\documentclass[a4paper,plainchapterheads,yschapters,oneside,truedoublelespace]{iitbthesis}
\usepackage{epsfig}
\usepackage[T1]{fontenc}
\usepackage[utf8]{inputenc}
\usepackage{xcolor}
\usepackage{mathptmx}
\usepackage{amssymb}
\usepackage{amsmath,epsfig}
\usepackage{graphicx,graphics}
\usepackage{url}
\usepackage{hyperref}
\usepackage[capitalise]{cleveref}
\usepackage{textcomp}
\usepackage{enumitem}
\usepackage{natbib}
\usepackage{pdflscape}
\renewcommand\bibname{References}
\usepackage{titlesec}

\usepackage{newtxtext}
\usepackage{newtxmath}
\usepackage{bm}
\usepackage{bm,bbm}
\usepackage{float,lscape}
\usepackage{subfig}

\usepackage{epstopdf}
\usepackage{tabu}
\usepackage{multirow}
\usepackage{array}
\usepackage{booktabs}
\usepackage{graphicx}
\usepackage{caption}
\usepackage{lettrine}
\usepackage{enumitem}
\usepackage{lscape}
\usepackage{rotating}
\usepackage{booktabs}
\usepackage{longtable}
\usepackage{mathptmx}
\usepackage{anyfontsize}
\usepackage{t1enc}
\usepackage{relsize}
\usepackage{placeins}
\usepackage{nomencl}
\usepackage{rotating}
\usepackage{glossaries}
\usepackage{scrlayer}
\usepackage{enumitem}

\usepackage[titletoc]{appendix}
\usepackage[nottoc]{tocbibind}
\newtheorem{theorem}{Theorem}[chapter]

\newtheorem{corollary}[theorem]{Corollary}
\newtheorem{definition}[theorem]{Definition}

\newtheorem{lemma}[theorem]{Lemma}

\usepackage{mathrsfs}
\usepackage{nicefrac}
\usepackage{dsfont}

\newcommand{\eop} {\hfill{$\blacksquare$}}

\renewcommand{\P}{{\mathcal P}}
\newcommand{\ulam}{{\underline \lambda}}
\newcommand{\x}{{\bf x}}
\newcommand{\xs}{ x}

\newcommand{\kb}{k_\beta}

\newcommand{\Cb}{C_\beta}

\newcommand{\um}{\bar{\lambda}}
\newcommand{\Q}{{\mathcal Q}}
\usepackage{multirow}
\newcommand{\ai}   {  {{\mathcal A}_I} }

\usepackage{algorithmic}
\usepackage{algorithm2e}
 \RestyleAlgo{ruled} 

\newcommand{\ignore}[1]{}

\newcommand{\Bggr}{\mbox{\fontsize{35}{2}\selectfont{$\}$}}}

\newcommand{\Bggl}{\mbox{\fontsize{35}{2}\selectfont{ $\{$}}}

\newcommand{\rbU}{{\bar{\bf u}}}

\newcommand{\rsbU}{\bar{u}}

\renewcommand{\H}{{\bf H}}  
\newcommand{\rH}{{\bf h}} 
\newcommand{\sH}{H}
\newcommand{\rsH}{h}

\newcommand{\A}{{\bf A}}  
\newcommand{\rA}{{\bf a}}

\newcommand{\G}{{\bf G}}
\newcommand{\rG}{{\bf g}}
\newcommand{\sG}{G}
\newcommand{\rsG}{g}

\newcommand{\ind}{\mathds{1}}
\newcommand{\bbeta}{{\bm \beta}}

\newcommand{\bgamma}{{\bm \gamma}}

\renewcommand{\S}{{\bf S}}
\renewcommand{\A}{{\bf A}}
\newcommand{\rS}{{\bf s}}
\newcommand{\rXn}{{\bf s}'}
\renewcommand{\rA}{{\bf a}}

\newcommand{\bd}{{\bf d}}
\newcommand{\bc}{{\bf c}}
\newcommand{\brd}{{\bf r}_\bd}
\newcommand{\tmu}{{\bm {\tilde \mu}}_\bd}
\newcommand{\tmud}{{\tilde \mu}_\bd}
\newcommand{\tP}{\mathbb{{\tilde P}}_\bd}
\newcommand{\rd}{{\bf r}_{\bd}}
\newcommand{\tr}{
{\bf {\tilde{r}}}_\bd}

\newcommand{\ns}{L}
\newcommand{\na}{M}
\newcommand{\bmu}{{\bm \mu}_\bd}

\newcommand{\mud}{{\mu}_{\bd}}
\newcommand{\uo}{{\bar u}^{(1)}}
\newcommand{\ut}{{\bar u}^{(2)}}
\newcommand{\Io}{i^{(1)}}
\newcommand{\It}{i^{(2)}}
\newcommand{\bet}{\beta}
\newcommand{\gam}{\gamma}
\newcommand{\eps}{\epsilon}
\newcommand{\del}{\delta}
\newcommand{\E}{\mathbb{E}}
\newcommand{\Pb}{\mathbb{P}}

\newcommand{\Bc}{\mathcal{B}}
\newcommand{\Fc}{\mathcal{F}}
\newcommand{\et}{\eta}
\newcommand{\Nc}{\mathcal{N}}
\newcommand{\T}{T}
\newcommand{\Sc}{\mathcal{S}}
\newcommand{\Ac}{\mathcal{A}}
\newcommand{\Cc}{\mathcal{C}}
\newcommand{\gs}{\gam^*_{ij}}
\newcommand{\Ps}{\Psi}

\newcommand{\Cv}{\zeta}

\newcommand{\gamv}{{\bm \gamma}}

\renewcommand{\i}{{\bm i}}

\newcommand{\rev}[1]{{\color{black} #1}}

\begin{document}

\abovedisplayskip=8mm
\abovedisplayshortskip=8mm
\belowdisplayskip=8mm
\belowdisplayshortskip=8mm



\pagenumbering{gobble}

\title{Navigating Resource Conflicts: Co-opetition and Fairness}
\author{Shiksha Singhal}
\date{June, 2023}

\rollnum{183190001} 

\iitbdegree{Doctor of Philosophy}

\reporttype{}

\department{Industrial Engineering \& Operations Research}

\setguide{Prof. Veeraruna Kavitha}

\maketitle


\begin{dedication}
\large{\textit{Dedicated to my beloved family}}
\end{dedication}


\chapter*{}
\thispagestyle{empty}
\vspace{-1in}
\begin{center}
{\Large  {\bf Thesis Approval}}
\end{center}
\vspace*{0.1in} \noindent This thesis entitled {\large \bf  Navigating Resource Conflicts: Co-opetition and Fairness} by {\bf \large
Shiksha Singhal} is approved for the degree of  {\large \bf Doctor of Philosophy}.\\\\
\hspace*{4 in} Examiners:\\\\
\hspace*{3.5 in} \ldots\ldots \ldots \ldots \ldots \ldots\ldots
\ldots \ldots \ldots\ldots\\\\
\hspace*{3.5 in} \ldots\ldots \ldots \ldots \ldots \ldots\ldots
\ldots \ldots \ldots\ldots\\\\
\hspace*{3.5 in} \ldots\ldots \ldots \ldots \ldots \ldots\ldots
\ldots \ldots \ldots\ldots\\\\
  \hspace*{3.5in} \ldots\ldots \ldots \ldots \ldots \ldots\ldots
\ldots \ldots \ldots\ldots\\\\
 Supervisor:\hspace*{3.5 in}  Chairperson:\\\\\\
\ldots\ldots \ldots \ldots \ldots \ldots\ldots
\ldots \ldots \ldots\ldots
\hspace*{1.0 in} \ldots\ldots \ldots \ldots \ldots \ldots\ldots
\ldots \ldots \ldots\ldots\\\\\\

\noindent
Date: \ldots\ldots \ldots \ldots\\\\
Place: \ldots\ldots \ldots \ldots


\clearpage
\thispagestyle{empty}

\begin{center}
\Large  {\bf Declaration }
\end{center}
I declare that this written submission represents my ideas in my own words and where others ideas or words have been included, I have adequately cited and referenced the original sources. I also declare that I have adhered to all principles of academic honesty and integrity and have not misrepresented or fabricated or falsified any idea/data/fact/source in my submission. I understand that any violation of the above will be cause for disciplinary action by the Institute and can also evoke penal action from the sources which have thus not been properly cited or from whom proper permission has not been taken when needed.
\vspace{4.5in}


%
 \begin{table}[h]
 \begin{flushleft}

\vspace{-3.2in} 
 \begin{tabular}{ccccc}
 \rule[5ex]{0pt}{-10ex}&& Date: && \\ 
 \end{tabular}
\end{flushleft}

\vspace{-0.5in} 
\begin{flushright}
 \begin{tabular}{ccccc}
 
 \hline 	\rule[5ex]{0pt}{-10ex}&& Shiksha Singhal&& \\ 
 \rule[5ex]{0pt}{-10ex}&& Roll No. 183190001&& \\ \\
 \end{tabular}
\end{flushright}
\end{table}

\pagebreak



\clearpage
\pagenumbering{roman}
\begin{abstract}
  \renewcommand{\thepage}{\roman{page}} \setcounter{page}{1}
In today's dynamic and interconnected world, resource constraints pose significant challenges across various domains, ranging from networks, logistics and manufacturing to project management and optimization, etc. Resource-constrained problems (RCPs) represent a class of complex computational \rev{problems} that require efficient allocation and utilization of limited resources to achieve optimal outcomes. This thesis aims to delve into such problems  involving multiple agents, where agents aim to enhance their own payoffs, or a neutral moderator aims to maximise the system revenue while distributing the resources appropriately among all agents. In the former type of problems, agents may seek collaboration to achieve higher individual shares, resulting in a cooperative game with competition, i.e., \textit{co-opetition}. Cooperative and non-cooperative game theory tools are utilized to analyze such games.  On the other hand, for the latter kind of problems, we use tools from optimization and Markov decision processes. 

In the first part of the thesis, we consider a \textit{coalition formation game} involving multiple agents (willing to cooperate) and a possible adamant agent (unwilling to cooperate) involved in resource sharing to identify the partitions (disjoint set of coalitions) that are stable against unilateral or coalitional deviations. Each player selects a set of agents to collaborate with (defining a strategy profile), forming a partition of coalitions. The agents in each coalition maximize their collective utilities leading to a non-cooperative resource-sharing game among the coalitions. The (unique) utilities at the resulting equilibrium are shared via an extended Shapley value concept (as Shapley value is defined only for grand coalition); these shares define agents' utilities for the given strategy profile in the coalition formation game. We also obtain the partition, which optimizes the social welfare of the system, and use it to estimate the price of anarchy, i.e., inefficiency that arises from selfish behavior in a game.
 We show that with a sufficient number of almost similar agents, no collaborative partition (agents working together) emerges at equilibrium. But the scenario reverses when the agents are significantly different: the system reaches a `lethargic state' where all partitions become stable against unilateral deviations. Surprisingly, in cases where the reputation factors of the agents are intermediate, successful collaborations are more likely to form between agents with higher and lower reputation factors. On the other hand, none of the partitions are stable against coalitional agents when the agents are similar while grand coalition is the only stable partition against coalitional deviations under certain conditions. We also show that the grand coalition optimizes social welfare. 

Next, we analyze a coalition formation game between strategic service providers of a congestible service, which has been modeled as an Erlang-B loss system. Each of the service providers has a certain, predefined number of servers. The key novelty of our formulation is that it is a constant sum game, i.e., the total
  payoff across all service providers (or coalitions of providers) is
  fixed, and dictated by the total size of the market. The game thus
  captures the tension between resource pooling (to benefit from the
  resulting statistical economies of scale) and competition between
  coalitions over market share,  i.e.,   \textit{market segmentation} based on the quality of service provided by each provider. In a departure from the prior literature on resource pooling for congestible services, we show that the grand coalition is in general not stable, once we allow for competition over market share. In fact, under classical notions of stability (defined via blocking by \emph{any} coalition), we show that no partition is stable. This motivates us to introduce more restricted (and practical) notions of blocking; interestingly, we find that the stable configurations under these novel notions of stability are \emph{duopolies}, where the dominant coalition exploits its economies of scale to corner a disproportionate market share.
Furthermore, we completely characterise the stable duopolies in heavy and light traffic regimes, and also study a
  dynamic variant of this game.

In the last part of the thesis, we consider a  neutral moderator who allocates resources appropriately among the agents, in the context of wireless networks. Towards this, a system with a base station  and multiple mobile/stationary users is considered. The base station uses millimeter waves (mmWaves) for data transmission and hence needs to align beams in the directions of the end-users.
The opportunistic schedulers that select  a `good' user in each time slot  are well 
known in the context of previous generation networks to achieve the best trade-off between the system efficiency (defined in terms of the sum of the user-utilities) and fairness (measured in terms of differences in utilities derived by individual users). Such schedulers require  good estimates of  the channel  conditions of individual users  in each  slot and hence require sufficiently accurate beam alignment towards each user in all slots. 
The idea here is to avail regular user-position updates, which help in accurate beam alignment towards multiple users, paving the way for opportunistic mmWave schedulers. We propose an  algorithm that uses a dual opportunistic and fair scheduler to allocate data as well as position-update channels, in each slot. The dual scheduler optimizes the  well-known alpha-fair objective function of the individual user-utilities,   after including the effects of the quality of   the user-position based  beam-alignment. The proposed schedulers have near-closed-form expressions -- one has to choose the best from  a finite set, each of which has a closed-form expression. 
The proposed opportunistic schedulers are also compared  with the previously proposed mmWave schemes; the latter schedulers choose one user in each slot and start data transmission only after accurate beam alignment.  We establish that  the opportunistic mmWave dual schedulers significantly outperform and have the versatility to achieve any required level of fairness. 

  \textbf{Keywords:}  Coalition formation game, Partition form game, Kelly mechanism, Erlang-B queueing system, Opportunistic and fair schedulers
\end{abstract}

\tableofcontents
\listoftables
\listoffigures



%
%



\setlength{\parskip}{2.5mm}
\titlespacing{\chapter}{0cm}{55mm}{10mm}
\titleformat{\chapter}[display]
  {\normalfont\huge\bfseries\centering}
  {\chaptertitlename\ \thechapter}{20pt}{\Huge}
  
  \titlespacing*{\section}
  {0pt}{8mm}{8mm}
  \titlespacing*{\subsection}
  {0pt}{8mm}{8mm}
\pagebreak
\pagenumbering{arabic}

\makeatletter
\def\cleardoublepage{\clearpage\if@twoside \ifodd\c@page\else
	\hbox{}
	\vspace*{\fill}
	\begin{center}
		This page was intentionally left blank.
	\end{center}
	\vspace{\fill}
	\thispagestyle{empty}
	\newpage
	\if@twocolumn\hbox{}\newpage\fi\fi\fi}
\makeatother

\newpage
\pagebreak
\cleardoublepage
\chapter{Introduction}
\label{chap_intro}

In today's dynamic and interconnected world, resource constraints pose significant challenges across various domains, ranging from networks, logistics and manufacturing to project management and optimization, etc. Resource-constrained problems (see \cite{minarolli2011utility,kumar2011resource,stoica1996proportional,tun2019wireless,koutsopoulos2010auction}) represent a class of complex computational puzzles that require efficient allocation and utilization of limited resources to achieve optimal outcomes. This thesis aims to delve into such problems: such problems might  involve multiple self-interested agents who must divide limited resources in order to attain their individual goals or a neutral moderator who strives to distribute the resources appropriately among all participants to achieve the required goals. In such scenarios, one may encounter several possibilities: (i) each agent may act independently, (ii) some agents may \textit{collaborate} in pursuit of improved resource allocations, (iii) neutral moderator may work to maximize the overall benefit for the system, or (iv) the moderator may strive to distribute the resources \textit{fairly} among all participants. 


\textit{Game theory} has been widely employed to examine scenarios where agents act in their own self-interest. It provides us with a set of mathematical tools to study the interaction among selfish agents and to analyze their behaviour. Even though agents may have self-serving motives, collaboration between them can be facilitated if they stand to benefit from working together. This brings forth the notion of cooperation among agents, leading to a \textit{cooperative game}. Such games can be analysed using tools from \textit{cooperative game theory} (\cite{narahari}).

In literature, cooperative games are primarily analyzed in \textit{characteristic form} and the stability of the grand coalition, comprising all players, is evaluated. This involves determining the existence of an allocation vector that discourages agents from deviating either independently or collectively from the grand coalition. However, it is also possible for a subset of agents to collaborate leading to a disjoint collections of agents, commonly referred to as partition. This gives rise to a \textit{coalition formation game} (\cite{saad}), where each coalition operates independently and competes with other coalitions, while agents within each coalition work together to maximize their coalition's welfare. This is a perfect example of \textit{co-opetition}.
In such situations, the welfare of a coalition may be influenced by both the members within it and the arrangement of players outside of it, leading to a \textit{partition form game} (\cite{saad}). One of the main aims of this thesis is to analyze these games in various contexts (for example, online auctions and lossy queueing systems) and determine the stable partitions that emerge from strategic interactions among the agents.

On the other hand, there may be situations where a moderator wants to distribute resources fairly among the agents. Fairness is a well-studied concept in literature (\cite{jain1984quantitative,lan2010axiomatic,kushner,cellular}). It is achieved by the moderator optimizing a certain concave function of the accumulated
utilities called $\alpha$\textit{-fair function} (\cite{lan2010axiomatic,kushner,cellular}) where the level of fairness is dictated by~$\alpha$. In many of these problems, the resources are shared among the agents, but the utility derived by the allocated agent depends on its individual state at the time of resource allocation. 
To ensure fairness, well-known opportunistic schedulers (see \cite{liu2001opportunistic,asadi2013survey} and the references therein) exploit these random variations in the states of competing agents and allocates resources to the `inferior agent', whenever it is in its `best' state. 

Specifically, we address the following problems in this thesis:



\noindent
\textbf{Coalition formation game in online auctions: }We examine an online auction where agents compete for larger spectrum shares using the proportional allocation algorithm (Kelly's mechanism). This allocates resources to each player based on their bid and the weighted sum of all players' bids, with weights reflecting reputation factors. We also consider a procuring cost proportional to their bids. Unlike previous research, we explore the potential for cooperation among agents to improve their shares. This leads to a coalition formation game of partition form. Interestingly, with more than four symmetric/identical agents (with same reputation factors), all agents being alone is the only partition stable against unilateral deviations; however, no partition is stable against coalitional deviations.  Asymmetry between agents (based on reputation factors) increases the number of stable partitions against unilateral deviations; beyond a threshold on the level of asymmetry, all partitions become stable; interestingly, grand coalition is the only  partition stable against coalitional deviations. We also explore the cost of not collaborating using the Price of Anarchy.

\noindent
\textbf{Coalition formation game in lossy queueing systems: }We examine a queueing system with multiple service providers, each with a fixed number of servers. Unlike prior research, we assume a (fixed/constant) shared customer pool divided among providers based on their server count via the well-known Wardrop equilibrium. This induces competition among the service providers for market share. We again consider the possibility of cooperation among the service providers. Our game turns out to be a partition form game and we analyze stable partitions with corresponding allocation vectors. Surprisingly, none are stable against coalitional deviations (i.e., the core is empty), motivating new, more realistic stability notions. Using these, we find that the predominant stable partitions are duopolies.

\noindent
\textbf{User position-based opportunistic fair schedulers for future generation networks: } We consider a system with a base station  and multiple mobile/stationary users. The base station uses millimeter waves (mmWaves) for data transmission and hence needs to align beams in the directions of the end-users. Departing from the existing literature, which considers selecting a user  for beam alignment and data transfer (i.e., single decision), we design a dual scheduler that ensures: (i) optimal dynamic update of information regarding user positions, and (ii) optimal dynamic assignment of channels to various users in different time slots. The idea here is to maintain sufficiently accurate position estimates of each of the users at the base station, which help in accurate beam alignment towards multiple users. The notion of fairness in optimality is also included.
The opportunistic schedulers that select  a `good' user in each time slot  are well 
known in the context of previous generation networks to achieve the best trade-off between the system efficiency (defined in terms of the sum of the user-utilities) and fairness (measured in terms of differences in utilities derived by individual users). Such schedulers require  good estimates of  the channel  conditions of individual users  in each  slot and hence require sufficiently accurate beam-alignment towards each user in all slots. 
%
Thus, we propose an  algorithm that uses a dual opportunistic and fair scheduler to allocate data as well as position-update channels, in each slot. This dual scheduler optimizes the  well-known alpha-fair objective function of the individual user-utilities,   after including the effects of the quality of   the user-position based  beam-alignment. The proposed schedulers have near-closed-form expressions -- one has to choose the best from  a finite set, each of which has a closed-form expression. The above is the  case with two users, while for (general) $N$ users one needs to solve $N-1$ dimensional deterministic equation for each choice. 
These schedulers are also compared  with the previously proposed mmWave schemes; the latter schedulers choose one user in each slot and start data transmission only after accurate beam alignment.  We establish that  the opportunistic mmWave dual schedulers significantly outperform and have the versatility to achieve any required level of fairness. 

\section{Contributions}
\label{sec:contributions}
This thesis contributes to  various domains. The key contributions of this thesis are as follows:
\begin{enumerate}[label=(\roman*)]
    \item The first contribution is towards cooperative game theory, where the solution concepts and stability concepts are extended to partition form games. Further, new and more meaningful notions of stability are introduced. We refer to these as stability against `Restricted Blocking'.
    \item Another contribution is towards queuing literature. The literature considers service providers with its own dedicated customer base. Instead, we consider a more realistic customer split based on the quality of service of each provider, introducing the concept of competition through market segmentation. We then study the possibility of providers cooperating and show results contrasting to the existing literature.
    \item The next contribution is towards Markov Decision Process literature, where we study a new kind of average cost Markov Decision Process. We optimize a function of finitely many average utilities, rather than directly optimising the single average utility.
    \item Further, this thesis extends the concept of opportunistic schedulers (well-known in the context of previous generation networks) to the future generation networks, where dual decisions are made in any time slot.
\end{enumerate}
	  
\section{Thesis outline}
\label{sec:outline}
	The subject matter of the thesis is presented in the following five chapters, 
\begin{enumerate}
\item	Chapter~\ref{chap_CGT} gives an overview of the cooperative games, their classification, various solution concepts, and notions of stability for each classification. The extension of solution and stability concepts to partition form games is also provided. It also includes the new stability notions introduced in this thesis. In the end, it also describes a general framework that allows converting any partition form game to a characteristic form game.
\item	Chapter~\ref{chap_PEVA} considers an online auction where agents compete for resources using the well-known Kelly's mechanism (proportional allocation) with an additional procuring cost. The partitions stable against unilateral as well as coalitional deviations are studied. 
\item	Chapter~\ref{chap_OR} obtains partitions stable against coalitional deviations for a constant sum coalition formation queueing game, under a more realistic customer split based on the quality of service of each provider. Under the classical notion of stability,  no partition being stable  is proved. Towards this, new stability notions are introduced and analyzed. Using these, we find that the duopolies are stable partitions.
\item	Chapter~\ref{chap_wireless} proposes a dual fair opportunistic scheduler for the future generation networks, while the thesis is concluded in Chapter~\ref{chap_conclusions}.  
\item The chapter-wise proofs are provided in Appendix~\ref{chap_appendixI}-\ref{chap_appendixIII}, at the end of the thesis.
\end{enumerate}

\chapter{Cooperative Game Theory}
\label{chap_CGT}

\section{Introduction}
Game theory is a branch of mathematics that studies how rational and intelligent decision-makers interact in various scenarios. The individuals who make the decisions are known as players or agents, and the interactions between them can involve both cooperation and conflict. Game theory provides mathematical tools for analyzing scenarios in which two or more players make choices that impact each other's well-being. A game can be viewed as a mathematical representation of a situation where each player strives to achieve the optimal outcome while being aware that every other player is also trying to achieve their own best result. This thesis will concentrate on games that involve cooperation.



A cooperative game (also known as a coalitional game) studies how rational agents collaborate  and make collective decisions to achieve mutual benefits; the collaborating agents (referred to as `coalitions') may have to compete with other groups.  Any coalition  acts as a single unit and makes a joint decision for all its members. The focus of these games is to foresee the coalitions that emerge `stable', the joint actions that these coalitions take, and their collective payoffs at some appropriate equilibrium. It differs from the conventional non-cooperative game theory, which focuses on anticipating the actions and outcomes of individual players. 

\section{Classification of Cooperative Games}
Cooperative games can be categorized based on various factors, with the first being whether the utility of the coalition can be transferred among its players or not (as in \cite{narahari}). While the second category takes into account the factors that determine the value of a coalition (see \cite{saad}). 
We begin with the former category.
\begin{enumerate}
    \item \textbf{Transferable Utility (TU) Games:}
    A coalitional game with transferable utilities can be described by a pair $(N,\nu)$ where $N$  is the set of players and $\nu: 2^N \to \mathbb{R}$ with $\nu_\emptyset = 0$, where $\nu_C$ for any $C \subset N$ represents the worth of coalition $C $. Basically, it is sufficient to describe the worth of a coalition by a single number because the utilities can be apportioned among coalition members in any desired manner, as the utilities are transferable. 

\item \textbf{Non-Transferable Utility (NTU) Games:} An NTU game is described by a pair $(N,V)$ where $N$ is the set of players and $V_C$ for any $C \subset N$ is the set of all possible payoff vectors that the players in $C$ can jointly achieve on cooperating. Thus,  the worth of any coalition $C$ is no longer a  real number, but rather is represented by a set of payoff vectors. In other words, $V_C$ is a closed and convex subset of $\mathbb{R}^{|C|}$. 

\end{enumerate}

\noindent
We now discuss the second classification of cooperative games (see \cite{saad}).
\begin{enumerate}
\item \textbf{Characteristic Form Games:} These are the most commonly studied versions of cooperative games in literature. Here, the worth of a coalition $C$ is solely determined by the members of that coalition and is independent of how the (outside) players in $N \backslash C$ are organized.

The primary aim of these games is to analyze the stability of the grand coalition (i.e., the coalition of all players), determine the benefits of cooperation, and assess how the gains from cooperation should be distributed among the players.

\item \textbf{Partition Form Games:} Unlike the characteristic form games, the value of a coalition $C$ in partition form games also depends on the arrangement of other players. The arrangement of the players is referred to as \textit{partitions}, which is a  set of mutually disjoint and exhaustive coalitions. For example,  $\P = \{C_1,\cdots,C_k\}$ is a partition of set of players $N$ into $k$ coalitions  if it satisfies the following,
$$
\cup_{l=1}^k C_l = N \text{ and } C_i \cap C_j = \emptyset \text{ for all } i \neq j.
$$
The partition form games are represented by $(N,\{\nu_C^\P\})$ where $\nu_C^\P$ is the worth of coalition $C$ under partition $\P.$

In Figure~\ref{fig:PFG}, we can see two partitions,~$\P_1 = \{C_1, C_2\}$ and~$\P_2 = \{C_1,C_3,C_4\}$, both of which partition the same set of players. In a characteristic form game, the worth of~$C_1$ is the same for both partitions, i.e.,~$\nu_{C_1}^{\P_1} = \nu_{C_1}^{\P_2}$. However, in general in a partition form game, the worth of~$C_1$ can differ depending on whether the remaining two players cooperate or not. Therefore, we have~$\nu_{C_1}^{\P_1} \neq \nu_{C_1}^{\P_2}$ in a partition form game.
\begin{figure}[ht]
    \centering
    \includegraphics[trim = {0cm 0cm 0cm 0cm}, clip, scale = 0.4]{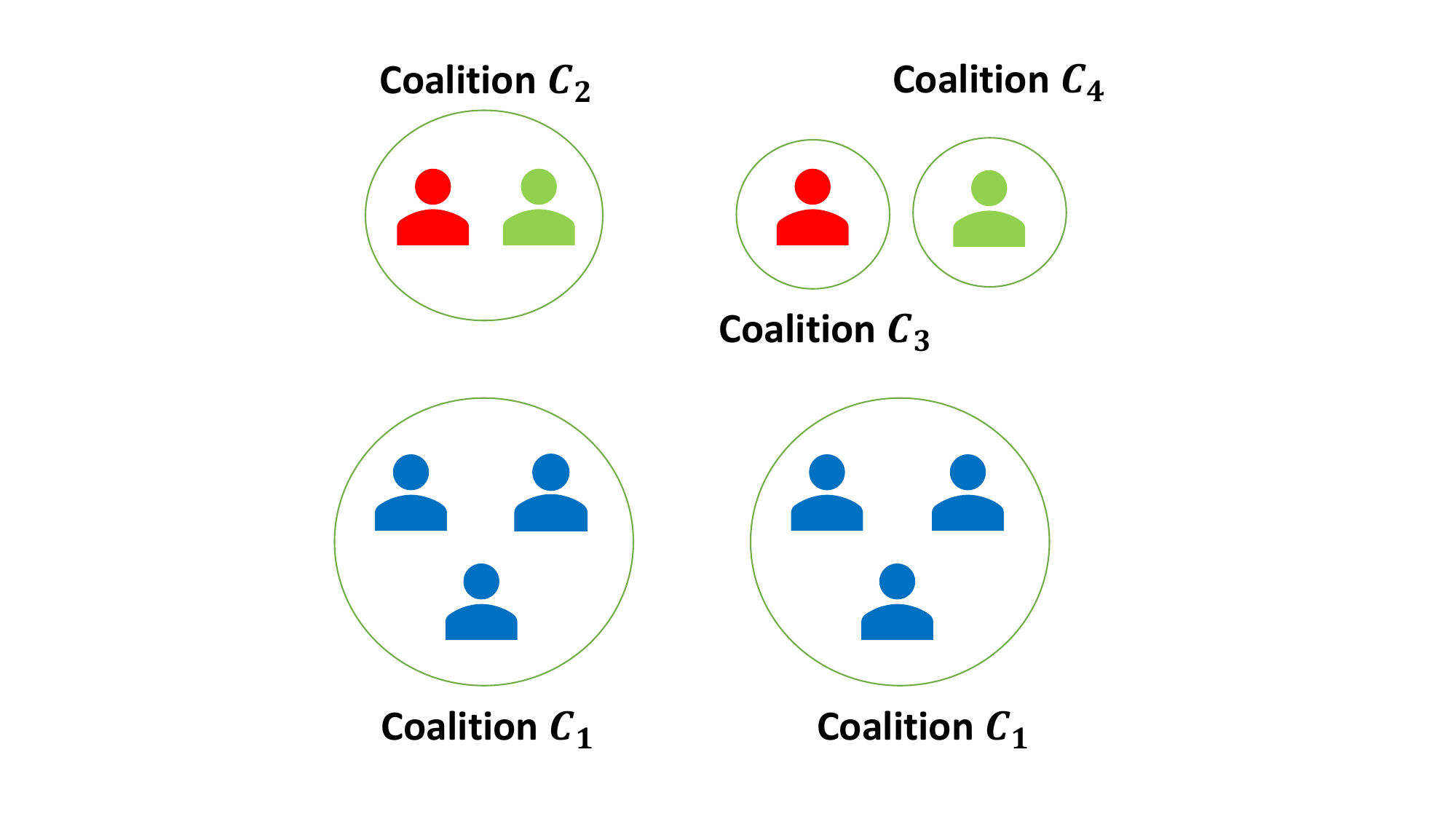}
    \caption{Characteristic v/s Partition Form Games}
    \label{fig:PFG}
\end{figure}

\item \textbf{Games in Graph Form:} In many coalitional games, the players are interconnected 
and communicate through pairwise links in a graph. However, in such situations, the characteristic form and partition form may not be appropriate as they fail to consider how the members of a coalition $C$ are connected. The worth of a coalition $C$ in a cooperative game $(N,\nu)$ with graph structure $G_C$ (directed or undirected) where vertices represent the members of $C \subset N$ is given by $\nu(G_C)$.
 \end{enumerate}

 \begin{figure}[ht]
     \centering
     \includegraphics[trim = {0cm 3cm 2cm 4.5cm}, clip, scale = 0.4]{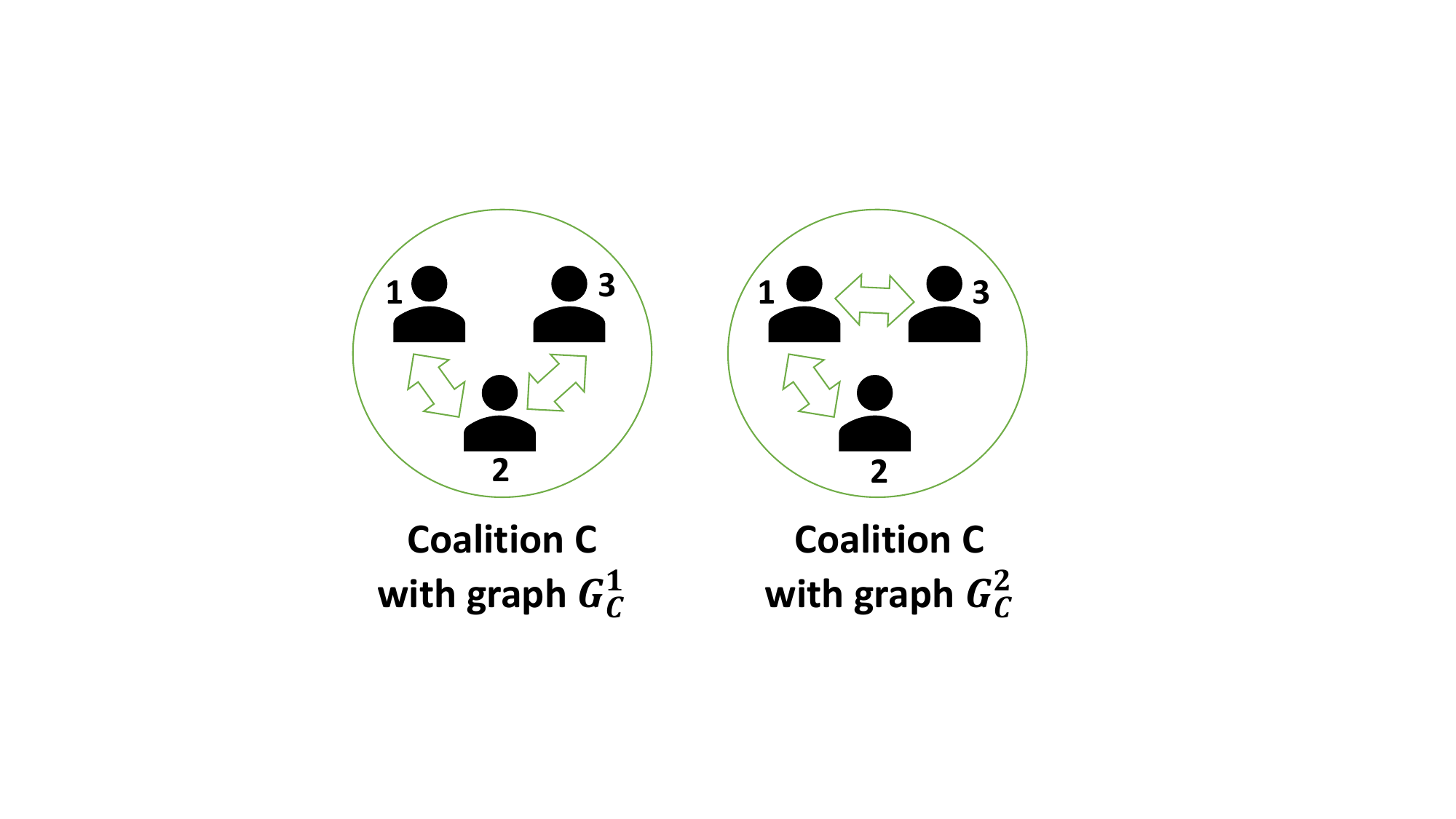}
     \caption{Games in Graph Form}
     \label{fig:graph}
 \end{figure}
 In Figure~\ref{fig:graph}, we are given two graphs~$G_C^1$ and~$G_C^2$ defined over coalition~$C = \{1,2,3\}$. In the former graph, player~$2$ is connected to both players~$1$ and~$3$, while in the latter, player~$1$ is connected to~$2$ and~$3$. Thus, the worth of the coalition $C$ further depends on the graph that represents the communication structure within the coalition, i.e., $\nu(G_C^1) \neq \nu(G_C^2).$

It is worth noting that all three forms of games mentioned above can be either of  transferable utility or non-transferable utility form. This thesis focuses on  transferable utility games. We next discuss the various solution concepts  and notions of stability for such games. We also introduce the new notions proposed in this thesis. 

 \section{Stability of Grand Coalition}
 \label{sec_GC}
Majority of the literature considers the emergence of grand coalition -- basically the stability of grand coalition, which depends upon the division of the total worth among its members. The payoff vector ${\bm \phi} = (\phi_1,\cdots,\phi_n)$ represents this division where  $\phi_i$ denotes the share to player $i$. Let the set of all possible payoff vectors ${\bm \phi}$ be denoted by $E^N$.  
The core and Shapley value are the two most frequently used solution concepts in this context. We begin by discussing the core, the payoff vectors of which render the grand coalition stable under the notions discussed below. Prior to this, \textit{any allocation/payoff vector ${\bm \phi}$ is said to be consistent} with GC (also referred as collectively rational in literature, e.g., \cite{narahari})   \textit{when $\sum_{i} \phi_i = \nu_N$}. Further (GC, ${\bm \phi}$) is referred to as \textit{configuration}. The \textit{excess} $e({\bm \phi},C)$ of the coalition $C \subset N$ is defined by,
\begin{equation*}
    e({\bm \phi},C) = \nu_C - \sum_{i \in C} \phi_i.
\end{equation*}
\begin{enumerate}
    \item \textbf{Core:} The core refers to a set of payoff vectors ${\bm \phi}$, in which no individual or coalition has an incentive to reject  the suggested payoff vector. In other words, core is defined as,
\begin{equation}
\mathcal{C} = \left\{ {\bm \phi} \in E^N: \sum_{i \in N} \phi_i = \nu_N \text{ and } e({\bm \phi},C) \le 0 \text{ for all } C \subset N \right \}.
\label{eq_core_classical}
\end{equation}
The payoff vectors of core under grand coalition satisfy certain stability properties, which are described below. 
\begin{enumerate}
    \item \textbf{Individual stability/rationality:} The grand coalition (GC) is considered to be individually stable with payoff vector ${\bm \phi}$ if no player can benefit by leaving the GC independently, i.e., $\nu_{\{i\}} \le \phi_i$ for all $i$, where $\nu_{\{i\}}$ is the worth that player $i$ can achieve independent of others. 
    \item \textbf{Coalitional stability/rationality:} A coalition $C \subset N$ is said to block the configuration (GC, ${\bm \phi}$) if the players in $C$ can obtain strictly better worth $\nu_C$ independent of others, this happens when $\nu_C > \sum_{j \in C} \phi_j$. The configuration (GC, ${\bm \phi}$) is considered coalitionally stable when there exists no  coalition that can block it, i.e., when for all $C \subset N, \sum_{j \in C} \phi_j \ge \nu_C$. 
\end{enumerate} 

To be more specific, core has all those payoff vectors ${\bm \phi}$ whose corresponding configurations (GC, ${\bm \phi}$) are individually, collectively and coalitionally rational.
However, despite its usefulness, the core also has some limitations: $(i)$ it is possible for the core to be empty, and $(ii)$ in some cases, the core can be large, making it challenging to choose a suitable allocation in core. To address these issues, Shapley (\cite{myerson1997game}) proposed an axiomatic approach, which we describe in the immediate below.
    \item \textbf{Shapley value:} 
    The Shapley value assigns a unique payoff to each player $i \in N$ as follows:
    \begin{equation}
        \phi_i = \sum_{C \subset N \backslash \{i\}} \frac{\left|C \right|!(|N|-|C|-1)!}{|N|!}\left[\nu_{C \cup \{i\}}-\nu_{C} \right]. \nonumber
    \end{equation}
    In the above, for any $C$, the value $\nu_{C \cup \{i\}}-\nu_{C}$ can be seen as the marginal contribution of player $i$ to coalition $C$, each of which contribute towards its payoff.
    The final Shapley payoff given above, is an appropriate weighted combination of such marginal contributions  and see for example \cite{narahari} for details of such a definition. For certain games, the Shapley value lies in the core, and hence satisfies all the corresponding stability properties; for all other properties and details, refer to  \cite{narahari}.
    
\end{enumerate}

\ignore{Once the worth of grand coalition is divided among its members, the next step is to study the stability of the grand coalition. \textit{Note here that the notions of stability of the grand coalition hinges on the payoff vector ${\bm \phi}.$}
In fact as seen later, the notions of stability of any general partition depends upon the payoff vector `consistent' with it.

 \subsection{Stability Concepts for Characteristic Games}

This sub-section presents various notions of stability for grand coalition;  here we also  adopt the nomenclature used in literature for partition form games (\cite{}) to pave way for the relevant  future discussions.

\textit{Thus, core is the collection of payoff vectors that are consistent with GC and are coalitionally as well as individually stable.}}
 
\section{Stability of any arbitrary partition}

In addition to a grand coalition, players have the option to organize themselves into coalitions. One can analyze the game confined to a given partition.  Such a  `constrained game'  is specified by $(N, \nu, \P)$, and considers only the \textit{payoff vectors ${\bm \phi}$ consistent with $\P$, i.e., those that satisfy  $\sum_{i \in C_j} \phi_i = \nu_{C_j}$ for all $C_j \in \P$} (see \cite{aumann1974cooperative}). Such `constrained games' can further be classified as either non-partition or partition form games, based on the dependency of $\nu$ on partition $\P$. The authors in \cite{aumann1974cooperative} define solution concepts and stability concepts for non-partition form games, which we discuss first. We later extend these ideas to partition form games  in Section \ref{sec:pfg_new_sc}.   

Recall $(N,\nu)$  is a \textit{game in characteristic form} where $N$ is the set of finite players and $\nu$ is a real-valued function on the family of subsets of $N$, with $\nu_\emptyset = 0$ and where $\nu_C$ represents the worth of coalition $C$.  
%
In non-partition form games, the worth of any coalition $C$, $\nu_C$ is independent  of partition $\P$.  Consider the following subset of consistent payoff vectors which are further constrained, as below:
\begin{equation}
X_{\P} = \left \{ {\bm \phi} \in E^N: \sum_{i \in C_j} \phi_i = \nu_{C_j} \text{ for all } C_j \in \P \text{ and } \phi_m \ge \nu_{\{m\}} \text{ for all } m \in N \right \}.   
\label{eqn_XP}
\end{equation}

We begin with the definition of core as provided in \cite{aumann1974cooperative}. 
\begin{enumerate}
    \item \textbf{Core:} The core is the set of payoff vectors where no coalition has the incentive to reject the proposed payoff allocation and depart from the current partition. In other words, the core can be defined as:
    \begin{equation}
    \mathcal{C}(N,\nu,X_\P) = \left \{ {\bm \phi} \in X_\P: e({\bm \phi},C) \le 0 \text{ for all } C \subset N \right \}.
        \label{eq_core_extended_arbitrary}
    \end{equation}
Note that as in the case of core corresponding to grand coalition in \eqref{eq_core_classical}, the payoff vectors in core for any arbitrary partition in \eqref{eq_core_extended_arbitrary} also satisfy the \textit{coalitional rationality} as well as \textit{individual rationality} (with $C = \{i\}$ for all $i$). However, the property of payoff vectors being consistent is now applicable with respect to each coalition in the partition, i.e.,     $\sum_{i \in C_j} \phi_i = \nu_{C_j}$ for all $C_j \in \P.$

    \item \textbf{Shapley value:} Under this extension (see \cite{aumann1974cooperative}), each coalition $C_j$ in the partition is treated as `grand coalition', 
    and then the usual definition of Shapley value is used to obtain the unique individual shares of the players in $C_j$. For any player $i$, the Shapley allocation is given by,
     \begin{equation}
        \phi_i = \sum_{C \subset C_j \backslash i} \frac{\left|C \right|!(|C_j|-|C|-1)!}{|C_j|!}\left[\nu_{C \cup \{i\}}-\nu_{C} \right] \text{ if } i \in C_j \in \P. 
        \label{eqn_shapley_non_PFG}
    \end{equation}
    Again, the unique payoff vector is now consistent with partition $\P$.
\end{enumerate}

Note here that when $\P = N$, the stability and solution concepts match with the ones described in Section \ref{sec_GC}. We now extend the above concepts to partition form games in the next section.



\section{Partition form games}
\label{sec:pfg_new_sc}

In these games, the worth of any coalition $C$ depends also on the arrangement of outside players, i.e., on partition $\P$, and hence we may have $\nu_C^{\P} \ne \nu_C^{\P'}$ in general when $\P \neq \P'$.
Thus, to extend the notions of previous section to partition form games,  any player or coalition that wishes to deviate from the current configuration (payoff vector and coalition) additionally need to take into account the retaliatory actions of other players; in other words, they need to predict the worth (call it $\nu^a_C$) of the new coalition $C$ that they are considering to form, based on anticipated reactions of other players; in the symbol  $\nu^a_C$, `$a$' represents anticipation. 
To assist with this prediction, researchers have studied several anticipation rules in the literature (e.g., \cite{pessimistic}), which are outlined below, i.e., we discuss $\{\nu_C^a\}$ for different anticipation rules. We describe these rules in our own wordings,   using our own notations and sometimes for more  simplified and sometimes for more general settings.  We have also described a new anticipation rule used in this thesis. 

\subsection{Anticipation Rules}
\label{sec_anticipation}

For all the discussions below consider any general partition $\P = \{C_1, \cdots, C_k\}$ and let $C \notin \P$ be a new coalition that is attempting to deviate from the arrangement in $\P$. To keep the explanations simple, we  consider deviation from a single coalition of $\P$, i.e., $C \subset C_j $ for some $ C_j \in \P$, but the ideas can readily be extended to the case when players from multiple coalitions attempt to deviate together, i.e., when $C\cap C_j \ne \emptyset$ for more than one $j$. 
\begin{enumerate}
    \item \textbf{Disintegration rule:} This rule was first introduced in~\cite{von1947theory}. According to this rule, coalitions can only be formed if all their members unanimously agree. Therefore, if a player or sub-coalition $C$ decides to deviate from their current coalition, say $C_j$, they can expect that the rest of the coalition they leave behind will disintegrate into individual players. 
    This anticipation helps in predicting the worth of the new coalition, as given below.
    $$
    \nu^a_{C} = \nu_C^{\P'} \text{ with } \P' = 
    \{ C\} \cup  \left  \{ \{i\} : i \in C_j \backslash C  \right  \} \cup \left \{ C_l \in \P: C_l  \neq C_j  \right \}.
    %
    $$
    When $C$ is formed from multiple coalitions then   
    $$\P' = \{ C\} \cup  \cup_{_{C_j: C_j \cap C \ne \emptyset}} \left  \{ \{i\} : i \in C_j \backslash C  \right  \} \cup \left \{ C_l \in \P: C_l \cap  C = \emptyset  \right \}.$$
    \item \textbf{Projection rule:} The authors in~\cite{hart1983endogenous} introduce~$\delta$ model of coalition formation -- all players announce the coalitions they prefer to participate in, and the players with matching interests end up in the same coalition. This leads to an anticipation where the players who choose to deviate from their current coalition can expect the coalition they leave behind to remain intact. In this case, the predicted worth of the deviating coalition $C$ is given by, 
    $$
    \nu^a_{C} = \nu_C^{\P'} \text{ with } \P' = \{ C\} \cup \{C_j \backslash C\} \cup \left \{ C_l \in \P: C_l  \neq C_j  \right \}. 
    $$
    \item \textbf{$\mathbf{\mathcal{M}}$-Exogenous rule~\cite{hafalir}:} This rule is characterized by an exogenous partition $\mathcal{M}$ of the player set $N$. Under this rule, players in deviating coalition $C$ expect other players to organize themselves according to the projection of $\mathcal{M}$ onto $N\setminus C$ -- they anticipate the partition after deviation to be ${\P}' := \{ S \backslash C : S \in{\cal M}\} \cup \{C\}$. We again have $\nu_C^a = \nu_C^{\P'}$. Two extreme special cases of exogenous rules are the $\underline{N}$-exogenous rule, where players expect all external players to form singletons (possible with ${\cal M} = \{ \{i\} : i \in N \}$), and the ${\bar N}$-exogenous rule, where players anticipate that all external players will join a single coalition $N\setminus C$ (when ${\cal M}  = N$). 
    
    Observe that under the disintegration rule, only the  left-over members of the coalition from which the players have deviated, are anticipated to disintegrate into singletons, while under $\underline{N}$-exogenous anticipation rule all players  in $N \backslash C$ are expected to disintegrate.
    
    \item \textbf{Optimistic rule:} The optimistic rule, introduced in~\cite{shenoy1979coalition}, assumes that the players in the deviating coalition expect the other players to select a partition that maximizes the worth of the deviating coalition, which is given by
    $$
    \nu^a_{C} = \max_{\P': C \in \P'} \nu_C^{\P'}. 
    $$
    \item \textbf{Pessimistic rule:} The pessimistic rule predicts the worth of the new coalition as the amount that it can guarantee for itself regardless of the arrangement of other players. This rule is inspired by the definition of the $\alpha$-core in~\cite{aumann1961} and is discussed in~\cite{hart1983endogenous}. In other words, according to the pessimistic anticipation rule,  the members of $C$ expect other players to select a partition that minimizes the worth of $C$ as given by,
    $$
    \nu^a_{C} = \min_{\P': C \in \P'} \nu_C^{\P'}. 
    $$
    \item \textbf{Max rule:} According to the max rule, as discussed in~\cite{hafalir}, players in the deviating coalition $C$ anticipate the rest of the  players to
    arrange in a way that maximizes the payoff of the external players. Thus, the worth of deviating coalition $C$ is given by,
    $$
    \nu^a_{C} =  \nu^{\P^*}_C \mbox{ where }  \P^* \in  \arg \max_{\P': C \in \P'} \sum_{S \in \P' \setminus C} \nu_S^{\P'}. 
    $$
    \item \textbf{Partial pessimistic rule:} In this thesis, we explore an additional form of anticipation that can be applied to the split of an  existing (single) coalition, say $C_j$. Here, the deviating coalition $C$ assumes that the players in $C_j \backslash C$ will strategically position themselves to inflict maximum harm upon the deviating coalition $C$, while the remaining players, i.e., those in   $N \backslash C_j$, maintain their positions. This specific anticipation rule is considered in Chapters \ref{chap_PEVA} and \ref{chap_OR} while defining the Shapley value for any arbitrary partition. Furthermore, it coincides with the disintegration rule for the model in Chapter \ref{chap_PEVA} and with projection rule in Chapter \ref{chap_OR}.
\end{enumerate}

Next, we present the extended versions of the classical solution concepts (core and Shapley value) for the partition form games. Now the payoff vectors consistent to a partition $\P$ (see \eqref{eqn_XP}) have the same meaning as before but with $\nu_C = \nu_C^P$, i.e., also depends on partition $\P$. Towards defining the stability of a given partition, as in \eqref{eqn_XP},  the following subset of the above constrained payoff vectors    is considered, but now  using the anticipated worths $\{\nu_{\{i\}}^a \}$:
$$
X_{\P}^a = \left \{ {\bm \phi} \in E^N: \sum_{i \in C_j} \phi_i = \nu_{C_j}^\P \text{ for all } C_j \in \P \text{ and } \phi_m \ge \nu_{\{m\}}^a \text{ for all } m \in N \right \}.
$$
In similar lines the excess is defined using the anticipated worths -- 
for any  ${\bm \phi}$, the anticipated \textit{excess} $e^a({\bm \phi},C)$ of the coalition $C$ in partition $\P$ can be defined as{\footnote{If $C \in \P$ one doesn't need anticipation as the worth is known. Also, for such $C$, $e^a({\bm \phi},C) = 0$ when one considers  payoff vectors consistent to partition $\P$.}},
\begin{equation*}
    e^a ({\bm \phi},C) = \begin{cases} \nu_C^a - \sum_{i \in C} \phi_i & \text { if } C \notin \P, \\
    \nu_C^\P - \sum_{i \in C} \phi_i & \text { else }.
    \end{cases} 
\end{equation*}
\noindent
A coalition $C$ that exhibits $e^a({\bm \phi}, C) > 0$ anticipates to achieve a superior worth in comparison to the current configuration $(\P,{\bm \phi})$. This type of coalition is commonly known as a `blocking coalition', and it serves as the foundation for defining the stability concepts presented below. 

\subsubsection*{Core: Stability under General Blocking}
The definition of the core can now be extended to partition games as done for non-partition form games in  \eqref{eq_core_extended_arbitrary}, but now using  the anticipatory quantities,  $\{X^a_\P\}$ and $\{e^a({\bm \phi}, C)\}$. Let $\mathcal{B}: = 2^N$ and then,
    \begin{equation}
    \mathcal{C}^a(N,\nu,X_\P^a) = \left \{ {\bm \phi} \in X^a_\P: e^a({\bm \phi},C) \le 0 \text{ for all } C \in \mathcal{B} \right \}.
        \label{eq_core_extended_arbitrary_anticipate}
    \end{equation}
In the above, $\nu$ includes $\nu^a_C$ and $\nu_C^\P$, and where $\{\nu^a_C\}$ can be computed for any given anticipation rule, once $\{\nu_C^\P\}$ is known. In contrast to the definition of core in \eqref{eq_core_extended_arbitrary} for non-partition form games which depends on the actual worths of the coalitions, core for partition form games depends on the worths of the coalitions under anticipation. However, it still satisfies the \textit{individual} and \textit{coalitional rationality} with $\nu_C$ being replaced by $\nu_C^a$, i.e., none of the coalitions anticipate to have an incentive to deviate from any configuration $(\P,{\bm \phi})$ with ${\bm \phi} \in \mathcal{C}^a(N,\nu,X^a_\P)$.

Considering a payoff vector ${\bm \phi} \in \mathcal{C}^a(N,\nu,X_\P^a)$, we observe that, under the given anticipation rule, no coalition outside of $\P$ has any incentive to break away from the configuration $(\P, {\bm \phi})$ (recall $e^a({\bm \phi},C) = 0$ for $C \in \P$). This is because for all coalitions $C \notin \P$, the inequality $\nu^a_C \le \sum_{i \in C} \phi_i$ holds true. In other words, such coalitions do not possess a higher cumulative value than what they would receive in the existing configuration. As a result, these coalitions will not \textit{block} or \textit{oppose} the current configuration.  In the above, $\mathcal{B}$ is representative of the set of coalitions that can block and $\mathcal{B} = 2^N$ implies every coalition has a potential to block (equivalently, coalitions from $2^N \backslash \P$). Later, we will discuss other notions where $\mathcal{B} \subsetneq 2^N \backslash \P.$  

A partition $\P$ that is not blocked by any coalition $C \notin \P,$ as described above, is considered stable against coalitional deviations (\cite{hafalir}). 
 We refer to this as stability under general blocking in Chapter \ref{chap_OR}. 
 
 The definition of the core as in \eqref{eq_core_extended_arbitrary_anticipate} matches with the `pessimistic core' and `optimistic core' in \cite{abe2017non} under the pessimal and optimistic anticipation rules respectively, described in sub-section \ref{sec_anticipation}.

\subsubsection*{RB-Core: Stability under Restricted Blocking }In certain scenarios, the conditions demanded by the core are too stringent, leading to an empty core. However, one can define more relevant notions of stability inspired by the practical rearrangements in the marketplace. Towards this, we introduce a new and more relevant notion of stability in Chapter \ref{chap_OR}, referred to as `Restricted Blocking (RB)'. Under this notion, the \textit{candidate blocking coalition} $C$ is restricted to only the mergers ($C = \cup_{C_{j_k}} C_{j_k} \text{ for some sub-collection } \{C_{j_k}\} \subset \P)$ or splits ($C \subset C_j \in \P)$ of the existing coalitions. Thus the restricted blocking core is defined as,
\begin{equation}
     \mathcal{C}^a_R(N,\nu,X_\P^a) = \left \{ {\bm \phi} \in X_\P^a : e^a({\bm \phi},C) \le 0 \text{ for all } C \in \mathcal{B}_R   \right \},
\end{equation}
where $\mathcal{B}_R = \{C : C\subset C_j \in \P \text { or  } C = \cup_{C_{j_k}} C_{j_k} \text{ for some sub-collection } \{C_{j_k}\} \subset \P \} \subset \mathcal{B}$.

The above definition is based on the assumption that the shares of the players are known to each other. However, this might not be the case always. For such scenarios, another solution concept namely, `Restricted Blocking Imperfect Assessment (RB-IA)'  has been discussed in Chapter \ref{chap_OR}.

\subsubsection*{U-Core: Stability against unilateral deviations}
In contrast to the stability against coalitional deviations, one can also consider stability against unilateral deviations (as in Nash Equilibrium in non-cooperative game theory).   The unilateral deviations only consider the potential movement/opposition by individuals, and hence the U-core is defined as follows,
\begin{equation}
    \mathcal{C}^a_U(N,\nu,X_\P^a) = \left \{ {\bm \phi} \in X^a_\P: e^a({\bm \phi},C) \le 0 \text{ for all } C \in \mathcal{B}_U \right \} \text{ with } \mathcal{B}_U = \{\{i\}: i \in N\}.
    \label{eq_core_extended_unilateral}
    \end{equation}
   
    This notion coincides with the notion of stability derived via the coalition formation game in Chapter \ref{chap_PEVA} (more details are in Section \ref{sec_CFG_core}).   Further, it is easy to observe that
    $$ \mathcal{C}^a(N,\nu,X_\P^a) \subset  \mathcal{C}^a_R(N,\nu,X_\P^a) \subset  \mathcal{C}^a_U(N,\nu,X_\P^a).$$
    Thus via the new notions $(\mathcal{C}^a_U, \mathcal{C}^a_R$) we could successfully derive solutions even for the cases where the original concepts provide no solution. Furthermore, these notions capture the practical tensions that exists in the marketplace. 
    
    \rev{It is easy to see from above that the above solution concepts are related as in Figure \ref{fig_nested}.
    \begin{figure}[ht]
        \centering
       \includegraphics[trim = {0cm 0cm 0cm 0cm}, clip, scale = 0.4]{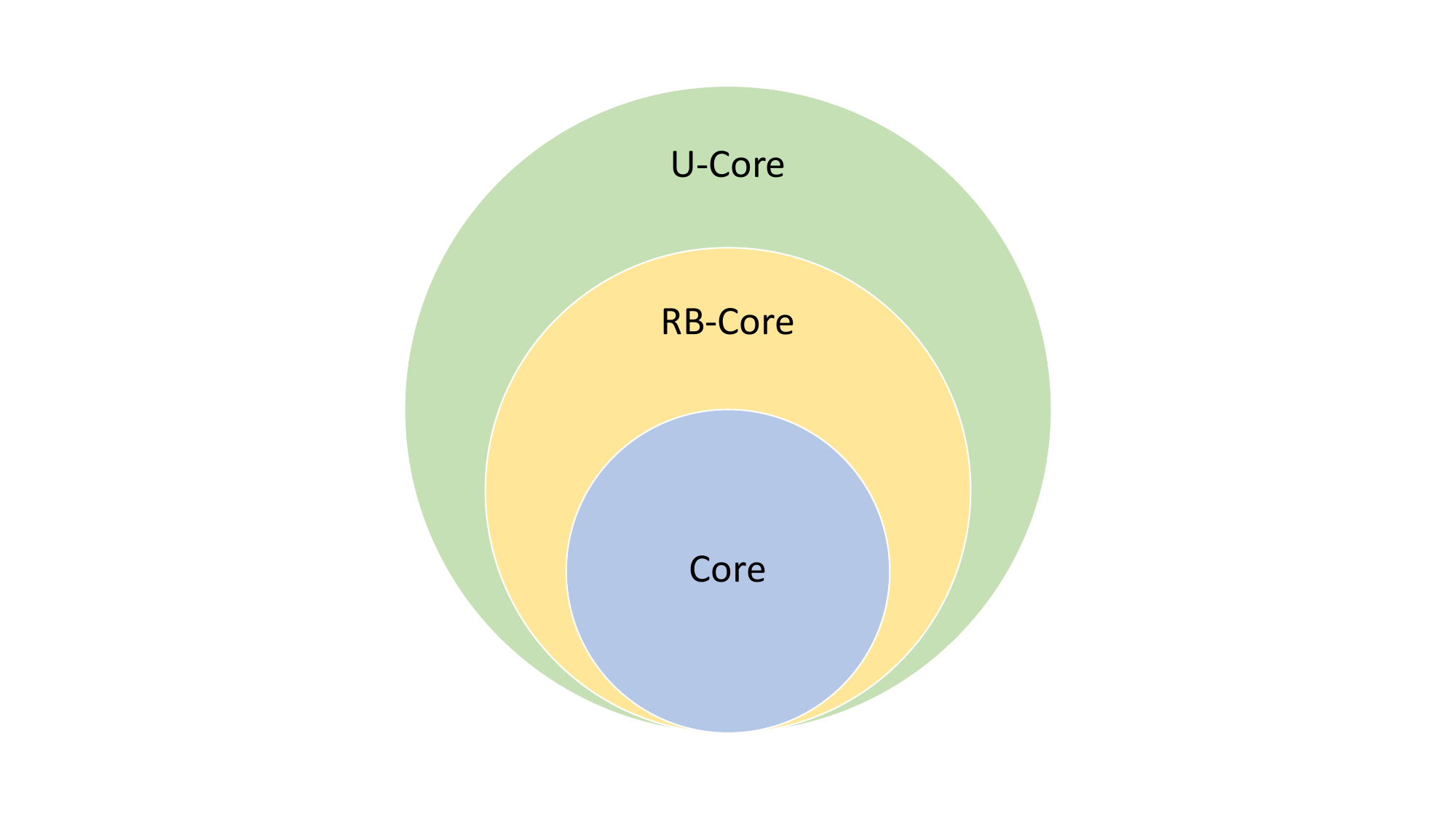}
    \caption{Relationship between different core definitions}
        \label{fig_nested}
    \end{figure}
    }

   \subsubsection*{Shapley value} Inspired by the extension of Shapley value to non-partition form games in \cite{aumann1974cooperative}, we extend it for partition form games  and use it in Chapters \ref{chap_PEVA} and \ref{chap_OR}. Under this extension, we treat each coalition $C_j$ in the partition as a `grand coalition', define a suitable `worth' $\nu^a_C$ for each $C \subset C_j$
    and then use the usual definition of Shapley value to obtain unique  individual shares of the players in $C_j$. Formally, for any $i \in C_j$,
    $$
    \phi_i = \sum_{C \subset C_j, i \notin C} \frac{|C|!(|C_j|-|C|-1)!}{|C_j|!}\left[ \nu^a_{C \cup \{i \}} - \nu^a_C \right] \text{ if } i \in C_j \in \P.
    $$

\noindent
Once again $\nu^a_C = \nu_C^\P$ whenever $C \in \P$. The anticipation here is based on the partial pessimistic rule described in sub-section \ref{sec_anticipation}. Under this rule, the pessimal anticipation is applied only to the players who deviate from their original coalition, while it is assumed that the other players remain in their original coalition. Consequently, when $C\cup \{i\} = C_j$ for some $C_j \in \P$ in the above equation, the worth of coalition $C\cup \{i\}$ is exact and has no anticipation. On the other hand, the worth of coalition $C$ is calculated under the partial pessimal anticipation and since the leftover coalition is a singleton, this anticipation leads to the following: $\nu^a_C = \nu_C^{\P'}$ where $\P' = \{C, \{i\}\} \cup \P \backslash C_j $.
Observe that similar to the core, Shapley value in this case depends on the anticipated worths in contrast to the actual worths in \eqref{eqn_shapley_non_PFG}.

\rev{Thus, this thesis extends the solution concepts for partition form games and proposes new stability notions as summarised in Figure \ref{fig_extension}.
\begin{figure}[ht]
    \centering
    \includegraphics[trim = {4cm 0cm 0cm 2cm}, clip, scale = 0.5]{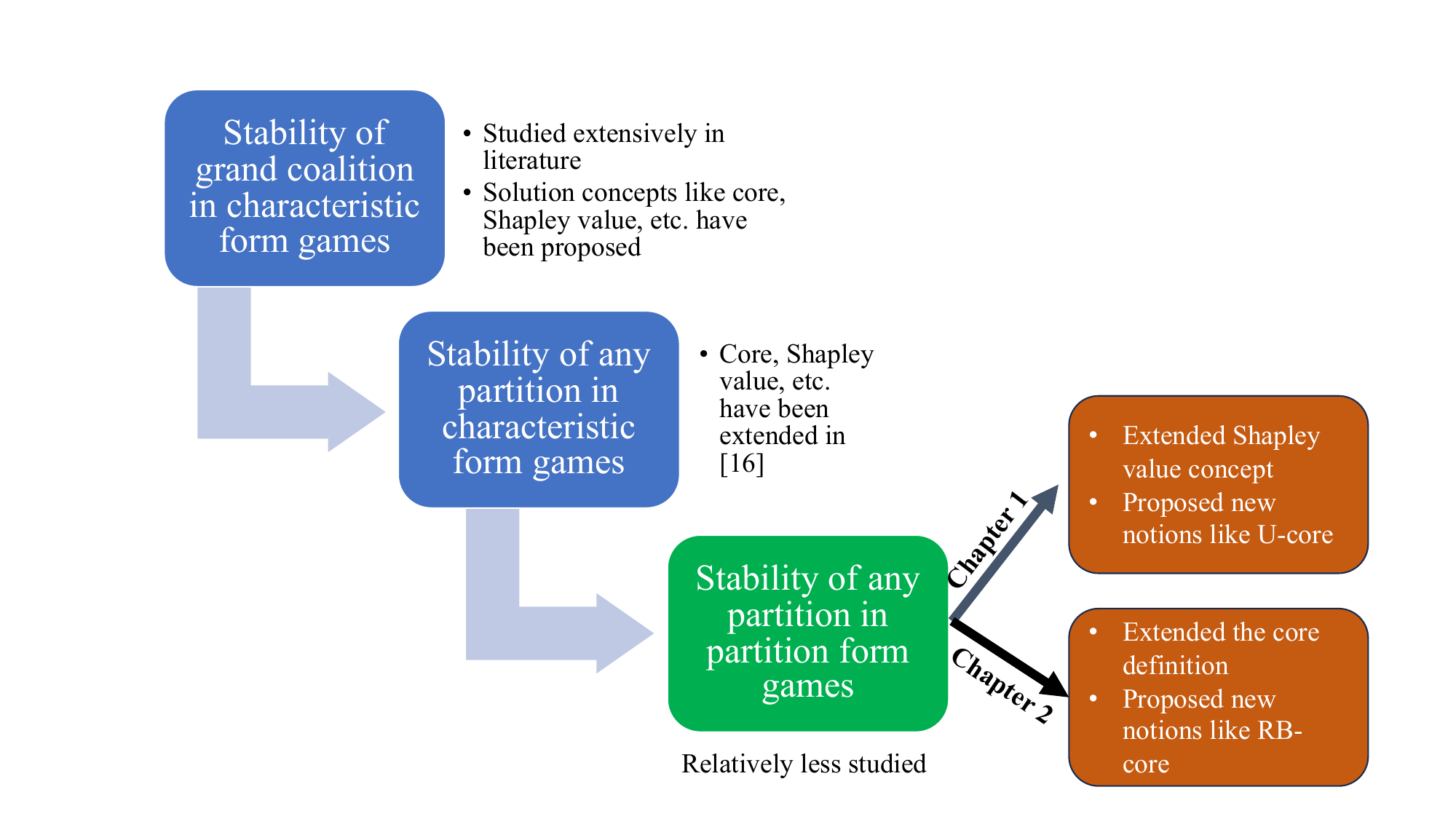}
    \caption{Overview of games and solution concepts}
    \label{fig_extension}
\end{figure}}


\section{General Characteristic (G-Ch) Form  Framework}
\label{sec_gch}

This is a more general framework that can handle games with transferable (TU) as well as non-transferable (NTU) utilities \cite{aumann1961}.  

Any such game can be described by a tuple~$(N,\nu,\mathcal{H})$ and we describe the ingredients one after the other.  Firstly, $N$ denotes the set of $n$ players. Secondly, $\nu$ is the \textit{characteristic function}. To cater to both forms of games, $\nu(C)$ for any $C \subseteq N$ is no longer a real number as in previous sections but is a set, representing the `worths achievable' by $C$. We call a payoff vector ${\bm \phi}$ to be \textit{effective for a coalition $C$} if players in $C$ can jointly ensure that every member $i \in C$ achieves at least $\phi_i$ (see \cite{aumann1960neumann}). Then, $\nu(C)$ is defined as the set of all payoff vectors of dimension~$n$ that are \textit{effective for coalition $C$},
$$
\nu(C) = \{ {\bm \phi}: {\bm \phi} \text{ is effective for coalition } C \}.
$$
Further, it is required that $\nu(C)$ is closed and convex.
Lastly, $\mathcal{H}$ is the set of all payoff vectors of dimension~$n$, that can be obtained by means of some correlated mixed strategy of the set of all players.  

Towards defining the stability concept, we first define the \textit{domination of a payoff vector}. We say a payoff vector ${\bm \phi}$ dominates payoff vector ${\bf y}$ via coalition $C$ if ${\bm \phi} \in \nu(C)$ and $\phi_i > y_i$ for all $i \in C$; ${\bm \phi}$ is said to  dominate  ${\bf y}$  if there exists a coalition $C$ such that $\phi_i > y_i$ for all $i \in C$. With these definitions in place, we now define a related solution concept called R-core, which is an extension of the classical definition of \textit{core}.

\subsubsection*{R-core}
R-core~$\mathscr{C}(\mathcal{H})$ is defined to be the set of vectors in~$\mathcal{H}$ which cannot be dominated by any other member of~$\mathcal{H}$. Once again, the payoff vectors in R-core are individually as well as group rational.

\medskip

\noindent
The authors in \cite{aumann1961} define two concepts of effectiveness (to define the characteristic function $\nu$), namely $\alpha$-effectiveness and $\beta$-effectiveness. We next discuss these concepts.
\begin{enumerate}
    \item ${\bm \alpha}$\textbf{-effectiveness: } A coalition $C$ is said to be $\alpha$\textit{-effective} for the payoff vector ${\bm \phi}$ if there is a correlated strategy $s^C$ of players in C such that for any correlated strategy of the opponents $s^{N \backslash C}$, we have the payoff of each player $i$ in coalition $C$, $U^i(s^C \times s^{N \backslash C})$ is at least as much as their payoffs in the current payoff vector ${\bm \phi}, i.e., U^i(s^C \times s^{N \backslash C}) \ge \phi_i$ for all $i \in C.$

    In other words, $\alpha$\textit{-effectiveness} means that $C$ can assure itself, independently of the actions of remaining players $N \backslash C$, that each of its members $i$ will receive at least $\phi_i$. 

    Given a game in partition form framework $(N,\{\nu_C^\P\})$ 
    the $\alpha$-effectiveness  coincides with \textit{pessimistic anticipation rule} when the characteristic function $\nu(C)$ of the latter is defined by, 
    $\nu(C) = \{ {\bm \phi}: \sum_{i \in C} \phi_i \le \nu^a_C\}$ 
     where $\nu^a_C$ is the worth of coalition $C$ under pessimistic anticipation rule.
     \item ${\bm \beta}$\textbf{-effectiveness: } A coalition $C$ is said to be $\beta$\textit{-effective} for the payoff vector ${\bm \phi}$ if for each correlated strategy of the opponents $s^{N \backslash C}$ there is a correlated strategy $s^C$ of players in C such that the payoff of each player $i$ in coalition $C$, $U^i(s^C \times s^{N \backslash C})$ is at least as much as their  payoffs in the current payoff vector ${\bm \phi}, \text{ i.e., } U^i(s^C \times s^{N \backslash C}) \ge \phi_i$ for all $i \in C.$
     
     In other words, $\beta$\textit{-effectiveness} means that $C$ can always act so that each of its members $i$ receives at least $\phi_i$, but the strategy that it must use to achieve this may depend on the strategy used by $N \backslash C$, in contrast to $\alpha $-effectiveness.
\end{enumerate}
Next, we explain how the partition form TU games can be fitted into the above  general characteristic form framework. 

\subsection{Transformation of partition form TU game to G-Ch form framework}
Now we are given $\{\nu_C^\P\}$ and need to convert to G-Ch framework.
Towards the required transformation, one needs to define $\nu$ and $\mathcal{H}$ appropriately.
Let~${\cal F} (\P) $ be the set of all feasible payoff vectors
under partition~$\P$, i.e.,\textit{ effective with respect to coalitions in partition}.  These are the vectors that satisfy the following:     the sum of payoffs of all  agents in any coalition~$C$ is less than or equal to that obtained by~$C$ under partition~$\mathcal{P}$,~$\nu_C^\mathcal{P}$ (which is a real number). Hence  
\begin{equation}
{\cal F} (\P) := \left \{{\bm \phi} = (\phi_i) : \sum_{i \in S} \phi_i \leq \nu_C^\mathcal{P} \text{ for all }  C \in \P   \right \}.
\end{equation}
Then $\mathcal{H}$, the set of all achievable/feasible payoff vectors is~$
\mathcal{H} = \cup_{\mathcal{P}} {\cal F} (\P) .
$ Observe that for grand coalition,~$\mathcal{F}(\mathcal{N}) = \mathcal{H}$ (this is because $\nu_C^\P \le \nu_N$ for all $C$ and $\P$) and hence is convex. Next, we define the characteristic function~$\nu$.

\textbf{Characteristic function: }The  characteristic function precisely describes the set of all possible divisions of the anticipated worth  of any coalition. One can define such a function for partition form games using an appropriate anticipation rule~\cite{pessimistic}, also defined in sub-section \ref{sec_anticipation} (similar to $\alpha$ and $\beta$-effectiveness in the previous section). Thus, the characteristic function for any coalition $C$ is defined below,
$$
\nu(C) = \left \{ {\bm \phi}: \sum_{i \in C} \phi_i \le \nu^a_C \right \},
$$
where $\nu^a_C$ is the anticipated worth of coalition $C$ (as explained in Section \ref{sec:pfg_new_sc}). 

With this transformation, one can define R-core $\mathscr{C}(\mathcal{H})$ as in Section \ref{sec_gch} for partition-form TU game. Further, these notions can be related to  the notions  described in Section \ref{sec:pfg_new_sc}. 
We in fact have that the R-core equals the union of all partition-specific cores defined in \eqref{eq_core_extended_arbitrary_anticipate}, i.e.,  
$$
\mathscr{C}(\mathcal{H}) = \cup_{\P} \mathcal{C}^a(N,\nu,X^a_\P).
$$


\section{Stable configurations} 

In the previous few sections, we discussed various types of games like partition form, non-transferable utility games, etc.,  and notions of solutions like core, Shapley value, etc. We also considered a study specific to a fixed partition $\P$. 
In this section, we use the above  results to discuss an alternate viewpoint. 

We now want to consider a completely partition-specific viewpoint. Many a times it is important to understand the `emergence of a particular type of partition'. Basically one would like to know if a given arrangement/partition of players can continue to operate forever. 
If the answer is yes, then one could probably say that the partition is stable/emerges.

Formally we say that a given configuration $(\P,{\bm \phi})$ with ${\bm \phi} \in \mathcal{F}(\P)$ is stable if no coalition $C \notin \P$ can `block/dominate' the payoff vector ${\bm \phi}$ (as   in Section \ref{sec_gch}), i.e., if:
$$
e^a({\bm \phi}, C) \le 0 \mbox{ for all } C \notin \P. 
$$
%
%
Blocking by a coalition implies that a coalition not in the original partition has  an anticipation of achieving a better worth and could break away from the configuration involving $\P$. So, one can call a configuration  stable only if no such tensions exist, i.e., if no coalition can block. Observe that the notion of stability once again depends on the anticipation rule.
A partition is said to be stable if there exists at least one configuration involving it, which is stable.




An alternative approach to examining cooperative games involves studying coalition formation games, which we describe next.

\section{Coalition Formation Game (CFG) leading to $U$-core}
\label{sec_CFG_core}
In Chapter \ref{chap_PEVA}, we consider a CFG where players chose their coalitions in a non-cooperative manner. Each player announces the set of players with which they prefer to form coalitions, leading to a strategy profile. Each strategy profile leads to at least one partition, based on certain rules as described in sub-section \ref{sec_partition_given_strategy_profile}. Depending upon the partition formed, agents derive utilities (via another internal non-cooperative game among the coalitions of the partition). The above is the description of the  non-cooperative coalition formation game whose Nash Equilibrium helps us in defining unilaterally stable or \textit{U-stable partitions} as below.

We further restrict ourselves to the coalition-selection strategy profiles leading to a unique partition, and referred them as \textit{natural strategy profile}, as others have inherent instability. We finally have the following definition. 
 Any partition is a U-stable partition if the corresponding natural strategy profile is a NE, i.e., no player can benefit by unilaterally deviating from the current partition.

Now it is easy to see that this concept is related to the U-core described in Section \ref{sec:pfg_new_sc}. 
Specifically, a partition $\P$ is stable if and only if  the corresponding U-core $\mathcal{C}^a_U(N,\nu,X^a_\P)$ is non-empty.

\chapter{Coalition Formation Resource Sharing Games in Networks}
\label{chap_PEVA}

In this chapter, we consider a  coalition formation game with players exploring cooperation opportunities in a non-cooperative manner, where the utilities of players/coalitions are resultant of a resource sharing game.

\section{Introduction}
\label{Intro}
Resource sharing problem is a well-known problem that aims to find an optimal allocation of shared resources.
  Wireless networks existing in the same region compete to obtain larger spectrum shares to cater for ever-growing traffic demands. It is well known that online auctions
   \cite{koutsopoulos2010auction} can be used to achieve optimal resource allocations, in particular can also be used for efficient and fair spectral allocations. These auctions majorly use a `proportional allocation  algorithm' (Kelly's mechanism \cite{kelly1997charging}) which is also  considered in a variety of other contexts;  e.g., \cite{stoica1996proportional} considers real-time performance in time-shared operating systems, \cite{kelly1998rate} considers rate allocation for communication networks, \cite{tun2019wireless} considers resource allocation  in  wireless  network  slicing, etc. In this mechanism (\hspace{-.005mm}\cite{kelly1997charging}), the resource allocated to any player is proportional to its bid and inversely proportional to the weighted sum   of  bids of all players, with the weights representing the influence factors. 

  We also consider Kelly's mechanism, but, with very important differentiating features:
i) possibility of cooperation among the willing players;  and,
ii) the possible presence of an adamant player, not interested in cooperation. For example, most of the literature related to spectrum auctions utilising Kelly's mechanism, considers non-cooperative players \cite{xu2019robustness, koutsopoulos2010auction}. As mentioned in \cite{koutsopoulos2010auction}, it would be interesting to see if the agents can buy spectrum together and divide the allocated spectrum amicably. However, one cannot rule out the existence of players who are not interested in this kind of collaborations, and these form the \textit{adamant player} of our work.
 As a second example, the market giants (e.g., e-retailers like Amazon, wireless network providers like EE in UK, Orange in France) 
tend to strive alone while smaller business entities (e.g., e-retailers - Flipkart, Walmart; network providers - Airtel, Vodafone) look for  collaboration opportunities. The acquisition of Flipkart by Walmart and Mannesmann AG by Britain's Vodafone AirTouch PLC are few examples of rational firms coming together to sustain competition.
 In this study, we consider  a relevant aspect for investigation: \textit{when and which subset of willing players find it beneficial to collaborate}. We consider such a study using `non-cooperative coalition formation games'.

Any   transferable utility cooperative game is defined by a set of players $N$ and the worth of each possible coalition  $\{\nu_S; S \subset  N\}$ (e.g, \cite{narahari}, \cite{saad}). Majority of the analysis  related to cooperative games discuss the  emergence of  grand coalition (includes all players) as a successful partition and then consider the division of worth among the players;  Shapley value, Core  etc., are some  such solution concepts   (e.g, \cite{narahari}).  But one can find many example scenarios, in which a partition of strict  coalitions (subsets) of $N$ might emerge at some appropriate equilibrium (\hspace{-0.005mm}\cite{saad},\cite{saad2008distributed}).   In this context, one of the key challenges is to generate a partition, i.e., an exhaustive and disjoint division of the set of agents, such that the performance of the system is optimized (see for example, \cite{saad} and references therein). This leads to a utilitarian solution.  In contrast, we consider a non-cooperative approach to generate partitions (e.g. as in \cite{Nevrekar2015ATO}, \cite{saad});  basically  the solution/partition would be stable against unilateral deviations. These are in general called as coalition formation games (CFGs) (\hspace{-0.005mm}\cite{saad}). 

Another important aspect  of cooperative games is \textit{characteristic} and \textit{partition form games} \cite{saad}. Majority of the literature focuses on the former type of games where the  worth of a coalition depends only on its members. 
In the latter form of games, the worth  is also influenced  by the partition of the  players outside the given coalition (\hspace{-0.005mm}\cite{saad}).
These inter-coalitional dependencies play a crucial role in many real-world   scenarios (e.g., \cite{yi2003endogenous}, \cite{hafalir}). Our problem falls into this latter category.

We consider a CFG  in the  possible presence of  an adamant player  and seek for  a non-cooperative solution.  
In  our    game,  the     strategy of a  player   is the set of players with whom  it wants to form coalition as in \cite{Nevrekar2015ATO}.
Given the strategies of all players,  basically the preferences of all the players,  an  appropriate partition of   \emph{coalitions}  is  formed;  and players in each coalition maximize their collective utilities. This leads to a non-cooperative resource sharing game (RSG) among the coalitions.  The utilities at the resulting equilibrium are shared via Shapley value (confined to each coalition);   these shares define  the utilities of individual players for the given coalition suggestive preferences of all players in CFG.  We used Shapley value as it is a widely used solution concept (e.g.,\cite{iturralde2011resource, kim2016asymptotic}). Further for the game with identical players (or when a coalition has only identical players), the Shapley value is equivalent to any other solution, so many of our results are independent of the solution concept.

 Kelly's mechanism is studied in various other  non-cooperative frameworks (e.g., with adversary in \cite{xu2019robustness}, attacker in \cite{vulimiri2012well}, misbehaving advertiser in \cite{reiffers2014game}, etc.). To the best of our knowledge, none of these papers consider the non-cooperative coalition formation games. Further, we have a \rev{brief initial study (with other players having same influence factors) on} non-rational adamant player whose presence influences the game and many real-world scenarios require such study.  At this point, we would like to compare our adamant player with the usual adversary in literature. The aim of the adversary player is to harm other players (\cite{xu2019robustness, vulimiri2012well}). The adamant player in our model can also have a deteriorating influence on the utilities of players, however it's aim is not to harm; it is just not interested in cooperating with others.
 
  Many applications can be modelled using the framework of this work. Some examples are 
 spectrum allocation \cite{koutsopoulos2010auction}, cloud computing \cite{wei2010game}, network slicing \cite{tun2019wireless}, allocation of advertisement space \cite{cui2011bid,reiffers2014game}, market share \cite{ma2018complexity}, etc. We present a detailed description of one of these applications (spectrum allocation) in Section \ref{sec_example}.

We first derive the solution of this non-cooperative CFG for   symmetric players (players with equal influence factor).  
For smaller number of players ($n \leq 4$), the  partitions at Nash Equilibrium (NE) depend upon the relative strength of the adamant player and that of the others, \rev{which is denoted by $\eta$}. 
The partitions at NE are not monotone with $\eta$:  coarser partitions result at lower and higher values of  $\eta$, while we have finer partitions for  intermediate values \rev{(see sub-section \ref{sec:small_no})}.  This non-monotone behaviour disappears  for $n > 4$;  the players remain alone  at equilibrium (irrespective of $\eta$), i.e., the finermost partition emerges \rev{ (Corollary \ref{corollary_ALC_NE})}. 
We also consider the  utilitarian solution (maximizes the sum of utilities of all players) \rev{in Lemmas \ref{Lemma_SO_Partition} and \ref{Lemma_SO_no_adv}} to derive the price of anarchy (\rev{PoA}), which captures the loss of players resulting due  to their rational behaviour. The \rev{PoA} increases with   $n$ and  $\eta$.  It also increases  with decrease in  $\eta$  to zero \rev{(see Tables \ref{tab:my-table2}, \ref{tab:my-table1}, and \ref{tab:my-table})}. 

We next derive the solution of the same non-cooperative CFG for asymmetric players (without adamant player). The following are our key findings: (i) near symmetry (almost equal influence factors) there is only one stable partition at which all are alone ($n>4$); (ii) as asymmetry (differences between influence factors) increases, more partitions emerge at equilibrium \rev{(see Theorem \ref{Thm_stable_big_asym})}; (iii) interestingly, the highest and the least influential players are the first ones to form successful coalitions \rev{(Theorems \ref{Thm_monotonicity}-\ref{Thm_monotonicity_new})}, intermediate players may participate in coalitions at equilibrium for higher levels of asymmetry \rev{(demonstrated through numerical examples in Section \ref{sec_numerical})}; (iv) on further increase in asymmetry, we in fact have absolute stability, where all partitions are stable against unilateral deviations \rev{(Theorems \ref{Thm_general}-\ref{Thm_general_sym}}; (v) some players with smaller influence factors become insignificant as they derive zero utilities at equilibrium; and (vi) the \rev{PoA} in this case increases with the increase in the number of significant players, it also depends on the relative influence factors \rev{(Theorem \ref{Lem_worst_NE_util})}. To summarize, strong players are seldom interested in collaborating with comparable enemies if there are other collaborative opportunities.

We briefly investigated the partitions that are stable against coalitional deviations. Interestingly, for the case of absolute stability, grand coalition (which is also the utilitarian solution) is the only partition which is coalitionally stable \rev{(Theorem \ref{Thm_GC_CS})}. On the other hand, for the scenario with identical players (for $n>4$), none of the partitions are coalitionally stable \rev{(Theorem \ref{Thm_CS_None})}. This aspect requires further future investigation to understand general scenario.

By one of our findings, collaborative partitions (at least one coalition with size more than one) emerge at NE, even in the presence of adamant player, for the case with fewer number of (similar) players. 
 As noted before, real-world mergers also occur when competition is among fewer, similar and smaller players, in presence of market giants. We also demonstrated that higher and smaller influential players have more affinity to form successful  coalitions. For example,  Britain's Vodafone Airtouch PLC has been continuously acquiring smaller telecoms since 1990s.

 \section{Spectrum Allocation via auction}
\label{sec_example}
Consider a wireless network where $n$ players want to buy spectrum. Let the bid of player $i$ be denoted by $a_i$. The spectrum allotted to any player is proportional to its bid and inversely proportional to the sum of \rev{all bids}, i.e., player $i$ receives $a_i/\sum_j a_j$ fraction of the available spectrum (\cite{johari2004efficiency, xu2019robustness} and references therein). Every player wants to buy maximum possible spectral share which is proportional to its bid; however, they also have an associated cost proportional to their bids, \rev{which equals} $\gamma a_i$ \rev{with $\gamma$ being the proportionality factor}. Thus, the overall utility of player $i$ is given by,
\begin{equation*}
\frac{ a_i}{\sum_j  a_j} - \gamma  a_i.
\end{equation*}
Now consider players with different strengths represented by factors $\{\lambda_i\}_i$. For a player with bigger strength (say $\lambda_1$), the negative value of the cost ($ \nicefrac{\gamma}{\lambda_1}x$) associated with bid amount $x$ is smaller than that valued by a weaker player (say $\lambda_2<\lambda_1$) for the same bid amount (value equals $ \nicefrac{\gamma }{\lambda_2}x$). Basically, a stronger player incurs a smaller cost as compared to the weaker player when both of them bid the same amount. If suppose, player $i$ bids $x_i = \lambda_ia_i$ amount to buy the spectrum then its utility function is given by,
\begin{equation*}
\frac{\lambda_i a_i}{\sum_j \lambda_j a_j} - \gamma a_i \mbox{ for each } i.
\end{equation*}
There is a considerable literature that considers spectrum allocation problems. Here our goal is to study the set of players that come forward to bid together for their own selfish reasons; the aim is to find collaborative configurations that are primarily stable against unilateral deviations. We also consider a brief study of configurations stable against coalitional deviations. To the best of our knowledge we have not seen any work that considers such a game-theoretic setting. However
as pointed out by the authors in \cite{koutsopoulos2010auction}, it would be interesting to see if the agents can buy spectrum together and divide the allocated spectrum amicably.  But one still cannot rule out the possibility of existence of `adamant players' who are not interested in cooperation. For example, Jio in India might not be interested in buying spectrum together with any other player. We exactly consider such a study in this work when the agents are selfish and agree to cooperate, only if it is beneficial for them. 

We explain the implications of our results to this example later in Section \ref{conclusion}.

\section{Problem Description and Background}
\label{Prob_Desc}
Consider a system with $(n+1)$  players  involved in a resource sharing game (RSG) and  let  $N = \{0, 1, 2, \cdots, n\}$  
denote the set of players  along with \rev{a possible} adamant player represented by index $0$.  The  $n$ players (other than the adamant player)   are 
interested in forming coalitions, and these  are referred to as C-players. These players are willing to cooperate with each other if they can obtain higher individual share while the adamant player is not interested in cooperation. 
The utility of players is  proportional to their actions  which also includes a proportional cost. 
Thus, when players choose respective actions $(a_0, a_1,  \cdots, a_n)$, the utility of player $i $ equals 
\begin{equation} 
\varphi_i = \frac{\lambda_i a_i}{\sum_{j=0}^n \lambda_j a_j} - \gamma a_i \; \forall \; i \; \in \; N ,
\label{eq:Util_NC}
\end{equation}
where $\gamma$ represents the cost factor, $\lambda_i$  represents the influence factor of $i^{th}$ player and $a_i $  represents the action of $i^{th}$ player with  
	$a_i \; \in \; (0,\hat{a}) \; \text{for some} \;  n /\gamma  < \hat{a} < \infty $  which ensures the
	 existence of a unique Nash Equilibrium (NE) (\cite{dhounchak2019participate}).

The first component of equation \eqref{eq:Util_NC} is the fraction of resource allocated to player $i$ and the other component represents the cost.
This resembles the utility of players in well known Kelly-mechanism for resource sharing and is relevant in various applications as mentioned in Section \ref{Intro} including the application considered in Section \ref{sec_example}. 
%

In the first few sections, we consider the case with symmetric C-players, i.e.,  $\lambda_i = \lambda$ for all $i \in N_C$ where $N_C := \{1,\cdots,n\}$ is the set of C-players and $\lambda_0 = \eta \lambda$ with $\eta \in [0,\infty)$. \textit{Here, $\eta =0$ implies the absence of adamant player (in Section \ref{no_adamant_player})}. The general case is considered from Section \ref{sec_asym_init} onwards.  Such a sequencing is considered for ease of exposition; this is   also because  the general case is considered in the absence of  the adamant player.  General case in the presence of adamant player  can be a topic of future research.

When the players choose their actions in a fully non-cooperative manner, i.e., when none of the  C-players are   interested in forming coalitions,  it results in a strategic form game with utilities as in \eqref{eq:Util_NC};
basically the rational and intelligent players choose their respective   actions to improve their own utility and 
the  utility derived by any player equals that  at NE{\footnote{ \label{footnote_NE} Nash Equilibrium (NE) is a well-known solution concept. For a strategic form game $\langle N,X,U \rangle$, a strategy profile $\underbar{x}^* = (x_1^*, \cdots, x_n^*)$ is called an NE, if
		$
		U_i(x_i^*,x_{-i}^*) \geq U_i(x_i,x_{-i}^*) \, \forall \, x_i \in X_i,  \ \forall \, i \in N.
		$ This solution is stable against unilateral deviations.}}.


When  the $C$-players are looking   for opportunities to form coalitions and work together, a set/collection of coalitions   emerge at an appropriate equilibrium (details in  later sections);  say   ${\mathcal P} = \{S_0,S_1, \cdots, S_k\}$   represents the partition of $N$ into different coalitions where $S_0 \hspace{-1mm} = \hspace{-1mm} \{0\}$ denotes  the adamant player. Observe that a partition \( \mathcal{P}  \) is a set of coalitions such that 
\begin{equation}
\cup_{i=0}^k \; S_i = N \; \text{and} \; S_i \cap S_j = \emptyset, \mbox{  null set, } \, \forall \, i \neq j.
\label{eq:partition}
\end{equation}
The players in \textit{coalition} $S_i$  choose their strategies together with an   aim to optimize  their social objective function (of their own coalition) and hence the utility of a coalition (for symmetric case) is given by:
\begin{eqnarray}
\label{Eqn_Util_coaltiion}
  \varphi_{S_m} ({\bf a}_m, {\bf a}_{-m} ) &=& \frac{\lambda\sum_{l \in S_m} a_l }{\lambda_0a_0+\lambda\sum_{l=1}^n a_l} - \gamma \sum_{l \in S_m} a_l; \; m \geq 1 \\
\label{Eqn_Util_coaltiion_adamant} \varphi_{S_0} ({\bf a}_m, {\bf a}_{-m} ) &=& \frac{\lambda_0a_0 }{\lambda_0a_0+\lambda\sum_{l=1}^n a_l} - \gamma a_0, \end{eqnarray}
 where, $\ {\bf a}_m = \{a_i, i \in S_m\}, {\bf a}_{-m} = \{a_i, i \notin S_m\},  \forall  S_m \in {\mathcal P}$,
which is the sum of their individual utilities.
The players will now try to derive maximum utility for their own coalition and hence  there would again be a non-cooperative game, 
but now  among coalitions. Thus we have  a reduced RSG (one for every $\mathcal{P}$) with each coalition representing one (aggregate) player and  the utilities given by \eqref{Eqn_Util_coaltiion} and \eqref{Eqn_Util_coaltiion_adamant}; utility of  any coalition  equals that at the corresponding NE. 
This utility is divided among the members of the coalition using the well-known  {\it Shapley value} (computed within the coalition), \textit{which simplifies to equal shares for symmetric players.}

This is the problem setting and our aim is to study the  coalitions/partitions that  emerge out successfully (at an appropriate equilibrium), when the C-players (at times we refer them as players) seek opportunities to come together in a non-cooperative manner.
There is a brief initial study of this problem in   \cite{dhounchak2019participate}, for the special case when players only form grand coalition, i.e., when ${\mathcal P} = \{ \{0\}, \{1,  \cdots, n\}\}$. For this case, it has been shown that: 

i)  The utility of grand coalition at CNE (Cooperative NE) is higher than the sum of individual utilities of players at the unique NCNE (Non-Cooperative NE) for majority of the scenarios. The paper also provides example scenarios for the 
case where the sum of utilities at NCNE is larger.

ii) Moreover, Shapley value (SV) does not always share this utility  in a fair manner;   the sum of utilities might be larger, but  the shares derived via SV by some  players  is smaller (especially ones with higher influence factors).  We also consider SV, as it is a widely accepted solution. Investigating with other solution concepts can be a part of future study.


The above study leads to new questions: a) can the players derive even  better utilities if they form strict sub-coalitions instead of grand coalition; b)  when  is it  beneficial for the players to cooperate; c) how stable are these resultant coalitions  (e.g., against unilateral or coalitional deviations); c) when players are asymmetric (with different influence factors), which type of players form `successful' coalitions? Is it all high influential or all less influential or a mix of them? etc.  

{\it We  build an appropriate non-cooperative framework to study these aspects. We also consider  solutions that optimize social objective function to  derive the Price of Anarchy.}

\section{Adamant Coalition Formation Games}

\label{ACFG}
We use non-cooperative framework to study this coalition formation  game (CFG) as in \cite{Nevrekar2015ATO}. 
For each C-player, i.e., for $i \in N_C$, strategy $x_i$  is defined as the  set of players with whom player $i$ wants to form coalitions, i.e., $x_i \subseteq N_C$ and the corresponding strategy set $X_i$  is defined as:\vspace{-2mm}
$$
X_i = \{x_i : i \in x_i \mbox{ and } x_i \subseteq N_C \}.
\vspace{-2mm}
$$
To construct a  strategic form game, we need to define the utility of all players for any given strategy profile, i.e., for any $\underbar{x} = (x_1,x_2,\cdots,x_n) \mbox{ with } x_i \in X_i \mbox{ for each }i \in N_C$.

As a first step, one needs to define appropriate partition(s) of coalitions (referred to as $\mathcal{P(\underbar{x})}$,  and made up of subsets of  $N$)  that can result for any   given  strategy profile $\underbar{x}$. 

\subsection{Partition for a given strategy profile \texorpdfstring{$\underbar{x}$}{x}}
\label{sec_partition_given_strategy_profile}

\rev{The majority of the literature focuses on grand coalition and analyses the corresponding solutions/stable payoff vectors. In other words, one can view this as a preference to bigger coalitions.  In view of this, we say a partition to be `better' when it is comprised of bigger coalitions, as defined below.} 

We say a {\it partition  $\mathcal{P}'$ is (strictly) better than partition $\mathcal{P}$, represented by the symbol $ \mathcal{P}' \prec \mathcal{P}$,}  if every coalition of the latter is a subset of a coalition of the former (with at least one of them being a strict subset), i.e.,   if
\begin{eqnarray}
\mathcal{P}' \hspace{-1mm} \ne \hspace{-0.5mm} \mathcal{P}  \mbox{, and,       for all } S \hspace{-0.5mm} \in \hspace{-0.5mm} \mathcal{P}    \
\exists \   S' \hspace{-1mm} \in \mathcal{P}'  \mbox{ such that } S \subset S'.  \hspace{-2mm} 
\label{Eqn_Max_Coalition_rule_0}
\end{eqnarray}
%
Note that the size   (number of coalitions)  of the better partition is strictly smaller than that of the other \rev{-- basically it is a coarser partition}; in other words, there exists at least two coalitions $S_1, S_2  \in {\mathcal P}$ such that 
$S_1 \cup S_2 \subset S$ for some $S \in {\mathcal P}'$.

{\it  Partition $\mathcal{P}(\underbar{x})$  formed  by $\underbar{x}$:}
We say ${\underbar x} \to  \mathcal{P} (\underbar{x})$, if  it satisfies the following two conditions as in \cite{Nevrekar2015ATO}:\\
i) {\it respects the preferences,} 
a coalition $S$ is an element of partition $\mathcal{P(\underbar{x})}$, i.e.,  $S \in \mathcal{P(\underbar{x})}$,  if  it satisfies: 
\begin{eqnarray}
i \in x_j  \mbox{ and }  j \in x_i  \mbox{   for all }  i, j  \in S;  \mbox{ and, }  \label{Eqn_Coalition_rule}  
\end{eqnarray} 
ii) {\it minimal partition,} there exists no other (see \eqref{Eqn_Max_Coalition_rule_0})
\begin{eqnarray}
\mbox{(better) partition  $\mathcal{P}' $ formed by } \underbar{x}  \mbox{, such that  } \mathcal{P}' \prec \mathcal{P}.    \label{Eqn_Max_Coalition_rule}
\end{eqnarray}

\noindent
Hence, a \emph{partition} formed by $\underbar{x}$ is a (minimal) subset of $2^N := \{S: S \subset N\}$  such that   \eqref{eq:partition} and  \eqref{Eqn_Max_Coalition_rule}  are satisfied and all its coalitions satisfy  \eqref{Eqn_Coalition_rule}.  Thus, \textit{preference is given to coarser partitions}. Using these rules, we may obtain multiple partitions for some strategy profiles (some examples are  in Tables \ref{tab:n=3} and \ref{tab:n=4}). 

To summarize, if 
$\underbar{x} = (x_1, \cdots, x_n)$ is the strategy profile, let $n(\underbar{x})$ represent the number of possible partitions   corresponding to $\underbar{x}$, and  let  the partitions formed be  represented by the following,
$
{\mathcal P}^1 (\underbar{x}), \, {\mathcal P}^2 (\underbar{x}) \,  \cdots \, {\mathcal P}^{n\text{\small{$(\underbar{x})$}}}  (\underbar{x}).
$
We now define the utilities derived by (all) the coalitions and then the individual players. We begin with  the case when $n(\underbar{x}) = 1$.
\subsection{Utilities of coalitions in a given partition}
\label{sec_util_coal}
Let $\mathcal{P(\underbar{x})}$ = $\{S_0,S_1,   \cdots, S_k \}$ be a  partition of $N$  with $k$ coalitions of C-players,   corresponding to $\underbar{x}$. We now  aim   to find the utility of 
coalitions in $\mathcal{P}(\underbar{x})$, represented by  $\varphi^*_{S_m}(\mathcal{P})$ for all   $ m \in \{0,1,2, \cdots, k\}$. We will see that  these utilities depend upon the strength of the adamant player, via $\eta := \lambda_0 / \lambda$, the relative ratio of the influence factors (recall $S_0 = \{0\}$ is the coalition with  only adamant player). 

As already mentioned, the resource sharing game (RSG) is now reduced to a \textit{(k+1)-(aggregate) player non-cooperative strategic form game} which is given by the tuple,
\begin{eqnarray}
\label{Eqn_reduced_RSG}
 \big \langle \{0, 1, \cdots, k\}, \{[0,\hat{a}]^{|S_0|} \times \cdots\times [0,\hat{a}]^{|S_k|}\},\mathbf{\varphi} \big \rangle,
\end{eqnarray}
where $|S_m|$ represents the cardinality of coalition $S_m$ and $\mathbf{\varphi}$ = $\{\varphi_{S_0}, \varphi_{S_1}, \cdots \varphi_{S_k} \}$, the vector of utilities is given by  \eqref{Eqn_Util_coaltiion} and \eqref{Eqn_Util_coaltiion_adamant}.
%
%
This kind of a game is analysed in \cite[Lemma 2]{dhounchak2019participate} for the  special case with grand coalition (GC) of C-players. 
Since, we consider all possible exhaustive and disjoint collection of players, i.e., all possible partitions (corresponding to various coalition suggestive strategy profiles), we  extend the above  result to a general partition. The complete theorem is available in Section \ref{sec_asym_init} as Theorem \ref{Thm_util_coal_general}. For the ease of notations, we present the result for game \eqref{Eqn_reduced_RSG} in the following for the special case with symmetric C-players  (proof is in Appendix~\ref{sec_AppendixA_PEVA}). 
\begin{theorem}{\bf [Utilities of coalitions]}
	\label{thm:Thm1}
	The game \eqref{Eqn_reduced_RSG} can have multiple NE, but the utilities at NE are unique and are ( for any $ 1\leq m \leq k$),
	 \begin{eqnarray}
		\label{Eqn_USm}	\varphi^*_{S_m}(\mathcal{P}) & = & 
		\frac{1}{  (1  +   k  \eta )^2} \mathds{1}_B 
		+ \frac{1}{ k^2   }  (1 - \mathds{1}_B),
\mbox{ and, }		\label{NE_util_adamant}
	\varphi^*_{S_0}(\mathcal{P}) 
	\	 = \  \left(\frac{1-k+k\eta}{  1  +  k  \eta }\right)^2 \mathds{1}_B,   
	\end{eqnarray}
	with indicator  $\mathds{1}_B := \mathds{1}_{\eta > \frac{k-1}{k}}$, $k =  | {\mathcal P} | -1$,  $\eta  := \lambda_0/\lambda$. 
 Further the optimal actions at any NE  satisfy:
	 \begin{eqnarray}
			{\bar a}^*_m :=  \sum_{j \in S_m} a_j^* =    \frac{k\lambda\lambda_0}{\gamma (\lambda + k \lambda_0)^2} \mbox{ \normalsize and, } {\bar a}^*_0 = \frac{k\lambda((1-k)\lambda+k\lambda_0)}{\gamma (\lambda + k \lambda_0)^2}.  \hspace{20mm}  
	\label{Eqn_alstar}
			\end{eqnarray}  
%
	
\end{theorem}

From \eqref{Eqn_alstar},  some (or all) players of a coalition can choose actions such that the sum of these actions equal corresponding $\bar{a}_m^*$; all such actions constitute NE; hence multiple NE exist. However,   utilities of coalitions are uniquely defined by  \eqref{Eqn_USm}.

\noindent
{\bf Significant Adamant Player:}
The adamant player gets non-zero utility at NE   when  $\mathds{1}_B = 1$,  i.e.,  when $\eta > 1 - 1/ k$ (see \eqref{NE_util_adamant}),  we then say the adamant player is significant, and otherwise it is insignificant.  However, it is always significant when grand coalition is formed, i.e., when $k=1$ (see \eqref{NE_util_adamant}).  This condition will play an important role in our CFG. 

To summarize, the utilities of any coalition of any given partition $\mathcal{P}$  are given by \eqref{Eqn_USm} and \eqref{Eqn_alstar}, which  are the utilities at NE of the reduced RSG with coalitions  as the players.

\subsection{Division of worth  within a coalition}
\label{sec_div_worth}
The next step is to divide the worth of a coalition among its members  using Shapley value confined to each coalition as in \cite{aumann1974cooperative}.
For  symmetric players, the utility of a coalition gets divided equally among its members because of equal influence factors (infact this is true for any reasonable solution/sharing concept); general case is again in Section \ref{sec_asym_init}. Hence from \eqref{Eqn_USm}, the utility of  player $i$  under  partition 
${\mathcal P} $ is given by (if $i \in S_m$ and $m \geq 1$):
\begin{equation}
\label{Eqn_USm_player}
\varphi^*_{i} ({\mathcal P})   = \frac{   \mathds{1}_B  }{  |S_m|  \left (1 +  k   \eta  \right )^2}
+ \frac{1- \mathds{1}_B }{ k^2  |S_m|  } \mbox{ with } k=|{\mathcal P}|-1.
\end{equation}

\subsection{Utility of a player}
\label{sec_util_player}
We define the utility of a player, say $i$ as the minimum utility among all the possible partitions for arbitrary $n(\text{{\small$ \underbar{x}$}})$  (see equation \eqref{Eqn_USm_player}),
\begin{equation}
\label{min_util_mult_partition}
U_i(\underbar{x})= \min_{\mathcal{P}(\text{{\small $\underbar{x}$}})} \varphi^*_i (\mathcal{P}(\underbar{x}) ).
\end{equation} 
This definition ensures minimum guaranteed utility to each player for the given strategy profile $\underbar{x}$ and is similar in concept to the \textit{security value} (\cite{narahari}), \textit{pessimistic rule} (\cite{bloch}) used in game theory. Basically, when a strategy profile (coalition suggestive strategies of all the players) can lead to multiple partitions, 
the eventual partition formed may depend on some further negotiations. 
Hence, it is appropriate to define the utility of each player as the worst possible  utility.

\subsection{Coalition Formation Game: Ingredients}
\label{sec_CFG_ingre}
\label{sec:CFG}
We now have a non-cooperative CFG with,  i) $N_C  $ as the set of players;
ii) $
X_i$ is the strategy set of player $i $; and 
iii) Utilities  of   players, $\{U_i ({\underbar{x} }) \}_{i,\text{{\small{$\underbar{x}$}}}}$ 
given by \eqref{min_util_mult_partition}. Recall  these utilities are defined via their Shapley value corresponding to the coalition that they belong to (based on their and others strategies), the worth of which is computed using NE of the reduced RSG. 
We study this game and consider two types of solution concepts: NE and Social Optima and also discuss the price of anarchy in the coming sections.

\section{Symmetric players} 
\label{init_analysis}
We begin with   symmetric game, i.e., where C-players have same influence factor $\lambda$.
We begin   with few definitions.

\subsection{Partition resulting from a unilateral deviation} 
\label{sec_partition_ud}
Recall a  {\it strategy profile}  $\underbar{x}$ {\it leads to  partition $\mathcal{P}$},  represented by $\underbar{x} \to \mathcal{P}$,  if    $\mathcal{P}$  results from   $\underbar{x}$ as explained in Section \ref{sec_partition_given_strategy_profile}, i.e., if  it satisfies \eqref{Eqn_Coalition_rule}, \eqref{eq:partition} and \eqref{Eqn_Max_Coalition_rule}. 
We say,     $\underbar{x}$ {\it leads to unique partition $\mathcal{P}$},  represented by $\underbar{x} \to !\mathcal{P}$,  further,  if  $\mathcal{P}$ is unique such partition, i.e., if $n(\underbar{x}) = 1$.

Consider  any  partition $\mathcal{P} = \{S_0, \cdots,S_k\}$ 
%
%
and say $ \underbar{x} \to !\mathcal{P}.$
Now consider a  unilateral deviation    of  player  $i$,  from  $x_i$  to  $\{i\}$ (strategy of being alone)  in   $ \underbar{x}$ and say $i \in S_l$.   
Then, the following lemma  shows that the new strategy profile ($\underbar{x}'$) also leads to a unique partition with $S_l$ coalition getting split into two; $\{i\}$ and $S_l/\{i\}$ (proof in Appendix~\ref{sec_AppendixB_PEVA}):
\begin{lemma}
	\label{Lem_partiton_uni_dev}
	Consider a strategy profile  $\underbar{x} \to !\mathcal{P}$, where  $i \in S_l$. 
	Let  $\underbar{x}' = (\{i\},  \underbar{x}_{-i})$ be the strategy obtained by the above unilateral deviation, then    $  \underbar{x}'  \to  !{\mathcal P}_{-i}$, where: 
	\begin{equation*}
	\mathcal{P}_{-i} :=   \{S_0,S_1, \cdots, \  S_{l-1}, \{i\},  \  S_{l} \backslash   \{i\},   S_{l+1}, \cdots, S_k\} . 
	\end{equation*}
\end{lemma}


%
We call partition  $\mathcal{P}_{-i} $ of the above lemma  as the {\it $i$-unilateral deviation  partition,  $i$-u.d.p.,}  of  the pair  $(\underbar{x}, \mathcal{P})$. 


\subsection{Weak Partition}
\label{sec_weak}
A \emph{partition is defined to be weak} if  for all $ \underbar{x} \to !\mathcal{P}$, there exists a  player $i$  which gets strictly better utility  at  its  $i$-u.d.p, i.e.,   if ($\mathcal{P}_{-i}$ defined  in Lemma \ref{Lem_partiton_uni_dev}),
$$U_i(\mathcal{P}_{-i})>U_i(\mathcal{P}). $$

With the above definitions, the following result characterizes the weak partitions (proof in Appendix~\ref{sec_AppendixB_PEVA}):
\begin{lemma}
	\label{Lemma_Partition_weak}
	Consider a partition $\mathcal{P}$ with {\small$|\mathcal{P}| = (k+1)$}.    Let  $m^* := \max_{S_i \in \mathcal{P}} |S_i|$, be the  size of the largest coalition. If $m^* >  (k+1)^2/ k^2 $, then   $\mathcal{P}$ is \emph{weak}. 
\end{lemma}
\textbf{Remarks: } (i) The above result identifies the weak partitions. It is easy to observe that as the size of the partition increases, it is more likely to be a weak partition, since $(k+1)^2/k^2$ is decreasing in $k$. For example, all partitions with size greater than two are weak if any of its coalitions contains more than one player.

 (ii) Say for all the strategy profiles $\underbar{x}$ leading to  $\mathcal{P}$ it is the unique such one (i.e., $\underbar{x} \to !\mathcal{P}$). Further, if it satisfies the above conditions, it  cannot be a partition at NE (because some player $i$ derives better at its $i$-u.d.p.). However, if there is a strategy profile leading to multiple partitions with one of them being $\mathcal{P}$, then $\mathcal{P}$ may  still emerge at an NE. We  investigate these aspects in the immediate following with an aim to derive the NE-partitions.

\subsection{Nash Equilibrium }
\label{sec_NE}
To study the CFG (see Section \ref{sec:CFG}), we again consider the solution concept \emph{Nash Equilibrium (NE)} \cite{narahari} (provided in footnote \ref{footnote_NE}).  The NE is now in terms of coalition suggestive strategy profile, but one might be more interested in
\textit{NE-partitions} (partitions emerging from an NE).
Lemma \ref{Lemma_Partition_weak} characterizes  weak partitions, and one may think weak partitions  cannot   result from an NE. However, as discussed before, if a weak partition is one amongst the multiple partitions emerging from  an NE, then it may also be a NE-partition. Thus we have (proof is straightforward):
\begin{lemma}{\bf [NE $\not \to$ Weak Partition]}
	\label{Lemma_weak_NE}
	Assume that the game does not have multiple partitions at NE. 
	Then, if a partition ${\mathcal P}$  is weak,  it cannot be a NE-partition. 
	
\end{lemma}

If for a  given set of parameters, it is known a priori that none of the NE lead to multiple partitions, then by the above lemma,     a weak partition can't emerge from an NE. We will then concentrate on  partitions that are not weak. 
We will use these intermediate results to derive the NE.  Before we proceed,   we discuss a relevant  social objective function.

\subsection{Social Optima}
\label{sec_SO}
In this work we are primarily studying the CFGs, in which the players choose their partners in a non-cooperative manner; \textit{basically the players are interested in coalition formation, so as to  improve their own objective function (selfishly)  and one requires a solution which is stable against unilateral deviations}. But if instead the players attempt to optimize a social/utilitarian objective (sum of utilities of all the players),  they would have achieved much better utilities; this aspect   is well understood in literature (~\cite{johari2004efficiency} and references therein) and we study the same in our context.   
A \emph{utilitarian solution}, referred to as SO (social optimizer),  is any  strategy profile  $\underbar{x}_S^
*$ that maximizes: 
$$
\sum_{i \in N_C}U_i(\underbar{x}_S^*) = \max_{\text{{\small{$\underbar{x}$}}} } \sum_{i \in N_C}U_i (\underbar{x})  . 
$$
In \cite{dhounchak2019participate} authors illustrated that the sum utility of the  C-players improve significantly, when all players come together to  form a grand coalition (as $n$ increases). However we will see in this study that for $n > 4$, the only  NE-partition is ALC (all alone). Because of the
selfish nature of the players, the efficiency of a system degrades and the utility received by players at NE is (much) lower than  that at SO.
We study  this loss  using the well known concept, \emph{Price of Anarchy}.

One might be interested in the NE or SO, basically the strategies that represent the solutions.  However in our context,  the {\it more interesting entities are the partitions at various equilibrium/optimal solutions;}   we are interested in  NE-partitions  and  the SO-partitions.
%
When one directly optimize\footnote{One can think of such an optimization,  as   the players are working together now. } using partitions;  it is easy to see that the  SO-partition, 
${\mathcal P}^*_S$   satisfies the following:
$$
U_{SO}^* := \sum_{S_i \in \mathcal{P}^*_S;i\neq0}U_{S_i}( \mathcal{P}^*_S) = \max_{\mathcal{P}'} \sum_{S_i' \in \mathcal{P}'; i \neq 0}U_{{S_i}'}( \mathcal{P}'),  
\vspace{-0mm}
$$
where $\mathcal{P}'$ includes all possible partitions 
(see \eqref{Eqn_USm}-\eqref{Eqn_USm_player}).  

\subsubsection*{\bf Some more notations}
{\it Let ${\mathcal P}_k$ represent any partition with $k$  coalitions of C-players}, i.e.,  $| {\mathcal P}_k | = 1+k$.
\rev{Consider the partition ${\mathcal P}_n = \{ \{0\}, \{1\}, \{2\}, \cdots, \{n\}\}$ where all the players operate alone. We refer it as the partition with {\it  All aLone Coalitions,} or  briefly as {\it ALC} partition.}
The  strategy $x_i = \{i\}$  is the  ALC strategy for any $i$, and 
{\it GC (Grand Coalition}) partition implies partition $\{ \{0\},  N_C\}$, while  GC strategy implies  $x_i = N_C$.

\textbf{Two groups of partitions:}
As seen in \eqref{NE_util_adamant},  at some equilibrium the {\it adamant player  becomes insignificant,} i.e., gets 0 utility.  
We distinguish  these  equilibrium partitions  from the others using superscript $^o$.  Thus, for example,  ALC is the NE-partition if adamant player is significant at that NE, otherwise, ALC$^o$ is the NE-partition. 

%

With the notations in place, we have the following result characterizing the  SO-partitions (proof  in 
Appendix~\ref{sec_AppendixB_PEVA}):
\begin{lemma}{\bf [SO-partitions]}
	\label{Lemma_SO_Partition}
	i) When $\eta \ge 0.707$ or when $\eta \le 0.414$, then GC is the   SO-Partition.\\
	ii)  When $0.414 \le \eta \le  0.5$,  any  ${\mathcal P}_2^o$  is the SO-partition.  \\
	iii)  Any  ${\mathcal P}_2$  is a  SO-partition for  {\small$ 0.5 < \eta \le 0.707$}. 
\end{lemma}

\subsection{Price of Anarchy and SO-partition}
\label{PoA}
Price of Anarchy (\rev{PoA}) is defined as the ratio between the sum utilities at  `social optima' and the sum utilities at the  `worst Nash Equilibrium', i.e., 
\begin{eqnarray*}
	\text{\rev{PoA}}  &\hspace{-1mm}=\hspace{-1mm}& \frac{\max_{\mathcal{P}} \sum_{S_i \in \mathcal{P}; i \neq 0} U_{S_i}  }{ \min_{\mathcal{P}^*} \sum_{S_i \in \mathcal{P}^*;i \neq 0} U_{S_i}  }  = 
	\frac{  U_{SO}^*  }{ \min_{\mathcal{P}^*} \sum_{S_i \in \mathcal{P}^*;i \neq 0} U_{S_i}  }, \text{ where $\mathcal{P}^*$ is any NE-partition}. 
\end{eqnarray*}

\subsection*{ALC/ALC$^o$  is always an NE-partition }
When all others choose to be alone, i.e., if $x_i = \{i\}$ for all $i \ne j$, then it is clear that the best response of $j$  includes $x_j = \{j\}$. This is true for any $j$.
This leads to an NE.  
From \eqref{NE_util_adamant}, the adamant player becomes insignificant at ALC when $\eta \le 1 - 1/n$, then the NE-partition is ALC$^o$, otherwise ALC is the NE-partition. Basically, this NE results when nobody is interested in collaborating with others.

\subsection{Large  number of players, \texorpdfstring{$n > 4$}{n>4}}
\label{sec_sym_large}
For the case with $n>4$, we have the following two results using Lemma \ref{Lemma_Partition_weak} (proofs in Appendix~\ref{sec_AppendixA_PEVA}):
\begin{corollary} {\bf [Weak Partitions]}
	All partitions other than ALC/ALC$^o$ are \textit{weak}. 
	\label{corollary_ALC_weak}
\end{corollary} 

\begin{theorem}{\bf [No Multiple Partitions at NE]}
	\label{Thm_No_MPs_for_grt_n}
	Any strategy profile leading to multiple partitions cannot  be an NE.  
\end{theorem}
In view of the above two results and Lemma \ref{Lemma_weak_NE}, only ALC/ALC$^{o}$ is the NE-partition. Further using \eqref{NE_util_adamant}, we have:   
\begin{corollary}
	\label{corollary_ALC_NE}
	{\bf [Unique NE for $n>4$]}
	ALC is the unique NE-partition if $\eta > (n- 1)/n$, else 
	ALC$^o$ is the unique NE-partition. 
\end{corollary}

In \cite{dhounchak2019participate}, authors defined BoC (benefit of cooperation) as the normalized improvement in sum of utilities that the players achieve at GC in comparison with that achieved when they  compete alone.  They showed that BoC increases significantly as $n$ increases (\cite[Lemma 3]{dhounchak2019participate}). 
Despite the fact that  BoC is large for large $n$, \textit{by the above Corollary  we have that  players prefer to remain alone at NE}. Thus the  price paid for anarchy (\rev{PoA})  can be significantly high.

{\bf  Price of Anarchy}
%
From Corollary \ref{corollary_ALC_NE}, 
we have ALC/ALC$^o$ as the only NE-partition and  from Lemma \ref{Lemma_SO_Partition},  GC is the SO-Partition  when  $\eta \ge  0.707$. Hence \rev{PoA} equals (see \eqref{Eqn_USm}): 
\begin{eqnarray*}
	\text{\rev{PoA}} \hspace{-2mm} &=& \hspace{-2mm} \frac{\frac{1}{(1+\eta)^2}}  {\frac{n}{(1+n\eta)^2} }
	\ = \  \frac{(1+n\eta)^2}{n(1+\eta)^2} \mbox{ when }  \eta \geq 0.707. 
\end{eqnarray*}
We compute \rev{PoA} for the  remaining  cases  in a similar way and the results are  in Table \ref{tab_PoA_large}.
Clearly as $n \to \infty$, \rev{PoA}   grows like   $n$, i.e., $\text{\rev{PoA}} = O(n)$;   this is another instance of strategic behaviour where the players pay high price for being strategic.

\begin{table}[h]
		\centering
				\begin{center}
		\begin{tabular}{|c|c|c|c|c|}
			\hline
			& Range                  & $\mathcal{P}$ at NE & $\mathcal{P}$ at SO & \text{\rev{PoA}}                                                                                                                  \\ \hline
			1 & $\eta > \frac{n-1}{n}$ & ALC                 & GC                  & $\frac{1}{n}\Big(\frac{1+n\eta}{1+\eta}\Big)^2$ \\ \hline
			2 &  $0.707 \le  \eta \leq \frac{n-1}{n}$                  &       ALC$^{o}$              &                    GC   &              $ \frac{n}{(1+\eta)^2}$                                                                                                       \\ \hline
			3 &       $0.5 < \eta  \leq  0.707$                 &           ALC$^{o} $           &     $\mathcal{P}_2$              &                                                                                                                      $ \frac{2n}{(1+2\eta)^2}$ \\ \hline
			4 &       $0.414 \le  \eta \leq 0.5$                 &          ALC$^{o} $           &      $\mathcal{P}_2^o$                & $\frac{n}{2}$                                                                                                                      \\ 
			\hline 
			5 &       $0 < \eta  \leq  0.414$                 &        ALC$^{o} $             &  GC                  &      \hspace{-3mm}$ \frac{{n}^{\hspace{1mm}}}{(1+\eta)^2}$                                                                                                                                                                                                                   \\ \hline
		\end{tabular}
		\vspace{1mm}
		\caption{NE-partitions, SO-Partitions and \rev{PoA} for $n > 4$
		\label{tab_PoA_large}}
		\vspace{-6mm}
			\end{center}	
	\end{table}
	
	\subsection{Small number of players, \texorpdfstring{$n \le 4$}{n<5}}
\label{sec:small_no}
In this section, we identify the  NE-partitions and derive the \rev{PoA},    for $n \leq 4$, by direct computations. 
\subsubsection*{When $n = 2$}
Here,  GC and ALC (or ALC$^o$) are the only possible partitions. 
Some strategy profiles and the corresponding partitions can be seen from Table \ref{tab:n=2}.

	\begin{table}[ht]
	\begin{minipage}{8cm}
	\begin{center}
		\begin{tabular}{|c|c|c|c|}
			\hline
			$x_1$   & $x_2$   & $\mathcal{P} ({\underline  x})$         \\ \hline
			GC& GC & GC     \\ \hline
			ALC   & GC &ALC \\ \hline
		\end{tabular}
		\vspace{-1mm}
		\caption{Partitions  at $n=2$
		\label{tab:n=2}}
		\end{center}
	\end{minipage}
	\begin{minipage}{8cm}
	\begin{center}
		\begin{tabular}{|c|c|c|c|c|}
			\hline
			$x_1$     & $x_2$     & $x_3$     & $\mathcal{P}({\underline  x})$                                       \\ \hline
			\{1,2\}   &GC&GC & \{\{1,2\},\{3\}\} \\
			&                 &                 & \{\{1\},\{2,3\}\}  \\ \hline
			ALC     &GC& GC &\{\{1\},\{2,3\}\}                            \\ \hline
		\end{tabular}
		\caption{Partitions at $n=3$}
		\label{tab:n=3}
		\end{center}
	\end{minipage}
\end{table}
We begin with  deriving  the best responses (BRs). Consider the case with    $\eta \ge 0.707$. Then from \eqref{Eqn_USm_player}, BR of player 2 against    player 1's  strategy, $x_1 = \{1,2\}$ is GC, because:
$$
\frac{1}{2}\left(\frac{1}{1+\eta}\right)^2   \geq   \left(\frac{1}{1+ 2 \eta}\right)^2.
$$    Thus  both GC and ALC  are  NE-partitions  when  $\eta \ge 0.707$. In a similar way one can verify that the only NE-partition is ALC for 
$0.5 < \eta \le  0.707 $ (see Table \ref{tab:my-table2}).

\begin{table}[h]
	\centering
	\hspace{1mm}
	{\small\begin{tabular}{|c|c|c|c|}
		\hline
		  $\mathcal{P}$ at NE                   &  Range &  $\mathcal{P} \text{ at SO}$    &  \rev{PoA}                                                        \\ \hline
		  GC, ALC        & $\eta \ge 0.707$ & GC & $ \frac{1}{2} \Big( \frac{1+2\eta}{1+\eta} \Big)^2$ \\ \hline
		 ALC &$0.5 < \eta \leq 0.707$  &  $\mathcal{P}_2 $ & 1 \\ \hline
		 ALC$^o$  & $0.414 \leq \eta \leq 0.5$  &   $\mathcal{P}_2 ^o$& 1 \\ \hline
		 GC, ALC$^o$& $0 < \eta \leq 0.414$  & GC  & $\frac{2}{(1+\eta)^2}$ \\ \hline
	\end{tabular}}
	\vspace{1mm}
	\caption{NE-partitions, SO-partitions and \rev{PoA} For  $n=2$ 
	\label{tab:my-table2}}
\end{table}

When  $0.414 \le  \eta \le  0.5$,   the adamant player is insignificant (gets 0 at NE) and ALC$^o$ is the  unique NE-partition. Interestingly, below $\eta \le 0.414$, 
the C-players find it beneficial  (again) to cooperate, note GC is also an NE. Thus we observe interesting  {\it non-monotone phenomenon} with ratio of influence factors, $\eta$. 

\subsubsection*{When $n = 3$} 
In this case, we can have three types of partitions:   GC,  ALC  and ${\mathcal P}_2$ type partitions.  In any ${\mathcal P}_2$ type partition,  two of the C-players are together  in one coalition, while  the remaining  one is   alone.   
Some strategy profiles and the resulting partitions are  in Table \ref{tab:n=3}.

We derive the analysis by directly computing the BRs as in the previous case. The results
are summarized in Table \ref{tab:my-table1} (some details are in Appendix~\ref{sec_appendixE_PEVA}).

\begin{table}[h]
\centering
		{\small 
		\hspace{-1mm}
		\begin{tabular}{|c|c|c|c|}
		\hline
		 $\mathcal{P}$ at NE                   &  Range&  $\mathcal{P}$ \text{at SO}    &\rev{PoA}                    \\ \hline
		 GC, $\mathcal{P}_2$, ALC             &           $\eta \geq 2.732$      &  GC            &    $\frac{1}{3} \Big( \frac{1+3\eta}{1+\eta} \Big)^2$       \\    \hline   
		  $\mathcal{P}_2$, ALC  & $ 2.414   \leq \eta \le  2.732 $                                        & GC                          & $\frac{1}{3} \Big( \frac{1+3\eta}{1+\eta} \Big)^2$ \\ \hline
		   ALC &   $ 0.707  \le   \eta  \le2.414$    &  GC & $\frac{1}{3} \Big( \frac{1+3\eta}{1+\eta} \Big)^2$   \\ \hline
		  ALC &   $   0.67 < \eta \leq 0.707 $    &   $\mathcal{P}_2$ & $ \frac{2}{3} \Big( \frac{1+3\eta}{1+2\eta} \Big)^2$ \\ \hline
		   ALC$^o$ &   $   0.56 \leq \eta \leq 0.67 $    &   $\mathcal{P}_2$ & $ \frac{6}{(1+2\eta)^2}$ \\ \hline
		$\mathcal{P}_2$, ALC$^o$  & $0.5 < \eta \leq 0.56 $  &  $\mathcal{P}_2$   & $ \frac{6}{(1+2\eta)^2}$  \\ \hline
		 $\mathcal{P}_2^o$, ALC$^o$  & $0.414 \le \eta \leq 0.5$  & $\mathcal{P}_2^o$   &  $\frac{3}{2}$ \\ \hline 
		 $\mathcal{P}_2^o$, ALC$^o$  & $0.15 \leq \eta \leq 0.414$  & GC   &  $\frac{3}{(1+\eta)^2}$ \\ \hline 
		GC, $\mathcal{P}_2^o$, ALC$^o$           &      $0 < \eta \leq 0.15$                     &      GC      & $\frac{3}{(1+\eta)^2}$  \\   \hline   
	\end{tabular}
	\caption{NE-partitions, SO-partitions and \rev{PoA} for  $n=3$ 
	\label{tab:my-table1}}
	}
	\vspace{-3mm}
\end{table}

Important observations are: a) If GC is an NE-partition, all others are also NE-partitions; b) recall  ALC/ALC$^o$ is always an NE-partition; 
c)  the utilities  of all the players  at GC are bigger than  those at ${\mathcal P}_2$ or ALC, when GC is a NE-partition, thus GC is the preferred NE  in row 1 and 9 of Table \ref{tab:my-table1}; d) the  
utilities  of all  players  at ${\mathcal P}_2$   are  bigger than  those at ALC,   when  ${\mathcal P}_2$   is an NE, in such cases,    ${\mathcal P}_2$ is the preferred one, etc. 

\subsubsection*{When $n = 4$}
In this case, we can have four types of partitions: GC, ALC, $\mathcal{P}_2$ and $\mathcal{P}_3$ type partitions.  In any ${\mathcal P}_3$ type partition,  two of the C-players are together  in one coalition, while  the remaining two players  are alone. While ${\mathcal P}_2$ type partition can have either two players in each coalition or three players in one coalition and the remaining one is alone. We refer the first one  as TTC (partition with Two-Two  coalitions). Some of the strategy profiles and the corresponding partitions can be seen from Table \ref{tab:n=4}.

\begin{table}[h]
		\vspace{-2mm}
		\begin{center}
		\begin{tabular}{|c|c|c|c|c|c|}
			\hline
			$x_1$       & $x_2$       & $x_3$       & $x_4$       & $\mathcal{P}$                                                   \\ \hline
			
			\{1,2,3\}     & GC & GC & GC & \{\{1,2,3\},\{4\}\}  \\
			&             &            &             & \{\{1\},\{2,3,4\}\}         \\ \hline
			\{1,2\}     & GC & GC & GC & \{\{1,2\},\{3,4\}\}  \\
			&             &            &             & \{\{1\},\{2,3,4\}\}         \\ \hline
		\end{tabular}
		\caption{Partitions at $n=4$
		\label{tab:n=4}}
		\end{center}
		\end{table}
	Once again BRs are computed directly  and the results are  in Table \ref{tab:my-table}.	Important observations are: a) GC is never an NE-partition; 
b)  the utilities  of all the players  at TTC are bigger than  those at ALC, when TTC is an NE-partition, thus TTC is the preferred NE.  
The non-monotone phenomena  observed for the case with $n=2$ can also be seen  for $n=3,4$.
			\begin{table}[ht]
		
		\centering
		\begin{tabular}{|c|c|c|c|c|}
		\hline
		& $\mathcal{P}$ at NE                 & Range   & $\mathcal{P}$ at SO & \rev{PoA} \\ \hline
		1     &  TTC, ALC                               &$\eta \geq 2.414$ & GC & $ \frac{1}{4} \Big( \frac{1+4\eta}{1+\eta} \Big)^2 $\\ \hline
		2     &ALC                   & $0.75 < \eta\leq 2.414$   & GC&  $ \frac{1}{4} \Big( \frac{1+4\eta}{1+\eta} \Big)^2 $ \\ \hline
		3     &     ALC$^o$      & $0.707 \leq \eta \le 0.75$  &GC &  $\frac{4}{(1+\eta)^2}$\\       \hline
		4     &    ALC$^o$       & $0.56 \leq \eta \leq 0.707$  &$\mathcal{P} _2$  &$\frac{8}{(1+2\eta)^2}$  \\  \hline
		5    &       TTC, ALC$^o$           & $0.5<  \eta \leq 0.56$   &  $\mathcal{P} _2$  & $\frac{8}{(1+2\eta)^2}$   \\ \hline
		6 &TTC$^o$, ALC$^o$ & $0.414 \le \eta \leq 0.5$  &  $\mathcal{P} _2^o$ & 2\\ \hline
		7 & TTC$^o$, ALC$^o$  & $0 < \eta \leq 0.414$  &  GC& $\frac{4}{(1+\eta)^2}$\\ \hline
	\end{tabular}
	\caption{NE-partitions, SO-partitions and \rev{PoA} for  $n=4$
	\label{tab:my-table}}
\end{table}

\textbf{\rev{PoA} and  Observations}
The \rev{PoA} for smaller $n$ is computed in the Tables \ref{tab:my-table2}, \ref{tab:my-table1}, \ref{tab:my-table}, and the overall observations are:

\begin{enumerate}[label=(\roman*)]
    \item From Table \ref{tab_PoA_large}, as the number of players increases the  \rev{PoA} also increases, and, $\text{\rev{PoA}} = O(n)$ when $n \to \infty$;  
    \item For  any $n$ as the adamant player grows strong (as $\eta \to \infty $),  the $\text{\rev{PoA}} \uparrow  n$ (see Tables \ref{tab_PoA_large}, \ref{tab:my-table2}, \ref{tab:my-table1}, \ref{tab:my-table});   and
   \item  Similarly  when the adamant player becomes weak ($\eta \to 0 $),   the \rev{PoA} again increases to~$ n$. 
\end{enumerate}

\section{Symmetric  players without Adamant Player}
\label{no_adamant_player}
We now consider the same model as in  previous sections, but without adamant player. 
Majority of the analysis goes through as in previous cases, we will only mention the differences. The utility of any  partition
$\mathcal{P}^o = \{S_1,\cdots,S_k\}$ and that of the individual players, using Theorem \ref{thm:Thm1}   and Shapley value (equal shares)  simplify to:  		
\begin{eqnarray}
\label{Eqn_USm_without_adv}
\label{Eqn_Ui_without_adv}
U_{S_m}   =   \frac{1}{ | {\mathcal P}^o|^2   } \forall m,  \mbox{ and }   U_i  =   \frac{1}{ | {\mathcal P}^o|^2  |S_m| }  \mbox{ if } i \in S_m.
\end{eqnarray}
These utilities are exactly the  same as those in the previous model with insignificant   adamant player, except for GC.

The results for the case with $n >4$ are exactly the same because of the following:
i) Lemmas \ref{Lem_partiton_uni_dev} and \ref{Lemma_weak_NE} are independent of adamant player;
ii) Theorem \ref{Thm_No_MPs_for_grt_n}  is also applicable,  since  only  ${\underline x}_G := \{N_C,  \cdots, \ N_C\}$ leads to   $GC$, and that too\footnote{Since in the previous case (in the presence of adamant player), adamant player was always significant; but this is not true here. } ${\underline x}_G \to! GC$; and 
iii) The proof of Lemma \ref{Lemma_Partition_weak}  can easily be adapted.	

\subsubsection*{Smaller n}
 One can compute NE for all these cases as before, and the results are in Table \ref{tab_without_adv}.  
In a similar way the SO-partition is GC$^o$  (proof in  Appendix~\ref{sec_AppendixB_PEVA}):

\begin{lemma}
	\label{Lemma_SO_no_adv}
	GC$^o$ is  the  SO-partition in the absence of adamant player, for all $n$. 
\end{lemma}
\begin{table}[ht]
\begin{center}
		\begin{tabular}{|c|c|c|c|}
		\hline
		$n$  &  $\mathcal{P}^o$ at NE&  $\mathcal{P}^o$ at SO & \rev{PoA} \\ \hline
		2    &  GC$^o$, ALC$^o$ & GC$^o$ & 2 \\ \hline
		3    & GC$^o$, $\mathcal{P}_2^o$, ALC$^o$ &GC$^o$  & 3  \\ \hline
		4    &GC$^o$,  TTC$^o$, ALC$^o$ & GC$^o$ & 4 \\ \hline
		$>4$ & ALC$^o$ & GC$^o$ & $n$  \\ \hline
	\end{tabular}
	\caption{NE-Partitions and \rev{PoA} without Adamant Player}
	\label{tab_without_adv}
\vspace{4mm}
\end{center}
	\end{table}

\section{Asymmetric players: initial analysis}
\label{sec_asym_init}
In this section, we consider the  general case with $n$ players having possibly different influence factors and without the adamant player. 
The results of the  previous sections (with symmetric and adamant players)  will be useful in  deriving some  results of this  section. 
Without loss of generality, assume  $\lambda_1 \geq \lambda_2 \geq \cdots \geq \lambda_n$. 
We discuss the problem formulation and some initial results here, while the main analysis is considered in the next section. 
We first recall and discuss further details of the CFG  (coalition formation game) under consideration.

\subsection{CFG for asymmetric players}
We again use non-cooperative framework to study the CFG as in Section \ref{ACFG}. The definition of strategy of a player, strategy set, strategy profile and the rules for the formation of corresponding partition under strategy profile remains the same (see Section \ref{ACFG} for details). The next step is \textit{defining utilities of coalitions in the partition $\mathcal{P}=\{S_1,S_2,\cdots,S_k\}$ involved in a RSG.} As already explained, the players in a coalition act together and hence the utility of any coalition $S_m$ for $ 1 \leq m \leq k$ in $\mathcal{P}$ is given by:
 \begin{equation}
\label{Eqn_Util_coaltiion_asym}
\varphi_{S_m} ({\bf a}_m, {\bf a}_{-m} ) = \frac{\sum_{l \in S_m} \lambda_l a_l }{\sum_{l=1}^n \lambda_l a_l} - \gamma \sum_{l \in S_m} a_l; \;   \text{ where actions, }
\ {\bf a}_m := \{a_i, i \in S_m\}, \mbox{ and, } {\bf a}_{-m} := \{a_i, i \notin S_m\}. 
\end{equation}
Moving forward, for a given partition $\P$, a player is said to be an \textit{active player} if it has the highest influence factor ($\um^\P_i := \max_{j \in S_i} \lambda_j$) in its coalition $S_i \in \P$.  {\it Without loss of generality, the coalitions in  $\P$  are arranged in the decreasing order of the influence factors of the corresponding active players, i.e., such that, $\um^\P_1 \ge \um^\P_2 \ge \cdots \ge \um^\P_k$.}
To determine the utilities of these coalitions (under RSG) at NE (without adamant player), we present an extended version of Theorem \ref{thm:Thm1} (proof in Appendix~\ref{sec_AppendixA_PEVA}).  
\begin{theorem} \textbf{[Utilities of coalitions]}
	\label{Thm_util_coal_general}
	The game \eqref{Eqn_reduced_RSG} with utilities $\{\varphi_{S_m}\}$ as in \eqref{Eqn_Util_coaltiion_asym},  can have multiple NE, but the utilities at NE are unique. There exists a $M^\P \leq k$ such that only the coalitions in $\mathcal{J}^* = \{S_1,\cdots,S_{M^\P}\}$ get non-zero utilities. The unique  NE-utility for any $m \le k$ is  given by  (recall  $\um^\P_m = \max_{i \in S_m} \lambda_i$),
	 \begin{eqnarray}
		\label{Eqn_USm_asym}
		\varphi^*_{S_m}(\mathcal{P})  =  \Bigg( \frac{s^\P-\frac{M^\P-1}{\um^\P_m}}{s^\P} \Bigg)^2    \mathds{1}_{ S_m \in \mathcal{J^*} } 
	, \mbox{ with, }	 M^\P  :=    \max  \left \{ m \le k :   \sum_{i=1}^m  \frac{1}{\um^\P_i}  -  \frac{m-1}{\um^\P_m} > 0      \right \}, \nonumber \\
	\mbox{ and,  }   s^\P = \sum_{m=1}^{M^\P} \frac{1}{\um^\P_m}.    
		\end{eqnarray}
 Further the optimal actions of any non-active player is $0$, while that of the active players at any NE  satisfy:	
 \begin{eqnarray}
			{\bar a}^*_m(\mathcal{P}) &:=&  \sum_{j \in S_m: \lambda_j = \um_m^\P} a_j^*(\mathcal{P}) =    \frac{(M^\P-1)\Bigg( s^\P-\frac{M^\P-1}{\um_m^\P} \Bigg)}{\gamma \um_m^\P (s^\P)^2}\mathds{1}_{ s^\P > \frac{M^\P-1}{\um_m}}  \text{ for all } m.
			\label{Eqn_alstar_asym}  \hspace{5mm} 
			\end{eqnarray}
\end{theorem}

In the above, only the first $M^\P$   number of coalitions obtain non-zero utility, and, $\mathcal{J}^*$  precisely is this set of  coalitions; also  the utility of any coalition depends  only upon the influence factor of its best/active player (and of course on the environment outside given by $\P$).  Observe here that any insignificant coalition (i.e., the ones with zero utility) is similar to the insignificant adamant player defined in \eqref{Eqn_USm} of Section \ref{sec_util_coal}. Further, one may have multiple insignificant coalitions in this (asymmetric) case depending upon the level of asymmetry. 
Next, we   consider the division of this coalitional worth among its members,  $\{\phi_i^\P\}_{i \in \mathcal{N}}$ (we use superscript $\P$ to explicitly denote the dependency on partition $\P$). 

\subsubsection{   Shares of individuals in a given partition }
\label{sec_sol_concept_pfg}

Consider any partition $\P = \{S_1, \cdots, S_k\}$.
 To define the individual shares of each player ($j$) in coalition $S_i$ and for each $i \le k$, we use a modified version of Shapley value (SV) (see \cite{aumann1974cooperative}). 
 Upon extending the well known concept of  SV    to compute the shares within a given coalition (in any $\P$),  we have:
	\begin{equation}
		\label{Eqn_SV}
		\phi_j^\P = \sum_{C \subseteq S_i, j \notin C} \frac{|C|!(|S_i|-|C|-1)!}{|S_i|!} \left[\nu_{C \cup \{j\}}^\P - \nu_C^\P \right]    \text{ and } j \in S_i \mbox{ and any } i \le k,
	\end{equation}where $\nu_C^\P$ is the worth of a  sub-coalition $C \subset S_i$ under $\P$.  
Thus, 
 in order to obtain SV for  all  players under any partition $\P$,  we need to define the (corresponding) worth of all the sub-coalitions, i.e.,  $\nu_C^\P$ for all $C \subset S_i $ and $\text{ for all } 1 \leq i \leq k$.
 Towards this,  we assume the environment, i.e., the coalition structure outside $S_i$  remains   fixed (as in \cite{aumann1974cooperative}),  while the players in $S_i\backslash C = S_i- C $ can be arranged in various ways leading to multiple partitions of  possibly different sizes. One can use \eqref{Eqn_USm_asym} to determine  the  utility  of $C$ at RSG-NE corresponding to each of these multiple partitions. The worth of  sub-coalition $C$ is defined as the minimum amongst the NE-utilities  of $C$ derived under these  multiple partitions  (see \cite{aumann1974cooperative}, \cite{bloch}). 
  Our immediate observation is  that the worth of any sub-coalition ($\nu_C^\P$) equals the  (RSG) NE-utility of the coalition $C$,  under `maximum-possible'   partition $  \Q^\P_C$  defined as below:
\begin{equation}
	\label{Eqn_partition_min_gen}
	\Q^\P_C = \Big  \{C,\underbrace{\{\{l\}\}_{l \in S_i \backslash C},}_{\text{Arranged as singletons}}  \underbrace{S_1,\cdots,S_{i-1},S_{i+1},\cdots, S_k}_{\text{Environment fixed} }\Big \}.
\end{equation}
Basically in $\Q^\P_C$, the environment is fixed  and  other players of $S_i-C$ are all alone (proof in Appendix~\ref{sec_appendixC_PEVA}). 
 \begin{lemma}
  	 Consider any partition $\P = \{S_1,\cdots,S_k\}$.
  		Then the worth of any sub-coalition,  $\nu_C^\P$ with $C \subset S_i$ for any  $ i \leq k$, is given by its value under the  partition, $\Q^\P_C$ (partition with maximum cardinality  as defined in  \eqref{Eqn_partition_min_gen}). Thus
  		$ 
  		\nu_C^\P  =  \varphi^*_C ({\Q^\P_C}), \mbox{ with }  \varphi^*_C (\cdot) \mbox{ as in \eqref{Eqn_USm_asym}. }
  		$
  	\label{min_worth_sub_coal_general}
  \end{lemma}
  
  \subsubsection{ Solution Concepts}

In the case with symmetric players (with $n > 4$), we observed that an NE does not lead to multiple partitions (see Theorem \ref{Thm_No_MPs_for_grt_n}). But this may not be the case with general players. 
Thus we include  a slightly modified  solution concept to study the general case.  

(i) \textit{U-stable partitions: }
For any given partition $\P$, let $\underbar{x}^\P = (x^\P_1, \cdots, x^\P_n)$ with $x^\P_j =  S_i$ for any $j \in S_i \in \P$, represent the natural strategy profile that uniquely leads to $\P$ ($\underbar{x}^\P \to !\P$); basically every player, under this natural strategy profile, precisely proposes to collaborate with all the players of the coalition to which it  belongs in $\P$.   
 A partition $\P$ is said to be a \underline{U-stable partition if  the corresponding natural  strategy profile $\underbar{x}^\P$ is a Nash Equilibrium.} Basically, \underline{there should be no strategy profile} ${\underbar x}'$ resulting from a unilateral deviation of one of the players (say player $j$) from $\underbar{x}^\P$, such that $j$ gets strictly better  utility, 
i.e., such that, 
\begin{equation}
	\label{Eqn_Ustable}
U_j (\underbar{x}^\P) < U_j (\underbar{x}' )  \text{ for any } j  \mbox{ and  after any $j$-unilateral deviation } \underbar{x}'.
\end{equation}

(ii) \textit{NE-partitions:} Recall that a  partition $\P$ is a NE-partition   if    $\underbar{x} \to \P$, where $\underbar{x}$ is an NE.

Observe that a partition  is a NE-partition if it is a U-stable partition or  
if it is  among the   partitions  (can also be multiple) resulting from a NE-strategy profile $\underbar{x}$.  
One can easily observe that {\it all U-stable partitions are NE-partitions but the vice-versa may not be true.} Our main focus will be on U-stable partitions since a partition which is not U-stable, may still qualify as a NE-partition, if it is one amongst the multiple partitions resulting from an NE. However, if it is formed it might still  not be stable, because, by \eqref{Eqn_Ustable} one among the players can deviate unilaterally to do strictly better.

We derive the analysis for the general case, which we also refer to as the case with asymmetric players, in the following steps. First we consider a case with `minimal number' of asymmetric players, i.e., case in which one player is different from all other symmetric players.  This case will provide some required insights into the results for completely general case which is considered in Section \ref{sec_full_asym}.

\subsection{One Asymmetric player}
\label{sec_one_asym}
We begin with the case where we have one asymmetric player with influence factor $\beta\lambda$ (with $\beta > 1$) and $n$ symmetric players with influence factor $\lambda$, henceforth referred to as A-player and S-players respectively. Thus, $N_C = \{
\beta \lambda, 1, \cdots, n\}$.

 We begin with few definitions some of which are specific to this sub-section. Any partition  in this special case  is of the form $\P = \{ C_\beta, \ S_1, \cdots, S_k\}$: the coalition containing A-player  $\Cb = \{\beta \lambda, 1, \cdots, \kb\}$, 
$\kb$ the number of S-players in $\Cb$ coalition and $k$ the number of coalitions containing only S-players.  We  will refer {\it them as $(\kb, k)$ partitions;}  we will soon see that the remaining details of the partition become irrelevant.  

We \underline{refer a $(\kb, k)$ partition as SS-$(\kb, k)$ partition,  when all coalitions ($k$ of them) other  } \\
\underline{ than $\Cb$ are SingletonS}, i.e., contain exactly one $S$-player.   In other words,
$\P = \{ \Cb, \{\kb +1\}, \cdots, \{n\}\}$ is an SS-$(\kb, n-\kb)$ partition.

Consider any $(\kb, k)$-partition. By symmetry and Theorem \ref{Thm_util_coal_general}, the SV of any player in any coalition $j \in S_i$ equals:
$$
\phi_j^\P = 
 \frac{ \varphi^*_{S_i} (\P) } {  |S_i| }  =  \frac{ 1 } {  |S_i| }   \left( \frac{1}{1+k\beta} \right)^2.$$
 It remains to derive the SVs for players of $\Cb$.  Recall 
 from Lemma \ref{min_worth_sub_coal_general}, the worth of any sub-coalition ($\nu_C^\P$) equals the NE-utility of the coalition $C$, in an appropriate   RSG played among the coalitions of (maximal) partition, $  \Q^\P_C$,   defined in \eqref{Eqn_partition_min_gen} and the worth of these sub-coalitions  are  instrumental in computing the required SVs (see \eqref{Eqn_SV}). 
Using Lemma \ref{min_worth_sub_coal_general} and equation \eqref{Eqn_SV}, the SVs of the players  in $\Cb$ coalition  are given by (proof in Appendix~\ref{sec_appendixC_PEVA}), 
 \begin{lemma}
The SV of A-player and any S-player of $\Cb$, denoted  respectively by $\phi_\beta^\P$ and $\phi_\lambda^\P$ is given by:

\vspace{-4mm}

{\small \begin{eqnarray}
	\phi_\beta^\P
	&=& \left \{   \begin{array}{llll}
	\frac{1}{\kb+1} \left[ \sum_{l=0}^{\kb-1}\frac{{(\kb-l+k)} (1-\beta) \left[ (\kb-l+k-2) -\beta(\kb-l+k) \right]}{\left( {1+(\kb-l+k)\beta} \right)^2} + \left( \frac{1-k+k\beta}{1+k\beta} \right)^2 \right]  & \text{ for } \kb \geq 1,   \\
		\left( \frac{1-k+k\beta}{1+k\beta} \right)^2 &  \mbox{ for }  \kb=0 \text{ and, } \\
		\label{Eqn_SV_A_player}
	\end{array} \right  .
\\	\phi_\lambda^\P & = &
		\frac{1}{\kb}\left[\left( \frac{1-k+k\beta}{1+k\beta} \right)^2 - \phi_\beta^\P \right] \mathds{1}_{\kb \geq 1}. \hspace{20mm} 
 \label{Eqn_SV_S_player} 
\end{eqnarray}}

\label{Lem_SV_A_player}
\end{lemma}

Our aim is to derive U-stable partitions (special NE-partitions)  and one can again observe that  ALC, i.e., each player alone, is always a U-stable partition and hence, an NE-partition.
In this direction, our first result presents all the possible U-stable partitions under this case (see Appendix~\ref{sec_appendixC_PEVA} for proof). 

\begin{theorem}
	Consider $n > 5$ and $\beta >1$.  A partition is \underline{not U-stable} if it is not an SS partition,    that is, if  any S-only coalition (i.e., any $S_j$)  has more than one player.   
	The  SS-$(\kb, k)$   is a U-stable partition if and only if 
		\begin{eqnarray}
			\phi_\beta^\P-\left( \frac{-k+(k+1)\beta}{1+(k+1)\beta} \right)^2 \geq 0 \quad  \text{ and } \quad \phi_\lambda^\P - \left( \frac{1}{1+(k+1)\beta} \right)^2 \geq 0,
			\label{Eqn_NE_partition_single}
			\end{eqnarray}
		where, SVs $\phi_\beta^\P$ and $\phi_\lambda^\P$ 
		are defined in \eqref{Eqn_SV_A_player} and \eqref{Eqn_SV_S_player} of Lemma \ref{Lem_SV_A_player}.
	\label{Thm_stable_big_asym}  
\end{theorem}

Thus once again   non-SS/non-Singleton partitions (with  two or more S-players  together in a coalition without A-player)    are not U-stable. As one may guess, this proof   follows 
from the results of the previous sections. However, more interestingly, 
  \textit{with  addition of just one asymmetric player, the number of stable partitions can increase from one} (from Table \ref{tab_without_adv} related to symmetric players, only ALC is the stable partition); some of the SS-partitions can become stable depending upon $\beta$.  Observe SS-partitions include (exactly) one coalition $\Cb$ with more than one C-player and are different from ALC. In fact, we showed the existence of a   threshold ${\bar \beta}_n$ such that SS-(1,(n-1)) partitions are stable for all $\beta \ge {\bar \beta}_n$. Similarly one can show the existence of $n_\beta$ such that  GC is U-stable if and only if $ n \le n_\beta$. One can also derive the results for the case with $\beta < 1$. We omitted these results due to lack of space.  We instead move on to the general case  with an aim to  understand if  `stability' can increase/decrease in some manner as the `asymmetry' increases. One first needs to understand what different levels of `stability' and `asymmetry'  mean and then the  two notions have  to be connected.   This is precisely the agenda of the next section, that provides real insights into the general case. 
  
  \section{Asymmetric players: Stability  Analysis}   
\label{sec_full_asym}


In Section \ref{sec_sym_large} on symmetric players, we observed that ALC is the only stable partition with $n > 4$. With a `small' introduction of asymmetry, in Section \ref{sec_one_asym}, we found that many more partitions are stable (U-stable).  We now consider a general case and investigate similar questions.  We will observe that as certain `measures of asymmetry' increase, more and more  partitions become U-stable and we would eventually have `absolute stability'.  

\subsection{Absolute Stability }
\label{sec_abs_stable}
 We say that the system  is \underline{absolutely stable, if each and every partition is stable against unilateral } \underline{deviations}, in the sense all the partitions are U-stable. 
 Our first interesting result is that there indeed exist conditions under which the system is absolutely stable. We  begin with some  assumptions 
(recall  $\lambda_1 \geq \lambda_2 \geq \cdots \geq \lambda_n$): 

{\bf A}.1 Assume,   \vspace{-3mm}
\begin{equation}
\label{Eqn_A.1}
 w_1 \le   \min_{ j \ge 2} \left( w_{j+1} - w_{j} \right) \text{ where } w_j := \frac{1}{\lambda_j}.
\end{equation}
   Under this assumption,  only the first two `big' coalitions derive non-zero utility  (basically $M^\P = 2$ in Theorem \ref{Thm_util_coal_general}) and this facilitates in the following simplification 
of the SVs for any partition (proof in Appendix~\ref{sec_AppendixD_PEVA}):
%
\begin{lemma}
Let $\lambda_1/(\lambda_1+\lambda_j) = \varrho_j$ for any $j \geq 2$ and set $\varrho_{n+1}=1$. Under {\bf A}.1,
\begin{enumerate} [label=(\roman*)]
\item Consider any partition $\P = \{S_1, S_2, \cdots, S_k\}$ with  $1 \in S_1$ and $2 \in S_2$. Then the SV of any player $i$ is given by:
\begin{eqnarray*}
\phi_1^\P = \varrho_2^2 ,  \ \ \phi_2^\P  = (1-\varrho_2)^2   \mbox{ and } \phi_i^\P  = 0   \mbox{ for any }  i >2 .
\end{eqnarray*}
\item Consider a partition with $\{1,2,\cdots,l\} \subseteq S_1$ and $l+1 \in S_2$, where $2 \leq l \leq n$ (here, $l=n$ implies GC). Then the SVs   are: 
\begin{eqnarray}
	\phi_i^\P = 
	\left \{   \begin{array}{llll}
		  \sum_{j=3}^{l} \frac{1}{j(j-1)} \left( \varrho_j^2-(1-\varrho_2)^2 \right) + \frac{1}{l}  \left( \varrho_{l+1}^2-(1-\varrho_2)^2 \right)+  \frac{1}{2} \varrho_2^2 & \text{ for } i=1, \nonumber \\
		\sum_{j=3}^{l} \frac{1}{j(j-1)} \left( \varrho_j^2- \varrho_2^2 \right) + \frac{1}{l}  \left( \varrho_{l+1}^2-\varrho_2^2 \right) +\frac{1}{2} \left(1-\varrho_2 \right)^2  & \text{ for } i=2, \nonumber   \\
	\sum_{m=1}^{l-i} \frac{1}{(i+m)(i+m-1)}  \left( \varrho_{i+m}^2-\varrho_i^2\right)+ \frac{1}{l} \left( \varrho_{l+1}^2-\varrho_i^2\right) & \text{ for } 3 \le i <l, \\
	\frac{1}{l} \left( \varrho_{l+1}^2-\varrho_l^2\right)  & \text{ for } i=l, \nonumber  \\
	\left(1- \varrho_{l+1}\right)^2  & \text{ for } i=l+1 \text{ and } i \le n,  \nonumber \\
	 0  & \text{ for } l+1<i<n.  \nonumber  \hspace{10mm}  
	\end{array} \right  .  	
	\label{Eqn_SV_Gen_A1}
\end{eqnarray}
\end{enumerate} 
\label{Lem_SV_general}
\end{lemma}
 We now identify the conditions for absolute stability of the system under {\bf A}.1 (with proof in Appendix~\ref{sec_AppendixD_PEVA}):


{\bf A}.2 For every $j \ge 2$, with  $c_j :=  1_{j = 2} + j 1_{j > 2}$,  assume (recall $\varrho_j =\lambda_1/(\lambda_1+\lambda_j) \text{ and } \varrho_{n+1}=1$), 

\vspace{-4mm}

{\small \begin{eqnarray}
 \left [ \varrho_{j+1}^2 - \varrho_{j}^2\right ]  &\geq & c_j  \left(1- \varrho_j \right)^2 \text{, that is assume, }  \ai \geq 1 \mbox{ where } \ai := \min_{2 \le j \le n}   \frac{ \varrho_{j+1}^2 - \varrho_j^2} { c_j (1-\varrho_j)^2},   \label{Eqn_Measure_Asym}
\end{eqnarray}}
%
\begin{theorem}{\bf [Absolute Stability]} Assume {\bf A}.1. Then the system is absolutely stable if and only if {\bf A}.2 is satisfied.    
\label{Thm_general}
\end{theorem}

{\bf Measure of Asymmetry (MoA):} We refer the index $\ai$ defined in  \eqref{Eqn_Measure_Asym} of assumption {\bf A}.2, as first measure of asymmetry (MoA).  The above theorem shows that the system is absolutely stable (i.e., all the partitions are U-stable) once MoA is greater than or equal to one (under {\bf A}.1). It also shows that some partitions are not stable once MoA is less than one.  We will see that, in general (even without assumption  {\bf A}.1), the  `level of stability', when  measured in terms of the number of U-stable partitions, depends upon  MoA and the same aspect  is studied in Section \ref{sec_numerical} on numerical analysis. 

By Corollary \ref{corollary_ALC_NE}, when players have equal influence factors  and when $n > 4$,  ALC is the only stable partition. 
This fact can easily be extended by simple continuity arguments to the case where players are of almost similar influence factors:
for example, one can find a neighbourhood  ${\mathcal B}:=\{ \lambda: |\lambda - \lambda_1| \le \delta\}$ of $\lambda_1$  (with $\delta$ depending upon $\lambda_1, n$) such that  ALC is the only stable partition, if  $\lambda_j \in {\mathcal B}$ for all $j$. 
 On the other extreme, when the agents are completely different from each other (the influence factor of player $j$ is sufficiently bigger than that of $(j+1)$ for each $j$,  e.g., as under {\bf A}.1 and with   MoA  greater than one), all the partitions are stable.  Absolute stability is possible even when two or more players are of similar influence with the rest of them being completely asymmetric.   
We now have another theorem (proof in Appendix~\ref{sec_AppendixD_PEVA}) which provides the conditions for absolute stability when some of the agents, say agents 3 and 4, have equal influence factors under the following modified assumptions:
{\bf A}.1$'$ Assume, $w_1 \le   \min_{ j \ge 2, j \ne 3} \left( w_{j+1} - w_{j} \right),$ $w_3=w_4$ (recall $w_j = 1/\lambda_j$).

{\bf A}.2$'$
We require same condition as in {\bf A}.2 for all $j > 4$, and for  others we need the following modified assumptions:
\begin{eqnarray}
\left ( \frac{2\lambda_1-\lambda_3}{2\lambda_1+\lambda_3} \right)^2  \geq    \varrho_2^2 + \left( 1- \varrho_2 \right)^2 \text{ for $j=2$, }  \hspace{10mm} \nonumber  \\
\left [ \varrho_4^2 - \left( \frac{2\lambda_1-\lambda_3}{2\lambda_1+\lambda_3} \right)^2 \right] \  \geq  \  c_j \left( \frac{\lambda_3}{2\lambda_1+\lambda_3} \right) ^2 \text{ for $j=3$, and,  }  
 \nonumber \\
  \left[ \varrho_5^2 - \varrho_4^2 \right] + \frac{1}{3} \left [  \varrho_4^2 - \left( \frac{2\lambda_1-\lambda_3}{2\lambda_1+\lambda_3} \right)^2 \right] \  \geq \   c_j  \left( 1-\varrho_3 \right) ^2 \text{ for $j=4$.  } \nonumber
\end{eqnarray}

\begin{theorem}{\bf [Absolute Stability]} Assume {\bf A}.1$'$. Then the system is absolutely stable if and only if {\bf A}.2$'$ is satisfied.  
\label{Thm_general_sym}  
\end{theorem}

  {\bf Remarks:}  By Theorems \ref{Thm_general}-\ref{Thm_general_sym}, we  have a surprising result:  there are conditions under which the system is absolutely stable. One may derive similar  conditions for absolute stability when other subsets of players are equal. There are several remarks in place in this regard. a) We observe  that more often only stronger players have significant contributions; b) the weaker players have non-zero contributions only when the stronger ones are together (e.g., when 1, 2 are in one coalition as given by Lemma \ref{Lem_SV_general}.(b)), but they can not impose the stronger ones to be together (using any unilateral deviation);  c) the stronger players can get better utilities at partitions in which they are together with other strong players, however a unilateral deviation by one strong player cannot result in bigger coalition; and d)   coalitions with bigger sizes  can result only when more players simultaneously propose bigger set of choices.

These observations also lead to a set of new questions, that of stability against `coalitional deviations';  we touch upon this topic briefly towards the end of this work and a more detailed analysis would be a part of future work.  
For now, we continue with the analysis under unilateral deviations.  

We next consider the case when the system is not very  `far away'  from symmetric case. We also study the stability patterns as the system becomes more and more `asymmetric',  measured using MoA \eqref{Eqn_Measure_Asym} and a second measure introduced in the next sub-section. 

\subsection{ With moderate asymmetry:}
\label{sec_moderate_asym}
Another important condition (apart from MoA) for absolute stability is either {\bf A}.1 or {\bf A}.1$'$. These conditions imply, 
\begin{equation}
\Delta_w : = \sum_{j \ge 1}^{n-1}  \left( w_{j+1} - w_j \right)  >  (n_d-1) w_1, \text{ where }  n_d \text{ is the number of distinct players}.
\label{Eqn_def_w_delta}
\end{equation}
Observe that $\Delta_w$ equals 0 for symmetric players, and would increase as the agents become more and more distinct. Thus {\it one can view $\Delta_w$ as  the second measure of asymmetry} and as seen above when this measure is high enough,  we have  absolute stability.  We now consider the  system with smaller  $\Delta_w$. As already mentioned, when $\Delta_w$ is close to 0 (e.g., all  $\lambda_i \in \mathcal{B}$, defined in Section \ref{sec_abs_stable}), ALC is the only stable partition.
But more   partitions become stable as one or some of the players have  very different influence factors (as we will soon see). This aspect is already observed in  the case  with one asymmetric player (as $\beta$  increases more SS-partitions become stable).  

In a given system, we say a \underline{player can form successful coalitions (FSC)}, if there exists a U-stable partition in which the player is not alone (i.e., atleast one more player belongs to its coalition). 
It would be interesting to find the players that can form successful coalitions, possibly depending on their strengths. As in Section \ref{sec_one_asym}, we call a partition $\P_C$ to be \underline{SS$(C)$ if  players in $N_C-C$ are alone}, i.e., if $\P_C = \{C,\{\{l\}\}_{l \notin C}\}$, i.e., if all coalitions other than $C$ are singletons. We immediately    have the following (see \eqref{Eqn_def_w_delta} for definitions and proof is in Appendix~\ref{sec_AppendixD_PEVA}):
\begin{theorem}
\label{Thm_monotonicity}{\bf [Bigger player, higher chance to FSC]} Assume $w_1 \ge  {\bar w}/(2 (n-2) )$, which implies $\Delta_w < w_1$.  
 Then, partition $\text{SS}(\{j-1,k\})$  is stable for any $j <k$, if partition $ \text{SS}(\{j,k\})$ is stable. 
\end{theorem}

{\bf Remarks:} The most important observation of the above result is that when a   player is able to form coalition with a weaker player (under stable partition and when the latter is weaker than the former), then a more influential  player (than the former) will also form coalition with the same weaker player; and this is possible when the rest of the players are arranged as singletons. This implies that the most influential player would be the {\it first one to start forming successful coalitions.}   

 Similarly, the next result shows that the least influential player (among significant players) would also be the first one to start forming successful coalitions among weaker players (proof in Appendix~\ref{sec_AppendixD_PEVA}): 
\begin{theorem}
\label{Thm_monotonicity_new}{\bf [Smaller player, higher chance to FSC]} 
The partition $\text{SS}(\{j,k+1\})$  is stable for any $j <k$, if partition $ \text{SS}(\{j,k\})$ is stable. 
 \end{theorem}
 \textbf{Remarks:} Theorems \ref{Thm_monotonicity} and \ref{Thm_monotonicity_new} suggest that the first successful coalition formed (as asymmetry increases) would consist of the strongest and the weakest players.
  The numerical examples of Section \ref{sec_numerical} reinforce this observation.

To explain the above concept more precisely, we consider a sequence of systems with `increasing asymmetry' and compare their stable partitions. 
Towards this, we begin with a system of almost symmetric players (such that all players are significant);  
 we  then consider a sequence of  systems where the influence factor of the strongest and the weakest player is increased and decreased simultaneously  by   same amounts (while keeping others influence factors' the same).  We consider all such resultant  systems, in  which  no player becomes insignificant. 
  Then, in the following we show, there always exists a threshold for 'minimum asymmetry' between players such that they find it beneficial to form coalitions (see Appendix~\ref{sec_AppendixD_PEVA} for proof). 

\begin{theorem}{\bf[Towards stable partitions]}
\label{Thm_Sym_to_Asym}
Start with a system of  $n$ players  with  $n > 4$, such that  partition $\text{SS}(\{1, n\})$  is not  U-stable.  Say  player 1 becomes more influential and player $n$ becomes less influential,  while maintaining ${\bar w}:= \sum_{j =2}^{n-1}  w_j + w'_n+ w'_1 $ constant, i.e., say $w'_1 = w_1 - \delta$  and $w'_n = w_n + \delta$ for some $\delta >0$ in the modified system. Then there exists a threshold  ${\bar \delta} $ 
such that $\text{SS}({\{1,n\}}) $ is U-stable  whenever $\delta > \bar{\delta}$ (with $n$-th player still being significant when $ \delta  <  {\bar w} / (n-1)-w_n $). 
\end{theorem} 
 {\bf Remarks:}  Through this choice of systems (one for each $\delta$) we established that 
 the highest and the least influence players  will start forming successful coalitions, as the asymmetry in the system  increases  (note $\Delta_w$ increases with $\delta$). 
 
 Next we consider a system with players of (possibly) different influence factors, with at least two symmetric players having maximum influence factor. The next result shows that any partition of the players with symmetric players together cannot be a U-stable partition under certain conditions given below (proof in Appendix~\ref{sec_AppendixD_PEVA}).
 
 \begin{theorem}
 \label{Thm_no_identical}
 Consider the case in which some of the top (at least two of them) players are identical, i.e., say $\lambda := \lambda_1 = \lambda_2$. Then, any partition $\mathcal{P}$ with at least three coalitions (i.e., $k \geq 3$) obtaining strictly positive utility (as given by Theorem \ref{Thm_util_coal_general}), cannot be a U-stable partition if at least two of the top identical players are together and the following is \underline{not satisfied} (recall  $\um^\P_m = \max_{i \in S_m} \lambda_i$ and $\um^\P_k \leq \um^\P_m$ for all $m$):
 \begin{equation}
 \frac{\sum_{l=1}^{k-1}\frac{\lambda}{\um^\P_l}}{k-2} > \frac{\lambda}{\um^\P_k} > \frac{\sum_{l=1}^{k-1}\frac{\lambda}{\um^\P_l}+1}{k-1}. 
 \label{Eqn_Counter}
 \end{equation} 
 \end{theorem}
\textbf{Remarks:} (i) We would first like to explain the relevance of equation \eqref{Eqn_Counter}. Observe that the partition is not U-stable (and so strong identical players do not find it beneficial to collaborate) when $\um^\P_k$ is not in the range specified by \eqref{Eqn_Counter}.

(ii) When the first $(k-1)$ active players are identical or near identical it is easy to verify that this range (approximately equals $\left (\frac{k-1}{k-2},\frac{k}{k-1} \right )$) 
  is small (more so with bigger $k$). Thus strong identical players may collaborate only when the $k$-th active player is in this small range.

(iii) The range in \eqref{Eqn_Counter} can become significant when the players are very different from each other (i.e., when $\lambda/\um^\P_m \gg 1$ for some $m$) and one may find cases where identical players prefer to collaborate. 

(iv) More so, it is not guaranteed that the identical players collaborate when $\um^\P_k$ is in the range given by \eqref{Eqn_Counter}; the arguments are too tedious and hence we skip them, but the actual range of $\um^\P_k$ for which identical players may find it beneficial to collaborate can be smaller than that shown by \eqref{Eqn_Counter}. Thus in all, \textit{strong identical players may collaborate (if at all) only in the presence of weaker players when the latter are not too weak to make a difference to the game}. In fact it is affirmed by Theorems \ref{Thm_monotonicity} and \ref{Thm_monotonicity_new}, the strongest player prefers to form coalition with the weakest player.

We now consider the SO-partitions and discuss the \rev{PoA}. 
\subsection{Price of Anarchy and SO-partition}
\label{POA_general}
Recall that the Price of Anarchy (\rev{PoA}) is defined as the ratio between the sum utilities at  `social optima' and the sum utilities at the  `worst Nash Equilibrium'. 
\begin{eqnarray*}
	\text{\rev{PoA}}  &\hspace{-1mm}=\hspace{-1mm}& \frac{\max_{\mathcal{P}} \sum_{S_i \in \mathcal{P}} U_{S_i}  }{ \min_{\mathcal{P}^*} \sum_{S_i \in \mathcal{P}^*} U_{S_i}  } = 
	\frac{  U_{SO}^*  }{ U_{NE}^*  }, \text{ where $\mathcal{P}^*$ is any NE-partition.}
\end{eqnarray*}
%
It is easy to see that GC is the SO-partition as the players obtain the maximum possible sum utility, which equals one.  
We show in the following that  $U_{NE}^* $ is achieved at  ALC and the  \rev{PoA} is given by the following (proof in Appendix~\ref{sec_AppendixD_PEVA}):
\begin{lemma}
\label{Lem_worst_NE_util}
For the general case, the \rev{PoA}  equals inverse of the sum utility at $ALC$, 

 \begin{eqnarray}
&& \text{\rev{PoA}}  =  \frac{1}{U^{A}}  \mbox{ with } U^{A} :=  \sum_{j=1}^{M^{A}} \left (   \frac{{\bar w}^A - (M^A-1) w_j}{{\bar w}^A} \right )^2, 
M^A := \sup  \left \{k:  \sum_{i=1}^k w_i - (k-1) w_k  > 0 \right \}, \nonumber \\
&& \hspace{110mm} {\bar w}^A := \sum_{j=1}^{M^A} w_j. \hspace{2mm}  
\end{eqnarray}  
\end{lemma}
Using simple algebra (from definition  of $M^A$) one can show that (as $w_j \ge w_1  $) 
\begin{eqnarray}
&& \hspace{-10mm} U^A \le M^A  \left (  1  - \frac{(M^A-1)  w_1 }{ {\bar w}^A}   \right )^2 = \frac{1}{M^A}  \left (    \frac{ {\bar w}^A  -  (M^A-1) w_1 }{  \frac{{\bar w}^A}{M^A}  }   \right )^2 \mbox{, and thus, }
\text{\rev{PoA}}  \ge  \frac{ M^A  } { \Omega^A }, \nonumber \\  
&& \hspace{80mm} \Omega^A :=  \left (    \frac{ {\bar w}^A  -  (M^A-1) w_1 }{  \frac{{\bar w}^A}{M^A}  }   \right )^2.    
\end{eqnarray}
Thus \rev{PoA} can  again increase with  the size of partition, however,  the growth rate is governed by  the number of significant players ($M^A$). 
Further it also depends upon relative   influence factors of the players;  the growth rate is more if the players are almost similar,   for symmetric case clearly, $\Omega^A = 1$.  

\subsection {Numerical results}
\label{sec_numerical}
In this section, we present the algorithm which can be used to obtain U-stable partitions for a  given set of influence factors. 
The algorithm is significantly simplified thanks to the observation made in footnote \ref{Foot_note1}.
We use this algorithm to validate our theoretical results and obtain insights when assumption {\bf A}.1/{\bf A}.1$'$ is not true.

\textbf{Algorithm}
\begin{enumerate}
\item \textbf{Input:} An array of $n$ influence factors and all possible partitions.
\item \textbf{For} each partition, do the following:
\begin{enumerate}[label=(\roman*)]
	\item Find the active player in each coalition of the partition and   the significant coalitions (using Theorem \ref{Thm_util_coal_general}).
	\item \textbf{For} each significant coalition do the following:
\begin{enumerate}[label=(\alph*)]
\item Generate all possible sub-coalitions of this coalition and calculate their worth (using Lemma \ref{min_worth_sub_coal_general}).
		\item \textbf{For} each player $j$ in the coalition, do the following:
		
		\begin{itemize}

		\item Use \eqref{Eqn_SV}  to obtain SV  of player $j$.
		\item Calculate the SV   after unilateral deviation to being alone\footnote{\label{Foot_note1}
		{\bf Comparison with unilateral deviation to being alone is sufficient:} Recall the definition of U-stable partition. When a player $j$ deviates unilaterally to any  $x'_j \subset S_m$ (say $j \in S_m$), then  the resulting strategy profile leads to two partitions, 
$$
\P_1 = \left \{ \{j\}, S_m - \{j\},  S_1, \cdots, S_{m-1} ,  S_{m+1},  \cdots, S_k \right \} \mbox{ and } \P_2 = \left \{  x'_j, S_m - x'_j,  S_1, \cdots, S_{m-1} ,  S_{m+1},  \cdots, S_k \right \}.
$$
	If the player obtains better after unilateral deviation to $x'_j = \{j\}$ (which  now leads to unique partition $\P_1$ by Lemma \ref{Lem_partiton_uni_dev}), then original  $\P$ is not stable.  If it obtains lesser, then
from \eqref{min_util_mult_partition}
 it obtains lesser with any other unilateral deviation $x'_j$, as $\P_1$ is always one of the partitions that result after the deviation.
		}, which leads to a unique partition again.
		\item  If  SV    is less than the  utility  after unilateral deviation,   declare the partition unstable and go to next partition. 
		\end{itemize}
\end{enumerate}
 
\end{enumerate}
\item \textbf{Output:} Stability/Instability of each partition.
\end{enumerate}

\begin{figure}[!h]
\begin{center}
\begin{minipage}{7.5cm}
\vspace{6mm}
\begin{center}
\includegraphics[trim = {1cm 7cm 0cm 0cm}, clip, scale = 0.4]{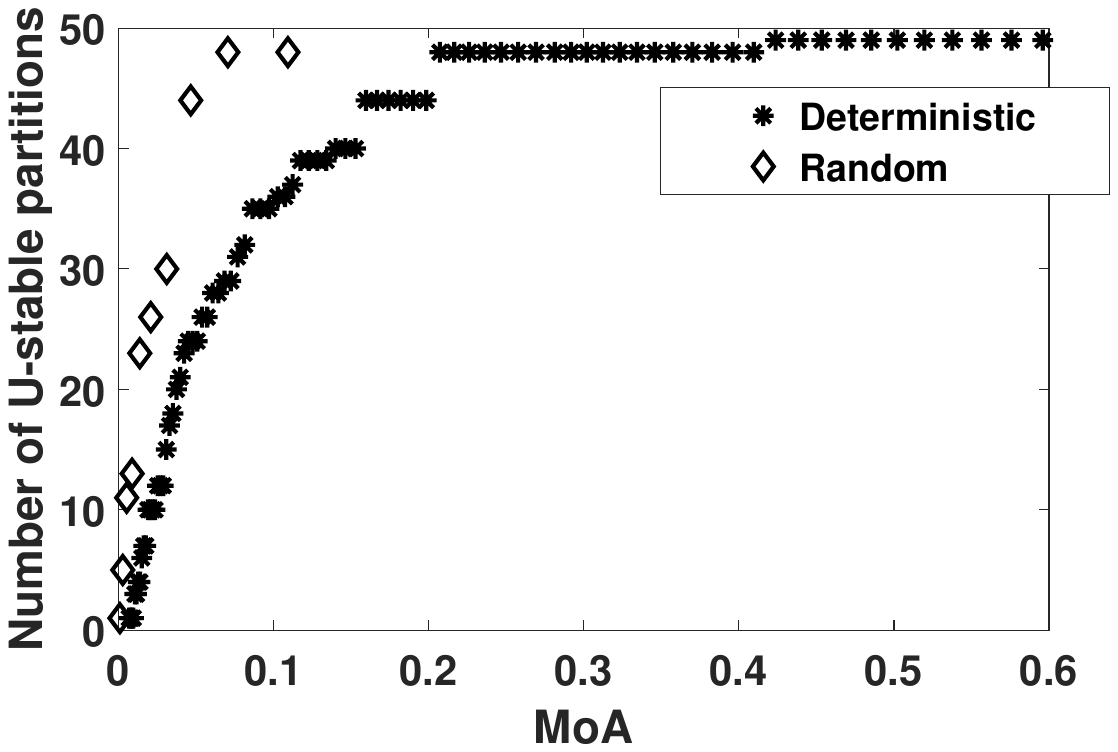}
\caption{ \textbf{Random:} $\alpha_j \sim U(0,1)$ and $\delta \in [1,20 ]$ \\ \textbf{Deterministic:} $\alpha$ = [0, 10, 12.5, 17.3, 21.5] and $\delta \in [0.1,0.93 ]$ 
   }
\label{MoA_vs_stable}
\end{center}
\end{minipage}
\begin{minipage}{7.5cm}
\vspace{-6mm}
\begin{center}
\includegraphics[trim = {0cm 5cm 0cm 0cm}, clip, scale = 0.4]{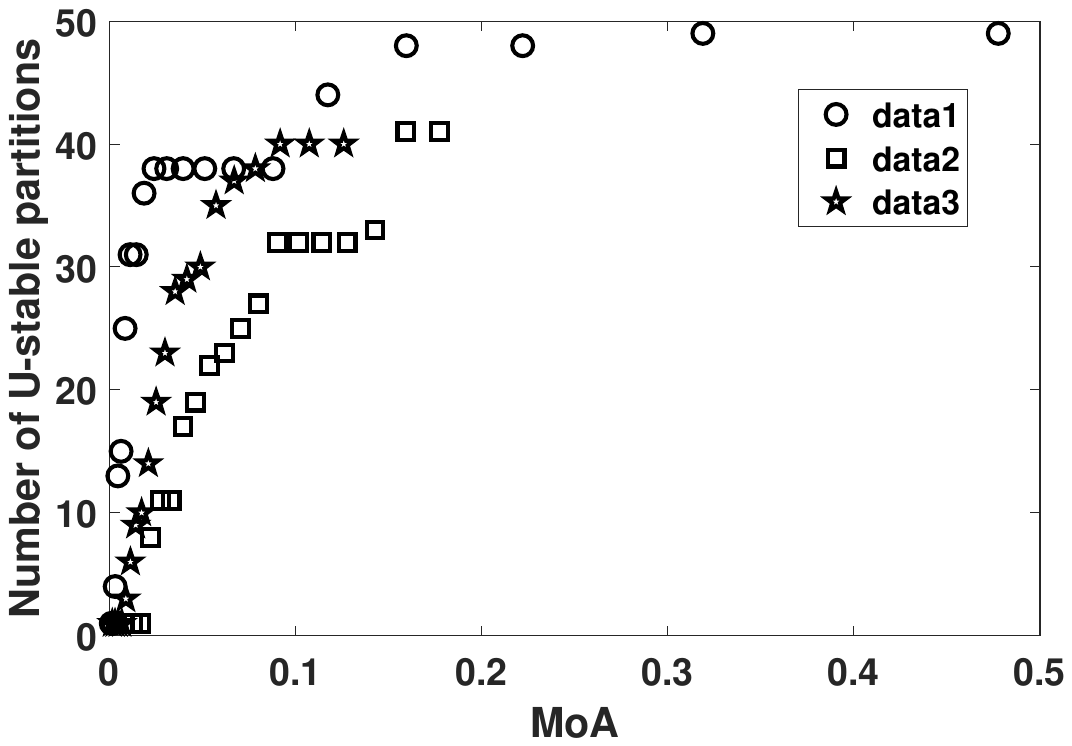}
\caption{
  \textbf{Random:} $\alpha_j \sim U(0,1)$ and $\delta \in [1,20 ]$ 
  }
\label{MoA_random}
\end{center}
\end{minipage}
\end{center}
\end{figure}

In our first case study, we begin with 5 symmetric players, with $\lambda=20$ as their influence factor.  
We obtain a sequence of systems (as explained in Section \ref{sec_moderate_asym}), one for each $\delta$, by  setting $\lambda_j =  \lambda - \alpha_j \delta$ for each $j$, where $\{\alpha_j\}$ is   a deterministic or a random  vector.  Basically   the asymmetry between players is increased by increasing $\delta$.  
As mentioned in Section \ref{sec_abs_stable}, the `level of stability' (represented by number of stable partitions) depends on MoA which can be seen in Figure \ref{MoA_vs_stable}. Also, one can observe that almost all partitions are stable for MoA around 0.4 (i.e., 49 out of 52). Figure \ref{MoA_random} considers random influence factors (for three samples) and we observe again that the number of stable partitions increases as MoA increases.

We consider another case study in Tables \ref{tab:case1} and  \ref{tab:case2}, with an aim to understand the players that are the  first to form successful coalitions, and the types of stable partitions etc., as `asymmetry' grows (via $\delta$).  \textit{We observe that the set of stable partitions only increases, no partition (stable for a lower $\delta$) becomes unstable for a larger $\delta$.} We thus tabulated only the new additions in the last columns. 

 \begin{table}[h]
\vspace{-2mm}
\centering
\hspace{-10mm}
{\small \begin{tabular}{|l|l|l|l|l|l|}
\hline
\multicolumn{1}{|c|}{\multirow{2}{*}{$\delta$} }                                       & \multicolumn{3}{c|}{No. of Stable partitions}                         & \multirow{2}{*}{Additional Stable Partitions}  \\ \cline{2-4}
 \multicolumn{1}{|c|}{} &     ALC & TTC & SS  & \\ \hline
 3                         &         1   & 0                 & 0 &     $\{\{1\},\{2\},\{3\},\{4\},\{5\}\}$                        \\ \hline
  3.2                            &     1    &    1                &  0 &     $\{\{5,2\},\{4,1\},\{3\}\}$				
                      \\  \hline 
  3.3                          &      1    &     2              & 0 &         $\{\{5,3\},\{4,1\},\{2\}\}	$ \\ \hline
  3.7                           &      1    &     2              & 1  &      $\{\{5,1\},\{4\},\{3\},\{2\}\}$ \\ \hline
                                &            &                   &    &   Additional SS Partitions            \\ \hline
  4.2                           &      1     &     4             & 2  &   $\{\{ 5, 2\},\{ 4\},\{ 3\},      \{ 1\}\}$ \\ \hline 
  4.95                          &      1     &      4            &  3  &  $\{\{ 5, 3\},\{ 4\},\{ 2\},      \{ 1\}\}$ \\ \hline
  5.7                            &      1     &      8            &  4  & $\{\{ 5, 4\},\{ 3\},\{ 2\},      \{ 1\}\}$ \\ \hline       
\end{tabular}}
\caption{$\lambda_j = 20-\alpha_j \delta$,  $\alpha$ = [0, 7, 11.8, 15.3, 19.3]/21.5
\label{tab:case1}}
\end{table}

Thus 
 in Tables \ref{tab:case1} and  \ref{tab:case2}, for any value of $\delta$, the partitions tabulated  in its row and  the rows  above are  stable.
We tabulate number of stable  TTC and SS  partitions, we refer a partition as TTC if there are two coalitions of two players while the remaining player is alone.  
  With further increase in $\delta$, as anticipated, 
lot more partitions are stable,   and hence to illustrate our results we mention only SS-partitions.  
 As we increase $\delta$ further, partitions other than ALC, TTC and SS start becoming stable. In Table \ref{tab:case2} we present such partitions under the column "Others" which also include TTC partitions.  We have the following observations (see Tables \ref{tab:case1} and  \ref{tab:case2}):
 \begin{enumerate}
\item The number of U-stable partitions increases as the asymmetry ($\delta$)  increases, i.e., as we move down the table.
\item Till $\delta = 3$ and $0.1$ respectively in the two tables, ALC is the only stable partition (first row). 
\item The highest and the least influential players are always the first ones to form successful coalitions (second row). 
They might form coalitions with each other (SS in Table   \ref{tab:case2}) or with others (TTC in Table   \ref{tab:case1}) depending upon $\alpha$. 
\item Let $j<k$. Then from  the tables   one can observe that (which explains the remarks after Theorems \ref{Thm_monotonicity}-\ref{Thm_monotonicity_new}),
\begin{enumerate}[label=(\roman*)]
\item if SS$(\{j,k\})$ is stable then SS$(\{j-1,k\})$   is also stable (e.g.,  row 4,   Table  \ref{tab:case2}, SS(\{5,2\})  and  SS(\{5,1\})  are stable).
\item if SS$(\{j,k\})$  is stable then SS$(\{j,k+1\})$  is also stable.
\item if SS$(\{j,k\})$  is stable then SS$(\{j-1,k+1\})$  is also stable.
\item if SS$(\{j,k\})$  is not stable, then it becomes stable as $w_j$ deviates from $w_k$ (by increasing $\delta$) as in Theorem \ref{Thm_Sym_to_Asym}.
\end{enumerate}
\end{enumerate}

\begin{table}[ht]
\centering
{\small \begin{tabular}{|l|l|l|l|l|l|}
\hline
\multicolumn{1}{|c|}{\multirow{2}{*}{$\delta$} }                                       & \multicolumn{3}{c|}{No. of Stable partitions}                         & \multirow{2}{*}{Additional Stable Partitions}  \\ \cline{2-4}
 \multicolumn{1}{|c|}{} &     ALC & TTC & SS  & \\ \hline
 0.1                         &         1   & 0                 & 0 &     $\{\{1\},\{2\},\{3\},\{4\},\{5\}\}$                        \\ \hline
  0.146                            &     1    &    0                &  1 &     $\{\{5,1\},\{4\},\{3\},\{2\}\}$				
                      \\  \hline 
 0.147                          &      1    &     2              & 1 &         $\{\{5,3\},\{4,1\},\{2\}\}	$ \\ 
                                &           &                    &   &         $\{\{5,2\},\{4,1\},\{3\}\}	$ \\ \hline
 0.18                           &      1    &     2              & 2  &      $\{\{ 5, 2\},\{ 4\},\{ 3\},      \{ 1\}\}$ \\ \hline
 0.19                           &      1     &     2             & 3  &       $\{\{ 5, 3\},\{ 4\},\{ 2\},      \{ 1\}\}$ \\ \hline 
                            &           &     Others             &    & Additional SS Partitions \\ \hline  
  0.21                          &      1     &      4            &  4 &  $\{\{ 5, 4\},\{ 3\},\{ 2\},      \{ 1\}\}$  \\ \hline     
  0.35                          &      1     &      17             &  5  &  $\{\{ 1, 4\},\{ 3\},\{ 2\},      \{ 5\}\}$  \\ \hline
   0.36                         &      1     &      17             &  6  &  $\{\{ 2, 4\},\{ 3\},\{ 1\},      \{ 5\}\}$  \\ \hline  
 0.37                           &      1     &      17             &  7  &  $\{\{ 3, 4\},\{ 2\},\{ 1\},      \{ 5\}\}$  \\ \hline         
\end{tabular}}
\caption{$\lambda_j = 20-\alpha_j \delta$, $\alpha$ = [0, 8, 11.5, 15.3, 21.5]
\label{tab:case2}}
\vspace{-7mm}
\end{table}
We studied many more examples (even with higher $n$) and our observations are exactly similar. In all examples, at the threshold of $\delta$, at which the number of stable partitions just increase from one, the highest and the least player are  always in some collaborative coalition in the new additions.
 \subsection*{Spectral sharing model}
 Next, we present a case study related to spectral sharing model described in Section \ref{sec_example} with $4$ players whose influence factors are given by $35$, $35$, $30$ and $30$. We also compute the individual spectral shares along with the individual utilities (obtained using SVs) in this example. To begin with, the spectral shares of any coalition is obtained using the utility and the optimal action of the same coalition at the NE of Theorem \ref{Thm_util_coal_general}. This coalitional spectral share is then divided among its members according to the same ratios as that of the the individual SVs of the coalitional utility. Figure \ref{Spectral_share} shows the utility and spectral shares obtained by each of the players at various stable configurations (i.e., U-stable partitions). It can be easily seen from this figure that the strongest player (i.e., player $1$) obtains the best spectral shares at one of the symmetric stable configuration (i.e., configuration $2$). The next best spectral share is achieved at ALC; however, the cost paid is also high and thus, the player obtains minimum utility at ALC. This case study also explains the observation made in Theorem \ref{Thm_no_identical} as SS-$(\{35,35\})$ partition (which satisfies the hypothesis of this theorem) is not U-stable.

\begin{table}[h]
\begin{minipage}{8.5cm}
\begin{center}
\begin{tabular}{|l|l|}
\hline
Stable Configurations & Partition \\ \hline
1                     & GC        \\ \hline
2                     &  $\{\{30,35\},\{30,35\}\}$         \\ \hline
3                     &  $\{\{30,30\},\{35,35\}\}$         \\ \hline
4                     &  $\{\{30,35\},\{30\},\{35\}\}$         \\ \hline
5                     & ALC       \\ \hline
\end{tabular}
\caption{Partitions described by stable configurations in Figure \ref{Spectral_share}}
\label{config}
\end{center}
\end{minipage}
\begin{minipage}{8.5cm}
\begin{center}
\includegraphics[trim = {1.5cm 0cm 0cm 0cm}, clip, scale = 0.23]{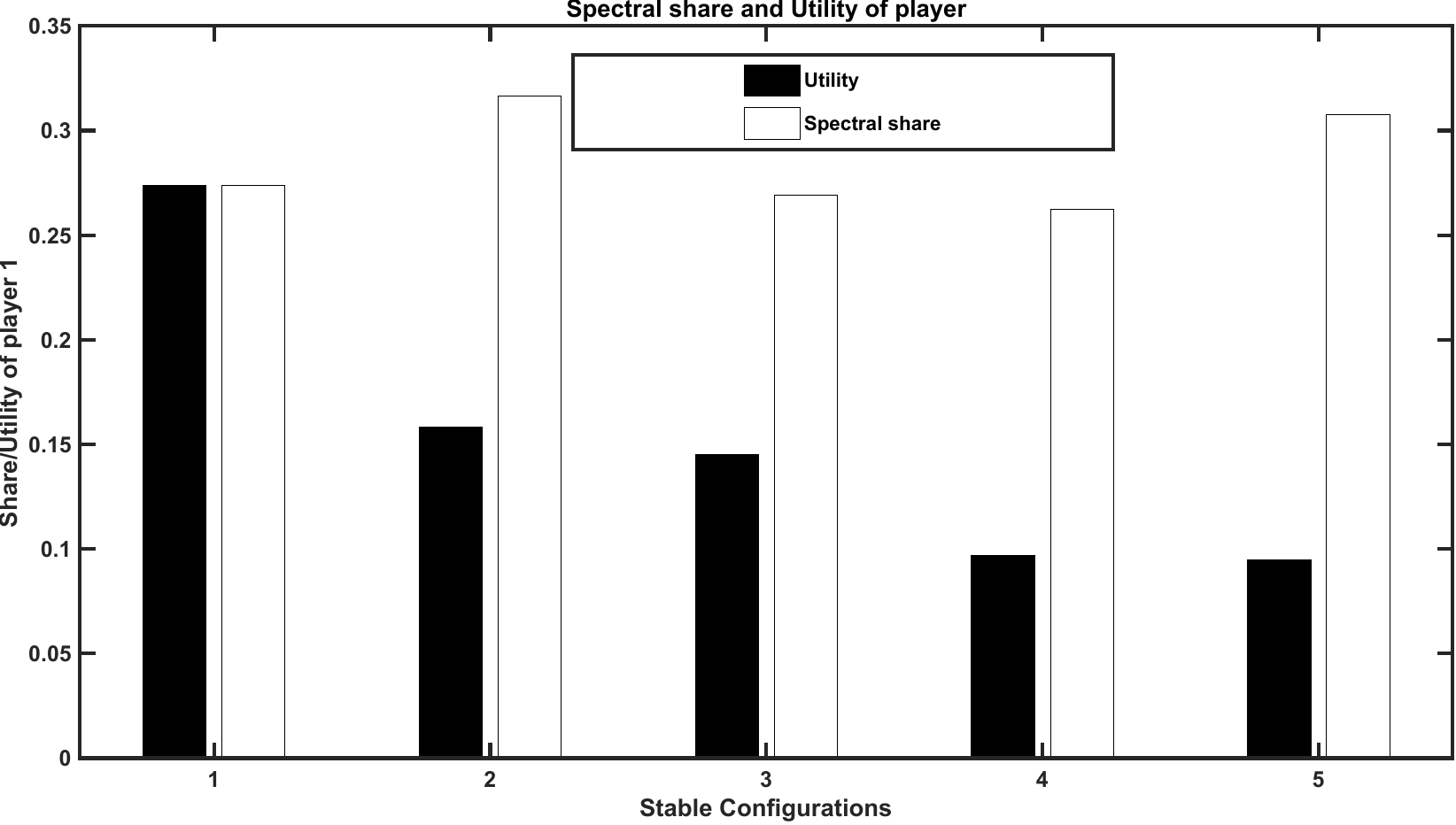}
\captionof{figure}{Spectral shares and utility  obtain- \\ ed under various stable configurations when \\
  $\lambda_j = $ [35, 35, 30, 30] and $\gamma = 1$
  }
\label{Spectral_share}
\end{center}
\end{minipage}
\end{table}

\section{ Stability against coalitional deviations}
In the previous sections, we studied the stability of a partition against unilateral deviations (i.e., any  player can change its strategy alone) under two divisions:  all  players being symmetric and players having a certain amount of asymmetry among them. Interestingly we have orthogonally different results in the two extremes: in the case with all identical players, only ALC is the stable partition (when $n>4$), while in the case with completely dissimilar players (i.e., under absolute stability), all partitions are stable. These  extreme results could be a result of our restriction on the type of deviations: what happens if some players deviate together to derive better shares for all of them.  This aspect requires further investigation and we precisely consider the same by studying the stability of a partition against coalitional deviations.


We begin with a few definitions that help explain \underline{coalitional blocking.} These concepts are inspired from \cite{aumann1961,CDC}, and are adaptations of those concepts to our partition form game.  
We basically need to start with {\it a partition $\P$ and check if it can be blocked by a coalition $S$.}  As a first step one needs to define {\it utility of the members of any given  coalition $S$} that attempts to block. 
Since our game is a partition form game, we must also consider the arrangement of outside players,  i.e., players  in $N_C-S$, while defining the above utilities.  
We again consider the \textit{pessimistic rule} (\cite{aumann1961,bloch}), where the worth of any coalition  $S$ is taken to be the worst among all possible partitions that contain $S$; observe here all possible partitions containing $S$ are allowed and there is no fixed environment.   Thus working as in Lemma \ref{min_worth_sub_coal_general}, the worth of blocking coalition $S$ is given by $\varphi_S^* ( \Q_S )$ with 
$\Q_S := \{S, \{ \{l\} \}_{l \in N_C -S} \}$ and \underline{the shares/utilities  of the members of $S$} equal the SVs  $\Phi_j^S :=  \phi_j^{\Q_S} $ for all $j \in S.$

We  say    \underline{a partition $\P$ is blocked by a coalition $S$}, if the vector of shares  $\{\phi_j^{\P} \}_{j \in S}$ of members of $S$ under $\P$ are strictly improved, i.e., if 
$\phi_j^\P < \Phi_j^S$ for all $j \in S$. 
 We say \underline{a partition to be C-sta-} \underline{ ble, i.e., stable against coalitional deviation, if there exists no coalition $S$ that can block it}.
Our next result presents  C-stable partitions  under {\bf A}.1-2 (proof in Appendix~\ref{sec_AppendixD_PEVA}).
\begin{theorem}
\label{Thm_GC_CS}{\bf [Coalitional Stability]}
Assume  {\bf A}.1-2. Then  GC is the only partition that is C-stable. 
\end{theorem}
Thus under this more robust form of stability (stable even under coalitional deviations),  the GC   is the only stable partition; this is true even when all partitions are stable against unilateral deviations. But for this, the players have to choose their collaborative strategies cooperatively.
Further this `more robust' partition  also achieves the social optimum.   

 We study the next case with symmetric players where no partition other than ALC is U-stable for $n >4$ (see Corollary \ref{corollary_ALC_NE}). Thus to find the C-stable partitions, it is sufficient to check the C-stability of ALC. The next theorem confirms the negative result (proof in Appendix~\ref{sec_AppendixD_PEVA}).
\begin{theorem}
\label{Thm_CS_None}
For symmetric players with $n >4$, none of the partitions are C-stable. 
\end{theorem}

As anticipated, the number of stable partitions against coalitional deviations are much lesser. However, the more interesting observation is that all the partitions were U-stable in the extreme case with dissimilar players while only GC is stable against coalitional deviations. In the other extreme, even ALC is not stable against coalitional deviations. This further reinforces our observation from the beginning that the identical players seldom want to collaborate: the system with all identical players (when $n>4$) is in `absolute C-instability', in that, none of the partitions are stable against coalitional deviations.

\section{Summary}
\label{conclusion}
We consider\rev{ed} a  coalition formation game with players exploring cooperation opportunities in a non-cooperative manner, where the utilities of players/coalitions are resultant of a resource sharing game.  We developed a framework 
to study the partitions  (non-overlapping and exhaustive set of  coalitions) that emerge at equilibrium. The  strategy of a player  is the set of players with whom it wants to form coalition, while the utilities of  players  are defined via (Shapley values of) the utilities of their coalitions and these coalitions/partition is formed based on the choice of all players;  the resulting coalitions     involve in a non-cooperative game along with a possible  adamant player (not willing to cooperate) and the utilities at the equilibrium define the utilities of the coalitions.

Our primary aim \rev{was} to identify the NE-partitions, we also derive\rev{d} the partitions that result at utilitarian solution (maximizes the sum of utilities).    We observe\rev{d} that  the agents 
derive much lower utilities at NE than that  at utilitarian solution,  and this loss is because of their strategic behaviour.  We considered \rev{PoA} (price of anarchy) to estimate the loss. 

With equal or almost equal players,  no one collaborate\rev{d} at equilibrium (if $n >4$) and  coarser partitions (some players  collaborate) emerge\rev{d} at NE  for smaller $n$; and the former  case does not depend upon  adamant player, while  latter case depends.   In all cases, the  \rev{PoA} increase\rev{d} with $n$ (as $O(n)$) and with increase/decrease in the strength of adamant player. Interestingly, none of the partitions are coalitionally stable for the symmetric case with $n>4$.

Surprisingly, when the players \rev{were} significantly different,  every partition \rev{was} stable against unilateral deviations.  One can view it as  lethargy of the system, where all players prefer to continue operating in their current configurations, as they do not  find it beneficial to deviate alone (or unilaterally).  However, many partitions \rev{were} challenged when players  deviate\rev{d} together. 
Interestingly, the only utilitarian partition, grand coalition, \rev{was} also the only partition stable against coalitional deviations (for a special case).

For the system with intermediate players, the number of stable partitions (stable against unilateral deviations) increase\rev{d} as asymmetry (a measure of differences in the influence factors of various players)
increases. Further and more interestingly, it \text{was} the highest and the lowest capacity players that first \rev{found} it beneficial to collaborate (form joint coalitions in some partitions that emerge at equilibrium).  When the number of players \rev{was}  sufficiently large, almost equal capacity (strong) players seldom \rev{found} it beneficial to collaborate.   

 When the players \rev{were} dissimilar, some smaller  players \rev{could} derive only zero utilities at equilibrium and  \rev{became} insignificant. The \rev{PoA} increase\rev{d} with the number of significant players, also depends upon relative strengths.

\subsection*{Spectrum auction model}
\label{sec_conclusions}
We applied our results to a spectrum auction model to understand the set of players that \rev{were} interested in bidding together for spectrum. Following are some important inferences:
\begin{enumerate}[label=(\alph*)]
\item  
None of the players with identical strengths buy the spectrum together, when their number is more than 4; this \rev{was} true irrespective of the way the allocated spectrum is divided among its members. This \rev{was} also true irrespective of the presence of adamant player. Identical players always \rev{found} better collaborative opportunities irrespective of the members with whom they \rev{were} currently collaborating/considering (as no partition \rev{was} coalitionally stable).
\item With a fewer number of identical players $(<5)$, a subset of players \rev{found} it beneficial to bid together for spectrum shares. 
This subset \rev{could} depend upon the presence and strength of adamant player.
\item When the strengths of players \rev{were} significantly different, any player would find it beneficial to bid together with any subset of players (when only unilateral deviations \rev{were} allowed). 
Further, when  all the  players share\rev{d} the available spectrum (i.e., form grand  coalition) according to  Shapley value, 
no subset of players \rev{found} it beneficial to deviate and derive better spectral chunks by bidding (only grand coalition \rev{was} coalitionally stable). 



\item  Another striking result \rev{was},  the bigger and smaller players more easily \rev{found} it beneficial to bid together for spectrum than the players of almost equal/intermediate strengths . 
\end{enumerate}
\chapter[On the Ubiquity of Duopolies in Constant Sum Congestion Fames]{On the Ubiquity of Duopolies in Constant Sum Congestion Games\raisebox{.3\baselineskip}{\normalsize\footnotemark}}
\footnotetext{Joint work also  with Prof. Jayakrishnan Nair, Electrical Engineering, IIT Bombay}
\label{chap_OR}

As in the previous chapter, we again focus on the stable partitions but now in a queueing system context, specifically the Erlang-B queueing system, with a more realistic customer behavior, i.e., where customers split based on the quality of service.

\section{Introduction}

Resource sharing is an efficient way of reducing congestion and
uncertainty in service industries. It
refers to an arrangement where service resources are pooled and used
jointly by a group (a.k.a., coalition) of providers, instead of each
provider operating alone using its own resources. Naturally, such a
coalition would be sustainable only if the participating providers
obtain higher payoffs than they would have obtained otherwise. The key
driver of coalition formation in congestion prone service systems is
the statistical economies of scale that emerge from the pooling of
service resources---this allows the coalition to offer a better
quality of service to its customers, and/or to attract more customers
to its service.

Not surprisingly, there is a considerable literature (for example, see~\citep{karsten} and the references therein) that analyses
resource pooling between independent providers of congestible services
via a cooperative game theoretic approach. In these papers, each
provider is modeled as a queueing system, with its own dedicated
customer base, that generates service requests according to a certain
arrival process. The payoff of each service provider is in turn
determined by the quality of service it is able to provide to its
(dedicated) customer base. In such a setting, the statistical
economies of scale from resource pooling typically drives the service
providers to pool all their servers together to form a \emph{grand
coalition}, which generates the greatest aggregate payoff across all
coalitional arrangements. Naturally, the resulting aggregate payoff
must be divided between the providers in a \emph{stable} manner, i.e.,
in such a way that no subset of providers has an incentive to `break
away' from the grand coalition. Such stable payoff allocations have
been demonstrated in a wide range of settings, including
single/multiple server environments, and loss/queue-based environments
(see~\cite{karsten,karsten2014}  and the references therein).

To summarize, the literature on coalition formation between providers
of congestible services suggests that a stable grand coalition would
emerge from the strategic interaction. However, a crucial aspect the
preceding literature fails to capture is \emph{user churn}. That is,
customers can switch service providers, if offered superior service
quality elsewhere. This aspect introduces \emph{competition} between the service providers (or coalitions of service providers) over market share, and turns the game into a \emph{partition form game} (described below). To the best of our knowledge, the interplay between resource pooling among service providers (aided by the associated economies of scale) and the competition between them, in the context of congestible services, has not been explored in the literature. This study seeks to fill this gap.

This chapter also contributes to the theory of coalition formation games
in terms of new notions of stability. In particular, we focus on partition form games; the main ingredients of such games are, a partition (an arrangement of players into disjoint coalitions), the worth of each coalition (which, crucially, also depends on the partition), and the anticipation rules by which a  blocking or opposing coalition estimates its new worth (depending upon the anticipated retaliation of the opponents). 
In such games, the classical notion of stability declares a partition to be stable if it is not blocked by \textit{any} coalition (\cite{aumann1961,narahari})---a coalition blocks a partition if it anticipates 
greater worth in the new arrangement. However, some case studies may have no stable partitions under such classical notions (e.g., the game studied in~\cite{Shiksha_Perf}, and the market-size driven coalition formation game of the present study). This necessitates a deeper study of such scenarios, possibly using new, more relevant notions of stability. \textit{In this study, we define novel notions of stability by suitably restricting the set of candidate blocking coalitions.} Indeed, in practice, rearrangements in the marketplace typically arise from mergers between, or the breaking up of, existing corporations---our new notions of stability restrict the focus only on such tensions in the marketplace. 

In this chapter, we analyse a coalition formation game  between a collection of service providers, each of which is modelled as an Erlang-B loss system. A key aspect of our model is that the total market size (captured via the aggregate arrival rate of customer requests) is fixed exogenously, and providers (or coalitions of providers) compete for market share---this leads to a \emph{constant sum, partition form game}. These aspects, as we show, dramatically alter the outcome of the strategic interaction between providers.
Interestingly, we find that under classical notions of stability\textit{, no arrangement of service providers into coalitions is stable, no matter how the payoff of each coalition is distributed across its members}. However, we demonstrate stable partitions when blocking coalitions are restricted to mergers and splits of the existing coalitions. Under our new notions of stability (we define two new notions, that differ on how a blocking coalition estimates its worth), \emph{the grand coalition is not} stable, except in a very specific corner case. Instead, the predominantly stable configurations are \emph{duopolies}, with the larger coalition exploiting economies of scale to corner a disproportionate portion of the market size. Our work also highlights several subtleties relating to different natural notions of stability in this context, the way the payoff of each coalition is divided between its members, and the degree of congestion in the system.

\subsection*{Our contributions}

\begin{enumerate}[label=(\roman*)]
\item We formally define a constant sum coalition formation
game between strategic service providers of a congestible
service (see Section~\ref{sec_model}). This model is the first, to the best of our knowledge, to
capture the interplay between resource pooling and
competition over market share.

\item Under the classical notion of stability for this \textit{partition form game} model (inspired by~\cite{aumann1961}), which we refer to as \textit{General Blocking-Perfect Assessment}, we show that no configuration is stable (see \rev{Theorem~\ref{thm_impossible}}). (A configuration specifies a partition of the set of providers into coalitions, and also the allocation of the total payoff of each coalition among its members.) This is because of the vast (specifically, all possible) range of deviations that can challenge any given configuration.

\item In view of this impossibility result, we define two novel \textit{restricted} notions of stability (see Section~\ref{stable_config}), where only coalitions arising from mergers or splits of existing coalitions can challenge the status quo.
The two notions differ with respect to the precision with which the coalition that seeks to `break' from the prevailing configuration can estimate the benefit from doing so. 

Interestingly, we show that our restricted notions of stability do admit stable configurations \rev{(Theorems \ref{Thm_two_partition} and \ref{Thm_stable_config_ruleB})}. Moreover, these stable configurations involve \emph{duopolies}, i.e., two competing coalitions (except for a certain corner case where the grand coalition is also stable). Intuitively, configurations involving three or more coalitions are unstable because economies of scale incentivize mergers of two or more (but not all) coalitions. On the other hand, the constant sum nature of the game dis-incentivizes the formation of a grand coalition (except in the corner case mentioned above).

\item Finally, we explore the impact of the overall congestion level on the stable duopolies, by analysing light and heavy traffic regimes (see \rev{Section~\ref{sec_two_partition}}). All duopolies are stable in heavy traffic \rev{(Theorem \ref{Thm_heavy})}, whereas only duopolies with nearly matched service capacities are stable in light traffic \rev{(Theorem \ref{low_traffic})}. We also present some initial ideas about the dynamic version of this game.
\end{enumerate}

\subsection*{Related Literature} 
This study is related to two distinct strands of literature: (i) the literature on coalition formation for resource pooling in queueing networks, and (ii) the literature on partition form games.

\emph{Resource pooling in queueing networks:} This literature is quite vast, and we only provide a brief survey here; a comprehensive review can be found in~\cite{karsten}. One line of this literature models each coalition as a single server queue. The service rate of each coalition is either assumed to be optimized by the coalition itself (see, for example,~\cite{gonzalez,garcia,yu}), or simply taken to be the sum of the intrinsic service rates of the members (see, for example,~\cite{anily2010,timmer,anily2011,anily2014}.
Another line of literature treats each coalition as a multi-server loss system--\cite{karsten2012} considers the case where the number of servers with each player is fixed apriori, and~\cite{ozen,karsten2014} consider the case where a coalition optimizes the number of servers it operates.
Finally,~\cite{karsten} analyses the setting where each coalition is an~$M/M/s$ queue (Erlang~C); they consider both the above mentioned models for the service capacity of a coalition.

All the above mentioned papers assume that each service provider has a dedicated customer base (modeled via an exogenously determined arrival rate of service requests). From a game theoretic standpoint, this simplification ensures that the worth/utility of each coalition depends only the members of that coalition. In contrast, in the present study, we explicitly model \emph{user churn}, which induces competition between coalitions, and turns the game into a \emph{partition form} game, wherein the worth/utility of a coalition also depends on the arrangement of players outside that coalition.

\emph{Partition form games:} The earliest work in this area can be found in~\cite{aumann1961}. The authors define a general definition of cooperative games which is applicable to both characteristic and partition form games (without using these names). The term ``partition form game" was first coined in~\cite{lucas}, where the authors further develop the theory of this class of games.~\cite{aumann1974cooperative} extends various existing stability notions for characteristic form games to partition form games.

Majority of the literature on cooperative games deals with the stability of the grand coalition in characteristic form games. In contrast, there is only a limited literature on partition form games.~\cite{hafalir} established the conditions under which the grand coalition is stable for convex partition form games. The authors in~\cite{saad_unilateral} (spectrum sensing and access), \cite{Shiksha_Perf} (Kelly's mechanism) show that certain finer partitions other than the grand coalition can be stable against unilateral deviations for partition form games, while the authors in~\cite{bloch,yi} show the same for the classical notions of stability against coalitional deviations. The authors in~\cite{Shiksha_Perf} also study stability against coalitional deviations to show that the  grand coalition is stable when players are significantly asymmetric, while  no partition is stable when the players are identical. Finally,~\cite{ray} considers a dynamic coalition formation game and shows that finer partitions can emerge at the sub-game perfect equilibrium.

\section{Model and Preliminaries}
\label{sec_model}

In this section, we describe our system model for coalition formation
between strategic service providers, characterize the behavior of the
customer base in response to coalition formation between service
providers, and introduce some background. 

\subsection{System model}
Consider a system with a set~$\mathcal{N} = \{1,\cdots,n\}$ of
independent service providers (a.k.a., agents), with provider~$i$
having~$N_i$ servers. Without loss of generality, we assume~$N_i \ge N_{i+1}$ for~$1 \leq i \leq n-1.$ All servers are identical,
and assumed to have a unit speed, without loss of generality. The
providers serve a customer base that generates service requests as per
a Poisson process of rate~$\Lambda.$ Jobs sizes (a.k.a., service
requirements) are i.i.d., with~$J$ denoting a generic job size, and~$\mathbb{E}[J] = 1/\mu.$

Service providers are strategic, and can form
coalitions with other service providers to enhance their
rewards. Formally, such coalition formation between the service
providers induces a partition~$\P = \{C_1,C_2,\cdots,C_k\}$ of~$\mathcal{N},$ where~$\cup_{i = 1}^kC_i = \mathcal{N}, \ C_i \cap C_j = \emptyset \text{ for all } i
\neq j.$ We refer to such a partition with~$k$ coalitions as a~$k$-partition. (Naturally, the baseline scenario where each service
provider operates independently corresponds to an~$n$-partition.)

In response to a partition~$\P$ induced by coalition formation between
service providers, the arrival process of customer requests gets split
across the~$k$ coalitions in~$\P$, with the arrival process seen by
coalition~$C$ being a Poisson process of rate~$\lambda^{\P}_{C},$
where~$\sum_{C \in \P} \lambda^{\P}_{C} = \Lambda.$ (We characterize
the split~$(\lambda^{\P}_{C},\ C \in \P)$ as a Wardrop equilibrium;
details below.) Each coalition~$C$ operates as an~$M$/$M$/$N_{C}$/$N_{C}$ (Erlang-B) loss system, with~$N_{C} = \sum_{j
  \in C} N_j$ parallel servers, and arrival rate~$\lambda^{\P}_{C}.$
This means jobs arriving into coalition~$C$ that find a free server
upon arrival begin service immediately, while those that arrive when
all~$N_{C}$ servers are busy get dropped (lost). Given the well known
insensitivity property of the Erlang-B system, the steady state
blocking probability associated with coalition~$C$ (the long run
fraction of jobs arriving into coalition~$C$ that get dropped),
denoted~$B_{C}^{\P},$ is given by the Erlang-B formula:
 \begin{align}
\label{Eqn_PB}
B_{C}^{\P} = B(N_{C},a^{\P}_{C}), \text{ where } a^{\P}_{C} := \frac{\lambda^{\P}_{C}}{\mu} \text{ and } B(M,a) = \frac{ \frac{a^{M}}{M!}   }{ \sum_{j=0}^{M} \frac{a^{j}}{j!} }. 
\end{align}

\subsection{User behavior: Wardrop equilibrium}

Next, we define the behavior of the customer base in response to
coalition formation across service providers, via the split~$(\lambda^{\P}_{C},\ C \in \P)$ of the aggregate arrival process of
service requests across coalitions. This split is characterized as a
Wardrop equilibrium (or WE; see~\cite{WE}).

In the context of our model, we define the WE split of the arrival
process of service requests across coalitions, such that the steady
state blocking probability associated with each coalition is equal.
Note that since the blocking probability associated with an `unused'
coalition would be zero, it follows that all coalitions would see a
strictly positive arrival rate. Thus, the WE (if it exists) is
characterized by a vector of arrival rates~$(\lambda^{\P}_{C},\ C \in \P)$ satisfying
\begin{equation}
  \label{Eqn_WE_properties}
  B^{\P}_C = B\left(N_C,\frac{\lambda_C^\P}{\mu}\right) = B^*  \ \forall\ C \in \P \text{ and }
 \sum_{C \in  \P } \lambda_C^\P  = \Lambda ,
\end{equation}
where~$B^*$ is the common steady state blocking probability for each
coalition. For any given partition~$\P,$ the following theorem
establishes the existence and uniqueness of the WE, along with some
useful properties (proof in Appendix~\ref{appendix_B}). 
\begin{theorem}
  \label{Thm_WE}
Given any partition~$\P$ between the service providers and market size~$\Lambda$, there is a
  unique Wardrop equilibrium~$(\lambda^{\P}_{C},\ C \in \P),$ where~$\lambda^{\P}_{C} > 0$ for all~$C \in \P,$ that satisfies~\eqref{Eqn_WE_properties}. Additionally, the following properties hold:
  \begin{enumerate}[label=(\roman*)]
      \item For each~$C \in \P, \lambda_{C}^\P$ is a strictly increasing
  function of the total arrival rate~$\Lambda.$
  \item  If the partition~$\P'$ is formed by merging two coalitions~$C_i$ and~$C_j$ in partition~$\P$ where~$C_i \cup C_j \neq \mathcal{N}$ (with all other coalitions in~$\P$
  remaining intact), then~$\lambda^{\P'}_{C_i \cup C_j} > \lambda^{\P}_{C_i} +
  \lambda^{\P}_{C_j}.$
  \item If~$\P = \{C_1,C_2\},$ with $N_{C_1} > N_{C_2},$
  then~$\frac{\lambda^{\P}_{C_1}}{N_{C_1}} > \frac{\Lambda}{N} >
  \frac{\lambda^{\P}_{C_2}}{N_{C_2}}, \text{ where } N = \sum_{i \in
    \mathcal{N}} N_i.$
  \end{enumerate} 
 \end{theorem}
 
\rev{The existence and uniqueness of the Wardrop split follows from the strict monotonicity of the blocking probability with the arrival rate (see Appendix~\ref{appendix_B}).} Aside from asserting the uniqueness and strict positivity of the
 Wardrop split, Theorem~\ref{Thm_WE} also states that equilibrium
 arrival rate of each coalition is an increasing function of the
 aggregate arrival rate~$\Lambda;$ see Statement~$(i).$ Additionally,
 Statement~$(ii)$ demonstrates the statistical economies of scale due
 to a merger between coalitions: the merged entity is able to attract
 an arrival rate that exceeds the sum of the arrival rates seen by the
 two coalitions pre-merger. \rev{The parts $(i)$ and $(ii)$ follow from the monotonicity of the blocking probability with the arrival rate and the WE constraint in \eqref{Eqn_WE_properties}.} Finally, Statement~$(iii)$ provides
 another illustration of statistical economies of scale for the
 special case of a 2-partition---the larger coalition enjoys a higher
 offered load per server than the smaller one. \rev{This follows because of the strict monotonicity of the probability with the number of servers and the WE constraint in \eqref{Eqn_WE_properties}. All these details are evident from the proof in Appendix~\ref{appendix_B}.}

 \subsection{Coalition formation game: Preliminaries}
 
 Having defined the behavior of the user base, we now provide some
 preliminary details on the coalition formation game between the
 service providers.

 Recall that each service provider is strategic, and only enters into
 a coalition if doing so is beneficial.
 Given a partition~$\P$ that describes the coalitions formed by
 the service providers, we define the value or payoff of each
 coalition~$C \in \P$ to be~$\beta \lambda_C^{\P},$ where~$\beta > 0.$
 This is natural when~$\lambda_C^{\P}$ is interpreted as being
 proportional to the number of subscribers of coalition~$C,$ with each
 subscriber paying a recurring subscription fee. Without loss of
 generality, we set~$\beta = 1.$
 
 The value~$\lambda_C^{\P}$ of each coalition~$C$ must further be
 apportioned between the members of the coalition. Denoting the payoff
 of agent~$i$ by~$\phi_i^{\P},$ we therefore have~$\sum_{i \in C}
 \phi_i^{\P} = \lambda_C^{\P} \text{ for all } C \in \P.$
 Since the providers are selfish, they are ultimately interested only
 in their individual payoffs.
 Thus, the coalition formation between providers is driven by the
 desire of each provider to maximize its payoff, given the statistical
 economies of scale obtained via coalition, and also the
 \emph{constant sum} nature of this game (the sum total of the payoffs
 of all providers equals~$\Lambda$). Thus, the relevant fundamental
 questions are:
\begin{enumerate}
\item Which partitions can emerge as a result of the strategic
  interaction between providers, i.e., which partitions are stable? 
  Indeed, a precursor to this question is: how does one define a
  natural notion of stability?
\item It is apparent that the answer to the above question hinges on
  how the value of each coalition is divided between its
  members. Thus, a more appropriate question is: which coalitional arrangement of agents and subsequent division of the coalitional shares results in  stable configurations?
\end{enumerate}
Our aim in this chapter is to answer these questions; such problems can
be studied using tools from cooperative game theory. 
In the next section, we begin with classical notions of stability and `blocking by a  coalition', available in the literature; we will observe that there exists no partition which is stable under these classical notions. In the later sections, we refine the notion of stability (using some form of restricted blocking) and study the configurations that are stable.

\section{Classical Notions of Coalitional Blocking and Stability }
\label{sec:classical}

It is well known that non-partition type transferable utility cooperative games are characterized by tuple~$(\mathcal{N},  \nu)$, where~$\nu(C)$ for any subset~$C \subset \mathcal{N}$ represents the utility of coalition~$C$.
However, this is not sufficient for a partition form game, where a coalition's utility depends not only on the coalition's players but also on the arrangement of other players.
In this case~$\nu(C)$ (more appropriately)  can be defined as the set of  payoff vectors (of dimension~$n$) that are  anticipated to be achievable by the players of the coalition~$C$ (e.g.,~\cite{aumann1961}); and this anticipation is based on their expectation of the reactions of the agents outside the coalition. 
The stability concepts (e.g., core) are extended to these type of games (e.g.,~\cite{aumann1961}), which are discussed at length in Appendix~\ref{appendix_A}. In this section we discuss the same ideas in our  notations, in particular, we consider the notion of $\alpha$-efficient  $R$-core defined in~\cite{aumann1961} (more details  are in Appendix~\ref{appendix_A}). 

This notion of stability  is interlaced with the notion of a partition
(more precisely, a configuration defined below) being \emph{blocked} by some
coalition. We begin with relevant  definitions.
Given a partition~$\P = \{C_1,\cdots,C_k\},$ the set of payoff
vectors consistent with~$\P$ is defined as:
$${\bm \Phi}^{\P} := \left  \{\Phi = [\phi_1,\cdots, \phi_n] \in \mathbb{R}^n_+:  \sum_{j \in C_i} \phi_j = \lambda^{\P}_{C_i}\ \forall \ i \right \}.$$
\textit{A \textbf{configuration} is defined as a tuple~$(\P,\Phi),$ such that~$\Phi \in {\bm \Phi}^{\P}.$} 
\newline
Note that a configuration specifies not just a
partition of the agents into coalitions, but also specifies an
allocation of payoffs within each coalition, that is consistent with
the partition.

\textit{\textbf{Blocking   by a coalition:} A configuration~$(\P,\Phi)$ is \emph{blocked} by a coalition~$C \notin \P$ if, for any
partition~$\P'$ containing~$C,$ there exists~$\Phi' \in {\bm \Phi}^{\P'}$
such that~$
  \phi_j' > \phi_j \text{ for all } j \in C.
$}

 Basically, a new coalition can block an existing configuration,
  if each one of its members can derive strictly better payoff from this
  realignment (irrespective of the responses of the opponents in~$C^c$).
Equivalently, $(\P,\Phi)$ is blocked by coalition~$C \notin \P$ if,
for any partition~$\P'$ containing~$C,$ we have
$\lambda^{\P'}_C > \sum_{j \in C} \phi_j.$
Note that the above equivalence hinges on the \emph{transferable
  utility assumption inherent in our cooperative game,  by virtue of  which (partial) utilities can be transferred across agents.} Intuitively, a
coalition~$C \subset \mathcal{N}$ blocks configuration~$(\P,\Phi)$, if
the members of~$C$ have an incentive to `break' one or more coalitions
of~$\P$ to come together and form a new coalition. In particular, it
is possible to allocate payoffs within the blocking coalition~$C$ such
that each member of~$C$ achieves a strictly greater payoff,
irrespective of any (potentially retaliatory) rearrangements among
agents outside~$C.$
This is referred to in the literature as a \textit{pessimistic
  anticipation rule} (see~\cite{pessimistic, Shiksha_Perf} and Appendix~\ref{appendix_A}) or $\alpha$-efficient rule in~\cite{aumann1961}.
 
We refer the above pessimal anticipation based blocking  as GB-PA (General Blocking--Perfect Assessment) rule,  we first provide the precise summary: 

\textit{\textbf{GB-PA rule:} Under this rule, a configuration~$(\P,\Phi)$ is
blocked by \emph{any} coalition~$Q \notin \P$  if
\begin{equation}
  \label{eq:blocking_PA}
  \ulam_Q > \sum_{i \in Q} \phi_i, \text{ where }
  \ulam_Q := \min_{\P': Q \in \P'} \lambda^{\P'}_Q.
\end{equation}A configuration is stable under the GB-PA rule if it is not blocked by any coalition.}

The term `General Blocking' is used for this notion, as any arbitrary coalition (mergers or splits of the existing coalitions or mergers of partial splits) can block; and the term `Perfect Assessment' is used as the players in blocking coalition are aware of the previous shares of all members of the blocking coalition, i.e., previous shares of players is `common knowledge' within~$Q$.

\textbf{Stability under GB-PA:} We establish a negative result for this classical notion of stability (proof in Appendix~\ref{appendix_C}):
\begin{theorem}
 For $n>2$, there exists no stable configuration under GB-PA rule.
\label{thm_impossible}
\end{theorem} 

 We establish the above result by showing that the configuration with the~$n$-partition (i.e., each agent operates alone) is blocked by a suitable merger, while for any other configuration, there exists a~$j \in \mathcal{N}$ such that either~$\{j\}$ or~$\mathcal{N} \backslash \{j\}$ blocks it. For $n=2$ it is trivial to observe that the only stable configurations are $\left(\P_2,\Phi^{\P}\right)$  and $\left(\{1,2\},\Phi^{\P}\right)$ where $\P_2:=\{\{1\},\{2\}\}$ and $\Phi^{\P}:= \left(\lambda_{\{1\}}^{\P_2},\lambda_{\{2\}}^{\P_2}\right)$.

Theorem~\ref{thm_impossible} states that no  configuration is stable under GB-PA for~$n>2$, in other words, the $\alpha$-core (R-core under $\alpha$-effectiveness) as defined in~\cite{aumann1961} is empty, for our game. This `impossibility' is due to the fact that under GB-PA, a configuration can be blocked by \emph{any} coalition that is not contained in it; this coalition can be formed via multiple mergers/splits of existing coalitions.  But in practice, either an existing coalition splits or two or more of the existing coalitions merge. Thus, to define more practical and relevant notions of stability, one may have to consider a more restricted set of blocking candidates. This is addressed in the next section.
 
 In the next section,  
we
also consider an alternate form of restricted blocking, where the
`prevailing worth' of the agents of the candidate blocking coalition
is assessed imprecisely. Prior to that, we conclude this section with a short discussion on other anticipation rules.

\textbf{Other Anticipation Rules:} There are many other anticipation rules considered in the literature, for e.g.,~$\beta$-effective rule in~\cite{aumann1961} (coalition~$C$ can block payoff vector~$\Phi$, if for every correlated strategy of players in~$\mathcal{N}-C$, there exists a correlated strategy of players in~$C$ which leaves them better-off) and max rule in~\cite{pessimistic} (the  opponents/players in~$\mathcal{N}-C$ are anticipated to arrange themselves in a partition that maximizes their own utilities).  
  Interestingly, the pessimistic rule coincides with the above mentioned anticipation rules for our constant sum game, mainly because of economies of scale established in Theorem~\ref{Thm_WE}.$(ii)$.
  
  There are other anticipation rules that do not coincide with the pessimal rule. For example, the optimistic rule (opponents are anticipated to arrange in such a way that the deviating coalition obtains the best utility) in~\cite{pessimistic}, the Cournot Nash Equilibrium (opponents are anticipated to remain in their old coalitions) in~\cite{alpha-core}, etc. However, the impossibility result established in Theorem~\ref{thm_impossible} also implies impossibility under these rules (if any coalition $Q$ anticipates a higher utility than what its members derive in the current configuration under the pessimal rule~\eqref{eq:blocking_PA}, it  would also anticipate higher utility using any other anticipation~rule). 

\section{Realistic Notions of Blocking and Stability} 
\label{stable_config}

Motivated by the impossibility of stable configurations under GB-PA (Theorem~\ref{thm_impossible}), in this section, we define weaker, more realistic notions of stability, that do admit stable configurations. Specifically, the proposed stability notions
differ from GB-PA on the class of candidate blocking
coalitions considered, as well as the precision with which the
`prevailing worth' of the members of the candidate coalition is
assessed and/or revealed. The former distinction is inspired by the observation that organisational rearrangements predominantly occur in practice via mergers or splits of existing coalitions. For each of these notions of stability, we characterize the
class of stable configurations. 

The main takeaway from our results is the following.
 Because of the interplay between statistical economies of scale and the
constant sum nature of the game, only configurations involving duopolies (i.e., partitions with two coalitions) are stable (except in a certain corner case, where the grand coalition is also stable). This is true for both the proposed notions of stability defined next.

\subsection{Restricted blocking and stability}

The first notion of stability we introduce simply restricts the set of
candidate blocking configurations to mergers and splits of prevailing
coalitions. Note that this is a natural restriction from a practical
standpoint, since complex rearrangements between firms in a
marketplace typically arise (over time) from a sequence of mergers and
splits. We refer to this as restricted blocking (RB). Further when one
assumes the precise knowledge of the worth of the blocking candidates,
it leads to the RB-PA (Restricted Blocking--Perfect Assessment)
rule. We begin with this rule.

\textit{\textbf{RB-PA rule:} Under this rule, a configuration~$(\P,\Phi)$ can be blocked only by a coalition~$Q$ that is formed either via~$(i)$ a merger of
coalitions in~$\P$ (i.e.,~$Q = \cup_{C \in \mathcal{M}} C$ for~$\mathcal{M} \subseteq \P$), or via ~$(ii)$ a split of a single coalition in~$\P$ (i.e.,~$Q \subset C$ for some~$C \in \P$). Further, such a~$Q$ blocks~$(\P,\Phi)$ if, for all
partitions~$\P'$ containing $Q,$ there exists
~$\Phi' \in {\bm \Phi}^{\P'}$ such that~$ \phi'_i > \phi_i \text{ for all } i \in Q.$}

\textit{Equivalently,~$Q$ (as described above) blocks the configuration~$(\P,\Phi)$ if
\begin{equation}
  \label{eq_def_RBPA}
  \ulam_Q > \sum_{i \in Q} \phi_i, \text{ where }
  \ulam_Q := \min_{\P': Q \in \P'} \lambda^{\P'}_Q.
\end{equation}
A configuration~$(\P,\Phi)$ is stable under the RB-PA rule if it is
not blocked by any merger or split.}

Note that like GB-PA, the RB-PA rule also involves pessimal anticipation; the
members of candidate blocking coalition are pessimistic in their
anticipation of the value of the new coalition. Moreover, it is
possible to allocate the payoff of the blocking coalition~$Q$ among its members such that each member is (strictly) better off, as discussed in the previous section. 

The next  notion  uses the same restriction on the set of
candidate blocking coalitions, but uses an imprecise assessment of
the prevailing worth of the members of the candidate blocking
coalition, resulting in an imprecise  assessment of the benefit/loss from blocking. We refer to this as the
RB-IA (Restricted Blocking--Imperfect Assessment) rule.

\textit{\textbf{ RB-IA rule:}
Under this rule, a configuration~$(\P,\Phi)$ is blocked by a coalition~$Q$ formed either via a merger or a split if:
\begin{align}
  \label{Eqn_unified_condition_S}
  &\ulam_Q := \min_{\P': Q \in \P'} \lambda^{\P'}_Q > \sum_{C \in \P} \frac{N_{C \cap Q}}{N_C} \lambda^{\P}_C, \\
  \label{Eqn_unified_condition_S_pt2}
  &\lambda^{\hat{\P}}_Q > \sum_{i \in Q} \phi_i, \text{ where }\hat{\P} =  \left( \bigcup_{C \in \P} \{ C\setminus Q\} \right) \bigcup \{Q\}.
 \end{align}
 A configuration~$(\P,\Phi)$ is stable under the RB-IA rule if it is not blocked by any merger or split.}
 
 Condition~\eqref{Eqn_unified_condition_S} can be
interpreted as a first stage check on the feasibility of the block, by
(imperfectly) assessing the total prevailing worth of the members
of~$Q$ (using the prevailing coalitional worths~$\{\lambda_C^\P\}$). This imprecise assessment is obtained as the sum of the proportional contributions of the members of~$Q$ to their respective parent coalitions; the imprecision stems from not using the actual payoffs~$\{\phi_i\}_{i \in Q}$.  
Note that this feasibility check is also under the pessimal anticipation rule, but with imperfect estimates. 

Condition~\eqref{Eqn_unified_condition_S_pt2} is the final validation of the block using precise estimates~$\{\phi_i\}_{i \in Q}$. This ensures that it is possible to
allocate the payoff of~$Q$ among its members such that each member is
(strictly) better off from the deviation. Here, the anticipation is that there would be no immediate retaliation from the leftover players, i.e., as seen from the definition of~$\hat{\P}$ in~\eqref{Eqn_unified_condition_S_pt2}, the opponents would remain in their
original coalitions (as in the Cournot Nash equilibrium~\cite{alpha-core}, or projection rule in sub-section \ref{sec_anticipation} of Chapter \ref{chap_CGT}). This is reasonable after the already pessimal feasibility check in~\eqref{Eqn_unified_condition_S}.

Let us now interpret the condition for blocking due to a split/merger separately under RB-IA. We begin with blocking due to a split.
By~\eqref{Eqn_unified_condition_S} and~\eqref{Eqn_unified_condition_S_pt2}, a configuration~$(\P,\Phi)$ is
blocked by a coalition~$Q$ that is formed by splitting a coalition~$C \in \P$ if: \vspace{-2mm}
\begin{align}
  \label{Eqn_condition_S}
  &\ulam_Q := \min_{\P': Q \in \P'} \lambda^{\P'}_Q > \frac{N_Q}{N_C} \lambda^{\P}_C, \\
  \label{Eqn_condition_S_pt2}
  &\lambda^{\hat{\P}}_Q > \sum_{i \in Q} \phi_i, \text{ where }\hat{\P} = (\P \setminus \{C\}) \cup \{Q, C\setminus Q\}.
 \end{align}
Condition~\eqref{Eqn_condition_S} estimates the total prevailing worth of the members
of~$Q$, as proportional to their fractional contribution towards the service capacity
of~$C$, i.e.,~$N_Q/N_C$. Condition~\eqref{Eqn_condition_S_pt2} is the final
stage check on split feasibility as discussed above. Note that~$\hat{\P}$ is the new partition that emerges after the split when opponents remain in their original coalitions. 

Applying~\eqref{Eqn_unified_condition_S} and~\eqref{Eqn_unified_condition_S_pt2} to a merger, 
a configuration~$(\P,\Phi)$ is
blocked by a merger coalition~$Q = \cup_{C \in \mathcal{M}} C, \text{ for some } \mathcal{M} \subseteq \P$, 
if 
\begin{equation}
  \label{Eqn_condition_M}
\ulam_Q  >  \sum_{C \in \mathcal{M}} \lambda^\P_C \mbox{, and } \lambda^{\hat{\P}}_Q > \sum_{i \in Q} \phi_i, \text{ where } \hat{\P} = \{Q, \P \backslash \mathcal{M} \}.
\end{equation}
Note that the first condition in~\eqref{Eqn_condition_M} is identical to~\eqref{Eqn_unified_condition_S}, the only difference being that the prevailing worth of all the deviating members~$(\sum_{C \in \mathcal{M}}\lambda_C^\P)$ is assessed precisely, given that full coalitions are deviating. The second condition in~\eqref{Eqn_condition_M} is the same as~\eqref{Eqn_unified_condition_S_pt2}.
However, observe   that~$\sum_{i \in Q} \phi_i =  \sum_{C \in \mathcal{M}} \lambda^\P_C,$ and hence the second condition in~\eqref{Eqn_condition_M} is implied by the first, as~$\ulam_Q  \le \lambda^{\hat{\P}}_Q$.

Note that RB-PA and RB-IA differ only in the condition for blocking
due to a split. This is natural, since the net worth of coalitions
 $\{\lambda_C^\P\}_{C \in \P}$ is often common knowledge,
whereas the internal payoff allocation within a coalition can often be
confidential.

Having defined our new notions of stability, we now consider each notion
separately, and characterize the resulting stable configurations. We
begin with RB-IA, which appears to admit a broader class of stable configurations.

\subsection{Stable configurations under RB-IA}

Our first result is that all configurations involving partitions
of size three or more are unstable. In other words, only monopolies or
duopolies can be stable (proof in Appendix~\ref{appendix_C}).

\begin{theorem}
\label{Thm_duo_mono}
\textit{Under the RB-IA rule, any configuration~$(\P,\Phi)$ with~$|\P| \geq 3$ is not stable. } 
\end{theorem}

The proof sheds light on why
configurations with~$|\P| \geq 3$ are unstable -- they are blocked by any merger leading to a~$2$-partition; this is because of the economies of scale arising from such a merger (as shown in Theorem~\ref{Thm_WE}.$(ii)$), and the pessimal anticipation rule.

Next, we move to the two remaining possibilities:
stable configurations involving the grand coalition, and those
involving 2-partitions.

 {\bf Grand Coalition:} Defining~$\P_G := {\cal N}$ as the
grand coalition, it is clear that any configuration of the
form~$(\P_G,\Phi)$ can only be blocked by a split. 
 We now show that
unless a single agent owns at least half the total service capacity
of the system, such a block is always possible. In other words, any
configuration involving the grand coalition is unstable, unless there
is a single `dominant' agent. On the other hand, if there is a single
agent who owns at least half the service capacity, we show that there
exist stable configurations of the form~$(\P_G,\Phi)$
(see Appendix~\ref{appendix_C} for proof).

\begin{theorem}
  \label{Thm_GC} 
Under the RB-IA rule: 
\begin{enumerate}[label=(\roman*)]
    \item If~$N_1 < \sum_{ i \neq 1}N_i$, then
  there exists no payoff vector $\Phi$ consistent with ${\P}_G$, such
  that $({\P}_G, \Phi)$ is stable.
  \item If $N_1 \ge \sum_{ i \neq 1}N_i$, then there
  exists at least one payoff vector $\Phi$ consistent with ${\P}_G$, such
  that $({\P}_G, \Phi)$ is stable. Specifically, any configuration $({\P}_G, \Phi)$  satisfying the following is stable:  
\begin{equation}
   \phi_1 \ge \max \left  \{ \ulam_C: C \subsetneq {\cal N} \mbox{ and } 1 \in C \right \}. 
   \label{Eqn_payoff_RB_GC}
\end{equation}
\end{enumerate} 
\end{theorem}
To prove part~$(i)$ of the above theorem, we show that for any payoff vector, there exists a coalition with~$n-1$ players that blocks the grand coalition (details in Appendix~\ref{appendix_C}).
For part~$(ii)$, note that only coalitions containing player~$1$  satisfy condition~\eqref{Eqn_condition_S} and hence are  potential blocking coalitions under RB-IA. Therefore, if player~$1$ is given a large enough allocation (as in~\eqref{Eqn_payoff_RB_GC}) in the grand coalition, it does not have an incentive to deviate, either alone or as part of a group.

{\bf Duopolies:} We are now left to examine the stability of duopolies, i.e.,~$2$-partitions, under the RB-IA rule.
Duopolies can, without loss of
generality, be represented as~$\P = \{C_1,C_2\},$ with~$k := N_{C_1} \geq
N_{C_2}.$ In the following, we identify a family of stable duopolies under RB-IA rule. An interesting property of the stable configurations we identify is that, the stability does not depend upon the payoff vector,~$\Phi$. Instead, it only depends upon the specifics of the partition (however this is not true for all partitions).  
\textit{This insensitivity 
to the payoff vector is not seen under the RB-PA rule.}
We begin by defining some preliminaries.

\textit{\textbf{Stable partition:}
  A partition~$\P$ is stable if all configurations
  involving it are stable, i.e., configuration~$(\P, \Phi)$ is stable
  for any~$\Phi \in {\bm \Phi}^\P.$}

 By
Theorem~\ref{Thm_WE},~$\lambda_{C_1}^\P = \lambda^\P_k$ is the unique zero of the following
function (see~\eqref{Eqn_WE_properties}):
\begin{equation}h(\lambda) := \frac  { \lambda^k } { k! } \sum_{j=0}^{N-k} \frac{
  (\Lambda-\lambda)^j }{j!}  -
\frac{{(\Lambda-\lambda)}^{N-k}}{(N-k)!}  \sum_{j=0}^{k}
\frac{\lambda^j} {j!}. \nonumber\end{equation} 
Now, we define~$\Psi(k; \Lambda):= \lambda_k^\P/k$ as the offered load (or market size) per
server of the larger coalition. Finally, define
\begin{eqnarray}
\label{def_k_star}
  k^{*}(\Lambda) := {\arg \max}_{k}
  \Psi(k ; \Lambda).  \label{Eqn_kstar}
\end{eqnarray}
Note that~$k^*(\Lambda)$ is the set of values of~$k$ that maximizes
the per-server offered load of the larger coalition among~$2$-partitions.

Let~${\mathbb C}^* := \{ C \subsetneq {\cal N}: N_C \in k^* (\Lambda) \}
$ be the set of coalitions~$C$, that can derive the maximum per-server
offered load among~$2$-partitions. In the following lemma, we provide a sufficient condition for
a class of partitions (recall any such partition is represented by $\P = \{C_1,C_2\}$) to be stable. 

\begin{lemma}
\label{Lemma_stable_two_RB-IA}
 Consider the RB-IA rule. A~$2$-partition~$\P$ is stable
  if there exists no coalition~$S \subset C_i \text{ for }
  i=\{1,2\}$ such that
 $  \frac{ \ulam_S } {N_S } > \frac{\ulam_{C_i}}{N_{C_i}}= \frac{\lambda_{C_i}^{\P}}{N_{C_i}}. $
\end{lemma}

The proof of the lemma follows directly from the definition of stability. Indeed, for~$2$-partitions that satisfy the hypothesis of the above lemma, none of the splits  are feasible (they violate~\eqref{Eqn_condition_S}); further, the merger of both coalitions (which leads to grand coalition) is also not feasible because of the constant sum nature of the game. A consequence of this lemma is the following (see Appendix~\ref{appendix_C} for the proof).

\begin{theorem}
  \label{Thm_two_partition}
Consider the RB-IA rule.
\begin{enumerate}[label=(\roman*)]
    \item There always exists a stable~$2$-partition.
  \item Any~$2$-partition~$\P$ with one
  of the coalitions from~$\mathbb {C}^*$ is a stable
  partition. 
  \item Additionally, any~$2$-partition~$\P= \{C_1,C_2\}$ (where~$N_{C_1} \ge N_{C_2}$) with no~$C \subsetneq C_1$ such that~$N_{C} > N/2$ is stable. 
    \end{enumerate}
\label{Thm_Prule_stability}
\end{theorem} 
Note that statement~$(i)$ directly follows from statement~$(ii)$ and the non-emptiness of $\mathbb{C}^*$. Statement~$(ii)$ follows as the duopolies identified here satisfy the hypothesis of Lemma~\ref{Lemma_stable_two_RB-IA}. A similar reasoning applies for statement~$(iii)$. 

From Theorem~\ref{Thm_Prule_stability}.$(iii)$, the duopolies with perfectly matched service capacities~($N_{C_1} = N_{C_2}$) are also stable; while from~$(ii)$ any duopoly with $N_{C_1} = k^*$ (see~\eqref{Eqn_kstar}) is stable.
Further, Theorem~\ref{Thm_two_partition} identifies a class of
\emph{stable partitions}, i.e., partitions that are stable for any consistent payoff vector. However, there can also exist duopolies that are stable only under certain consistent payoff vectors and unstable for others (see Section~\ref{sec_numerical_OR}).

In Section~\ref{sec_two_partition}, we provide a complete
  characterization of the class of stable partitions under RB-IA, in
 the heavy and light traffic regimes.

 \subsection{Stable configurations under RB-PA}
\label{sec_PA}

Next, we consider stable configurations under the RB-PA rule. Under this rule, 
we show that only configurations involving~$2$-partitions can be stable,
i.e., configurations involving the grand coalition, or
involving~$k$-partitions with~$k \geq 3$ are always
unstable. In contrast, for the RB-IA rule, recall that the grand coalition  is stable   under certain conditions. Moreover, also in contrast to RB-IA, the stability/instability of duopoly configurations under RB-PA appears to depend on the associated payoff vector.

We begin by characterising the space of stable allocations under RB-PA.
From \eqref{eq_def_RBPA},
it is easy to see that a stable payoff vector lies in the polyhedron~\eqref{eqn_polyhedra} defined below.
\begin{lemma}
  {\bf[Polyhedral Characterisation]} Given any partition~$\P$, stable allocations lie in the polyhedron defined by
  \begin{equation}
  \label{eqn_polyhedra}
      \sum_{i \in Q} \phi_i \geq \ulam_Q \mbox{ for all } Q \subseteq C_j \in \P \text{, and for all } j.
  \end{equation}
\end{lemma}

It is clear from the above lemma that RB-PA does not admit stable partitions (unlike RB-IA). In other words, stability under RB-PA is tied to the payoff vector. Interestingly, stable partitions under RB-IA, paired with a special payoff vector (defined next) are stable; see Theorem~\ref{Thm_stable_config_ruleB}.

The \emph{proportional} payoff
vector $\Phi^{\P}_p,$
associated with any partition~$\P$, assigns to each member a payoff in proportion to the number of servers they bring to the
coalition: 
\begin{eqnarray}
\label{Eqn_PSA}
\phi^{\P}_{p, i}  =  \frac{N_i}{ \sum_{j \in C} N_j }   \lambda_C^\P \mbox{ for any } i  \in C \in \P.
\end{eqnarray}

Our results for the RB-PA rule are summarized as follows (see Appendix~\ref{appendix_C} for the proof).
\begin{theorem}
  \label{Thm_stable_config_ruleB}
Under the RB-PA rule:
\begin{enumerate}[label=(\roman*)]
    \item No configuration involving the grand coalition is
  stable.
  \item No configurations involving~$k$-partitions, for~$k \geq 3$ are stable.
  \item There exists at least one~$2$-partition~$\P$ such that~$\left(\P,\Phi_p^\P\right)$ is stable. Specifically, consider any stable~$2$-partition~$\P$ under the RB-IA rule. Then~$\left(\P,\Phi_p^\P\right)$ is stable under the RB-PA rule. 
Further, there exists a neighbourhood~${\cal B}^\P_p$ of the payoff vector~$\Phi_p^\P$ such that~$(\P,\Phi)$ is stable for all~$\Phi \in {\cal B}^\P_p$.
\end{enumerate}
 \label{Thm_R_rule_stability}
\end{theorem}
Like RB-IA, RB-PA also does not admit any stable configurations involving~$3$ or more coalitions. Moreover, under RB-PA, the grand coalition is  also unstable for all payoff vectors (unlike RB-IA, which admits payoff vectors that stabilise the grand coalition under certain conditions).
Finally, turning to duopolies, Theorem~\ref{Thm_stable_config_ruleB} conveys that partitions that are
stable under the RB-IA rule (irrespective of the associated payoff
vector), are also part of stable configurations under RB-PA, but under
a restricted class of payoff vectors. Specifically, the payoff vectors
we identify are `close' to proportional allocations.

Next we investigate  
other natural payoff structures  that also induce stability under RB-PA. In particular, we consider a payoff vector inspired by the classical Shapley value.  

\textbf{Shapley value: }Shapley value  is one of the well-known sharing concepts used in cooperative game theory~(\cite{narahari}). 
We begin by defining an extended version of Shapley value for partition form games, to divide a coalition's worth among its members~(\cite{aumann1974cooperative}). Under this extension, we treat each coalition~$C_i$ in the partition as a `grand coalition', define a suitable `worth'~$\nu_C$ for each~$C \subset C_i$, and then use the usual definition of Shapley value to obtain individual shares of the players in~$C_i$. Formally, for any~$j \in C_i$ , 
\begin{eqnarray}
		\label{Eqn_SV_OR}
		\phi_{s,j}^\P := \sum_{C \subseteq C_i, j \notin C} \frac{|C|!(|C_i|-|C|-1)!}{|C_i|!} \left[\nu_{C \cup \{j\}} - \nu_C \right], 
	\end{eqnarray} where $\nu_C$ is defined using \textit{pessimal anticipation} as below: 
 \begin{equation}
    \nu_C = \lambda_C^{\P'},  \text{ where } \P' = \P \backslash \{C_i\} \cup \{C, C_i \backslash C\}.   \label{eq_subcoalition_worth}
\end{equation}
Note that~$\nu_C$ is defined as the payoff obtained by~$C$ when (i) players outside of~$C_i$ remain attached to their original coalitions (as in the Cournot equilibrium), and (ii) the players in~$C_i \setminus C$ form a single competing coalition (in the spirit of pessimal anticipation).

Next, we present some contrasting results (compared to Theorem~\ref{Thm_R_rule_stability}) for a small number of service providers, for any~$2$-partition~$\P = \{C_1,C_2\}$ (proof in Appendix~\ref{appendix_C}).
  
\begin{theorem}
   \label{Thm_SV_small}
 Under the RB-PA rule, with the Shapley payoff vector~$\Phi^\P_s$ as defined in~\eqref{Eqn_SV_OR} and~\eqref{eq_subcoalition_worth},
  \begin{enumerate}[label=(\roman*)]
      \item for~$n=3$, the configuration~$\left(\P,\Phi_s^\P\right)$ is stable for any~$2$-partition~$\P$, and
      \item for~$n=4$, the configuration~$\left(\P,\Phi_s^\P\right)$ is stable for any~$2$-partition~$\P$ such that~$|C_1| = |C_2| = 2$. 
  \end{enumerate}
  \end{theorem} 
 Note that Theorem~\ref{Thm_SV_small} establishes the stability of certain~$2$-partitions under the Shapley payoff vector that are not covered in Theorem~\ref{Thm_R_rule_stability} under the proportional payoff vector (for~$n = 3,4$). Specifically, under the Shapley payoff vector, any~$2$-partition for~$n=3,$ and any~$2$-partition with equal-sized coalitions for~$n=4,$ is stable. In contrast, recall that the~$2$-partitions that are shown to be stable under the proportional payoff vector depend on the number of servers within each coalition (see Theorem~\ref{Thm_R_rule_stability}). We present a few examples in Section~\ref{sec_numerical_OR} to demonstrate these contrasts numerically.

 \section{Stable Duopolies: Heavy and Light Traffic }
\label{sec_two_partition}

In this section, we provide a complete characterization of stable partitions under RB-IA, and stable configurations under RB-PA with the proportional payoff vector, in heavy and light traffic regimes. Specifically, we provide the necessary and sufficient conditions for stability, as~$\Lambda \uparrow \infty$ (heavy traffic) and~$\Lambda \downarrow 0$ (light traffic), with other system parameters remaining unchanged.
 
 Our analysis presents interesting contrasts between the heavy and light traffic regimes. In heavy traffic, we find that \textit{all} duopolies form stable partitions under RB-IA and stable configurations (with the proportional payoff vector) under RB-PA. Intuitively, this is because economies of scale discourage splits in heavy traffic; as we show in Lemma~\ref{lem_psi_inc} in Appendix~\ref{appendix_C}, the per server utility of the larger coalition \textit{increases} with the number of servers it possesses (Interestingly, this is a `second order' effect; per server scales as~$\nicefrac{\Lambda}{N}$ for both coalitions in heavy traffic (see Lemma~\ref{lem_accuracy_heavy_tight} in Appendix~\ref{appendix_C}).). In contrast, in light traffic, we find that only duopolies where the two coalitions are `closely matched' in the number of servers they possess, are found to be stable. Intuitively, this is because economies of scale get significantly diluted in light traffic, discouraging any coalition from becoming `too large'.

\subsection{Heavy Traffic}  
Our main result in heavy traffic is the following (proof in Appendix~\ref{appendix_C}).
\begin{theorem} 
\label{Thm_heavy}
There exists  a ~${\bar \Lambda}$ such that for all~$\Lambda \ge {\bar \Lambda}$, the following holds: given any~$2$-partition~$\P,$
 \begin{enumerate}[label=(\roman*)]
     \item $\P$ is a stable partition under RB-IA, and
     \item $\left(\P,\Phi_p^\P\right)$  is a stable configuration under RB-PA. 
 \end{enumerate} 
 \end{theorem}
 This result can be interpreted as follows. Note that due to the constant sum nature of the game, duopolies can never be blocked due to a merger. Thus, our stability analysis hinges on the feasibility of splits. Specifically, we prove Theorem~\ref{Thm_heavy} by showing that given any~$2$-partition~$\P$,
 \begin{enumerate}[label=(\alph*)]
     \item for any consistent payoff vector~$\Phi,$ the configuration~$(\P,\Phi)$ cannot be blocked by a split under RB-IA, and
     \item the configuration~$\left(\P,\Phi_p^\P\right)$ cannot be blocked by a split under RB-PA.
 \end{enumerate}
 These statements in turn follow from the fact that in heavy traffic, the per-server offered load~$\Psi(k)$ of the larger coalition increases monotonically with the number of servers~$k$ it possesses, i.e., its service capacity (see Lemma~\ref{lem_psi_inc} in Appendix~\ref{appendix_C}). In other words, economies of scale persist in heavy traffic. Indeed, the above monotonicity property, which is proved by exploiting the analytical extension of the Erlang-B formula to real-valued service capacities (see~\cite{jagerman}), renders condition~\eqref{Eqn_condition_S} for a split under RB-IA, and condition~\eqref{eq_def_RBPA} for a split under RB-PA, invalid.
	
\subsection{Light Traffic}
Next, we consider the light traffic regime and our main result here is  (proof in Appendix~\ref{appendix_C}):
\begin{theorem}
 Let~${\mathfrak P}$ denote the space of~$2$-partitions~$\P = \{C_1,C_2\}$ (where~$N_{C_1} \ge N_{C_2}$) satisfying the following condition:
 there does not exist~$C \subset C_1$ such that~$N_{C} > N/2.$ There exists~${\underline \Lambda} >0$, such that for all~$\Lambda \le {\underline \Lambda}$,
  \begin{enumerate}[label=(\roman*)]
  \item $\P$ is a stable partition under RB-IA  if and only if~$\P \in \mathfrak{P}$, and
  \item $\left(\P,\Phi_p^\P\right)$ is a stable configuration under RB-PA  if and only if~$\P \in \mathfrak{P}$.
  \end{enumerate}
   \label{low_traffic}
     \end{theorem}
Theorem~\ref{low_traffic} highlights that the~$2$-partitions that are stable under RB-IA and form stable configurations (with the proportional payoff vector) under RB-PA are those where the service capacities of the two coalitions are nearly matched. Formally, the larger coalition~$C_1$ should not have a sub-coalition~$C$ with more than half the total service capacity. In particular, note that duopolies with perfectly matched service capacities~($N_{C_1} = N_{C_2}$) also satisfy this condition. Intuitively, the result holds because in light traffic, the larger coalition corners almost the entire offered load (i.e.,~$\nicefrac{\lambda_{C_1}^\P}{\Lambda} \to 1 \text{ as } \Lambda \to 0$); see Lemma~\ref{lem_light_new} in Appendix~\ref{appendix_C}. 

Our results in the heavy and light traffic regimes shed light on the impact of congestion (via the total offered load~$\Lambda$, a.k.a., the \textit{market size}) on coalition formation. In light traffic, the per-server utility of the larger (by service capacity) coalition~$\Psi(k)$ \textit{decreases} with its service capacity~$k$ (as the larger coalition captures almost the entire~$\Lambda$, irrespective of~$k$). This in turn encourages duopolies where the service capacities of the two coalitions are closely matched (even though the larger coalition corners most of the total utility). On the other hand, in heavy traffic, the per-server utility of the larger (by service capacity) coalition~$\Psi(k)$ \textit{increases} with its service capacity~$k.$ These economies of scale induce stability in all duopolies, including those that have coalitions with highly asymmetric  service capacities. This suggests that in general, at moderate congestion levels, the per-server utility of the larger (as before, by service capacity) coalition peaks at an intermediate value of~$k$ between~$\nicefrac{N}{2}$ and $N,$ encouraging the formation of moderately asymmetric duopolies. This is consistent with what we find in our numerical experiments  (see~\ref{fig:stable partitions}).

Finally, it is important to note that we are able to provide necessary and sufficient conditions for stability under RB-I and RB-PA in heavy and light traffic regimes; in contrast, we could only   provide sufficient conditions for stability (see Theorems~\ref{Thm_two_partition} and~\ref{Thm_stable_config_ruleB}) outside of these limiting regimes.

\section{Dynamic coalition formation game}
\label{sec:dynamic}
In this section, we consider a dynamic version of the game discussed in
the previous sections. We begin with a queueing system and agents
operating in some configuration.  The agents are constantly on the
lookout for greener pastures, and would stop their
quest only if they are satisfied with the existing configuration.

Agents may consider joining existing collaborations or may consider
splitting from some of them.  The (new) payoffs derived by the agents
after the new collaborations (if any), depend upon the previous
payoffs and the value of the new operational
arrangement/coalition. Depending upon the new payoffs, some of
the agents might again consider another movement.  On the other hand,
the system might settle, if all the agents are satisfied with the
configuration.  We study this aspect by considering a sequence of
dynamic coalition formations.


{\bf Dynamics:} The system starts with some operational
arrangement given by $\P_0$ and with a payoff vector
$\Phi^{0} = [ \phi_1^{0}, \cdots, \phi_n^0]$. If the configuration
$\left(\P_0, \Phi^{0}\right)$ is stable as defined in previous sections, it is
not beneficial for any member to consider any (coalitional) deviation
and hence the system does not undergo any change.  If that is not the
case, some members of the partition merge/split.
 
There could be more than one movement (merger/split)  that may be successful,  under both the assessment
rules (RB-PA and RB-IA). We assume that
  any such blocking coalition~$Q$ is equally likely to form, causing
  the system to evolve to a new partition, say $\P_1.$ In case of the
RB-IA rule, any new payoff vector $\Phi^{1}$ that satisfies
 $\phi^1_i > \phi^0_i$ for all $i \in Q$ would
suffice.  We discuss the RB-PA rule towards the end of this section.

The system stops if the new configuration $\left(\P_1, \Phi^{1}\right)$ is
stable.  If not, it switches to yet another configuration
$\left(\P_2, \Phi^2\right)$ randomly (and equally likely among all possible
movements) in a similar way. This  evolution continues until
stopped by a stable configuration. Our aim is to understand if such a
limit stable configuration exists.

By the results of the previous section we have stable configurations
only with 2-partitions or grand coalition and we immediately have the
following result under the following assumption:

 \begin{figure}[h]
 \vspace{-4mm}
 \centering
   \centering 
    \includegraphics[trim = {2cm 0cm 0cm 0cm}, clip, scale = 0.4]{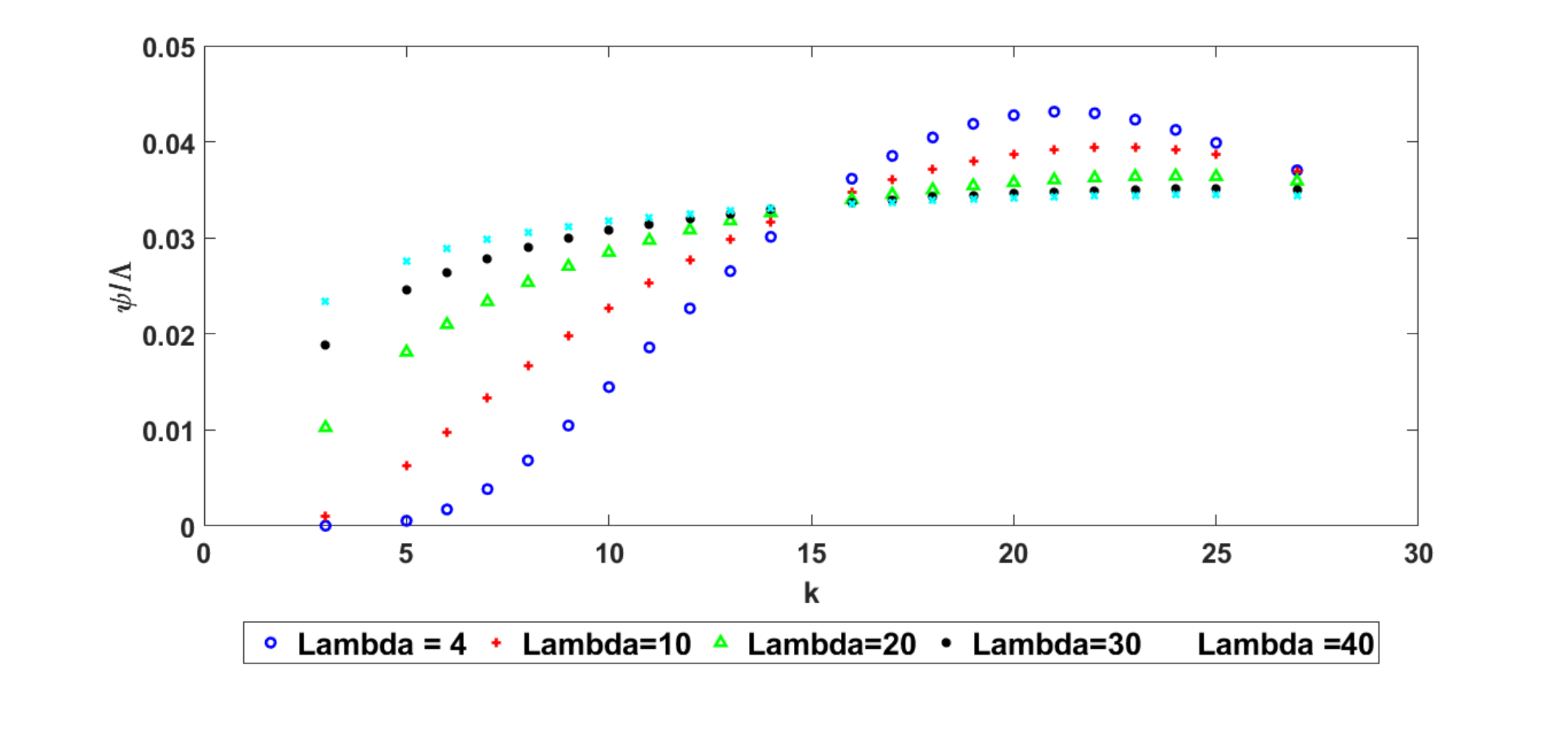}
    \vspace{-4mm}
    \caption{$\Psi(k;\Lambda)/\Lambda$ v/s $k,$  with $[N_i] = [9, 7, 6, 5, 3].$ 
      \label{fig:psi_vs_k}}
   \vspace{-3mm}
 \end{figure}

{\bf A.1)} 
%
 %
If $C$ is any coalition that does not contain any element of  ${\mathbb C}^*$,
i.e., if $C \cap C^* \ne C^*$ for all $C^* \in {\mathbb C}^*$, then we
have the following:
$$
\frac{ \ulam_S }{N_S}  < \frac{ \ulam_C }{N_C}  \mbox{ for all strict subtsets,  }  S \subset C.
$$

Basically, this assumption ensures that any 2-partition that is not stable, necessarily contains a coalition that is a strict superset of an element from ${\mathbb C}^*$.  From simulations, we
have seen that this assumption is satisfied by our queuing system for
all the cases that we considered (for example, see
Figure~\ref{fig:psi_vs_k}) and further by Theorems~\ref{Thm_heavy} and~\ref{low_traffic}  can be shown to hold  under heavy
and light traffic conditions. Under this assumption, we can show that
the dynamics stops after finite number of movements (proof in Appendix~\ref{appendix_C}).
\begin{theorem} {\bf [Convergence]}
  Assume {\bf A}.1. Then the random dynamics under RB-IA rule
  converges to one of the stable configurations under RB-IA rule in
  finite number of steps, with probability one.
  \label{dynamic_conv}
\end{theorem}

The above theorem proves that the random dynamics under RB-IA rule is
stopped in finite number of steps with probability one, and the limit
is a stable partition.
However under this imprecise anticipation rule, it is important to observe that the payoff vector at the stopped configuration can be  arbitrarily skewed (as also indicated in Theorem \ref{Thm_two_partition}).  

\subsection*{Dynamics under RB-PA rule}
Under RB-PA rule, the random dynamics behaves exactly similar to RB-IA
rule (as described in the proof), however it may not stop even after
touching a stable 2-partition, stable under RB-IA. As seen from Theorem
\ref{Thm_stable_config_ruleB} for RB-PA rule, the payoff vector is
equally important in the definition of stable configuration.

This shows the importance of appropriate reallocation of individual
shares after the new move towards the stability of the new system; it
is not sufficient to only ensure all members of the new
coalition derive positive increments, rather we will require that the
new allocation matches the payoff vector in the corresponding stable
configuration.  As seen from Theorem \ref{Thm_stable_config_ruleB},
one of the payoff vectors that provides stable configurations is the proportional  payoff vector given by equation \eqref{Eqn_PSA}.  Thus one probably has to design reallocation
policies that converge towards the proportional payoff vectors for the dynamics
under the RB-PA rule to stop.

Alternatively there might be other payoff vectors which would also
form a part of the stable configurations and they could be the ones at
limit.  We would study this aspect in the future, but for now we could
say that one can't have partitions of size greater than 2 or the grand
coalition (when none of the agents dominate) to be a part of the limit
(stable) configuration (if one exists), in view of Theorems
\ref{Thm_duo_mono} and \ref{Thm_GC}.  We can also say that the
dynamics stops if it hits upon a configuration with stable 2-partition
and the corresponding proportional payoff vector \eqref{Eqn_PSA}.

\subsection*{Dynamics under GB-PA rule}  It is not difficult to show
that the dynamics does not stop  even if it starts
with  or hits a stable configuration under RB-PA rule identified in Theorem \ref{Thm_R_rule_stability}.  It is interesting
to observe that the dynamics toggles between stable configurations of
RB-PA rule, even when it starts with one of them.

\section{Numerical Case Studies}
\label{sec_numerical_OR}
		
In this section, we present some numerical case studies that illustrate our key findings. 
Importantly, we also consider examples for which the conditions of our theorems are not satisfied; these provide additional insights. 
We numerically compute~${\lambda}_C^\P$  for various~$C$ and~$\P$ using zero finding algorithms and then compute~$k^*$ of~\eqref{Eqn_kstar} or  use equations~\eqref{Eqn_condition_S}-\eqref{Eqn_condition_S_pt2} or~\eqref{eq_def_RBPA} to determine the stable configurations.

 \textbf{RB-IA rule:} Recall that Theorem~\ref{Thm_two_partition} provides sufficient condition for a \textit{stable partition} under RB-IA, i.e., a partition that is stable under any consistent payoff vector. Here, we illustrate that RB-IA also admits stable configurations that are not supported by stable partitions. Consider the example with~$\Lambda = 13$ and~$4$ service providers having service capacities:~$N_1 = 10,$ $N_2 = N_3 = N_4= 2$. Note that the partition~$\P = \{\{1,2,3\},\{4\}\}$ does not satisfy the hypothesis of Theorem~\ref{Thm_two_partition} (in this case,~$k^* =\{12\})$. Moreover, configuration~$\left(\P,\Phi_p^{\P}\right)$ is blocked by~$Q = \{1,2\}$ as split~$Q$ satisfies~\eqref{Eqn_condition_S}, while,~$\Phi_p^\P$ and~$Q$ satisfy~\eqref{Eqn_condition_S_pt2}. Thus, $\P$ is not a stable partition. However, the configuration $(\P,\Phi)$ is stable for the following set of payoff vectors:
\begin{eqnarray}
  \left \{\Phi : \phi_1 \ge \lambda_{\{1\}}^{\{\{1\},\{2,3\},\{4\}\}}, \ \phi_1+\phi_2 \ge \lambda_{\{1,2\}}^{\{\{1,2\},\{3\},\{4\}\}}, \  \phi_1 + \phi_3 \ge \lambda_{\{1,3\}}^{\{\{1,3\},\{2\},\{4\}}, \right. & \nonumber \\ 
 \left. \phi_2 + \phi_3 \ge  \lambda_{\{2,3\}}^{\{\{1\},\{2,3\},\{4\}\}}, \ \phi_1+\phi_2+\phi_3 = \ulam_{\{1,2,3\}}   \right \}. \nonumber 
\end{eqnarray}
It can be checked that this set is indeed non-empty.
This demonstrates that it is possible for a partition to be stable under some but not all consistent payoff vectors.
 
\begin{figure}[!ht]
\begin{minipage}{.55\textwidth}
  \centering
    \begin{tabular}{|c|c|c|}
    \hline 
\multicolumn{1}{|c|}{\multirow{2}{*}{$N_1$} }                                       & \multicolumn{2}{c|}{Unstable $2$-partitions}            \\ \cline{2-3}
 \multicolumn{1}{|c|}{} &     Proportional & Shapley   \\ \hline
         $2-9$  & None & None \\ \hline
         $10-17$ &  $w \in \{14,\cdots,21\}$  &   None   \\ \hline
         $18-40$ & $w \in \{20,\cdots,44\}$ & None \\ \hline
    \end{tabular}
    \vspace{6mm}
    \captionof{table}{Unstable partitions under RB-PA for \\ different allocation rules with~$N_2=N_3=N_4=2, \\ \Lambda = 13$}
    \label{tab:my_label}
\end{minipage}
\begin{minipage}{.45\textwidth}
     \centering     \includegraphics[trim = {3cm 8cm 0cm 8cm}, clip, scale = 0.5]{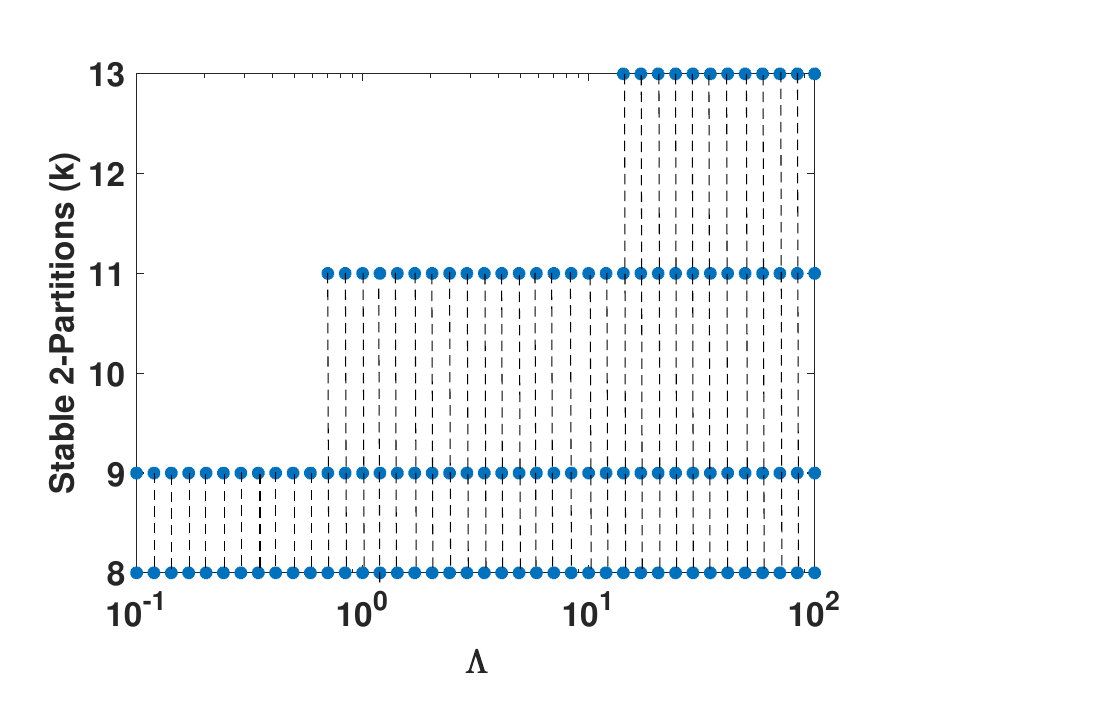}
     \captionof{figure}{Set of stable partitions (under RB-IA) v/s~$\Lambda$  (on log scale) for~$N_1 = 7, N_2 =  N_3 = N_4 = N_5 = 2 $}
     \label{fig:stable partitions}
\end{minipage}
\end{figure}

 \textbf{RB-PA rule:} Next, we study the RB-PA rule. Our aim is to first compare the stability of two allocation mechanisms---proportional allocation and Shapley value. Consider the following example with~$\Lambda = 13$ and~$4$ service providers. Here~$N_1$ is varied from~$2-41$, while  the remaining service capacities are fixed at~$N_2 = N_3 = N_4= 2$. 
Table~\ref{tab:my_label} presents the set of $2$-partitions that are unstable under each allocation mechanism. Here,~$w$ denotes the number of servers in the coalition that includes provider~$1$. For example, the second row considers the cases where~$N_1$ lies between~$10$ and~$17$. In all these cases, the proportional payoff vector renders   those~$2$-partitions with~$w \in \{14, 15, \cdots, 21\}$ unstable, whereas all two partitions are stable under Shapley value. This suggests that Shapley value renders more partitions stable in comparison to the  proportional payoff vector.

We consider another such example with 3 agents,~$\Lambda=100,$~$ N_1=80,$~$ N_2=20$ and~$N_3=5$. By Theorem~\ref{Thm_SV_small}.$(i)$,~$\left(\P,\Phi_s^\P\right)$ is stable for~$\P = \{\{1,2\},\{3\}\}$. However, we find (numerically) that~$\left(\P, \Phi_p^\P\right)$ is not stable (it is blocked by~$Q = \{1\}$). (Numerically, we find that~$k^*=\{80\}$, implying~$\P$ does not satisfy the hypothesis of Theorem~\ref{Thm_R_rule_stability}, as expected.) 


\textbf{Impact of congestion: }In Figure~\ref{fig:stable partitions}, we consider a final example that demonstrates how the set of stable partitions under RB-IA varies with the market size~$\Lambda.$ Here, we consider five service providers with service capacities~$N_1 = 7, N_2= N_3 = N_4 = N_5 = 2.$ 
Note that the left and right extremes in the figure are consistent with the light-traffic and heavy traffic results (Theorems~\ref{low_traffic} and~\ref{Thm_heavy} respectively). In particular, in light traffic, the only stable duopolies are those that are nearly matched with respect to service capacity---one where the dominant coalition is composed of agents~$2, 3, 4,$ and~$5$ ($k = 8$) and another the dominant coalition is composed of agent~$1$ and one of the remaining agents ($k=9$). In heavy traffic, all duopolies are stable. Importantly, the figure shows that the set of stable duopolies grows monotonically with~$\Lambda.$

\section{Summary}

We consider\rev{ed} a Erlang-B (lossy) queueing system with several strategic service providers with different server capacities.  Each service provider is on lookout for collaboration opportunities that improve their individual payoffs. The customer base responds to any operational arrangement formed by such collaborations, the customer arrivals are split across various operational units according to the well known Wardrop equilibrium that equalizes the steady state blocking probability of all the units. Any operational configuration is challenged by new coalition, and the former is dissolved if the new coalition finds it beneficial. A configuration is stable if there is no coalition to challenge it.  We have an `impossibility result' where we show\rev{ed} that no partition is stable under this classical notion of stability.  We defined  new and more meaningful notions of stability where only blocking via mergers or splits is allowed. Using these notions, we show\rev{ed} that the duopolies are the predominantly stable partitions, which highlight\rev{ed} that in competitive service systems enjoying statistical economies of scale, coalition formation games have very distinct equilibria when the total payoff across agents is a constant. We also explore\rev{d} the impact of the overall congestion on the stable partitions, by analyzing light and heavy traffic regimes. Finally, we present\rev{ed} some initial ideas on the dynamic version of the same game. 





\chapter{Dual Opportunistic Fair mmWave Scheduler: Position-Aided Beam Alignment and User Assignment }
\label{chap_wireless}

In this chapter, we consider a central moderator (the Base Station) whose aim is to allocate the resources among the users, while ensuring a prescribed level of fairness, in the context of 5G/6G wireless networks.

\section{Introduction}
\label{sec_intro_CN}

With the emergence of new technologies that support applications like the internet-of-things (IoT), high-definition (HD) 3D video, virtual
and augmented realities, etc., data traffic has increased significantly. There are two alternatives to achieve higher data rates, either by using large   bandwidth  or by increasing the transmit power. However, the transmit power cannot be increased arbitrarily  due to health guidelines. Thus the only feasible option is to increase the bandwidth. Towards this, the new generation networks propose to use  Millimeter Waves (mmWaves) in the spectral range of 24GHz to 40GHz (\cite{5g}).   

In mmWave communications, the base station (BS) needs to align the beam in the  direction of the end user.  Beam alignment is a challenging task as the location of the users may not be known apriori and further could be varying continuously. Hence it is a time-consuming process, leading to the degradation of the system performance \cite{mmwave,orikumhi2018location,shokri2015beam,lee2019beam}. Authors in \cite{mmwave} argue  that  completely relying on the beam direction from the previous slot may not be effective, however starting the current beam search algorithm using the previous estimate can significantly reduce the time spent in aligning the beam. Towards this, we further propose to maintain fairly accurate estimates of  the user positions at BS. The previous alignment information and the current user position can significantly improve the alignment process. But of course, the improvement depends on the age of the available information. The main focus of this work is to study this precise aspect  and design a dual scheduler that ensures:  (i) optimal dynamic update of information regarding user positions, and (ii) optimal dynamic assignment of channels  to various users in different time slots. Our aim is also to include the notion of fairness in optimality. 

Opportunistic schedulers (\hspace{-1.4mm}~\cite{tejas,kushner,cellular,debayan}) are widely used in wireless networks to take advantage of  `diversity gain'; the channel conditions are sufficiently independent across slots and users. They exploit the fluctuations in users' channel conditions and allocate data channels to some of those in `good' condition.  In every slot, the BS seeks  channel estimate of each of the users and say selects the `best' user for data transmission. Such allocations can be referred as efficient decisions as they  result in an `efficient' solution that maximises  the total utility gained  across all users and time slots (e.g., see \cite{info_efficient}). However, this approach can regularly deprive the users with `inferior' channels (bad channel conditions with high probability), resulting in minimal utility accumulations for them. To 
incorporate fairness, 
some sub-optimal allocations (deviations from the efficient decisions) may need to be made for users with inferior channels. It is also important to balance efficiency and fairness to achieve optimal channel allocation, and hence deviations from efficient decisions should only be made when the loss incurred is minimal. Generalized $\alpha$-fair opportunistic  schedulers are designed precisely for this purpose: allocate a channel to deprived users when the opportunities are the `best', to an extent depending upon the `required' level of fairness. In fact the authors in \cite{mayur}  proved that the Price of Fairness (\rev{PoF}) reduces to a negligible value as the number of users increases. Thus, we aim to design a Dual opportunistic Fair  Scheduler (DoFS) that includes user-position update schedules towards optimal beam alignment. 

With position information based beam alignments (as in \cite{mmwave,orikumhi2018location}), the channel conditions of the same user across different slots may not be as diverse as in previous generation networks (where beam alignment was not required). So we also investigate if the available diversity can once again provide significant improvement via opportunistic schedulers. 

\begin{figure}[!h]
\vspace{-2.5mm}
    \centering
   \includegraphics[trim = {1cm 3cm 0cm 3cm}, clip, scale = 0.6]{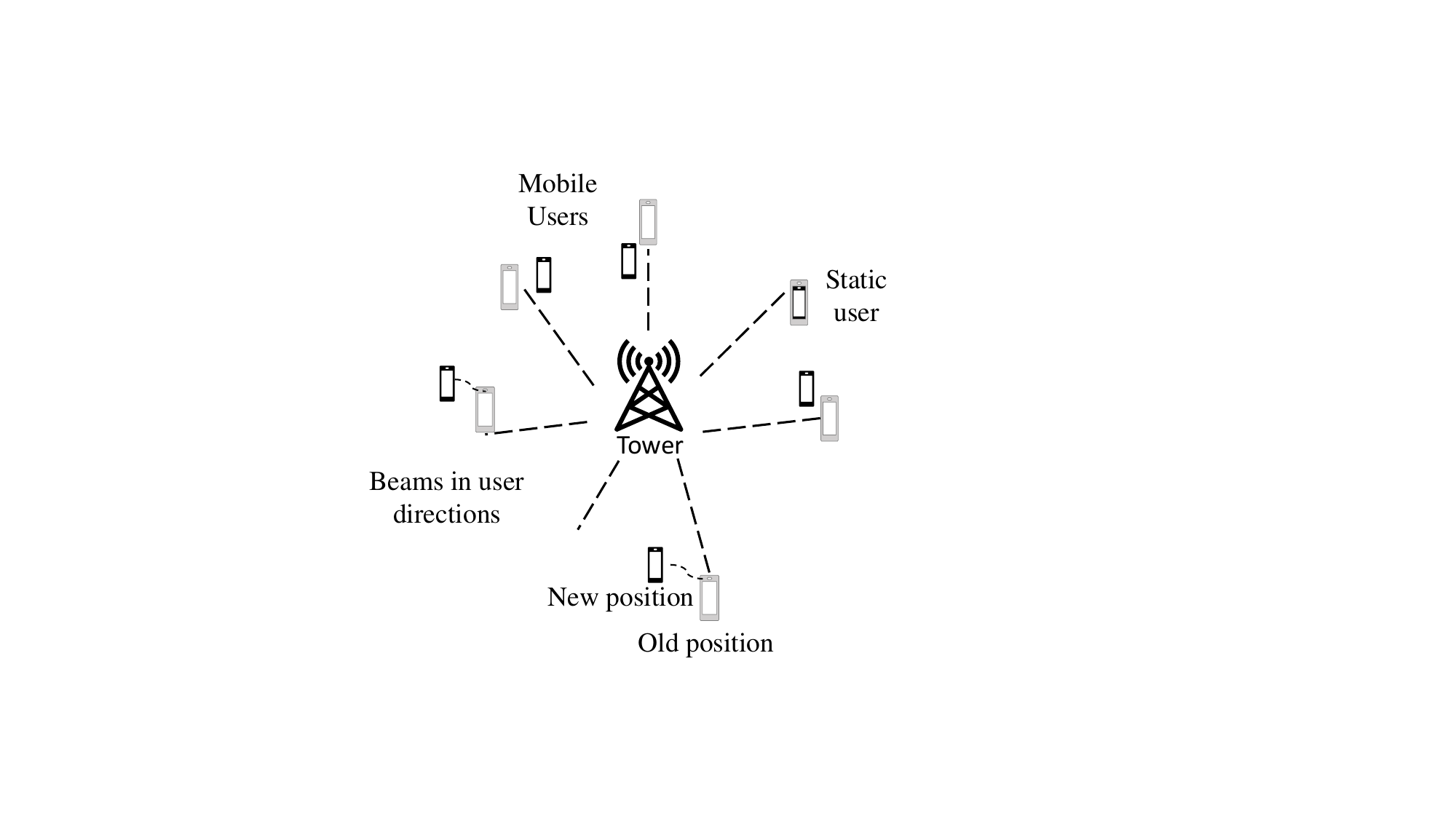}
    \caption{Beam alignment for multiple mobile/static users \cite{itc}}
    \label{fig:AoI}
    \vspace{-1mm}
\end{figure}

With mmWave transmissions and its desired accurate beam alignment, it is difficult to obtain the channel estimates from all the users, as desired for opportunistic schedulers.  However, to the best of our knowledge, the literature (e.g., \cite{ming,qureshi,irmak,mmwave,orikumhi2018location,shokri2015beam,he2022cross}) considers selecting one user in each slot and accurately aligning the beam towards the selected user prior to data transfer. The primary reason for our proposal of user-position-estimate aided beam alignment is to pave way for mmWave-opportunistic schedulers - \textit{we propose to maintain sufficiently accurate user position updates of each user at BS (see Figure \ref{fig:AoI})}, that  enables derivation of sufficiently good channel estimates for all the users in all the time slots. The quality of the position update estimates are recorded using the recently introduced metric Age of Information (AoI) (\cite{kaul}), which captures the time elapsed since the last position-information update. Our basic assumption is that the beam alignment times are negligible with accurate position updates.
Further, the users can instead transmit the alignment directions, if there are privacy concerns. 

 The wireless users are often mobile,
the BS needs  regular updates of the positions of all the users. Further, the frequency of the position-information updates can be different for different users, based on their individual mobility patterns. Thus the proposed dual scheduler should  consider various aspects like, opportunities provided by diverse channel conditions, fairness, and the influence of age of information and mobility patterns.  The precise dual tasks of the scheduler is to assign a user for position-update and another/the same user for data transfer in each time-slot in an optimal manner. 
In contrast to the existing literature on AoI that optimize average or peak AoI (e.g., \cite{yuan,kaul,kavitha}), we directly optimise the well-known $\alpha$-fair objective function (e.g., \hspace{-1.4mm}~\cite{kushner,cellular}) which is constructed using time-average utilities of all the users, which in turn are influenced by individual AoI trajectories and mobility patterns. 

In an initial conference paper \cite{itc}, towards this direction we consider a sub-optimal solution for the same problem--Markov Decision Process (MDP) based framework is applied only for age decisions, while the  data choices were according to  the  $\alpha$-fair scheduler of previous generation networks. In this study, we consider a dual scheduler in the true sense, that makes combined decisions in a dynamic fashion to optimise the given fairness-based (average cost) objective.

In Third Generation Partnership Project Technical Specification Group Radio Access Network (3GPP TSG RAN) meetings (for example, see \cite{3Gmeeting}) there has been an increased interest in the cellular community working towards beyond 5G and 6G networks, to provide the BSs with a more accurate user position -- our solution precisely banks on utilizing such updates. They propose  to make provisions in the next generation networks  to facilitate accurate  estimation of user position using alternate techniques (for example, they have included the Positioning Reference Signal (PRS) pilots which are specifically meant to improve the accuracy of the user position)  -- once user position estimates are readily available, the overhead in maintaining the position-updates 
 in our proposed schedulers will also be eliminated.



\subsection*{Our Contributions}
We derive solution of the proposed dual scheduler by  modeling it as an average cost Markov decision process (MDP). Departing from the regular solution approaches available in the literature, we propose a gradient based offline iterative algorithm under uni-chain assumption. The dual scheduler is parameterised by $\alpha$ which indicates the   level of fairness achieved by the scheduler. We provide near-closed-form expressions for the dual schedulers for the case with two   users, using an alternate approach as in \cite{debayan}. Finally, the opportunistic and non-opportunistic schedulers  are compared. We illustrate significant improvements and much smaller  Price of Fairness (\rev{PoF}) with opportunistic mmWave schedulers.

\subsection*{Related Literature}
Accurate  beam forming  requires  a beam search algorithm which incurs tens to hundreds of milliseconds overhead if all possible directions are scanned \cite{mmwave}. To reduce the overhead, current standard activities \cite{ieee2007ieee,wang2009beam} suggest a two-stage beam form technique. For a fixed bandwidth (given granulity of searching),  \cite{li2012efficient} suggest a new technique to replace the two-stage technique and reduce the beam alignment overhead. In \cite{patra2015smart}, a smart beam steering algorithm is proposed under user mobility, which uses knowledge of the previous feasible antenna sector pair to narrow the sector search space. Many other beam forming algorithms based on various techniques (for example, Kalman filtering in \cite{zhang2016tracking}, deep learning model in \cite{zhang2021deep,alkhateeb2018deep}, etc.) have been suggested in the literature for the case of single user. This set of papers focus on  beam forming techniques and not on resource allocation aspects.

There is a relatively limited  literature for mmWave scenarios with multiple users. In \cite{khalili2020optimal}, authors propose a non-interactive beam alignment procedure, that does not require feedback from the users,  to optimise the beam alignment overhead. The energy efficient beam alignment protocols are designed for the case of two users in \cite{hassan2018multi}; their  goal  is to minimise the power consumption during data transmission, subject to rate constraints for individual   users.
The authors in \cite{mmwave} consider the problem of minimizing the long run beam alignment overhead cost for the case with large number of mobile users. 
This set of papers discuss resource allocation aspect, however they assume one  user to be selected for data transfer in each slot, and beam alignment  in any slot is achieved only towards the selected user. Hence they do not consider  opportunistic aspects while allocating the resources. 

In a recent paper \cite{he2022cross}, the authors solve the joint optimisation problem of  user scheduling and beamforming subject to the requirement of per-user quality of service and the maximal allowable transmit power for multi-cell multi-user joint transmission networks. They also provide a table (\cite[Table 1]{he2022cross}) with recent related references, none of which consider the opportunistic aspect.

Conclusively, to the best of our knowledge, none of the papers consider opportunistic mmWave schedulers. Also, none of them discuss elaborately about   fairness aspects, like the spectrum of schedulers one for each level of fairness, measures of degradation of efficiency with fairness levels, etc.  

\section{Problem Definition and MDP Formulation}

Consider a system with a Base Station (BS) and a set of $N$ users, both stationary and mobile, labeled by~$n \in \Nc = \{1,\cdots, N\}$. Our focus is on the scenario where only one user can be served at a time. Due to the use of millimeter waves (mmWaves) for transmission, it is necessary to utilize narrow beams to establish a connection.  Further, the reliable connection can only be achieved when the narrow beam is sufficiently well-aligned with the users. Moreover, as mentioned in Section \ref{sec_intro_CN}, we would like to design \textit{opportunistic} schedulers, which rely upon channel estimates from all the users in any time slot. This requires sufficiently good alignment of beams towards all the users in any time slot. 

\begin{figure}[h]
\vspace{-3mm}
    \centering
    \includegraphics[trim = {0cm 7.5cm 0cm 2.5cm}, clip, scale = 0.5]{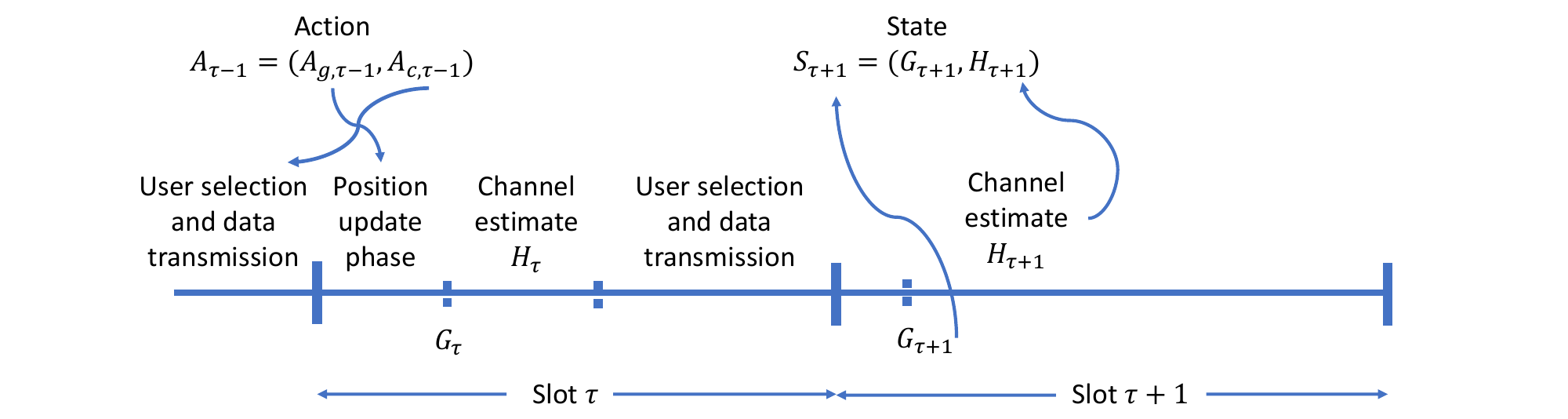}
    \caption{Various phases in a  slot, state and action description }
    \label{fig:state_desc}
\end{figure}

To address this issue, we propose maintaining precise estimates of the positions of all the users at the BS. It would be a large overhead to update position information of all the users in all time slots. We instead propose to update position of one of the users, and use the previous estimates for  others. 
These updates can be used by BS to align the beams in users' directions, collect their channel estimates, and assign data channel to one of them based on the  estimates.
 Thus each time slot has two scheduling decisions as described below,

\medskip
\textbf{Position update scheduler:} During the initial phase of each slot, the position of one of the users is updated, while maintaining the previous estimates for the remaining users. The quality of the position information is gauged using the recently introduced metric, Age of Information (AoI), which denotes the time elapsed since the user's last position update. Towards this, we utilize a vector \textit{$\G_\tau = (\sG^{(1)}_{\tau}, \cdots, \sG^{(N)}_{\tau})$ to represent the age of  position updates of each user in slot $\tau$}, where $\sG^{(n)}_{\tau} = 1$ implies that the position of user $n$ has just been updated -- basically, the beam alignment is perfect for that user.
Thus the age vector of previous slot $\G_{\tau-1}$ modifies to $\G_\tau$ as below, when the position of user $n$ is  updated in slot $\tau$ (see Figure \ref{fig:state_desc}),
\begin{eqnarray}
\sG^{(n)}_{\tau} = 1, \text{ and, } 
\sG^{(w)}_{\tau} = \min\left\{\sG^{(w)}_{\tau-1}+1, \ \bar{g}\right\}
\text{ for all } w \neq n \text{ with } \bar g > 1. 
\label{eqn_age_evolution}
\end{eqnarray}
Here, $\bar{g}$ represents the upper bound on the age of the position updates beyond which the beam alignment does not serve any purpose (channel conditions/estimates with such misalignment are close to zero). \textit{We briefly refer the position update decisions  as age decisions.} 
It is obvious that the chances for poor beam alignment increases with age, which in turn results in inferior channel conditions. Finally the set $\mathcal{G}$ consists of all possible realisations of age  vector $\rG$, which is described as below,
\begin{equation}
\label{eqn_age_set}
\begin{aligned}
    \mathcal{G} = \left \{ \rG =(g^{(1)},\cdots,g^{(N)}): \text{there exists } i_1,\cdots,i_N \text{ s.t. } 1=g^{(i_1)} <  g^{(i_2)}\le \cdots \le g^{(i_N)} \le {\bar g} \right. \\
   \left. \text{ with } g^{(i_k)} = g^{(i_{k+1})} \text{ iff } g^{(i_k)} = {\bar g} \right \}.
    \end{aligned}
\end{equation}

\medskip
\textbf{Data scheduler:} After the position update phase, the remaining time in every slot is utilized for data transmission, during which one of the users is allocated the data channel. In order to take advantage of the opportunities, as mentioned before, 
the BS aligns beams in the direction of the position estimates of each user to obtain their respective channel estimates, represented by $\H_{\tau} = (\sH^{(1)}_{\tau},\cdots,\sH^{(N)}_{\tau})$ for slot $\tau$. The transmission occurs at one of the  rates depending on the available code-books (\cite{zhao2015resource}) -- in fact we represent these rates directly by $\H_{\tau}$  and $\mathcal{H}$ represents the set of all such possible transmission rate vectors. Accordingly, we assume    $\{\sH_{\tau}^{(n)}\}$ to be 
independent across users and slots  (once conditioned on appropriate age vectors)  with support on finitely many values. The age of any user dictates the distribution of these channel estimates\footnote{The channel conditions depend upon the accuracy of alignment and then the estimation procedure provides the estimates of these channel conditions. We  assume that the estimation errors are almost negligible,   i.e.,  the channel estimates almost precisely represent the channel conditions. This assumption is reasonable (such assumptions are common in wireless literature \cite{zhao2015resource,chen2014distributed}) given the fact that we only need to have a coarse estimate (or discrete nature) on which code-book can be used for the given channel condition.  }.  In other words, when age $\sG^{(n)}_{\tau}$ is high,  $\sH^{(n)}_{\tau}$ takes smaller values with higher probability. Further, this probability depends upon the mobility of the user, i.e., increases with  speed  of the user.
Furthermore   the  rate/channel estimates of the same user are i.i.d. across time slots in which their age is the same.

 Next, we describe the evolution of the system (see Figure \ref{fig:state_desc}), and the resultant utilities derived by the users.

\medskip
\textbf{System Evolution:} 
Thus we have a dual scheduler, which at any time includes two decisions: (i) age decisions, and (ii) data decisions, denoted by $(a_g,a_c)$ respectively. 
In any time slot $\tau$, once the age decision is taken, the age vector from previous time slot, $\G_{\tau-1}$ transforms to $\G_{\tau}$ as in \eqref{eqn_age_evolution}. 
As mentioned before, based on the position information, the BS aligns the beams towards each user and collect their channel/rate estimates $\H_{\tau}$. The distribution of vector $\H_{\tau}$ depends upon $\G_\tau$ but its components (corresponding to different users) are conditionally independent. Depending on $\H_{\tau}$ vector, one of the users (say user $n$) is allocated the data channel by BS for data transfer. Thus, user $n$ derives $\sH^{(n)}_{\tau}$ instantaneous utility while others obtain zero utility in time slot $\tau$. 
The overall utility of any user equals the average of such instantaneous utilities over the entire time horizon, and hence
the problem can be modeled using average-cost Markov Decision Process (MDP).

Towards typecasting the problem into
  MDP framework, first observe that the   distribution  of   $\H_{\tau}$ (the channel vector estimated in slot $\tau$) depends upon age vector $\G_\tau$ (the vector after position update phase of slot $\tau$),  which in turn depends upon  $\G_{\tau-1}$ and the age decision component of $\A_{\tau-1}$ that prescribes the user whose age is updated in   slot $\tau$; observe here that $\A_{\tau-1}$ depends on previous time slot components $\G_{\tau-1}$ and $ \H_{\tau-1}$. Accordingly,  we gather together age and channel estimates of the    slot $\tau$  as state $\S_\tau = (\G_\tau, \H_\tau)$ and action $\A_{\tau-1} $ includes age decision that dictates the initial position update phase corresponding to slot $\tau$ (see Figure \ref{fig:state_desc}). However, the second component $A_{c,\tau-1}$ indicates the  user allocated with data channel in previous slot $(\tau-1)$ --- such a pairing  of actions facilitates MDP modeling and also allows channel allocation $A_{c,\tau-1}$ of slot $(\tau-1)$ to depend on   estimates $\H_{\tau-1}$ corresponding to the same slot. 
We now describe the ingredients of the MDP  in the following:
\begin{enumerate}[label=(\roman*)]
    \item \textbf{Decision Epochs:} Each time slot constitutes a decision epoch.
    \item \textbf{State Space:} The state of system in slot $\tau$ is denoted by $\S_\tau = (\G_\tau,\H_\tau)$ and hence the state space $\Sc = \mathcal{G} \times \mathcal{H}$, let  $|\Sc| = \ns.$ The  realization of the state is represented by $\rS_\tau = (\rG_\tau, \rH_\tau)$ where for example, $\rH_\tau = (h_\tau^{(1)},\cdots,h_\tau^{(N)}).$
    \item \textbf{Action Space:}  The action  $\A_{\tau} = (A_{g,\tau},A_{c,\tau})$  comprises of dual scheduling decisions, and the action space is represented by $\Ac$ where $\na := |\Ac|$. The realization of  the action is represented by $\rA_{\tau} = (a_{g,\tau}, a_{c,\tau}),$ -- here $\rA_{\tau}=(n', n)$ represents that the data channel is allocated to the user $n$ in the last (data) phase of  slot $\tau$, while the position of user $n'$ is updated in the initial phase of slot $(\tau+1)$. 
    \item \textbf{Utilities:} Depending on the state $\rS_{\tau}$ and action $\rA_{\tau}$ user $n$ obtains the following instantaneous utility in slot $\tau$, 
    \begin{equation}
        r^{(n)}(\rS_{\tau},\rA_{\tau})= \rsH^{(n)}_{\tau}\mathds{1}_{\{a_{c,\tau}= n\}}.    \label{eqn_instant_reward} 
    \end{equation}
    Thus the time average utility obtained by player $n$ in the long run equals (with initial condition  $\rS_0$),
\begin{equation}
\label{eqn_time_avg_util}
    \rsbU^{(n)}_{\infty} (\rS_0) =  \limsup_{\T \to \infty} \frac{\sum_{\tau \le T}r^{(n)}(\rS_{\tau},\rA_{\tau})}{T} 
\end{equation}
    \item \textbf{Transition Probabilities:} Once an age decision is made, the state component corresponding to age (briefly referred to as age-state), transitions as in \eqref{eqn_age_evolution} depending upon decision $ a_{g,\tau}$ -- let $q(\rG_\tau, a_{g,\tau},\rG_{\tau+1})$ represent this transition.  The remaining components ($\H_\tau$) evolve probabilistically based on the new age components. Thus the transition probability to  new state $\rS_{\tau+1}$ from  current state $\rS_\tau$ and under action  $\rA_\tau$ is given by:
   \begin{equation}
   \label{Eqn_trans_prob}
    p(\rS_{\tau+1} = (\rG_{\tau+1},\rH_{\tau+1} ) \mid  \rS_\tau,\rA_\tau) = q(\rG_\tau,    a_{g,\tau-1},\rG_{\tau+1}) p(\rH_{\tau+1}| \rG_{\tau+1}).   
   \end{equation}
\end{enumerate} 
Observe here that $a_{c,\tau}$ has no impact on  the above transitions while $a_{g,\tau}$ does not directly reflect in the immediate rewards given in \eqref{eqn_instant_reward}.

Hence our aim is to obtain an optimal policy,   which describes the age and data decisions for all time slots. Towards this, we restrict ourselves to Stationary Markovian Randomised (SMR) policies, denoted by $\bd^{\infty}$ (as in \cite{Putterman}). For the ease of notations, we refer $\bd^{\infty}$ as $\bd$ where $\bd : \Sc \to \mathcal{P}(\Ac)$ and $d(\rS,\rA)$ denotes the probability of choosing action $\rA = (a_g,a_c)$ in state $\rS$. 

As the state space is finite, for every SMR policy $\bd$, there exist at least one stationary distribution. Further, one may have more than one (but finitely many) stationary distributions  with support on  disjoint closed communicating classes in $\Sc \times \Ac$ (see e.g., \cite{hoel}). 
Thus under any  SMR policy $\bd$, and for any initial condition\footnote{For any given SMR policy $\bd$, any such $\rS_0$ fixes the initial distribution (i.e., distribution of $(\rS_0, \rA_0)$) of controlled Markov chain $\{(\S_\tau, \A_\tau)\}$.} $\rS_0$, as $\tau \to \infty$,
 the controlled Markov chain $\{(\S_\tau, \A_\tau)\}$ converges weakly to the  unique stationary distribution (an appropriate convex combination of the previously mentioned finitely many), call it  $\bm \mu_{\bd,\rS_0}$. Let $(\S_\infty, \A_\infty)$ with $\S_\infty = (\G_\infty,\H_\infty)$ and $\A_\infty = (  A_{g, \infty}, A_{c,\infty}) $ represent the corresponding limiting random variables, whose distribution is governed by $\bm \mu_{\bd,\rS_0}$; and let the corresponding utility of user $n$ (and then for all users) under the  distribution $\bm \mu_{\bd,\rS_0}$ be represented by,
 \begin{equation}
\label{Eqn_ensemble_multi}
  \rsbU^{(n)}_{\bm \mu_{\bd,\rS_0}} = \E_{\bm \mu_{\bd,\rS_0}} \left[  \sH_\infty^{(n)}\mathds{1}_{\{A_{c, \infty}={n}\}}\right] \mbox{ and }  \rbU_{\bm \mu_{\bd,\rS_0}} :=  \left [ \rsbU^{(1)}_{\bm \mu_{\bd,\rS_0}}, \cdots, \rsbU^{(N)}_{\bm \mu_{\bd,\rS_0}}\right ],    
 \end{equation}
 where~$\E_{\bm \mu_{\bd,\rS_0 }}\left[\cdot \right]$ represents the expectation with respect to distribution~$\bm \mu_{\bd,\rS_0}$.
By   well-known Ergodic theorems (see e.g.,~\cite{balter}), the time average utilities in~\eqref{eqn_instant_reward}-\eqref{eqn_time_avg_util},  under~$\bd$  and with initial condition~$\rS_0$  converge to the ensemble average in~\eqref{Eqn_ensemble_multi}, i.e.,
\begin{equation}  \label{eqn_ensemble_avg_util}
 \rsbU^{(n)}_{\infty}(\rS_0) =  
\rsbU^{(n)}_{\bm \mu_{\bd,\rS_0}} \text{ for all }n \in \Nc, \text{ and let } \rsbU^{(n)}_{\bd }(\rS_0) := \rsbU^{(n)}_{\infty}(\rS_0); 
\end{equation}
the above definition is to explicitly indicate the dependence on SMR policy $\bd$.

\ignore{{\color{red}We now provide more details of the SMR policy $\bd$ dependent time/ensemble averages of \eqref{Eqn_ensemble_multi}.
If the Markov chain $(\S_\tau,\A_\tau)_\tau$  under some SMR policy $\bd$   has  unique stationary distribution $\bmu$  (i.e., has single closed communicating class)  then the time-average utilities are the same  for all initial conditions $\rS_0$, i.e., 
\begin{equation}
\label{eqn_final_util}
 \rbU_{\bd}(\rS_0) = \rbU_{\bmu}  \mbox{ for all } \rS_0  \mbox{ with vector of utilities,  } \rbU_{\bd}(\rS_0) := \left [\rsbU^{(1)}_{\infty}(\rS_0), \cdots, \rsbU^{(N)}_{\infty}(\rS_0) \right ] .
\end{equation}

On the other hand, say there exist multiple stationary distributions under some $\bd$ -- let $\{C_i\}$ represent the corresponding closed communicating classes and $\bm \mu_{\bd,C_i}$ represent the distribution with support over class $C_i$ -- \textit{we refer $\{\bm \mu_{\bd,C_i}\}$ as distinct stationary distributions under $\bd.$}  Let $\sigma_{\bd,C_i}(\rS_0)$ represent the absorption probability to class $C_i$ when Markov chain starts in $\rS_0.$ 
For each initial condition, there exists a unique limiting  distribution given by   a convex combination
of distinct stationary distributions, using which one can describe the time averages  as   below:


 
\begin{equation}
\label{eqn_util_mult}
 \rbU_{\bd}(\rS_0) = \sum_{ C_i} \sigma_{\bd,C_i}(\rS_0)\rbU_{\bm \mu_{\bd,C_i}}.   
\end{equation}

For simplicity, we again represent  $\rsbU_{\bd}^{(n)}(\rS_0)$, etc., by $\bmu$, $\rsbU_\bd^{(n)}$ when there is clarity.}}

Thus there are $N$ users and hence there are $N$ distinct objectives (for any initial condition). Further, there is a central controller (or BS) that allocates the resources to one of them in every time slot. Hence, it is a multi-objective optimization problem. The BS would like to maximize the sum utility $\left(\sum_{n} \rsbU^{(n)}_\bd (\rS_0)\right)$ but might also want to be fair to individual users. 
\textit{The main aim of this study is to propose an opportunistic and fair scheduler for beam alignment and data transmission}. This problem is well understood in the context of previous generation wireless networks with the help of $\alpha$-fair schedulers, which we briefly discuss next. We begin with a similar approach in the next section where we propose an algorithm to design dual scheduler. In the later sections, we derive more insights for a special case with $N=2$, using a constrained optimization approach as in \cite{debayan}, which results in an algorithm with significantly lesser computational complexity. Using these insights, we propose a dual opportunistic scheduler for any $N$, which can easily be implemented. 

We now conclude the section with a theorem (proof in Appendix \ref{sec_AppendixB_CN}) which shows that optimization over all SMR policies is equivalent to the optimization over SMR policies with unique stationary distribution, referred to as uSMR policies.  
%
%
%
We derive this result under  minimal reachability assumption (observe we just need positive probability) on channel states, which we assume throughout. 



\begin{theorem}
\label{lem_multichain}
Assume $p( \rH|\rG) > 0$  for all $\rG$, $\rH$ in \eqref{Eqn_trans_prob}.   Let 
$f$ be any objective function. Then for any initial condition $\rS_0$, optimisation over SMR policies is equivalent to optimisation over uSMR policies, i.e., 
    \begin{equation}
    \label{eqn_sup_obj}
        \sup_{\bd \in D^{SMR}} f(\rbU_{\bd}(\rS_0)) 
        = \sup_{\bd \in D^{uSMR}} f(\rbU_{ \bd}).
    \end{equation}
    
\end{theorem}

\medskip 

\noindent
\textbf{Remark:} Observe that the above result assumes the possibility of being in any of the channel conditions (from $\mathcal{H}$), irrespective of the age vector. While (for example) it is true that the probability of being in certain  channel conditions may decrease with age, it is still reasonable to assume that such probabilities are non-zero.

\medskip

In view of the above result, optimizing over uSMR policies is sufficient. 
At this point, we would like to draw attention to several intriguing contrasts from the Markov Decision Process (MDP)  literature:
\begin{enumerate}[label=(\roman*)]
    \item It is well-known that the value, i.e., the optimal value of $f$ is the same for all initial conditions, if the model is unichain (i.e., if the controlled Markov chain has unique stationary distribution for all SMR policies). However, in our case because of the structure of the problem, the same is true in spite of the model being multi-chain; 
    \item The second contrast relates to the optimization approach. In the MDP literature, the objective typically focuses on optimizing a cumulative  or  an average reward/cost over a sequence of actions. However, in this particular context, we consider \textit{optimization of  a function of several average utilities.} The approach developed in Section \ref{sec_offline_CN} can also be applied to other problems of this kind. 
\end{enumerate} 
\ignore{{\color{red}Take any $\bm \mu_{\bd,C_i} \in \mathcal{B}$--basically, the stationary distribution has support only on $C_i$. 
As in the proof of Lemma \ref{lem_multichain}, one can construct a policy $\tilde \bd$ such that all states in $C_i^c$ are transient and $C_i$ is the unique closed communicating class under $\tilde \bd$. Further, the utility vector $\rbU_{\tilde \bd}(\rS_0)$ (for any $\rS_0$) under SMR policy $\tilde \bd$ exactly equals $\rbU_{\bm \mu_{\bd,C_i}}.$ Thus for ease of explanation and with slight abuse of notation, we represent $\bm \mu_{\bd,C_i}$ by corresponding $\bm \mu_{\tilde \bd}$.}}




\subsection{Background On $\alpha$-Fair Opportunistic Schedulers}
The concept of fairness has been extensively studied in the context of previous generation wireless networks, as evidenced by the research cited in \cite{tejas,kushner,cellular,debayan}, and other related works. Users are located at different positions with respect to the BS and hence some users may have inferior channel conditions with higher probabilities than others. As a result, efficient schedulers that maximize the sum of user-utilities may starve some users, for example, those away from the BS. To address this issue, fair schedulers have been proposed to cater to the needs of these deprived users.

An opportunistic and fair scheduler observes the channel estimates of all the users in any time slot, and allocates the channel in a controlled manner to a deprived user at the best possible opportunity. They aim to maximise the sum of the user-utilities, while maintaining a `required' level of fairness. The well-known generalized $\alpha$-fair schedulers (see \cite{tejas,kushner,cellular,debayan}) achieve this by optimizing a certain parameterized concave function of the average utilities obtained by each user, offering varying levels of fairness indicated by parameter $\alpha$:
\begin{equation}
\begin{aligned}
  \sup_{\bd = (d^{(1)}, \cdots, d^{(N)})} \sum_{n \in \Nc}  \Gamma_\alpha \left ( \rsbU^{(n)}_\bd \right ) \text{ with } \rsbU^{(n)}_\bd:=\E[\sH^{(n)} d^{(n)} (\H)], \text{ and, } \\
 \Gamma_\alpha(\rsbU^{(n)}) := \frac{\left(\rsbU^{(n)}\right)^{1-\alpha} \ind_{ \left \{ \alpha \neq 1 \right \}} }{1-\alpha} + \log\left(\rsbU^{(n)}\right) \ind_{\left \{ \alpha =1 \right \}},  
\end{aligned}  
 \label{Eqn_alpha_fair}
\end{equation}
here $d^{(n)}$ denotes the probability of data channel being allocated to user $n$, and $\rsbU^{(n)}$ denotes the average utility of user $n$. In the previous generation networks, either $\{\H_\tau\}$ are assumed to be i.i.d. or Markovian channels and importantly were not dependent on beam alignment. We now propose dual schedulers that achieve $\alpha$-fairness as well as optimal beam alignment in the next.

\section{Dual $\alpha$-Fair Opportunistic Scheduler}
\label{sec_offline_CN}
The natural extension of $\alpha$-fair schedulers in \eqref{Eqn_alpha_fair} to the case with beam alignment  is to  include age decisions in scheduler $\bd$, leading to a \textit{dual scheduler} --   in view of Theorem \ref{lem_multichain}, it is sufficient to   work with   uSMR policies.  The remaining details in \eqref{Eqn_alpha_fair} are exactly the same, except that   individual utilities $\{\rsbU^{(n)}_\bd\}$ now also depend on age decisions.  These time-average or stationary utilities of individual users now have the form as in \eqref{Eqn_ensemble_multi}-\eqref{eqn_ensemble_avg_util}.
Under any uSMR policy, quantities like $\rbU_{\bmu,\rS_0}$, $\rbU_\bd(\rS_0$) etc., do not depend on initial condition $\rS_0$, and hence we \textit{drop the notation $\rS_0$ henceforth.}

Under any uSMR policy $\bd$, the time-average utility in \eqref{eqn_time_avg_util} equals that in \eqref{eqn_ensemble_avg_util} and can be re-written as:
%
\begin{equation}
\rbU_{\bd} = \bmu \brd^T, 
\label{Eqn_optim}
\end{equation}
where:   
(i) with  $\mud(\rS)$ representing the stationary probability of being in state $\rS$ under $\bd$, the $\ns$-dimensional row vector $\bmu = \{ \mud(\rS) \}_{\rS \in \Sc}$ represents the stationary distribution; 
(ii) $\rbU_\bd = (\rsbU_\bd^{(1)},\cdots,\rsbU_\bd^{(N)})$ is an $N$-dimensional row vector of individual utilities (see \eqref{eqn_ensemble_avg_util}); and 
(iii) $\brd$ is a $N \times \ns$-dimensional reward matrix under policy $\bd$ whose components, for user $n$  and state $\rS$,  are given by (see \eqref{eqn_instant_reward} and \eqref{Eqn_ensemble_multi}),
\begin{equation}
\label{eqn_rd}
     r^{(n)}_{\bd}(\rS) := \sum_{\rA \in \Ac}r^{(n)}(\rS,\rA)d(\rS,\rA) \text{ for all } \rS \in \ \Sc \text{ with }  \ r^{(n)} (\rS, \rA) = h^{(n)} \ind_{\{a_{c} = n\}}. 
\end{equation}
\ignore{Under uSMR policy $\bd$, it is easy to observe  by conditioning that 
$$
 \E_{\bmu} \left [  \E [ \sH^{(n)} | \sG^{(n)}_\infty, A_{g,\infty} ] \ind_{\{A_{c,\infty} =  n\}}\right ]  = \E_{\bmu} \left [  \sH^{(n)} \ind_{A_{c,\infty} =  n}\right ], 
$$where it has to be understood that the distribution of $H^{(n)}$ conditioned on  $(\sG^{(n)}_\infty, A_{g,\infty})$ is independent of $A_{c, \infty}$ and that it is different from  the limiting variable $H^{(n)}_\infty$ -- basically if $H^{(n)}_\infty$ (and $\G_\infty, \A_\infty$) corresponds to the current time-slot then $H^{(n)}$ corresponds to the next time-slot. In view of this one can replace the immediate rewards of \eqref{eqn_instant_reward} with a less complicated version
Recall from \eqref{eqn_instant_reward} and \eqref{Eqn_ensemble_multi},
\vspace{-2mm}
\begin{equation}
    
    \label{eq_reward_less}
\end{equation}}

 \subsection{Algorithm}
We  begin with deriving the unique stationary distribution   $\bmu$ under any uSMR policy $\bd$, which satisfies: 
\begin{eqnarray}
\bmu \Pb_\bd = \bmu \text{ and }
{\bm \mu}_\bd {\bf e} = {\bf e}, \text{ where, }   \Pb_\bd = \{\Pb_\bd(\rS,\rXn)\} \mbox{ with }\Pb_\bd(\rS,\rXn) =  \sum_{\rA \in \Ac} p(\rXn|\rS,\rA)d(\rS,\rA),
\label{eqn_stationary_orig}
\end{eqnarray}
 is an $\ns \times \ns$-dimensional transition probability matrix under policy $\bd$ constructed using the probabilities given in \eqref{Eqn_trans_prob}, and ${\bf e}$ is the $\ns$-dimensional column vector of all $1$'s.
 We use gradient based approach to obtain the optimal policy $\bd$, that maximises the $\alpha$-fair function \eqref{Eqn_alpha_fair} of average utilities in \eqref{Eqn_optim}.
 
 To this end we optimise $\bc = (c(\rS,\rA))$, an $\ns \times \na$-dimensional matrix,
 which  defines  the decision rule (or uSMR policy) $\bd$  as below, similar to the technique  considered in MDP-LP literature \cite{Putterman},
 \begin{equation}
 \label{Eqn_d_term_c}
 d(\rS,\rA) := \frac{c(\rS,\rA)}{\sum_{\rA' \in \Ac} c(\rS,\rA')}.    
 \end{equation}
Observe this ensures that policy $\bd$ satisfies $\sum_{\rA \in \Ac} d(\rS,\rA) = 1$ for any $\rS$. We now update $\bc_{\tau+1}$ at time $\tau+1$ based on its estimate at time $\tau$, using step size $\psi_\tau$   and various gradients as in the following:
\begin{equation}
\label{eqn_c_update}
    \bc_{\tau+1} = \bc_\tau + \psi_\tau \sum_{n \in \Nc} \left( \frac{1}{ \left(\rsbU^{(n)}_{\bd}\right)^\alpha} \nabla_{\bc}^{\rsbU^{(n)}_{\bd}} (\bc_\tau) \right),
\end{equation}
as the partial derivative of  $\Gamma_\alpha$ with respect to $\rsbU^{(n)}_\bd$ is
$ 1/ \left(\rsbU^{(n)}_\bd \right)^\alpha $ and $\nabla_{\bc}^{\rsbU^{(n)}_{\bd}} (\bc_\tau)$ represents the $L\times M$ dimensional partial derivative of $\rsbU^{(n)}_{\bd}$ with respect to $\bc$ at $\bc_\tau$. 
An iterative procedure using the relevant gradients and the simplification steps derived  in Appendix \ref{sec_AppendixA_CN} is provided in  Algorithm \ref{algo_offline}, which computes optimal $\bc$ and thereby $\bd$ -- we refer this algorithm by $\alpha-$DoFS. 
The remaining relevant partial derivatives used in Algorithm \ref{algo_offline} are as in Table \ref{tab_nabla_notations}.
 We further require the projection, $\bc_{\tau+1} = \max \{0, \bc_{\tau+1}\}.$

\medskip 

\noindent
 {\bf Removing Irreducibility:} 
During the initial warm-up period, it is possible to reach a stage where age of only one user is updated.
As in Q-learning, to explore the optimality of reaching other states, one  needs to incorporate a leakage factor $\delta_\tau$ in the decisions (the probability of which reduces as time progresses); for example, the decision is updated as below (when $\S_\tau = \rS_\tau$):
$$
\bc_{\tau+1} (\rS_\tau)  = (1-\delta_\tau) \left( \bc_\tau(\rS_\tau) +  \psi_\tau \sum_{n \in \Nc} \left( \frac{1}{ \left(\rsbU^{(n)}_{\bd}\right)^\alpha} \nabla_{\bc_\tau}^{\rsbU_{\bd}} (n) \right) \right) + \delta_\tau K, \text{ where $K$ is a positive constant.}
$$




\begin{algorithm}[H]
\SetAlgoLined
\KwIn{$N, \ {\bar g}, \ \alpha, \ \Sc, \ \Ac, \ p(\rS'|\rS,\rA)$ for each $\rS=(\rG, \rH) \in \Sc, \ \rA \in \Ac$},
\KwOut{$\bd, \ \bmu, \ \rbU_\bd, \ \Gamma_\alpha(\rbU_\bd)$}
\noindent
\hspace{5mm} Initialise $\bc, \ \bd$ at $\tau=0$ \\
\hspace{5mm} \textbf{While} \\
\hspace{10mm} $(i)$ Using   \eqref{eqn_rd}, and \eqref{eqn_stationary_orig}  estimate $r^{(n)}_{\bd}(\rS)$ and $\Pb_{\bd}(\rS,\rS')$ as given below,
{\small \begin{eqnarray*}
    r^{(n)}_{\bd}(\rS) = \sum_{\rA \in \Ac}\rsH^{(n)}\mathds{1}_{\{a_{c}= n\}}d(\rS,\rA) \text{ and }  \Pb_\bd(\rS,\rXn) = \sum_{\rA \in \Ac}   q(\rG, a_{g},\rG') p(\rH| \rG)d(\rS,\rA) \text{ for all } \rS,\rXn \in \Sc, \rA \in \Ac 
\end{eqnarray*}}

  \hspace{10mm} $(ii)$  Calculate stationary distribution $\tmu$ (the first $\ns-1$ components, see   \eqref{eq_reduced_sd})  as in the following:
  \begin{equation}
  \begin{aligned}
      \tmu = {\bf b}^T_{\bd} \left(I - \tP \right)^{-1} \text{ where }   {\bf b}_{\bd} = \begin{bmatrix}
\Pb_{\bd}(\ns,1) \\
\vdots \\
\Pb_{\bd}(\ns,\ns-1)
\end{bmatrix}\text{ and, } \\
{\tilde \Pb}_{\bd}(\rS,\rXn) = \Pb_{\bd}(\rS,\rXn)-\Pb_{\bd}(\ns,\rXn) \text{ for all } \rS,\rXn \in \Sc'
\end{aligned}
  \end{equation} 
   \hspace{11mm} $(iii)$ Obtain $\bmu$ from $\tmu$ using \eqref{eq_bmu_from_tmu} \\
  \hspace{13mm} $(iv)$ Update $\bc$ using \eqref{eqn_c_update} (details in Appendix \ref{sec_AppendixA_CN})  
%
%
%
%
%
           %
\begin{eqnarray}
    \bc &\leftarrow & \bc + \psi_\tau \sum_{n \in \Nc}\frac{1}{\left(\rsbU^{(n)}_\bd\right)^\alpha} \nabla_{\bc}^{\rsbU_\bd}(n) \text{ with }  \nonumber \\ 
    \nabla_{\bc}^{\rsbU_\bd}(n) &\leftarrow & \sum_{\rS \in \Sc} \mud(\rS) \nabla_{\bc}^{r_\bd}(n,\rS)  \nonumber \\
    && +  \sum_{ \rS \in \Sc', \rA \in \Ac} \left( r^{(n)}(\rS,\rA)-r^{(n)}(\ns,\rA)\right)d(\rS,\rA)\nabla_{\bc}^{{\tilde \mu}_\bd}(\rS) \label{eq_grad_u_wrt_c} \\
\nabla_{\bc}^{r_\bd}(n,\rS) &\leftarrow &\sum_{\rA \in \Nc}r^{(n)}(\rS,\rA)\nabla_{\bc}^{\bd}(\rS,\rA) \text{ and } \label{eq_grad_r_wrt_c}  \\
\nabla_{\bc}^{{\tilde \mu}_\bd}(\rS) &\leftarrow & \sum_{\rXn \in \Sc'} \nabla_{\bc}^{{\tilde \mu}_\bd}(\rXn) \tP(\rXn,\rS) + \begin{bmatrix}
\mud(1) \Cv(\rS,1) \\
\vdots \\
\mud(\ns-1) \Cv(\rS,\ns-1)
\end{bmatrix} \label{eq_grad_tmu_wrt_c} 
\end{eqnarray}
 \caption{$\alpha$-DoFS - Dual opportunistic fair scheduler}
\label{algo_offline}
\end{algorithm}

\begin{eqnarray}
\nabla_{\bc}^{\bd}(\rS,\rA) & \leftarrow & \begin{bmatrix}
\left(  \nabla_{sym}^{\bd}(\rS,\rA) \right)^T \ind_{\{\rS=1\}} \\
  \vdots \\
  \left( \nabla_{sym}^{\bd}(\rS,\rA) \right)^T \ind_{\{\rS=\ns\}}
\end{bmatrix} \label{eq_grad_d_wrt_c}  \\
\nabla_{sym}^{\bd}(\rS,\rA) & \leftarrow &  
\frac{1}{
\sum_{\rA'} c(\rS,\rA')} \begin{bmatrix}
\ind_{\{\rA=1\}} \\
\ind_{\{\rA=2\}} \\
\vdots  \\
\ind_{\{\rA=\na\}}  \\ 
\end{bmatrix} - \frac{1}{
\left(\sum_{\rA'} c(\rS,\rA')\right)^2} { \bf e}
\label{eq_grad_dsym_wrt_c} \\
\Cv(\rS,\rS') & \leftarrow & \sum_{\rA \in \Ac} \left\{\left(\nabla_{sym}^{\bd}(\rXn,\rA)\right)^T p(\rS|\rXn,\rA)\right\} \label{eq_zeta}
\end{eqnarray} 
 \hspace{11mm} $(vi)$ Projection: $\bc  \leftarrow \max \left \{ 0,\bc  \right \}$   \\
\hspace{5mm} \textbf{end} \\


\begin{table}[h]
\centering
\begin{tabular}{|c|c|c|}
\hline
Symbols                                & Dimension & Explanation \\ \hline
$\nabla_{\bc}^{\rsbU_\bd}(n)$              &  $\ns \times \na$     &  Partial derivative of $\rsbU^{(n)}_\bd$ wrt matrix $\bc$    \\ \hline
$\nabla_{\bc}^{r_\bd}(n,\rS)$          & $\ns \times \na$   &   Partial derivative of $r_\bd^{(n)}(\rS)$ wrt matrix $\bc$      \\ \hline
$\nabla_{\bc}^{{\tilde \mu}_\bd}(\rS)$ &  $\ns \times \na$     &  Partial derivative of ${\tilde\mu}_\bd(\rS)$ wrt matrix $\bc$    \\
&& $\tmu = (\mud(1),\cdots,\mud(\ns-1))$\\ \hline
$\nabla_{\bc}^{\bd}(\rS,\rA)$          & $\ns \times \na$       &  Partial derivative of $d(\rS,\rA)$ wrt matrix $\bc$   \\ \hline
$\nabla_{sym}^{\bd}(\rS,\rA)$          & $\na \times 1$       &  Partial derivative of $d(\rS,\rA)$ wrt vector $\bc(\rS)$   \\ \hline
\end{tabular}
\caption{Various partial derivatives used in Algorithm \ref{algo_offline}}
\label{tab_nabla_notations}
\end{table}

\subsection{Non-opportunistic Fair Scheduler (NoFS) } 
\label{sec_NoFS}
In \cite{mmwave}, the authors consider  an optimal user scheduling problem to minimize the beam alignment overhead in mmWave networks, while maintaining the desired QoS (rewards related to data transmission) of each user.  In each time slot  the BS selects one user, and the beam search algorithm finds the most appropriate beam towards the selected user - it starts the search in the direction of the last beam alignment used for the same user -- 
the alignment is faster if the time elapsed since the last position update (which we again call as the age of  information) is smaller.   
  Their scheduler is non-opportunistic (does not make decisions based on channel estimates from all users), as  opposed to the ones  discussed in this work and further  they do not consider fairness.

We compare our opportunistic schedulers with the schedulers of \cite{mmwave}, after incorporating fairness into their framework. Towards this, we suggest an appropriate algorithm in the immediate next, which we refer to as NoFS (Non-opportunistic Fair Scheduler). 

To implement such  non-opportunistic scheduler after incorporating  optimization of the fair objective as in \eqref{Eqn_alpha_fair}, one just needs to consider weighted  expected conditional channel estimates of the users $\{O(\sG^{(n)}_\tau) {\bar h}^{(n)}\}$  in place of (actual age dependent) exact channel estimates $\{\sH^{(n)}_{\tau}\}$ -- here 
weight $O(\sG^{(n)}_\tau)$ depends on the age $\sG^{(n)}_\tau$ of user $n$  in slot $\tau$ and   characterises the time lost in aligning the beam to the user, and ${\bar h}^{(n)}:= \E[\sH^{(n)} |\rsG^{(n)} = 1]$
 is the expected channel estimate of user $n$ when its age is one.  Such a scheduler can be implemented using Algorithm \ref{algo_offline} after changing state space with  $\Sc = \mathcal{G}$, dual scheduler $\rA $ with just age scheduler $a = a_g $ and replacing Step $(i)$ with  
\begin{equation}
\label{eq_NoFS_reward}
\begin{aligned}
    r^{(n)}_{\bd}(\rG) = \sum_{a \in \Ac} O (g^{(n)})
    {\bar h}^{(n)} \mathds{1}_{\{a= n\}}d(\rG,a), \text{ and }  \\
    \Pb_\bd(\rG,\rG') =   \sum_{a \in \Ac} d(\rG,a)  q(\rG, a,\rG')  \text{ for all } \rG,\rG' \in \Sc=\mathcal{G}, a=a_g \in \Ac. 
  \end{aligned}
\end{equation}

\subsection{Numerical Examples}
\label{sec_num1}
\begin{figure}[h]
    \centering   
    \begin{minipage}{.45 \textwidth}
    \includegraphics[trim = {2.9cm 7cm 0cm 6cm}, clip, scale= 0.4]{Figures/PoF_fig1.pdf}
    \end{minipage}
     \begin{minipage}{.45 \textwidth}
     \includegraphics[trim = {1cm 7cm 0cm 8cm}, clip, scale = 0.4]{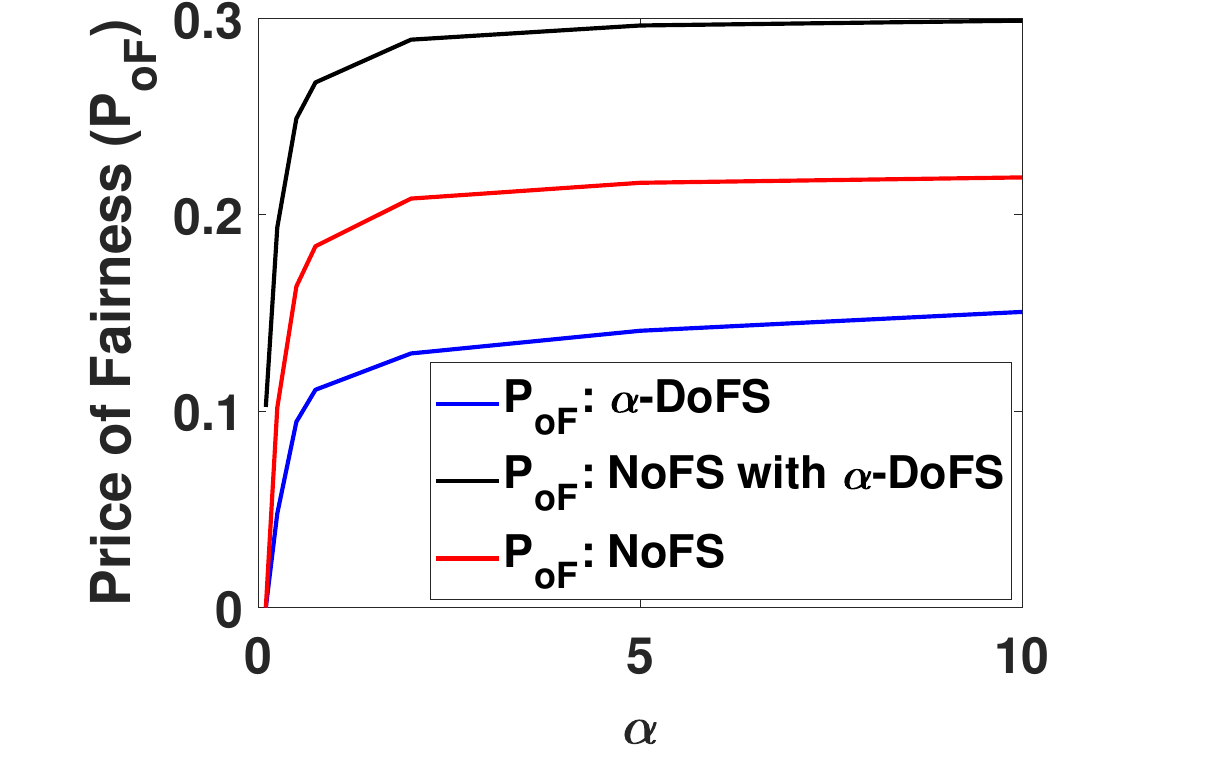}
    \end{minipage}
    \caption{Price of Fairness \rev{PoF} for example 1 (left sub-figure) and example 2 (right sub-figure)}
    \label{fig:pof12}
\end{figure}

In this sub-section, we numerically analyse and compare $\alpha$-DoFS as implemented in Algorithm \ref{algo_offline} and the algorithm to implement NoFS  suggested in sub-section \ref{sec_NoFS}. It is clear that higher the time lost in alignment given by $1-O(g^{(n)})$ in \eqref{eq_NoFS_reward},
inferior will be the performance of NoFS. For all the examples considered in this sub-section, we assume  $O(g^{(n)}) = 1$ for all $g^{(n)} \le \bar g$ and $n \le N$, i.e., no time is lost in aligning the beam for NoFS and illustrate that NoFS is still significantly inferior. We also plot  the Price of Fairness (\rev{PoF}) in  Figure \ref{fig:pof12}, 
$$
\text{\rev{PoF}} = \frac{\sum_{n \in \Nc} \left(\rsbU_e^{(n)}-\rsbU_\alpha^{(n)}\right)
}{\sum_{n \in \Nc}\rsbU_e^{(n)}},
$$
where $\rsbU_e^{(n)}$ is the utility of user $n$ under efficient scheduler, i.e., at $\alpha=0$ 
while $\rsbU_\alpha^{(n)}$ is the utility of user $n$ at $\alpha$. We plot two types of \rev{PoF} for NoFS: first is the usual one where $\rsbU_e^{(n)}$ is the utility of user $n$ under efficient scheduler with NoFS, while the second one (which we refer to as Global \rev{PoF} or G\rev{PoF}) is the utility of user $n$ under efficient scheduler with $\alpha$-DoFS.  The sum $  \sum_{n \in \Nc}\rsbU_\alpha^{(n)}$ is also referred to as efficiency  at fairness level $\alpha$ and hence \rev{PoF} captures the normalized percentage loss in efficiency. 

In Figure  \ref{fig:mobile1}, we consider an example with $3$ users, each with $2$ channel conditions. Each of these channel conditions are realised with a certain probability based on the age of that user. The details are as below,
\begin{equation}
\text{ channel condition matrix, } \mathbb{C} = \begin{bmatrix}
    2.4 & .8 \\
    3 & .5 \\
    1.2 & .3
    \end{bmatrix} \text{ and conditional probabilities, } \mathbb{Q} =  \begin{bmatrix}
.7 & .4 & .3 & .2 \\
.5 & .4 & .3 & .2 \\
.8 & .7 & .6 & .5            \end{bmatrix}
\label{eqn_ch_condtions1}
\end{equation}
\ignore{\begin{figure}[h]
    \centering   
    \begin{minipage}{.6 \textwidth}
    \includegraphics[trim = {1cm 6.5cm 2.5cm 6cm}, clip, scale= 0.5]{Figures/sd_fig1.pdf}
    \end{minipage}
     \begin{minipage}{.35 \textwidth}\includegraphics[trim = {2.5cm 7.5cm 0cm 6cm}, clip, scale = 0.4]{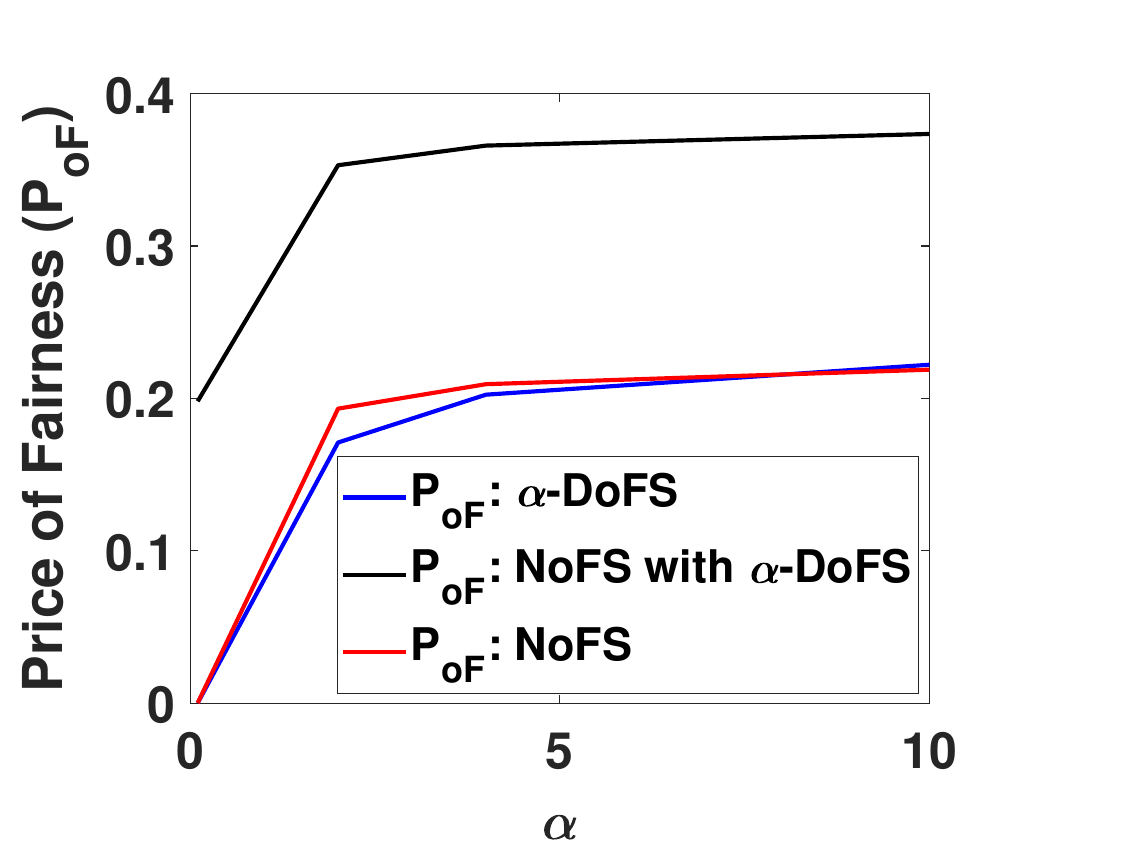}
    \end{minipage}
    \caption{Price of Fairness \rev{PoF} and age update probabilities}
    \label{fig:pof_mu_1}
\end{figure}}
Each row in $\mathbb{C}$ and $\mathbb{Q}$ matrices corresponds to one of the users. The first column in matrix $\mathbb{C}$ provides the utility under good channel conditions for respective users. The entry in the $n^{th}$ row and $m^{th}$ column of $\mathbb{Q}$ matrix depicts the probability of user $n$ having the best channel condition  when its age is $m$.  Both the schedulers are implemented using the algorithms as explained before and the results  are plotted in the left (individual user-utilities) and the right (sum of user-utilities) sub-figures  of Figure \ref{fig:mobile1}. 
The observations are as follows:

\begin{figure} [h]
  \begin{minipage}{.55 \textwidth}
  \centering
    \includegraphics[trim = {.5cm 8cm 0cm 5cm}, clip, scale = 0.4]{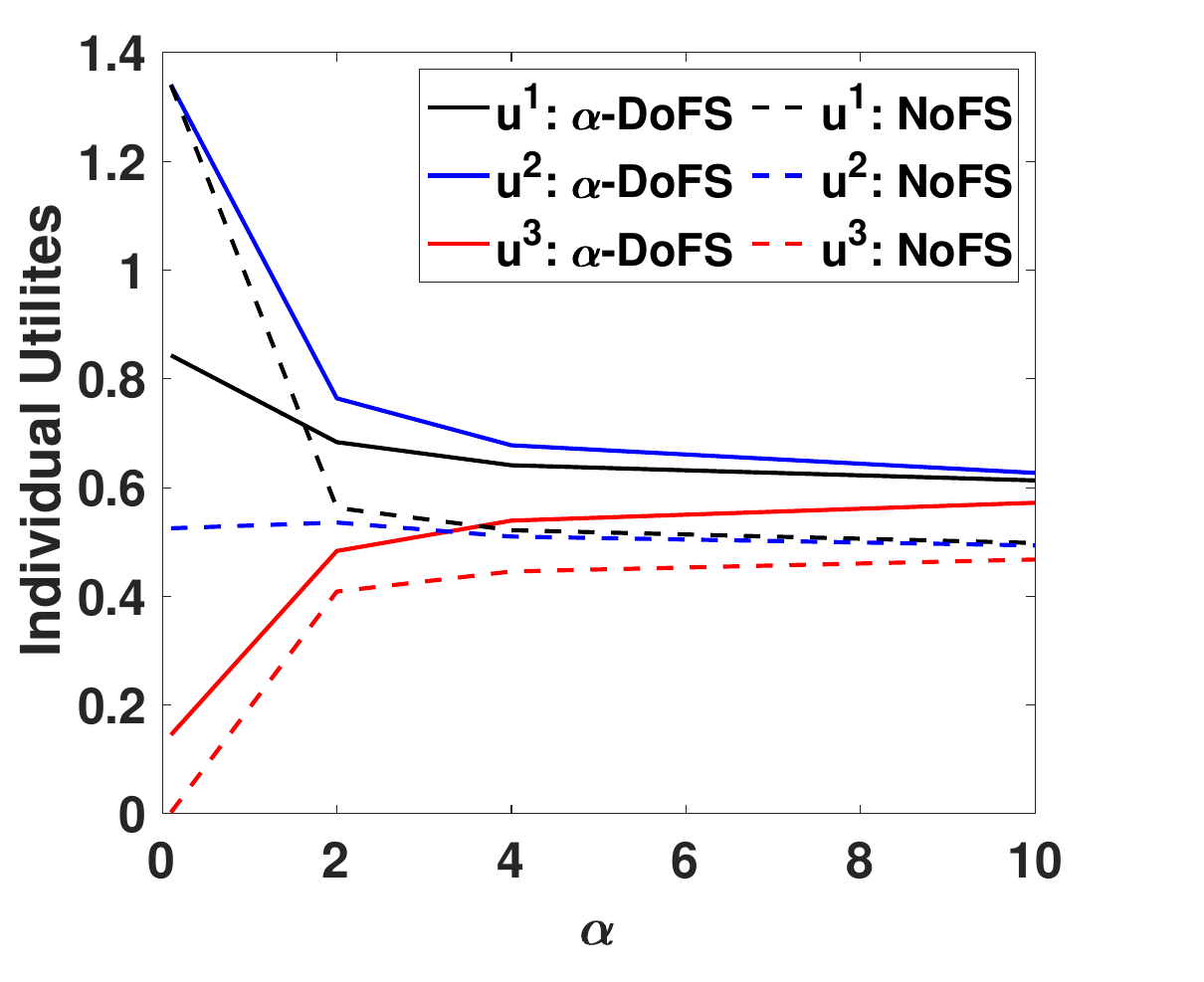}
        \end{minipage}
     \begin{minipage}{.5 \textwidth}
     \vspace{-2mm}
       \includegraphics[trim = {2cm 9.8cm 2cm 6cm}, clip, scale = .4]{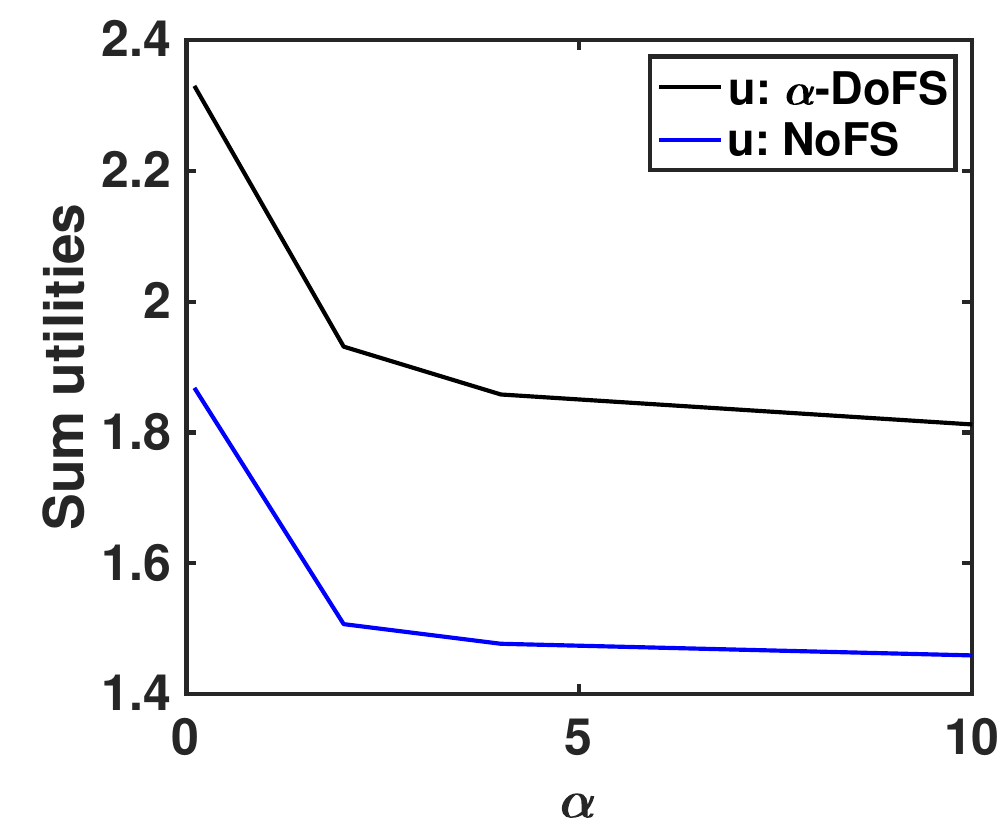}
     \end{minipage}
    \caption{  $\alpha$-DoFS significantly outperforms NoFS: individual utilities in the left and sum utilities in the right  sub-figures. }
    \label{fig:mobile1}
\end{figure}

$\bullet$ The $\alpha$-DoFS      is significantly better --    the sum-utilities in  right sub-figure (black for $\alpha$-DoFS) are significantly higher, even after setting the  time lost in aligning the beam  to  zero.
 
  $\bullet$ As $\alpha$ increases the difference in the individual user-utilities reduces (solid lines for $\alpha-$DoFS and dotted lines for NoFS). At $\alpha = 10$, the individual  utilities are almost equal for both the schedulers (indicating max-min fairness).  
 
   $\bullet$ Interestingly, user $1$ with higher expected channel conditions (see \eqref{eqn_ch_condtions1}) obtains  maximum utility under NoFS while user $2$ with the best utility 
(among all users) under respective good channel conditions obtains the maximum  under $\alpha$-DoFS. This becomes obvious when one observes that  the decisions of $\alpha$-DoFS depend upon instantaneous channel conditions, while the same under NoFS depend on the expected channel conditions. Thus near $\alpha=0$, user $1$ is more starved in comparison to user $2$ under $\alpha$-DoFS while it is the vice-versa under NoFS. 


\begin{figure} [h]
  \begin{minipage}{.56 \textwidth}
    \includegraphics[trim = {.5cm 9cm 0cm 4cm}, clip, scale = 0.4]{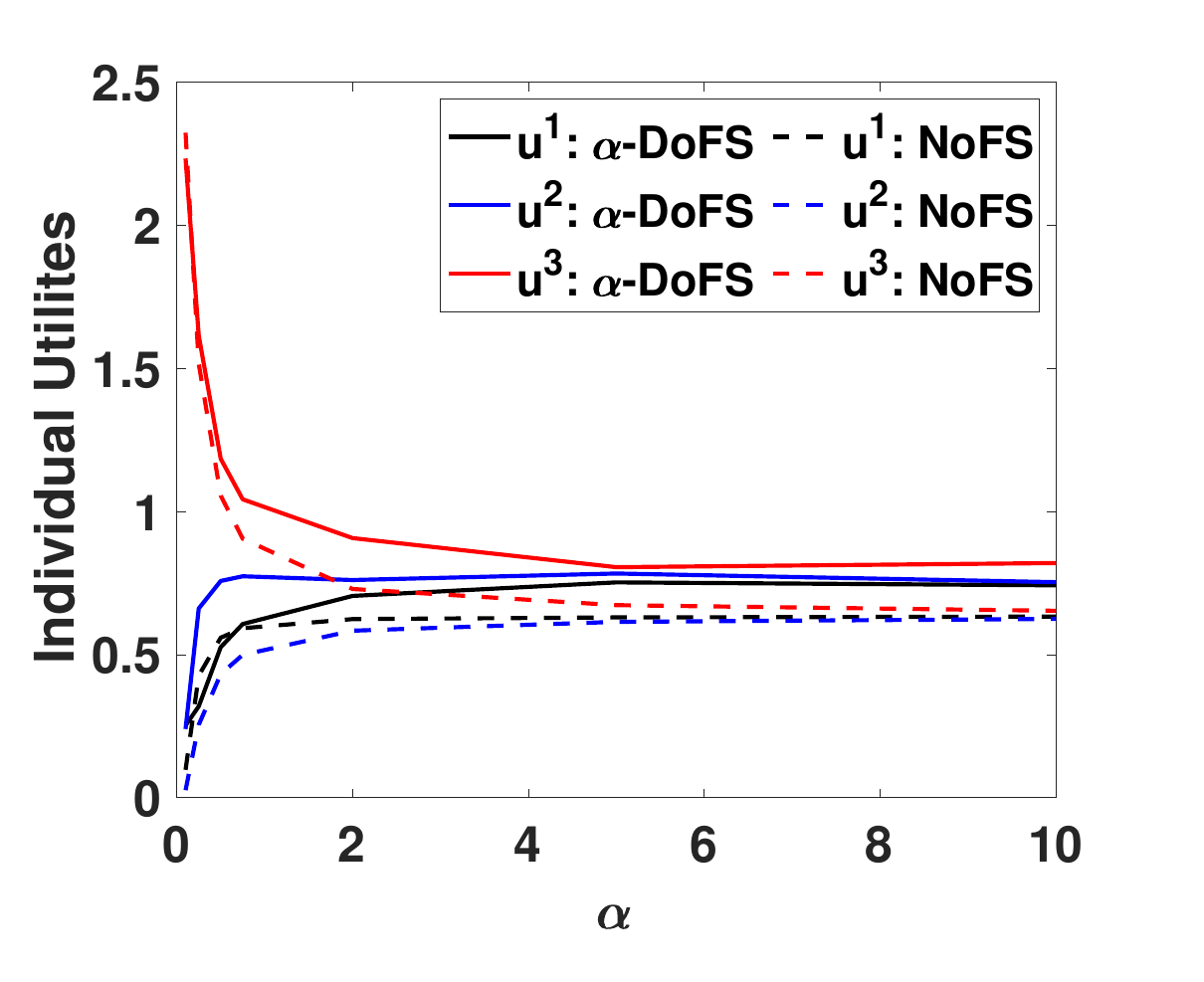}
        \end{minipage}
     \begin{minipage}{.34 \textwidth}
     \vspace{-2mm}
       \includegraphics[trim = {2cm 10cm 0cm 5.5cm}, clip, scale = .4]{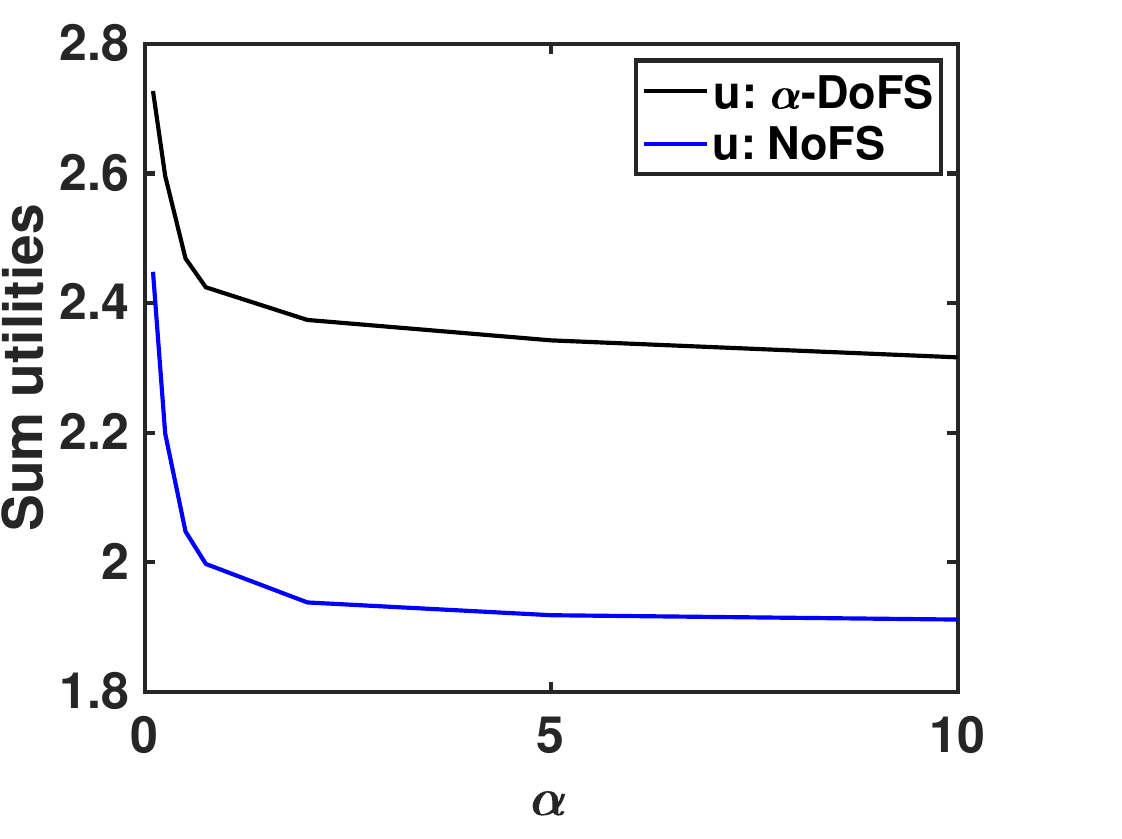}
     \end{minipage}
    \caption{Comparison of $\alpha$-DoFS v/s NoFS}
    \label{fig:mobile2}
\end{figure}

Our aim now is to compare the price of fairness, \rev{PoF}. Towards this we consider  a second 
example in Figure \ref{fig:mobile2})  with 
$
\text{ channel condition matrix } \mathbb{C}= \begin{bmatrix}
2 & 1.4 \\
    2.4 & .8 \\
    3 & .5 
\end{bmatrix} $ and conditional probabilities,  $\mathbb{Q}  = \begin{bmatrix}
    .7 &.4 & .3 & .2 \\
            .5 & .4 & .3 & .2 \\
            .8 & .7 & .6 & .5
\end{bmatrix}.
$
The observations are similar to that in the previous example, except that both the schedulers prefer user 3 near $\alpha = 0$ -- user 3 is better in terms of `channel condition at good state' and `expected utilities'. These differences in the two examples  implies a huge disparity in \rev{PoF} of the two schedulers as seen in Figure \ref{fig:pof12}. 


 Next, we present the comparison of \rev{PoF} between $\alpha$-DoFS and NoFS. As we have already seen, the sum-utility and the disparity in the individual utilities (near $\alpha = 0$) under $\alpha$-DoFS are significantly higher and lower (respectively) than those derived under NoFS. Interestingly, \rev{PoF} of $\alpha$-DoFS is never inferior to NoFS (even in other examples). This implies that the \rev{PoF} under NoFS can still be on par with $\alpha$-DoFS (see left sub-figure of Figure \ref{fig:pof12}) or significantly different from $\alpha$-DoFS (see right sub-figure of Figure \ref{fig:pof12}). However, when the utilities under efficient scheduler obtained with $\alpha$-DoFS are considered as a benchmark even for NoFS, then the \rev{PoF} is significantly inferior with NoFS. This also represents the cost of not using opportunistic schedulers. 

We now consider an alternate approach based on constrained optimisation as in \cite{debayan} to derive the entire spectra of $\alpha$-fair schedulers using a different set of parameters  $\Theta$,  for the special case with $N=2$.

\section{Dual Decoupled  Fair Opportunistic Scheduler (DDoFS) }

The purpose of this section is to derive possibly approximate `near closed-form expressions' (throughout the chapter, we say the solution has near-closed-form expression  if it is the best that optimises the sum of user-utilities over a given finite set  of schedulers, each of them having closed-form expressions), for any given `level of fairness', using an alternate approach. Towards this we consider $N=2$ and 
a special class of dual schedulers which we refer as dual decoupled schedulers (or briefly by DDoFS) -- here age decision $a_{g_\tau}$   depends only upon previous age vector $\G_\tau$ and  the data allocation decision $a_{c_\tau}$ depends only upon channel vector $\H_{\tau}$ (note however that the quantities $\H_\tau$ and $\G_\tau$ are correlated). 
Such decoupled SMR policies are considered for mathematical tractability as well as to design less complicated and practically viable decisions -- furthermore we will observe that in almost all the examples considered the loss of optimality by considering this sub-class of DDoFSs is  negligible.

The scheduler with $\alpha = 0$ is called the   efficient scheduler (when exists), and is represented by $\bd_e^*$.
From \eqref{Eqn_alpha_fair}, such a scheduler   maximises the sum of user-utilities $O(\bd) :=  \sum_{n \in \Nc}   \rsbU^{(n)}_\bd$, i.e., $$ \sup_\bd  O(\bd)  = O(\bd_e^*) = \sum_{n \in \Nc} \rsbU^{(n)}_{\bd^*} . $$
The sum of user-utilities,  $O^*_e:= O(\bd^*_e)  $ represents the total utility derived by the BS (or the central agent) under efficient scheduler $\bd^*_e$ and hence denotes the efficiency of the scheduler.   
 When one deviates from   efficient scheduler to provide a required level of fairness, such a  sum of user-utilities    is reduced. 
Every $\alpha$ corresponds to a level of fairness and defines an optimization problem - 
we refer the sum of user-utilities,  $O(\bd^*) =  \sum_{n \in \Nc} \rsbU^{(n)}_{\bd^*}$, at the optimal scheduler $\bd^*$  of any such relevant optimization problem as   \textit{efficiency} of that problem. 
One can alternatively achieve required levels of fairness by maximizing this efficiency under certain fairness constraints (as in \cite{debayan}).
Towards this, we introduce 
  the following notion of $\Theta$-fairness (for any given vector $\Theta := (\theta_1,\cdots,\theta_{N-1})$ with each $\theta_i \ge 1$), via  the following constrained optimisation:
\begin{equation}
\label{eqn_general_theta_fair}
 \sup_{\bd} \  O(\bd)  \ \ 
 \text{ such that }  \rsbU_\bd^{(N)} - \theta_{n}\rsbU^{(n)}_\bd = 0 \text{ for all } 1 \le n < N.
\end{equation}
In the above, without loss of generality, the users are arranged in the order of their utilities at efficient scheduler, i.e.,  $\rsbU_{\bd_e^*}^{(1)}  \le \rsbU_{\bd_e^*}^{(2)} \le \cdots \le \rsbU_{\bd_e^*}^{(N)}.$
Note that the above optimisation problem is equivalent to the constrained problem in \cite{debayan} (because of \cite[Theorem 3]{debayan}) if the optimal policy $\bd$ is comprised of only data decisions, and when the constraints  are suitably modified. We now derive the solution of \eqref{eqn_general_theta_fair} that  includes optimal age decisions for the case with two users. In this case, $\Theta = \theta_1 = \theta$, hence we refer it as $\theta$-fairness.

We begin with showing the existence of a feasible point of \eqref{eqn_general_theta_fair} for two users (proof in Appendix \ref{sec_AppendixB_CN}). 

\begin{theorem}
    \label{thm_existence}
    For $N=2$ and $\theta \ge 1$, there exists a feasible point of \eqref{eqn_general_theta_fair} for $\theta-$fair DDoFS. 
\end{theorem}

The above theorem implies the existence of an optimal or $\epsilon$-optimal solution.

\begin{figure}[h] 
     \centering
     \vspace{-3mm}
     \begin{minipage}{.48\textwidth}
        \includegraphics[trim = {4cm 4cm 6cm 4cm}, clip, scale = 0.4]{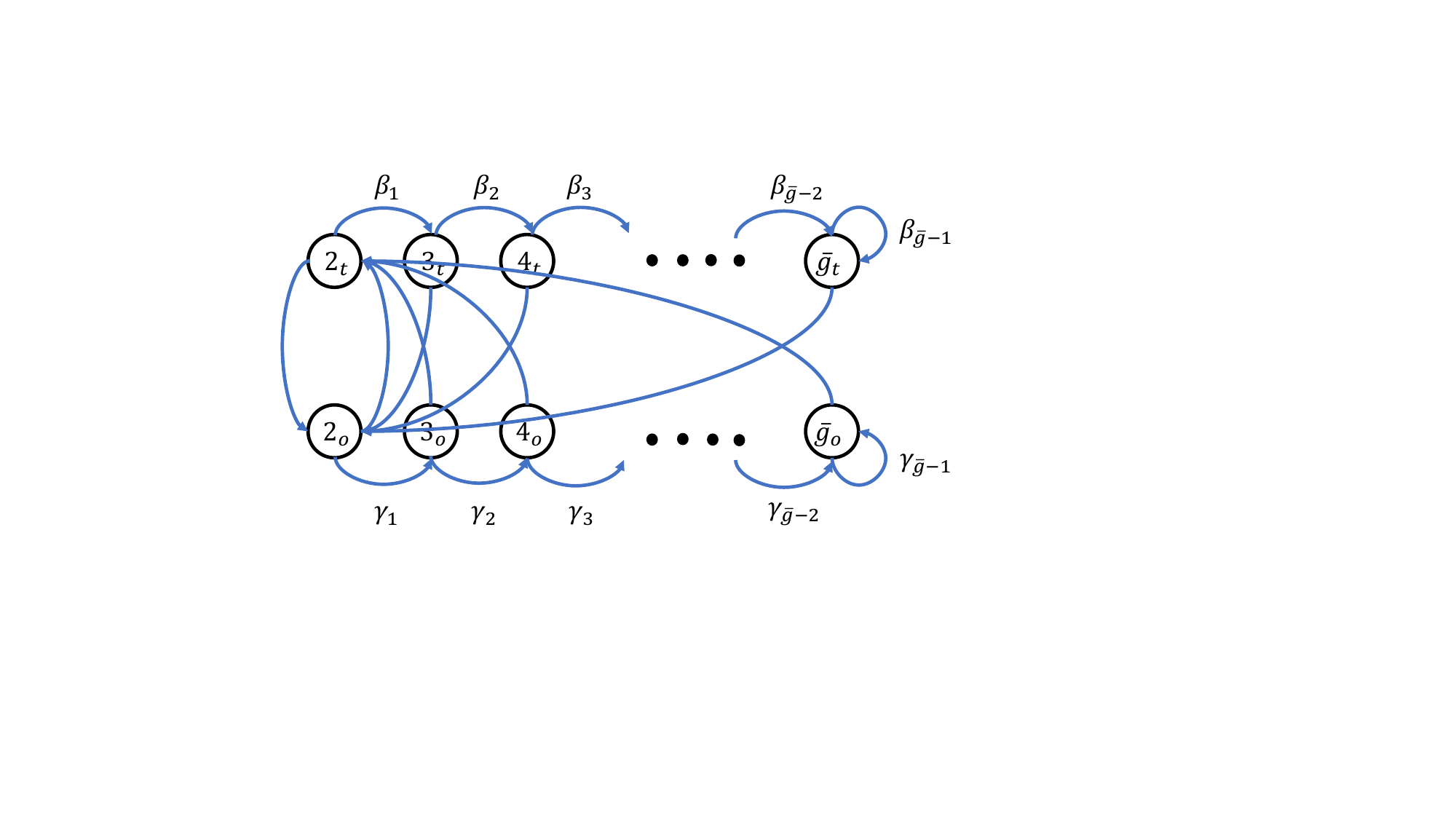} 
     \vspace{-12mm} 
      \caption{Transition Diagram of age Markov Chain}
           \label{fig:TD_MC}
     \end{minipage}
     \begin{minipage}{.48\textwidth}
         \includegraphics[trim = {4cm 4cm 0cm 5cm}, clip, scale = .4]{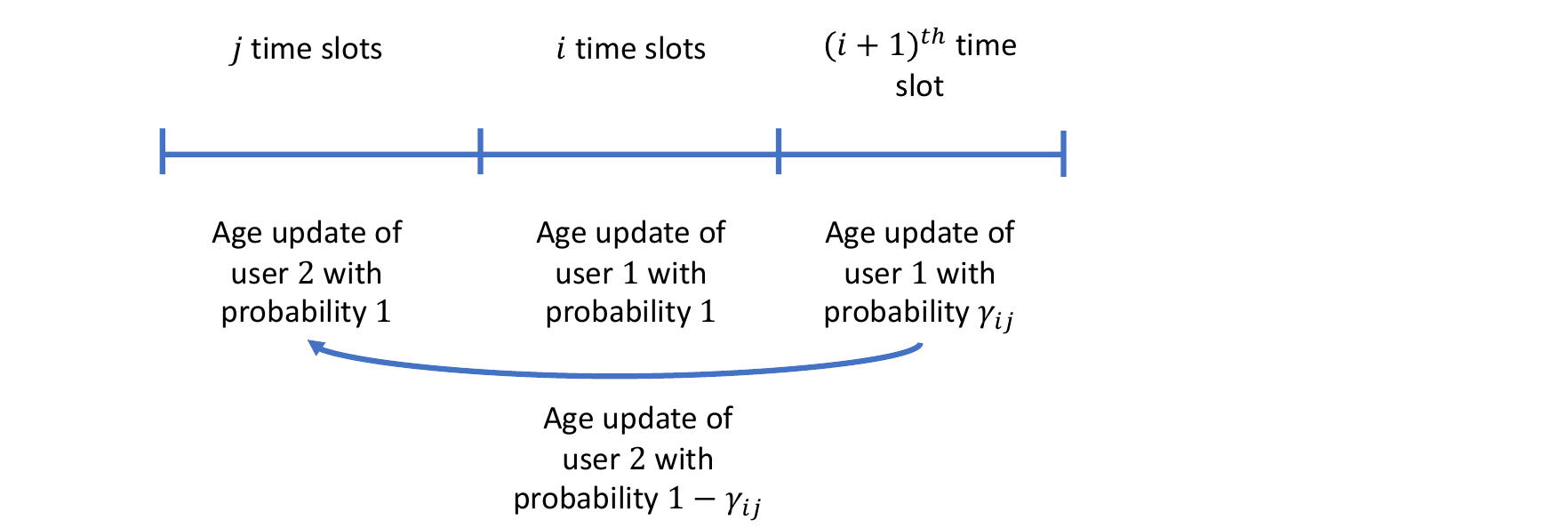}
         \caption{Random periodic age scheduler \textbf{RPA}$(i,j,\gamma_{ij}$) for $i,j < \bar g$}
         \label{fig:RPA}
     \end{minipage}    
 \end{figure}

Next, we describe the notations specific to the case of two users.   \textit{Recall  (w.l.g.)  user $1$ is inferior under efficient scheduler, i.e., that $\rsbU^{(1)}_{\bd^*_e} \le \rsbU^{(2)}_{\bd^*_e}$.}
The set of age vectors with $N=2$ simplifies to the following,
 $$
 \mathcal{G} = \left \{ (g,1) \text{ and } (1,g) \text{ such that } g \in \{1,\cdots,{\bar g}\}  \right \}.
 $$
 We briefly denote the age-states of the form $(g,1)$ and $(1,g)$ by $g_t$ (user two has age 1, while user one has age $g$) and $g_o$ (user one has age 1) respectively. Let \textit{$\bet_{j}$  represent the probability of  updating the position of user 2 when age-state is $(j+1, 1)$ -- this is the probability that the age-state transitions from $(j+1)_t$ to $(j+2)_t$.} Similarly   $\gam_i$ denotes the transition probability from age-state $(i+1)_o$ to  $(i+2)_o$ (see Figure \ref{fig:TD_MC}).
 It is easy to verify that any decoupled SMR policy for $N=2$ can alternatively be represented by parameters $\left(\bbeta, \bgamma, \{\eps(\rH)\}, \{\del(\rH)\}\right)$ where  
the corresponding data decisions are given by: \vspace{-2mm}
\begin{equation}
  d_{\eps,\delta}(\rS, a_c=1) =1-  d_{\eps,\delta}(\rS, a_c=2) = 
    \begin{cases}
    1 & \text{ if } \rsH^{(1)}-\rsH^{(2)}  > \eps(\rH), \\
    (1-\del) & \text{ if } \rsH^{(1)}-\rsH^{(2)}   = \eps(\rH), \\
    0 & \text{ else.}
    \end{cases}
\end{equation}
Such parametric dependency is captured via special notation, $\bd = d(\bbeta,\gamv,\{\eps(\rH)\},\{\del(\rH)\})$ is an SMR policy. We first show that  it is sufficient to consider optimal   among a
simplified   class  of policies  where the mapping
 $\rH \mapsto (\eps(\rH),\del(\rH))$  defining the DDoFS is represented by a  single point $(\eps,\del)$ (the proof is in Appendix \ref{sec_AppendixB_CN}).
\begin{theorem}
\label{thm_common_eps_del}
For $N=2$, the optimizer of \eqref{eqn_general_theta_fair} among the DDoFSs is within a simplified class of schedulers  $\Cc$:   every scheduler in $\Cc$ is parameterised by vectors $\bbeta, \bgamma$ and scalars $\eps,\del,$ and is represented by $d(\bbeta, \bgamma, \eps, \del)$,  where  the data decisions are as defined below using special functions $(\eps(\cdot),\del(\cdot))$
{\small \begin{equation}
\label{eqn_ep_del_h}
  d_{\eps,\delta}(\rS, a_c=1) =1-  d_{\eps,\delta}(\rS, a_c=2) = 
    \begin{cases}
    1 & \text{ if } \rsH^{(1)}-\rsH^{(2)}  > \eps(\rH) , \\
    (1-\del(\rH)) & \text{ if } \rsH^{(1)}-\rsH^{(2)}   = \eps(\rH), \\
    0 & \text{ else,}    
    \end{cases}
    \hspace{8mm}
    \begin{array}{l}
     \eps(\rH) := \eps(\theta h^{(1)}+h^{(2)}) \\ 
     \del(\rH) := \del(\theta h^{(1)}+h^{(2)}) \\  
    \end{array}
\end{equation}}
\end{theorem}

Under SMR policy $\bd = d(\bbeta,\bgamma, \eps,\delta)$ of  the above special class $\Cc$, the (decoupled) data decisions follow a certain order 
 as described below and hence we refer to data components of such dual schedulers as ordered data schedulers:
 
 {\bf Ordered Data Scheduler \textbf{OD}$(\eps, \del)$: }
\begin{eqnarray}
\label{Eqn_orderDataSchedule}
    \bd (  a_c = 1 \mid  \rsH^{(1)}, \rsH^{(2)} ) =     \begin{cases}
    1 & \text{ if } \rsH^{(1)}-\rsH^{(2)}  > \eps (\theta \rsH^{(1)}+\rsH^{(2)} ),  \\
    (1-\del) & \text{ if } \rsH^{(1)}-\rsH^{(2)}  = \eps (\theta \rsH^{(1)}+\rsH^{(2)} ) , \\
    0 & \text{ else.}
    \end{cases} 
\end{eqnarray}
 
It is not difficult to deduce that the  age components $\{\G_\tau\}$ under decoupled SMR scheduler themselves form a Markov chain and  have  transitions as in Figure \ref{fig:TD_MC}. Further analysis completely  depends upon this Markov chain. Also,  define the conditional utilities  under the  ordered data scheduler \textbf{OD}$(\eps,\del)$, conditioned that the  age-state of the Markov chain equals $g_t$,
\begin{equation}
\begin{aligned}
  \rsbU_{g_t}^{(1)} (\eps,\del) &:= \E \left [ \left . \sH^{(1)} \ind_{  \sH^{(1)} - \sH^{(2)} > \eps  \left (\theta  \sH^{(1)} + \sH^{(2)} \right ) } +  \sH^{(1)}  (1-\delta)  \ind_{  \sH^{(1)} - \sH^{(2)} =  \eps  \left (\theta  \sH^{(1)} + \sH^{(2)} \right ) } \right | \G = g_t  \right ]    \\ 
   \rsbU_{g_t}^{(2)}(\eps,\del) &:=  \E \left [ \left . \sH^{(2)} \ind_{  \sH^{(1)} - \sH^{(2)} < \eps  \left (\theta  \sH^{(1)} + \sH^{(2)} \right ) } +  \sH^{(2)}  \delta  \ind_{  \sH^{(1)} - \sH^{(2)} =  \eps  \left (\theta  \sH^{(1)} + \sH^{(2)} \right ) } \right | \G = g_t  \right ]     \mbox{ and }    \\ 
   \rsbU_{g_t}  (\eps,\del) &:= \rsbU_{g_t}^{(1)}(\eps,\del)+\rsbU_{g_t}^{(2)}(\eps,\del). 
   \end{aligned}
\end{equation}
Define similar quantities for each $g_o$. 
\textit{We use brief notations like $\rsbU_{g_t}^{(1)} $ by suppressing $(\eps,\delta)$ when there is clarity.} 
We begin with deriving the efficient scheduler using the above Markov chain.

\subsection{Efficient scheduler}

We now derive $\bd_e^*$, the efficient scheduler, which is obtained at $\alpha  =0$ in \eqref{Eqn_alpha_fair} with dual scheduler $\bd$ or  at $\theta = \theta_e := \rsbU^{(2)}_\bd/\rsbU^{(1)}_\bd$ in \eqref{eqn_general_theta_fair}, and which  maximises the sum of user-utilities.
Towards this define the following quantities at $(\eps,\del) = (0,1)$:
\begin{equation}
    \label{eqn_ubar_avg_eff}
    \Omega_m 
    = \frac{\sum_{g =2}^m \rsbU_{g_t}(0,1) + \rsbU_{2_o}(0,1)}{m} \text{ for any } m \le {\bar g}.
\end{equation}



 Now, the next result obtains the optimal dual decoupled efficient scheduler (proof in Appendix \ref{sec_AppendixB_CN}) under a natural assumption that that the average utilities reduce/remain the same with age of position update, i.e., ${\bar u}_{g_t} \ge {\bar u}_{{(g+1)}_t}$ and  ${\bar u}_{g_o} \ge {\bar u}_{{(g+1)}_o}$ for all $2 \le g \le \bar g$.

Let $\bd_{(\eta)}$, with $\eta < {\bar g}$, represent a scheduler  that updates the position of user $2$ for $(\eta-1)$ times and then updates once the position of user $1$ (irrespective of $\rH$ component of the state $\rS$) and data decisions are according to {\bf OD}$(0,1)$ -- observe this is an SMR  scheduler, which is in fact a dual scheduler.
Let $\bd_{(\eta)}$, with $\eta = {\bar g}$ be the scheduler that updates only position of user $2$ while using {\bf OD}$(0,1)$ for data scheduler.

 \begin{theorem}
\label{thm_efficient}
Say ${\bar u}_{g_t} > {\bar u}_{g_o}$ for $g = 2,3$. Let 
$\et \in \arg \max \limits_m \Omega_m$.  Then $  \bd_{(\et)} = \bd^*_e$, i.e.,  $  \bd_{(\et)} $ is an  efficient scheduler and the  optimal value,  $O(\bd^*_e)   = \Omega_\et$. Further, 
  \begin{enumerate}[label=(\roman*)]
     \item     the  stationary distribution at optimality  with $\et <\bar g$, is given by:
     \begin{eqnarray} \mu_{\bd_{(\et)}}(2_o)= \mu_{\bd_{(\et)}}({g_t}) = \frac{1}{\et} \  \text{ for any } \ 2 \le g \le \et, \  \ 
  \mu_{\bd_{(\et)}}({g_t}) = 0 \text{ for all }   g  > \eta,  \mbox{ and } \nonumber \\ \mu_{\bd_{(\et)}}(g_o) = 0 \mbox{ for all } g \ge 3, \nonumber
  \end{eqnarray}
 \item   if $\et = \bar{g}$, then the optimal
 $
 O(\bd_{(\et)}) = \Omega_\et = \bar{u}_{{\bar g}_t}  \text{ with stationary distribution, } \mu_{\bd_{(\et)}}({{\bar g}_t}) = 1
 $ and all others $0$. \eop
 \end{enumerate}
 \end{theorem}



{\bf Remarks:} $(i)$ Thus for the efficient scheduler the data decisions are according to {\bf OD}$(0,1)$ -- one can refer these as efficient data decisions as the system in any slot allocates the channel to the user with the highest instantaneous channel rate. The age scheduler is periodic with period $\eta$ given in Theorem \ref{thm_efficient}.

$(ii)$ Further  the age scheduler continuously updates the information of that user which contributes majorly towards the efficiency. For example, in  one extreme case as in Theorem \ref{thm_efficient}.(ii), the scheduler always update the age of user $2$.

$(iii)$ The above result shows that the efficient scheduler mostly tries to update the age of user $2$, which is superior. Further, depending on the degradation of the channel estimates, it sometimes also updates the age of the user $1$, once in a while.

 When one diverges from the efficient scheduler for fairness, either one has to diverge from efficient data decisions or the age scheduler should diverge from updating the major user, or a combination of the two. We precisely investigate this while deriving the $\theta$-fair schedulers with $\theta < \theta_e$ in the following.

 \subsection{$\theta$-fair schedulers}

Towards deriving $\theta$-fair schedulers for any given $\theta$, we first consider a convenient sub-class of schedulers and obtain the optimal among them.
To this end, fix $i,j,\eps,\del$ and consider the sub-class of schedulers:
$$
\Cc(i,j, \eps, \del) = \left\{d(\bbeta,\bgamma,\eps,\del): \bet_g = 0 \text{ for any $g \ge j$, $\gam_g = 0$ for any $g \ge i$, and $\bet_g, \gam_g \in [0,1]$ for all $g$} \right \}. 
$$
The theorem below identifies the sub-optimizers  among the above sub-class   (proof  is in Appendix \ref{sec_AppendixB_CN}). 
\begin{theorem}
\label{thm_SD}
Let $N=2$ and define 
\begin{equation}
\label{eqn_gamma_star}
    \gs =  -\frac{ \sum_{g=2}^{j} \left(\ut_{g_t}-\theta\uo_{g_t} \right)+  \sum_{g=2}^{i-1} \left(\ut_{g_o}-\theta\uo_{g_o} \right)}{\left(\ut_{i_o}-\theta\uo_{i_o} \right)} \mbox{ and } {\cal R}_\gamma (i) = \left \{ \begin{array}{ll}
      (0, 1]   &  \mbox{ if }   i < \bar g  \\
     (0, \infty)    &  \mbox{ if }  i = \bar g
    \end{array} \right . .
\end{equation}
i) If  $\gs \in  {\cal R}_\gamma (i) $   at least one sub-optimizer of \eqref{eqn_general_theta_fair} 
  exists in the class $\Cc(i,j,\eps,\del)$.  \\
  ii) The solution is unique if and only if $i < \bar g$ and 
the corresponding set of  sub-optimizers 
  is given by:
 \begin{equation}
\begin{aligned}
  \Cc^*(i,j, \eps, \del) = & \Bggl d(\bbeta,\bgamma,\eps,\del) \in \Cc(i,j, \eps, \del):  \left( \gamma_{i-2}, \gamma_{i-1} \right)  \in \Upsilon_i,  \\
 &    \bet_g  = 1  \mbox{ for }  \mbox{all }   2 \le g \le j-2, \ \bet_{j-1} = 0,  \text{ and }  \gam_g = 1 \text{ for all } 2 \le g \le i-3  \Bggr.
    \end{aligned}
    \label{eqn_gam_last}
    \end{equation}
where
    $$   
     \Upsilon_i =  \left \{ \begin{array}{ll}
       \left\{ \left(  \gs,0 \right ) \right \}   &  \mbox{ if }  i < \bar g  \\
      \left \{ (\varrho_1, \varrho_2) \in[0,1]^2:  \varrho_1 = (1-\varrho_2) \gs \right  \}   & \mbox{ if } i = \bar g.
    \end{array}\right. \\
    $$
iii)
One can have multiple sub-optimizers, but the stationary distribution of age-states (see Figure \ref{fig:TD_MC}) under any of those sub-optimizers is  the same. The unique stationary distribution is given by,
\begin{equation}
\label{eqn_optimal_sd}
\begin{aligned}
 \mud(g_t) = \mud(w_o) = \frac{(1-\gam_{i-1})}{(1-\gam_{i-1})(i+j-3)+\gam_{i-2}} \text{ for all } g,w \le j, i-1, \text{ and, } \\
 \mud(i_o) = \frac{\gam_{i-2}}{(1-\gam_{i-1})(i+j-3)+\gam_{i-2}}.   
\end{aligned}
\end{equation}
\eop
\end{theorem}
For the purpose of completion we 
 set $\Cc^*(i,j, \eps, \del) = \emptyset$, when there is no solution. 

From \eqref{eqn_gam_last}, the age decisions corresponding to the optimizer (one of them when $i=\bar g$) are  random and periodic   as defined below:
\begin{definition} {\bf Random periodic age scheduler -- RPA($i,j, \gamma_{ij}$):} 
This age scheduler has a random periodic sequence of age decisions: i) it first updates the position information of user 1 in   $i$ consecutive time slots; ii) it then updates the position information of user 2 in   $j$ consecutive time slots; 
iii) in the next slot it updates the position of user 2 with probability $\gamma_{ij}$; iv) after that it starts updating information of user 1 and continues as in previous steps.

 \end{definition}

 \ignore{\begin{figure}[h]
     \centering
     \includegraphics[trim = {0cm 4cm 0cm 5cm}, clip, scale = .5]{RPA.pdf}
     \caption{Random periodic age scheduler \textbf{RPA}$(i,j,\gamma_{ij}$)}
     \label{fig:RPA}
 \end{figure}}

Thus by Theorem \ref{thm_SD} and  from \eqref{eqn_gam_last}, the optimal DDoFS among  $\Cc(i, j, \eps, \delta)$ (if one exists) has data decisions given by ordered scheduler \textbf{OD}$(\eps,\del)$ as in \eqref{Eqn_orderDataSchedule} and the age decisions given by random periodic scheduler as below,
\begin{equation}
\bd^*_{i,j,\eps,\del} =    \begin{cases}
  \left(\text{\textbf{OD}}(\eps,\del),\text{\textbf{RPA}}(i, j, \gs)\right )   & \text{ if } i < {\bar g} \vspace{1mm}  \\ 
 \left(\text{\textbf{OD}}(\eps,\del),\text{\textbf{RPA}} \left (i, j,1-\frac{1}{ \gs}  \right  ) \right)  & \text{ if } i = \bar g.
\end{cases} 
\label{eqn_DDoFS_scheduler}
\end{equation}

\subsection{$\theta$-fair $\rho$-approximate Dual decoupled Scheduler ($\theta,\rho$-DDoFS) }
\label{sec_approx}
By Theorem \ref{thm_common_eps_del},   the  optimizers of  \eqref{eqn_general_theta_fair} among DDoFS is from the following set
\begin{equation}
\bigcup_{i,j}  \bigcup_{\eps \in {\cal E}, \  \del \in [0,1] } \Cc (i, j, \eps, \del), 
\mbox{ where }  {\cal E} := \left \{ \frac{h^{(1)}-h^{(2)}}{\theta h^{(1)}  + h^{(2)}} : h^{(1)}, h^{(2)} \in {\cal H} \right \}.
\end{equation}
Further by Theorem \ref{thm_SD} and  \eqref{eqn_gamma_star} it is sufficient to consider the following set   of (\textbf{OD},\textbf{RPA}) schedulers (of form  as in \ref{eqn_DDoFS_scheduler})  
\begin{equation}
\label{eqn_decoupled_set}
{\cal D} := \left \{ \bd^*_{i,j, \varepsilon , \del} : \ \del  \in [0,1],  \varepsilon \in  {\cal E}, \   \gs  \in {\cal R}_\gamma (i), \ i, j \le \bar g,  \right \}, \mbox{ where }  {\cal E} := \left \{ \frac{h^{(1)}-h^{(2)}}{\theta h^{(1)}  + h^{(2)}} : h^{(1)}, h^{(2)} \in {\cal H} \right \}.
\end{equation}
The cardinality $|{\cal E}|$ is finite, as that of 
  $|{\cal H}|$ is finite, and hence the following is a finite subset of ${\cal D}$ for any $\rho \in (0,1)$,
$${\cal D}_\rho := \left \{ \bd^*_{i,j, \varepsilon , \del} : \del = k\rho \mbox{ for some integer } k \ge 0,   k\rho \le 1,  \varepsilon \in {\cal E},  \  \gs  \in {\cal R}_\gamma (i), \ i, j \le {\bar g}  \right \}.$$
One can clearly anticipate that the optimal among ${\cal D}_\rho$ approaches the best among ${\cal D}$, there by towards a DDoFS that optimizes  \eqref{eqn_general_theta_fair}. 
Thus we define  a $\theta$-fair $\rho$-approximate DDoFS as:
\begin{eqnarray} 
\label{eqn_rho_scheduler}
\bd^*_{\theta, \, \rho} := \arg \max_{ \bd \in  {\cal D}_\rho  }  \left ( \rsbU^{(1)}_\bd  +  \rsbU^{(2)}_\bd  \right ).
\end{eqnarray}
Observe that $\bd^*_{\theta, \, \rho}$ has near-closed-form-expression.
One requires a technical proof to show that the above schedulers approach $\theta$-fair DDoFS, as $\rho \to 0.$ We would instead focus on illustrating the same using numerical examples, while the technical proof is deferred to future work.

\ignore{Under certain conditions, one can obtain a better approximation among a much smaller set of schedulers. 
Towards this
  fix $j, \eps, \del$ and consider a bigger sub-class
$$
\Cc(j,\eps,\del) =  \left\{d(\bbeta,\bgamma,\eps,\del) \in \Cc(i,j,\eps,\del): 2 \le i \le {\bar g}  \right \}, 
$$to identify  the corresponding sub-optimal schedulers  
  under certain conditions (proof is in Appendix B)
\begin{thm}
    \label{thm_i_eps_del_scheduler}
    If  $\Cc(j,\eps,\del) \neq \emptyset$, then the optimality among $\Cc(j,\eps,\del)$ is achieved at a unique point with `minimal' $i$ if $\rsbU_{g_t}>\rsbU_{{\bar g}_o}$ for all $g$:
    \begin{equation}
    \label{eqn_gam_last_two}
        \Cc^*(j,\eps,\del) =  \bigg \{d(\bbeta,\bgamma,\eps,\del) \in \Cc^*(i,j,\eps,\del): \gamma^*_{(i-1)j} \notin (0, 1]     \bigg \},
    \end{equation}
    else set $\Cc^*(j,\eps,\del) = 0.$ \eop
\end{thm}
Thus    from \eqref{eqn_gam_last_two}, the optimal DDoFS among  $\Cc(j, \eps, \delta)$ (if one exists) is given by, 
$$\bd^*_{j,\eps,\del} =  \bd^*_{i^*(j), j,\eps,\del}, \mbox{ where } 
i^*(j) := \min \{i : \gamma^*_{(i-1)j} \notin (0,1]\}
$$

Define $\Cc(i,j)$ to be the class of SMR schedulers such that $\bet_g= 0$ for any $g > i$ (i.e., when the position information of user 1 has age above $g$ and that of user 2 is $1$), and $\gam_g = 0$ for any $g \ge j$. We first obtain the sub-optimisers of \eqref{eqn_general_theta_fair} that optimise among schedulers of $\Cc(i,j).$}

\subsection{$\theta$-fair Dual decoupled Scheduler ($\theta$-DDoFS) }
\label{sec_algo}
We now find $\theta$-DDoFS, 
  the optimizer  of \eqref{eqn_general_theta_fair}  among DDoFSs or equivalently from among ${\cal D}$. As in previous sub-section one can find the best among a sub-class with fixed $(\eps,\del)$; this  again has `near-closed-form-expression' as it is the best among the following set of finitely many schedulers:
  $$
  \bd^*_{\eps,\del} = \arg \max_{\bd \in {\cal D}_{\eps,\del}  }   \left ( \rsbU^{(1)}_\bd  +  \rsbU^{(2)}_\bd  \right ), \mbox{ where  }  {\cal D}_{\eps,\del} := \{ \bd^*_{i, j, \eps,\del} : \gs \in {\cal R}_\gamma (i) \}.
  $$
  One can then obtain $\theta$-DDoFS by optimizing over  $(\eps, \del)$.  
We provide such an iterative procedure in Algorithm \ref{algo_theta_Fair} by using a gradient based method.  

\begin{algorithm}[]
\caption{Gradient based $\theta$-fair DDoFS}
{\bf Inputs:} $\rho > 0$, $\theta$,  ${\cal E}$ of \eqref{eqn_decoupled_set}  \\
{\bf Initialization:} Order the set ${\cal E}$ in decreasing fashion  and let the ordered set  ${\cal E}_o $ \\Initialize $\bd$, $\eps \in {\cal E}$ and $ \del$ using the corresponding values defining  $ \bd^*_{\theta, \, \rho}$ of \eqref{eqn_rho_scheduler}. Set $E= \rsbU_\bd ^{(1)} + \rsbU_\bd ^{(2)}$.\\   Set gradient $\nabla_\delta $ with a small value and $k=1$. \\
{\bf Step k:}
i) {\bf Update} $\delta_o \leftarrow \delta$, $E_o \leftarrow E$ \\
ii) {\bf Update}  
$
\delta \leftarrow \delta + \frac{1}{k+1} \nabla_\delta 
$\\
 {\bf Projection:} If $\delta < 0$, then set $\delta = 1$  and update $\eps$ to the previous value in ${\cal E}_o$ (if there is one).\\
if $\del > 1$ then set $\delta = 0$  and update $\eps$ to the next value in ${\cal E}_o$ (if there is one).
\\
 If no such $\eps$ exists, then STOP. \\
iii) {\bf Update} $\bd \leftarrow \bd^*_{\eps,\del}$,  $E \leftarrow \rsbU_\bd ^{(1)} + \rsbU_\bd ^{(2)}$ \\
iii) {\bf Update gradient } $\nabla_\delta \leftarrow (E-E_o)/(\delta-\delta_o)$
\\
Update $k \leftarrow k+1 $ and go back to {\bf Step k} until the algorithm converges.

\label{algo_theta_Fair}

\end{algorithm}

\ignore{\begin{algorithm}[H]
\SetAlgoLined
\textbf{Input:}~$N, K, {\bar g}, \Theta, \text{ and}$ set of difference between the channel estimates of user~$1$ and~$2$ (in decreasing order),~$ \mathcal{J}$ \\
\textbf{Output: }Optimal $(i,j,\Ps,\eps,\del)$ \\
\textbf{Initialise} iter $= 0$ \\
\noindent
 \hspace{5mm} \textbf{For} $\eps \in \mathcal{J}$  \textbf{do} \\ 
 \hspace{10mm} \textbf{For} $\del = 0:0.01:1$ \textbf{do} \\
 \hspace{15mm} $i,j,\gs,\rsbU^{(1)}_\bd,\rsbU^{(2)}_\bd,O^*_{\eps, \del}$ = 
BestForGivenEpsDel$(\eps,\del,{\bar g})$ \\
 \hspace{10mm} \textbf{end} \\
 \hspace{5mm} \textbf{end} \\
 \hspace{5 mm} Store $i,j,\gs,\rsbU^{(1)}_\bd,\rsbU^{(2)}_\bd,O^*$
 \\
 \hspace{5mm} \textbf{While} $\eps, \del$ converges \\
 \hspace{10mm} $i,j,\gs,\rsbU^{(1)}_\bd,\rsbU^{(2)}_\bd,O^*_{\eps, \del}$ = BestForGivenEpsDel$(\eps,\del,{\bar g})$ \\
 \hspace{10mm} \textbf{If} BestForGivenEpsDel$(\eps,\del,{\bar g}) = \emptyset$ \\
 \hspace{15mm} \\
 \hspace{10mm} \textbf{end} \\
 \hspace{5mm} \textbf{end}

\medskip

 \textbf{function} [$i,j,\gs,\rsbU^{(1)}_\bd,\rsbU^{(2)}_\bd,O^*_{\eps, \del}$] = BestForGivenEpsDel($\eps,\del,{\bar g}$) \\
 \hspace{5mm} \textbf{For} $i = 1,2,\cdots,{\bar g}$ \textbf{do} \\
\hspace{10mm}  \textbf{For} $j = 1, 2, \cdots,{\bar g}$ \textbf{do} \\
 \hspace{15mm} iter = iter + 1 \\
 \hspace{15mm} Use Theorem \ref{lem_SD} to obtain $\bmu$ \\
 \hspace{15mm} \textbf{If} $j = {\bar g}, $ and $\gs > 0$ or $j < {\bar g}$ and $0 < \gs \le 1$ \\
\hspace{20mm} Store $i,j,\gs,\rsbU^{(1)}_\bd,\rsbU^{(2)}_\bd,O_{\eps,\del,\text{iter}}$ with $O_{\eps,\del,\text{iter}}$ denoting the objective value \\
\hspace{20mm} \textbf{If} 
$O_{\eps,\del,\text{iter}} > O_{\eps,\del,\text{iter-1}}$ \\
\hspace{25mm} Replace values of previous iteration \\
\hspace{20mm} \textbf{end} \\
\hspace{15mm} \textbf{end} \\
 \hspace{10mm} \textbf{end} \\ 
\hspace{5mm} \textbf{end} \\
 \textbf{end} \\
 \caption{Dual Fair Algorithm 2 with $\Ps = \frac{\gam_{j-2}}{1-\gam_{j-1}}$}

\end{algorithm}}


\ignore{\subsection {Max-min scheduler}
In this section, we assume that the data decisions are same as efficient



  \begin{thm}
    \label{thm_existence}
    For $\theta = 1$, there exists an optimal dual $\Theta$-fair scheduler $\bd^*$ of \eqref{eqn_general_theta_fair} if $\rsbU_{\It_o}^{(2)}-\theta\rsbU_{\It_o}^{(1)} < 0$ when data decisions are efficient.
\end{thm}
\noindent
\textbf{Proof: } From \eqref{eqn_stationary_orig}, observe that $\mud(2_t) = \mud(2_o)$.

\begin{equation}
\label{eq_con_expand}
    \left(\ut_{2_t}-\theta\uo_{2_t}+\ut_{2_o}-\theta\uo_{2_o} \right)\mud(2_t) + \sum_{g=3}^{\Io} \left(\ut_{g_t}-\theta\uo_{g_t} \right)\mud(g_t) + \sum_{g=3}^{\It} \left(\ut_{g_o}-\theta\uo_{g_o} \right)\mud(g_o)  = 0.
\end{equation}

 Since the value of \eqref{eq_con_expand} is $>0$  at a $\bd$ which allocates data channel always to user $2$, irrespective of the age decisions. Similarly, the value of \eqref{eq_con_expand} is $<0$  at a $\bd$ which allocates data channel always to user $1$, irrespective of the age decisions, and thus the existence of the solution now follows from Intermediate Value Theorem.
\eop

\begin{thm}
\label{thm_max_min_states_cut}
If the Markov chain visits states $g_t$ for $g \le \et$ under efficient, then the
 Markov chain will never visit state $g_t$ for any $g>\et$ under max-min constraint with efficient data decisions.
\end{thm}

\noindent
\textbf{Proof: }We have the optimisation problem as in \eqref{eqn_theta_fair_optimisation} with $\theta=1.$
    We know ${\bar u}_{g_t} > {\bar u}_{w_t}$ for all $g+1 \le w \le \Io$  and ${\bar u}_{2_o} > {\bar u}_{w_t}$ for all $w > \et$. Further, $\ut_{g_t}-\uo_{g_t} < \ut_{w_t} - \uo_{w_t}$ since $\ut_{g_t} < \ut_{w_t}$ and $\uo_{g_t} > \uo_{w_t}$ for all $g+1 \le w \le n$. 
    Consider the optimal stationary distribution with $\mud^*((g+1)_t), \mud^*(g_t)>0$ for any $g\ge \et.$ Thus, if we move mass from $\mud^*((g+1)_t)$ to $\mud^*(g_t)$, the objective function is improved. Also, $\ut_{(g+1)_t} - \uo_{(g+1)_t} > \ut_{g_t}-\uo_{g_t}$ because of which constraint is now negative. Thus, we need to move mass from $\mud(g_o)$ if $\ut_{g_o}-\uo_{g_o}<0$ (such a $g$ is bound to exist as $\ut_{g_t}-\uo_{g_t}>0$ for all $g$ and the constraint is negative). Further, we also know ${\bar u}_{l_t}>{\bar u}_{g_o}$ for all $g \ge 3$, which improves the objective. \eop }

    \section{ Numerical examples}

In this section, we demonstrate the superiority of the proposed algorithms (i.e., Algorithm \ref{algo_offline}, \ref{algo_theta_Fair}) in this chapter. We further demonstrate that the $(\theta,\rho)$-DDoFS of sub-section \ref{sec_approx} (which is easier to implement) converges to the solutions of the other two other algorithms. We also show that these algorithm outperforms the non-opportunistic schedulers and the online algorithm of \cite{itc} (which we refer as SDoFS). Towards this, we again assume  $O(g^{(n)}) = 1$ for all $g \le \bar g$ and $n \le N$, i.e., no time is lost in aligning the beam for NoFS.

\ignore{We first consider an example with $N=2$. Each user has $5$ channel estimates. The details are as below,
\begin{eqnarray*}
\text{ channel condition matrix, } \mathbb{C} &=& \begin{bmatrix}
    1.2 & 0.8 & .7 & .5 &.2 \\
            1.6 & 1.3 & 1.2 & .7 & .2
\end{bmatrix}\text{ and, } \\
\text{ channel probabilities, } \mathbb{Q} & = & \begin{bmatrix}
    0.4545  &  0.2727  &  0.0909  &  0.0909  &  0.0909 \\
    0.4444  &  0.2222 &   0.1111    & 0.1111  &  0.1111 \\
    0.3333  &  0.2222 &   0.2222   & 0.1111  &  0.1111 \\
    0.2000  &  0.2000  &  0.2000   & 0.2000  &  0.2000
\end{bmatrix}
\end{eqnarray*}

Each row in $\mathbb{C}$ and $\mathbb{Q}$ matrices corresponds to one of the users.  The entry in the $n^{th}$ row and $m^{th}$ column of $\mathbb{Q}$ matrix depicts the probability of user $k$ having the  channel condition in the $k^{th}$ row and $m^{th}$ column of $\mathbb{C}$ matrix  when its age is $m$.
The performance under the $\alpha$-fair, $\theta$-fair and in fact even under $(\theta,\rho)$-fair and approximate scheduler is almost the same (see Figure \ref{fig:new_schedulers}). In fact the three sets of curves are indistinguishable from each other and hence we used markers of 3 different sizes (and black lines) to illustrate that they are almost the same. 

\begin{figure} [h]
    \centering
    \begin{minipage}{.56 \textwidth}
\includegraphics[trim = {0cm 6cm 0cm 5cm}, clip, scale = .4]{Fig1.pdf}  
    \end{minipage}
     \begin{minipage}{.34 \textwidth}
\includegraphics[trim = {1cm 11cm 0cm 2cm}, clip, scale = .4]{Fig2.pdf}
\end{minipage}
  \caption{ Comparison between $\alpha$-DoFS, $(\theta,\rho)$-DDoFS, and $\theta$-DDoFS
}  
    \label{fig:new_schedulers}
\end{figure}}

\begin{figure} [h]
  \begin{minipage}{.56 \textwidth}
    \includegraphics[trim = {0cm 4cm 0cm 4cm}, clip, scale = 0.36]{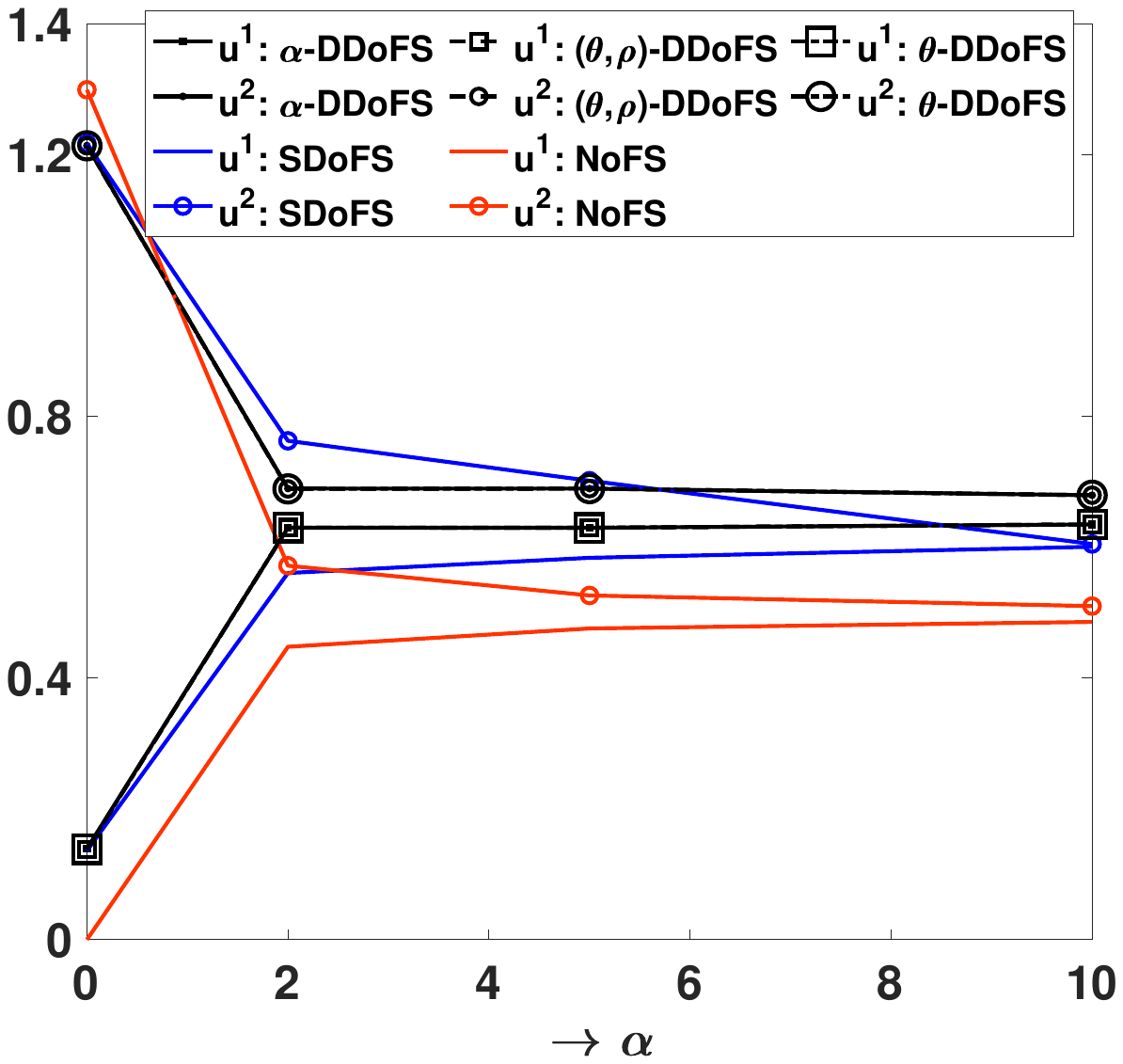}
        \end{minipage}
     \begin{minipage}{.34 \textwidth}
     \vspace{-4mm}
       \includegraphics[trim = {0cm 4cm 0cm 4cm}, clip, scale = 0.36]{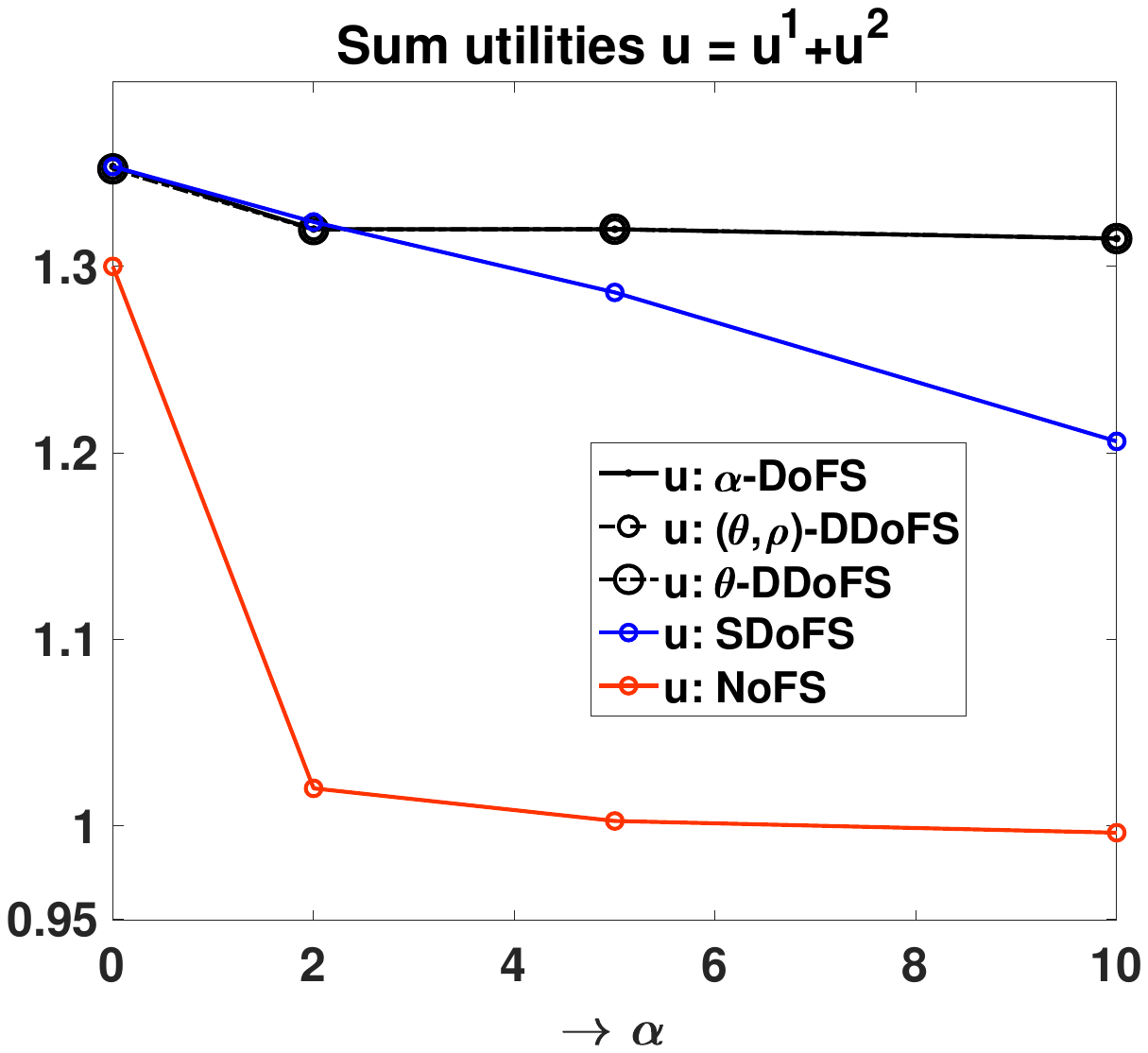}
     \end{minipage}
    \caption{Comparison of all schedulers: $\mathbb{C} = \begin{bmatrix}
    1.4 & .2 \\
    1.5 & .5
    \end{bmatrix}$ and $\mathbb{Q}
 = \begin{bmatrix}
 .5 & .4 & .3 & .2 \\
            .8 & .7 & .6 & .5
            \end{bmatrix}$, and $L_g = 1$ for all $g$}
    \label{fig:all_schedulers}
\end{figure}

In Figure \ref{fig:all_schedulers}, we consider  an example that compares the performance of all the schedulers studied in this chapter. In this example, we have $2$ users each with $2$ channel estimates (see caption in Figure \ref{fig:all_schedulers}). The notations are same as explained in sub-section \ref{sec_num1}.  
The left sub-figure has individual utilities while the right one has corresponding sum of user-utilities. The utilities under first three schedulers plotted using black lines with varying sizes of markers are the utilities under $\alpha$-fair, $(\theta,\rho)$-fair-approximate and $\theta$-fair schedulers respectively, while the red lines correspond to non-opportunistic scheduler (NoFS) of sub-section \ref{sec_NoFS}. While plotting $\theta$-fair schedulers, we first derived a value of $\theta$ that corresponds to the given value of $\alpha$ as in the previous example. 
In our initial conference paper \cite{itc}, we consider a sub-optimal solution for the same problem--MDP framework was applied only for age decisions, while the  data choices were according to  the  $\alpha$-fair scheduler of previous generation networks. We refer it as sub-optimal DoFS (SDoFS) in this study. The red-lines in both sub-figues of   Figure \ref{fig:all_schedulers}  illustrate the utilities under the sub-optimal DoFS of \cite{itc}. 

One can make several observations from Figure \ref{fig:all_schedulers}. 
\begin{enumerate}[label=(\roman*)]

    \item The performance under the $\alpha$-fair, $\theta$-fair and in fact even under $(\theta,\rho)$-fair and approximate scheduler is almost the same. In fact the three sets of curves are indistinguishable from each other and hence we used markers of 3 different sizes (and black lines) to illustrate that they are almost the same. 

    \item These three schedulers significantly out-perform both the non-opportunistic scheduler of subsection \ref{sec_NoFS}, as well as the sub-optimal DoFS of \cite{itc}.
    
    \item  By varying $\alpha$ from $0$ to 10, one can cover all levels of fairness. At one end,  when $\alpha = 0$, we have efficient schedulers with maximum sum of user-utilities, but also with maximum difference between the utilities of the two users - thus we have maximum efficiency but are `minimal' with respect to fairness.  At the other end, when $\alpha =10$, we almost  have max-min fairness where both the users derive equal utility; however the efficiency or the sum of user-utilities is the least.
    \item Furthermore the efficiency is significantly less for non-opportunistic schedulers. The sub-optimal DoFS of \cite{itc} performs on par with the dual schedulers of this chapter for small values of $\alpha$ (i.e., for small levels of fairness), however the efficiency is sufficiently reduced for larger $\alpha$; nonetheless its efficiency is significantly more than the non-opportunistic scheduler. 
\end{enumerate}

\section{Summary}

We investigate\rev{d} a system comprising a base station and multiple mobile/stationary users, where data transmission employ\rev{ed} millimeter waves (mmWaves) necessitating beam alignment towards the end-users.
In the context of earlier generation networks, opportunistic schedulers have been widely recognized for achieving an optimal balance between system efficiency (sum of user-utilities) and fairness (minimizing differences in individual user utilities). These schedulers rely on accurate channel condition estimations for each user in every time slot, hence requiring precise beam alignment.

In this study, we propose\rev{d} leveraging regular updates of user positions to facilitate accurate beam alignment with multiple users, thus paving the way for opportunistic mmWave schedulers.  We propose\rev{d} an  algorithm that uses a dual opportunistic and fair scheduler to allocate data and position-update channels in each slot. By incorporating the impact of user-position-based beam alignment, the dual scheduler optimizes the well-known alpha-fair objective function of the individual user-utilities.  Notably, the proposed schedulers have near-closed-form expressions -- one has to choose the best from  a finite set, each of which has a closed-form expression for the case of two users. 

Furthermore, we compare\rev{d} the performance of the proposed opportunistic schedulers against previously suggested mmWave schemes. The latter schedulers select\rev{ed} one user per slot and commence\rev{d} data transmission only after achieving accurate beam alignment. Our results demonstrate\rev{d} that the opportunistic mmWave dual schedulers outperform\rev{ed} the previous approaches significantly and exhibit remarkable versatility in achieving any desired level of fairness.

\chapter{Conclusions}
\label{chap_conclusions}

This thesis contributes towards various domains. The first contribution is towards cooperative game theory, where we extended various solution concepts like Shapley value, and core to partition form games, and provided a method to transform any partition form game to a characteristic form game. We also introduced new and more meaningful notions of stability, which we refer to  as the stability against `Restricted Blocking'.

We also contribute towards queueing literature. We consider an elaborate study where the  customer base can shift across the service providers, depending upon the quality of service delivered by the individual providers. This introduces competition among providers through \textit{market segmentation} and we show that the only stable partitions are the duopolies. This is in contrast to the existing literature where each provider has its own dedicated customer base and the grand coalition is proved to be stable.
    
    Next, we consider a new variant of the average cost Markov Decision Process, which optimizes a concave function of finitely many average utilities, rather than directly optimizing a single average utility.
   Here we also extend the concept of opportunistic schedulers (well-known in the context of previous generation networks) to the future generation networks, where dual decisions are proposed to be made in any time slot.

   We now discuss the chapter-wise contributions.

   In Chapter~\ref{chap_PEVA}, we considered a  coalition formation game with players exploring cooperation opportunities in a non-cooperative manner, where the utilities of players/coalitions are resultant of a resource sharing game.  We developed a framework 
to study the partitions  (non-overlapping and exhaustive set of  coalitions) that emerge at equilibrium. 

We observe that no one collaborates at equilibrium (if $n >4$) with equal or almost equal players.  Further,   none of the partitions are coalitionally stable for the symmetric case with $n>4$.

However, when the players are significantly different,  every partition is stable against unilateral deviations, while grand coalition is the only partition stable against coalitional deviations (for a special case).   
%
For the system with intermediate players, the number of stable partitions (stable against unilateral deviations) increase as asymmetry (a measure of differences in the influence factors of various players)
increases. Further and more interestingly, it is the highest and the lowest capacity players that first find it beneficial to collaborate (form joint coalitions in some partitions that emerge at equilibrium). 


 Next, we consider an Erlang-B (lossy) queueing system with several strategic service providers with different server capacities in Chapter~\ref{chap_OR}.  Each service provider is on lookout for collaboration opportunities if it improves their individual payoffs. As a resultant of this, the customer base responds to these collaborations, and the customer arrivals are split across various operational units (or coalitions) according to the well known Wardrop equilibrium (WE). WE equalizes the steady state blocking probability of all the units. We first have an`impossibility result' where we show that no partition is stable under classical notions of stability.    We then defined  new and more meaningful notions of stability where blocking is allowed only by mergers or splits. Our major findings are: (a) configurations with more than two coalitions are never stable; (b) grand coalition can be stable  only if there exists a single dominant player with more than half the server capacity of the system; and (c) some configurations with two coalitions are stable and there always exists at least one such configuration. We also study the impact of overall congestion in the system, under heavy and light traffic regimes. Finally, we present the dynamic version of this game.

 In Chapter \ref{chap_wireless}, we investigate a system comprising a base station and multiple mobile/stationary users, where data transmission employs millimeter waves (mmWaves) necessitating beam alignment towards the end-users.
In the context of earlier generation networks, opportunistic schedulers have been widely recognized for achieving an optimal balance between system efficiency (sum of user-utilities) and fairness (minimizing differences in individual user utilities). These schedulers rely on accurate channel condition estimations for each user in every time slot, hence requiring precise beam alignment.

In this study, we propose leveraging regular updates of user positions to facilitate accurate beam alignment with multiple users, thus paving the way for opportunistic mmWave schedulers.  We propose an  algorithm that uses a dual opportunistic and fair scheduler to allocate data and position-update channels in each slot. By incorporating the impact of user-position-based beam alignment, the dual scheduler optimizes the well-known alpha-fair objective function of the individual user-utilities.  Notably, the proposed schedulers have near-closed-form expressions -- one has to choose the best from  a finite set, each of which has a closed-form expression for the case of two users. 

Furthermore, we compare the performance of the proposed opportunistic schedulers against previously suggested mmWave schemes. The latter schedulers select one user per slot and commence data transmission only after achieving accurate beam alignment. Our results demonstrate that the opportunistic mmWave dual schedulers outperform the previous approaches significantly and exhibit remarkable versatility in achieving any desired level of fairness.

\subsection*{Future Directions}
We proposed new notions of coalitional stability for partition form games  in Chapter~\ref{chap_OR}. One can explore this new notion to study many other applications. 

In Chapter~\ref{chap_PEVA}, one can seek to derive answers to  several other interesting questions. Some of them are: how does an adamant player influence the equilibria with dissimilar players? 
  What if the players have partial information about the strengths of their opponents? What happens if the worth of coalitions is divided according to some other solution concept than Shapley value, etc.?  

  The work in Chapter~\ref{chap_OR} highlights that in competitive service systems enjoying statistical
economies of scale, coalition formation games have very distinct
equilibria when the total payoff across agents is a constant. In
particular, we demonstrate that \emph{duopolies} emerge, with the
dominant coalition exploiting economies of scale to corner a
disproportionate fraction of the total payoff.

This work motivates future work along several directions. Firstly, one could explore alternative models for a coalition's utility. For instance, one could define the utility of a coalition to be the rate of customers \emph{served} (rather than the rate of customer \emph{arrivals}); this is meaningful in scenarios where providers only earn revenue when a customer is successfully served. Preliminary analysis suggests that this modification of the utility structure alters the nature of stable equilibria. More generally, this work motivates a systematic understanding of how payoff structures influence
the nature of equilibria in partition form games. 
Another potential direction of inquiry involves exploring the effect of different queueing models, including models where customers can wait for service with/without balking or reneging.
Finally, it would also be interesting to explore dynamic variants of coalition formation games. This would entail examining whether any limiting behaviors emerge (particularly when stable equilibria do not exist). 

In Chapter \ref{chap_wireless}, we introduced dual decoupled schedulers for two users, which offer much-simplified solutions. It would be interesting to derive such schedulers for the case with  general  number of users. 

Currently, our approach allows for the update of only one user's position during each time slot, with only one user being assigned the data channel at a time. Alternatively, a dynamic decision-making process can be employed to determine if a position update is necessary and to determine the number of users requiring such an update. The remaining time slot can then be utilized for data transmission.

Another promising direction to consider is enabling the updating of the transmitting user's position. 
This action resets the age of the scheduled user to one, potentially resulting in a significantly improved dual scheduler, even in spite of the loss of some time for a second age update in the same slot.


\begin{appendices}
\renewcommand{\thechapter}{\Roman{chapter}}
	\chapter{Proofs Related to Chapter~\ref{chap_PEVA}}
\label{chap_appendixI}

\section{Appendix A}
\label{sec_AppendixA_PEVA}

{\bf Proof of Theorems \ref{Thm_util_coal_general} and \ref{thm:Thm1}:} Theorem \ref{thm:Thm1} is a special case of Theorem \ref{Thm_util_coal_general} and hence, we provide only the proof of Theorem \ref{Thm_util_coal_general}. For uniformity of notations assume adamant players is indexed by $1$,  others by $j \ge 2$ and consider $n-1$ S-players  for Theorem \ref{thm:Thm1}.
Consider any partition $\P = \{S_1,\cdots,S_k\}$. 
We omit superscript $^\P$ in some notations in the proof for ease of explanation. 
From equation \eqref{Eqn_Util_coaltiion_asym}, the utility of a coalition $S_m \in \P $  without cost factor $\gamma$ is given by,
\begin{eqnarray*}
\frac{\sum_{j \epsilon S_m} \lambda_ja_j}{\sum_{l=1}^n \lambda_la_l} \ = \ 1- \frac{\sum_{j \notin S_m}\lambda_ja_j}{\sum_{l=1}^n \lambda_la_l} \ < \  1- \frac{\sum_{j \notin S_m}\lambda_ja_j}{\bar{\lambda}_m \sum_{j \epsilon S_m}a_j + \sum_{j \notin S_m}\lambda_ja_j  } \ = \  1- \frac{\sum_{j \notin S_m}\lambda_ja_j}{\bar{\lambda}_m\bar{a}_m + \sum_{j \notin S_m}\lambda_ja_j  } \\ \text{ where } {\bar a}_m :=  \sum_{j \in  S_m}a_j.
\end{eqnarray*}
The utility of coalition $S_m $, for any $  m \leq k$,  can be upper bounded by:

\begin{eqnarray*}
\frac{\sum_{j \epsilon S_m} \lambda_ja_j}{\sum_{j=1}^n \lambda_ja_j} - \gamma \sum_{j \epsilon S_m}a_j \ < \
 1- \frac{\sum_{j \notin S_m}\lambda_ja_j}{\bar{\lambda}_m \bar{a}_m + \sum_{j \notin S_m}\lambda_ja_j} -\gamma \sum_{j \epsilon S_m}a_j \ = \  1- \frac{\sum_{j \notin S_m}\lambda_ja_j}{\bar{\lambda}_m \bar{a}_m + \sum_{j \notin S_m}\lambda_ja_j} -\gamma \bar{a}_m.
\end{eqnarray*}
The bound $\hat{a}  > n / \gamma$ on the actions is sufficiently large (as in \cite{dhounchak2019participate}), hence we can take $\bar{a}_m = \sum_{j \in S_m}a_j$ as the action of active player alone in $S_m$. Thus we  have,
\begin{equation}
    \max_{a_g \;\in \; [0, \hat{a}]^{l_m} } \varphi_{S_m} (a_g) \; \leq  \max_{a_g \;\in \; [0,  \hat{a}] \; \times \{0\}^{l_{m}-1} } \varphi_{S_m} (a_g). 
\label{eq:Eq3.21}
\end{equation}
\begin{tabular}{l l l}
  where & $\varphi_{S_m}$ & represents the utility of $S_m$ coalition, \\
  & $l_m$ & represents $|S_m|$, i.e., number of players in $S_m$, and  \\
&  $a_g$ & represents the $l_m$-dimension vector of actions for players in $S_m$.
\end{tabular}

\medskip   
Thus, in every coalition the player with the maximum influence factor remains active.
Hence, the utility of a coalition $S_m $, given by equation  \eqref{Eqn_Util_coaltiion_asym} 
 can be re-written as (for all $1\leq m \leq k$):
\begin{eqnarray*}
	\varphi_{S_m} = \frac{  \bar{\lambda}_m{\bar a}_{m} }{  \sum_{l =1}^k \bar{\lambda}_l{\bar a}_l} - \gamma {\bar a}_m , \mbox{ where aggregate actions, } 
	{\bar a}_m :=  \sum_{j \in  S_m}a_j.
\end{eqnarray*}
Now, our game is reduced to a similar game as studied in \cite{dhounchak2019participate} (with action of each coalition given by the aggregate action) and the result follows from \cite[Theorem 1]{dhounchak2019participate}; the utilities at NE are given by equation \eqref{Eqn_USm_asym}. 

Further by \cite[Theorem 1]{dhounchak2019participate},  any action profile, in which the aggregate actions  of each coalition equals \eqref{Eqn_alstar} (for the case of symmetric players with adamant player) forms a NE for RSG. One can similarly derive expression for the aggregate actions with asymmetric players (see \eqref{Eqn_alstar_asym}). Thus one can  have multiple NE, but the aggregate actions  and utility of each coalition  are the same at  all NE. 
\eop

\textbf{Proof of Theorem \ref{Thm_No_MPs_for_grt_n}:} 
Consider a strategy profile $\underbar{x}$ which leads to multiple partitions. Then as in \eqref{min_util_mult_partition} we define utility of a player to be the minimum utility among all the possible partitions emerging from $\underbar{x}$.

Let $(k_m+1)$ be the size of the biggest partition{\footnote{It can be seen from equation \eqref{Eqn_USm_player} that the utility of coalition decreases when the partition size, i.e., $k$ increases. \label{utilty_vs_coalition}}}  emerging from $\underbar{x}$ (call it ${\mathcal P}^*$), i.e.,  $k_m+1  =  \max_{{\mathcal P} (\text{{\small{$\underbar{x}$}}}) } |{\mathcal P} (\underbar{x})|$. 
$$
{\mathcal P}^* = \{S_0, S_1, \cdots,   S_l , S_{l+1}  \cdots\}, \mbox{ with } |{\mathcal P}^*|  = k_m+1.
$$
Now, if suppose player $i$ in coalition $S_l$ of size $m_m>1$ (we can always find such a player since otherwise all players are alone in this partition and we cannot have multiple partitions  because  of \eqref{Eqn_Coalition_rule} and \eqref{Eqn_Max_Coalition_rule})  deviates unilaterally to the strategy of being alone, i.e.,  to $\{i\}$ (changing strategy profile to $\underbar{x}'$), then we can have a partition with size at maximum  $k_m + 2$, call it $\mathcal{P}^*_{-i}$ (after splitting as in Lemma \ref{Lem_partiton_uni_dev}; since remaining players in $S_l/\{i\}$ may merge with some other coalition keeping the partition size intact).
$$
{\mathcal P}^*_{-i} \hspace{-0.5mm} = \hspace{-0.5mm} \{S_0, S_1, \cdots, \{i\}, S_l / \{i\}, S_{l+1}  \cdots\} \mbox{ with } |{\mathcal P}^*_{-i}| \hspace{-1mm}  = k_m+2.
$$ We have three cases based on adamant player:

\textbf{Case 1: When $\eta > 1 - 1/ (k_{m}+1)$:}  In this case the  adamant player gets non-zero utility in both the partitions, i.e.,  partition with $k_m+1$ as well as $k_m+2$ coalitions (see \eqref{NE_util_adamant}).
Then, utility of player $i$ with strategy profile  $\underbar{x}'$(using \eqref{Eqn_USm}),
	\begin{equation}
	U_i (\underbar{x}') \ge  \frac{\lambda^2 }{ (\lambda + (k_m+1) \lambda_0)^2} = \frac{1}{(1+(k_m+1)\eta)^2}.
	\label{min_x'}
	\end{equation}
The inequality above follows because the utility of a player decreases with increasing number of coalitions (see footnote \ref{utilty_vs_coalition}).

The utility of same player $i$ under strategy profile  $\underbar{x}$ equals,
$$
U_i ( \underbar{x}) \le   \frac{\lambda^2 }{ m_m  (\lambda + k_m  \lambda_0)^2} = \frac{1}{m_m(1+k_m\eta)^2},
$$
since  the utility of a player is defined to be the minimum utility among all possible partitions. 
From \eqref{min_x'} and 
Lemma \ref{Lemma_Partition_weak},  we have   (as $m_m \hspace{-1mm}> \hspace{-0.5mm} 1$ and because one can't have $m_m =2 $ and $k_m= 2$ simultaneously  for $n > 4$){\footnote{also see Corollary \ref{corollary_ALC_weak} for more details;}}:
$$
U_i (\underbar{x}) \leq \frac{1}{m_m(1+k_m\eta)^2} <    \frac{1}{(1+(k_m+1)\eta)^2} \leq  U_i (\underbar{x}'),
$$

\textbf{Case 2: When $ 1 - 1/ k_{m} < \eta \le  1 - 1/( k_{m}+1)$:}  From \eqref{NE_util_adamant},  the adamant player gets non-zero utility in partition with $k_m+1$ coalitions but zero utility with $k_m+2$ coalitions.
Once again, from \eqref{Eqn_USm} the
utility of player $i$ with strategy profile $\underbar{x}'$ (adversary insignificant),
\begin{equation}
\label{adv_zero_one}
 U_i(\underbar{x}')  \ge   \min \left \{ \left( \frac{1}{k_m+1}\right)^2, \left( \frac{1}{(1+ k_m\eta)^2}\right) \right \},
\end{equation}
since the minimum utility obtained after unilateral deviation may also be in one amongst the partitions with smaller size (observe this was not possible in Case 1).
Thus the utility of player $i$ with strategy profile  $\underbar{x}$ equals (inequality as explained in Case 1), 
\begin{equation*}
U_i(\underbar{x})  \leq \frac{1}{m_m(1+ k_m\eta)^2}.
\end{equation*}
By the conditions of Case 2, we have
$k_m\eta > (k_m-1)$ and  as in Lemma \ref{Lemma_Partition_weak} we have  $ \sqrt{m_m} k_m  >  k_m+1$ (as in Case 1):
\begin{equation*}
\sqrt{m_m}(1+ k_m\eta) >  \sqrt{m_m} k_m  >  k_m+1. 
\end{equation*}  
Hence 
from \eqref{adv_zero_one}: 
\begin{equation*}
U_i(\underbar{x})   \leq  \frac{1}{m_m(1+ k_m\eta)^2} \ < \   \min \left \{ \left( \frac{1}{k_m+1}\right)^2, \left( \frac{1}{(1+ k_m\eta)^2}\right) \right \}  \le U_i(\underbar{x}') .
\end{equation*}

\textbf{Case 3: When adamant player gets zero utility in both the partitions:}
Once again, using similar arguments as in Case 2 and conditions of Case 3, the
utility of player $i$ with strategy profile $\underbar{x}'$,
\begin{equation}
\label{adv_zero_two}
U_i(\underbar{x}')  \ge   \left( \frac{1}{k_m+1}\right)^2.
\vspace{-1mm}
\end{equation}
%
As before:
$
U_i(\underbar{x})  \leq 1/(m_mk_m^2). 
$
As in Lemma \ref{Lemma_Partition_weak}, $\sqrt{m_m} k_m  >  (k_m + 1)$ and hence  
\begin{equation*}
U_i(\underbar{x})   \leq  \frac{1}{m_mk_m^2}   <   \left( \frac{1}{k_m+1}\right)^2  \le  U_i(\underbar{x}') .
\end{equation*}
Thus, player $i$ finds it strictly better to deviate. 
\eop

\textbf{ Proof of Corollary \ref{corollary_ALC_weak}:} Consider any partition ${\mathcal  P}$ other than ALC. Let $m^*$ be the size of the biggest coalition  of ${\mathcal  P}$.  Then $m^* \geq 2$.  If $m^* = 2$, then $k := |{\mathcal P}| - 1 \ge  \lceil n/2 \rceil $ (lower bound achieved  when maximum coalitions are exactly of size  2). Thus   $m^* k^2  > (k+1)^2$, as $n > 4$  (note as $k$ increases, $(k+1)^2/k^2$ decreases).  If  $2 <  m^* \le n/2$, then  $k := |{\mathcal P}| - 1 \ge  \lceil n/m^* \rceil$ and hence:
 \begin{eqnarray*}
		\left ( \frac{k+1}{k}   \right )^2  =  \left (  1 + \frac{1}{k}  \right )^2 & \le &  \left (  1 +  \frac{m^*}{n} \right )^2 \le  (1.5)^2  <  3 \le m^*.
\end{eqnarray*}
For $n/2 < m^* < n$, we have $k \ge 2$\footnote{$k=1$ refers to the grand coalition of players which is never possible for $n>4$ (check Lemma \ref{Lemma_Partition_weak} conditions)} and hence 
	$$
	\left(1+\frac{1}{k}\right)^2 \le  \left(1+\frac{1}{2}\right)^2 = 2.25  <  \frac{n}{2}  < m^*.
	$$
	since $n>4$. 
	
	For $m^* = n$, we have $k=1$ and hence,
	$$
	\left(1+\frac{1}{k}\right)^2 = 4 <  n  = m^*.
	$$
	Thus conditions of Lemma \ref{Lemma_Partition_weak} are satisfied for all partitions other than ALC/ALC$^o$ and hence the result. \eop
	
	\section{Appendix B}
	\label{sec_AppendixB_PEVA}
	
	\textbf{Proof of Lemma \ref{Lem_partiton_uni_dev}:} We prove it in two steps: i) $\underbar{x}' \to {\mathcal P}_{-i}$ and  ii) $\underbar{x}' \to !\mathcal{P}_{-i}$. 
	
	To prove $\underbar{x}' \hspace{-1mm} \to \hspace{-1mm} \mathcal{P}_{-i}$:  it is clear by definition that  every coalition of $ \mathcal{P}_{-i}$ satisfies the    requirement \eqref{Eqn_Coalition_rule}  (with  $\underbar{x}'$). Hence, it suffices to prove that it is minimal  as in \eqref{Eqn_Max_Coalition_rule}.
	
	If possible consider a (better)  partition $\mathcal{P}'$ which satisfies \eqref{Eqn_Coalition_rule} and  such that $ \mathcal{P}'  \prec \mathcal{P}_{-i}$.  This means, from \eqref{Eqn_Max_Coalition_rule},   there exist at least a pair of coalitions $S_1, S_2  \in  \mathcal{P}_{-i}$ and an $S \in \mathcal{P}'$ such that $S_1 \cup S_2 \subset S$. Observe that $\{i\} \in \mathcal{P}'
	\cap \mathcal{P}_{-i}$, 
	as player $i$ deviates unilaterally to $\{i\}.$
	
	If  all such merging coalitions in $\mathcal{P}_{-i}$ are not equal to  $S_{l}\backslash \{i\}$ the merging coalitions will  also belong  to $\mathcal{P}$  (i.e., for example  if 
	$S_1 \neq S_2 \neq S_{l}\backslash \{i\}$, then $S_1, S_2 $ also belong  to $\mathcal{P}$),    then  one can  construct a  better partition\footnote{Partition ${\mathcal P}''$ 
		contains all coalitions of  ${\mathcal P}'$, except that $\{i \}$  and $S_{l}\backslash \{i\}$ are merged in  ${\mathcal P}''$.
	} ${\mathcal P}'' \prec {\mathcal P}$ and   $\underbar{x} \to {\mathcal P}''$,   which contradicts  $\underbar{x}  \to \mathcal{P}$.  
	
	On the other hand, if one of the merging coalitions equal $S_{l}\backslash{\{i\}}$,  then  ${\mathcal P}'$ is not comparable  with  ${\mathcal P}$ as in \eqref{Eqn_Max_Coalition_rule_0}  (i.e., neither is better than the other), as  $\hspace{-.5mm} \{i\} \hspace{-1mm} \in \hspace{-1mm} {\mathcal P}'$. 
	Further  ${\mathcal P}' \hspace{-1mm}$ satisfies  \eqref{Eqn_Coalition_rule} with  $\underbar{x}$ and hence
	$\underbar{x} \hspace{-1mm}\to \hspace{-1mm} {\mathcal P}'$. That means $\underbar{x}$   leads to multiple partitions and this contradicts the hypothesis that    $\underbar{x} \hspace{-1mm}\to  !{\mathcal P}$.
	This proves (i). 
	
	Next we prove uniqueness in (ii). 
	If possible  
	$\underbar{x}'$ leads to multiple partitions, say $\mathcal{P}_{-i}$ (defined in hypothesis) and   $\mathcal{P}'$. 
	This implies   $\mathcal{P}_{-i}$ is not comparable to  $\mathcal{P}'$. Further  observe  $ \{i\} \in \mathcal{P}'$ and hence  $ \mathcal{P}'$ is not even comparable to $  \mathcal{P}$. 
	Further more,   it is easy to verify that any coalition  that satisfies \eqref{Eqn_Coalition_rule} with   $\underbar{x}'$ also satisfies    \eqref{Eqn_Coalition_rule} with   $\underbar{x}$.  In all we have that 
	$\underbar{x} \to {\mathcal P}'$,   which again contradicts the uniqueness of  $\underbar{x}  \to ! \mathcal{P}$.   \eop

\textbf{Proof of Lemma \ref{Lemma_Partition_weak}:}
	W.l.o.g. we can assume that the $n$ C-players (i.e., with influence factor $\lambda$) form $k$ coalitions where $k\leq n$ , i.e., 
		\begin{eqnarray*}
			\mathcal{P}= \{\{0\},\{1,\cdots,m_1\},\{m_1+1,\cdots,m_2\},\cdots,
			\{m_{k-1}+1,\cdots,n\}\}.
	\end{eqnarray*}
	Consider the best response of (say $m_1 \hspace{-1mm}=\hspace{-1mm} m^*$) player $1$ against any strategy profile $\underbar{x} \hspace{-1mm} \to !\mathcal{P}$; player $1$ could either choose to remain alone (i.e., $x_1 \hspace{-1mm}= \hspace{-1mm} \{1\}$) or could form coalition with all or a subset of $(m_1-1)$ players (i.e., $x_1 \subset  \{1,\cdots,m_1\}$) resulting into a new strategy profile $\underbar{x}'$. In particular, we would show that forming coalition with all  players (as given by $\underbar{x} \to !\mathcal{P}$)
	is strictly inferior to remaining alone, i.e., player $1$ could get higher utility by unilaterally deviating to $\{1\}$.
	
	\textbf{Case 1: When} $\eta > 1-1/(k+1)$: In this case the adamant player gets non-zero utility in both the partitions, i.e., partition with $k+1$ as well as $k+2$ coalitions.
	
	Then, from \eqref{Eqn_USm} utility of player 1 when it chooses to remain alone (with strategies of the others remaining the same),
	\begin{equation}
	U_1(\underbar{x}') = \left(\frac{\lambda}{\lambda+(k+1)\lambda_0}\right)^2  = \left(\frac{1}{1+(k+1)\eta}\right)^2.
	\label{eq:Eq4.21}
	\end{equation}
	Similarly, utility of  player 1 when it proposes to form coalition with all $(m_1-1)$ players,
	\begin{equation*}
	U_{1}(\underbar{x})=   \frac{1}{m_1}\left(\frac{\lambda}{\lambda+k\lambda_0}\right)^2 =  \frac{1}{m_1}\left(\frac{1}{1+k\eta}\right)^2.
	\label{eq:Eq4.22}
	\end{equation*}
	
	\noindent 
	Since $m_1 >1$, 
	from \eqref{eq:Eq4.21}, player $i$ finds it better to deviate: 
	 \begin{equation*}	U_1(\underbar{x}')   =   \left(\frac{1}{1+(k+1)\eta}\right)^2 \ > \  \frac{1}{m_1}\left(\frac{1}{1+k\eta}\right)^2 = U_1(\underbar{x}) .
		\end{equation*}
	if $ \sqrt{m_1}k  >  k+1$.
	
	\textbf{Case 2: When $ 1 - 1/ k < \eta \le  1 - 1/( k+1)$:}  From \eqref{NE_util_adamant},  the adamant player gets non-zero utility in partition with $k+1$ coalitions but zero utility with $k+2$ coalitions.
Now, utility of player 1 when it proposes to form coalitions with $(m_1-1)$ players,
	\begin{eqnarray}
	\label{case2_coa}
	U_1(\underbar{x})& = & \frac{1}{m_1}\left( \frac{1}{1+k\eta}\right)^2.
	\end{eqnarray}
	When player 1 chooses to remain alone (with strategies of other players remaining the same) then, utility of player 1 is given by,
	\begin{eqnarray}
	U_1(\underbar{x}') & = & \left( \frac{1}{k+1}\right)^2.
	\label{case2_alone}
	\end{eqnarray}
	
	\noindent By the conditions of Case 2 we have $k\eta > (k-1)$,  
	\begin{eqnarray*}
		\sqrt{m_1}(1+k\eta) >  \sqrt{m_1}(1+k-1) =\sqrt{m_1}k .
	\end{eqnarray*}  
	Hence 
	from \eqref{case2_alone}: 
	 \begin{eqnarray*}
			U_1(\underbar{x}')   = \left( \frac{1}{k+1}\right)^2 \ > \   \frac{1}{m_1}\left( \frac{1}{k}\right)^2 > \   \frac{1}{m_1}\left( \frac{1}{1+k\eta}\right)^2 = U_1(\underbar{x}),
	\end{eqnarray*}
	if $ \sqrt{m_1}k  >  k+1$.

		\textbf{Case 3: When adamant player gets zero utility in both the partitions}
	
\noindent	Once again the utility of player 1 when it proposes to form coalitions with $(m_1-1)$ players,
	\begin{eqnarray}
	\label{case3_coa}
	U_1(\underbar{x}) & = & \frac{1}{m_1}\left( \frac{1}{k}\right)^2.
	\end{eqnarray}
	When player 1 chooses to remain alone (with strategies of other players remaining the same) then, utility of player 1 is given by,
	\begin{eqnarray}
	\label{case3_alone}
	U_1(\underbar{x}') & = & \left( \frac{1}{k+1}\right)^2.
	\end{eqnarray}

\noindent	Hence 
	from \eqref{case3_alone}: 
	\begin{eqnarray*}
		 U_1(\underbar{x}')   = \left( \frac{1}{k+1}\right)^2 \ > \ \frac{1}{m_1}\left( \frac{1}{k}\right)^2  = \ U_1(\underbar{x}) . 
	\end{eqnarray*}
	if $ \sqrt{m_1}k  >  k+1$. Thus, we have the result. \eop
	
	{\bf Proof of Lemma \ref{Lemma_SO_Partition}:}
		Let $\eta \ge 1$ (adamant player gets non-zero utility in all such partitions). 
		From \eqref{Eqn_USm}, one can verify  ($\{m_i\}$-sizes of coalition, $k+1$-size of partition):
			\begin{eqnarray}
			U_{SO}^* &:=& \max_{\mathcal{P}} \sum_{S_i \in \mathcal{P};i \neq 0} U_{S_i}   =   
			{  \max_{ \{ \{ m_i \}_{i \le k} , k  \} }  \sum_{i=1}^k \sum_{j=1}^{m_i}  \frac{ \lambda^2 }{m_i (\lambda +  k \lambda_0)^2 }  } =    {  \max_{ \{ \{ m_i \}_{i \le k} , k  \} }  \sum_{i=1}^k \frac{ \lambda^2 }{ (\lambda +  k \lambda_0)^2 }.  } \nonumber \\ &=& 
			{   \max_{1\le   k   \le n}   \frac{ k  \lambda^2 }{ (\lambda +  k \lambda_0)^2 }  }.  \label{Eqn_USO_initial}
			\end{eqnarray}
		One can equivalently  minimize: \vspace{-4mm}
		\begin{eqnarray}
			\min_{1\le   k   \le n}   \frac{ (\lambda +  k \lambda_0)^2 }{ k  \lambda^2 }  \mbox{ or } \min_{1\le   k   \le n}   \frac {  \lambda^2   } {k }   + k  \lambda_0^2.
				\label{Eqn_Convex_function}
		\end{eqnarray}
		By relaxing $k$ to real numbers, and equating the derivative to zero  (verify the second derivative is positive) we obtain:\vspace{-2mm}
		\begin{eqnarray*}
			-\frac{\lambda^2}{k^2} + \lambda_0^2   = 0   \mbox{ or }  
			k^*   =    1/\eta. 
		\end{eqnarray*} 
		This implies (by convexity) that the optimizer among integers  is  $k^* = 1$ when $\eta \ge 1$, i.e., GC is the SO-partition. 
		When $\eta \le  1- 1/2 = 0.5$,  from \eqref{NE_util_adamant} the adamant player gets insignificant in all partitions other than GC, and one needs to maximize 
		\vspace{-4mm}
	 \begin{eqnarray*}
				U_{SO}^* &=& \max \left \{  \frac{\lambda^2}{(\lambda+\lambda_0)^2} ,      \max_{ \{ \{ m_i \}_{i \le k} , k >  1 \} }  \sum_{i =1}^k  \frac{ 1 }{  k^2 }   \right \}  \\
				&& \hspace{-16mm}= \max \left \{  \frac{\lambda^2}{(\lambda+\lambda_0)^2} ,      \max_{ \{ \{ m_i \}_{i \le k} , k >  1 \} }   \frac{ 1 }{  k }   \right \} \ = \ 
				\max \left \{  \frac{\lambda^2}{(\lambda+\lambda_0)^2} ,     \frac{1}{2}   \right \}.
		\end{eqnarray*}
		When $\eta \leq   \sqrt{2}-1= 0.414 $,  GC is the  SO-partition  and for $ 0.414   \le \eta \leq 0.5$ any  ${\mathcal P}^o_2$ is an SO-partition, which completes the proof of part (ii).  
		Similarly when $\eta  \le  1- 1/3$ we have
		\vspace{-1mm}
		\begin{eqnarray*}
			{\bar U}^*_{SO}
			= \max \left \{  \frac{\lambda^2}{(\lambda+\lambda_0)^2} ,   \frac{2\lambda^2}{(\lambda+2\lambda_0)^2} ,      \frac{1}{3}  \right \} .
		\end{eqnarray*} 
		Progressing this way, for any $\eta \le 1 - 1/k$  (as in \eqref{Eqn_USO_initial}): 
		\begin{eqnarray*}
			{\bar U}^*_{SO}
			= 
			\hspace{-2mm}  \begin{array}{llll}
				\max \left \{ \max_{k' < k } \frac{ k' \lambda^2}{(\lambda+ k' \lambda_0)^2}  , \   \frac{1}{k}  \right \}   & \mbox{for any }  k \le  n,   \\
				%
			\end{array} 
		\end{eqnarray*}
		%
		%
		%
		But with $\eta  > 1/2$, the relaxed $k^* = 1/ \eta  < 2$.  Thus  by convexity of \eqref{Eqn_Convex_function} the maximizer of the first term among integers is either at 1 or 2, i.e., when $\eta \le 1 - 1/k$,
		
		\vspace{-4mm}
		\begin{eqnarray*}
			{\bar U}^*_{SO}
			=
				\max \left \{  \frac{\lambda^2}{(\lambda+\lambda_0)^2} ,   \frac{2\lambda^2}{(\lambda+2\lambda_0)^2} ,   \frac{1} {  k }  \right \} & \mbox{ \normalsize if }  
				k \le  n.
		\end{eqnarray*}
		We have GC is best among the first two if 
		\vspace{-2mm}
		$$
		1+ 2\eta  \ge    \sqrt{2} (1+\eta)  \mbox{ or if  } \eta \ge   1/ \sqrt{2}  = 0.707.
		\vspace{-2mm}
		$$Further in this range for all $n$,  GC is better than the third possibility also (if it is feasible). In a similar way one can prove that ${\mathcal P}_2$ is optimal for all other values of $\eta$.
		Thus we proved the lemma. \eop
		
		{\bf Proof of Lemma \ref{Lemma_SO_no_adv}:}
	For each $i$, $m_i:= |S_i|$, the size of coalition and $k$ be the size of partition $\mathcal{P}^o$, i.e. , $|\mathcal{P}^o|$.
	\vspace{-3mm}
	\begin{eqnarray}
	\hspace{3mm} U_{SO}^*:= \max_{\mathcal{P}^o} \sum_{S_i \in \mathcal{P}^o} U_{S_i}  \hspace{-2mm} = 
	{  \max_{ \{ \{ m_i \}_{i \le k} , k  \} }  \sum_{i=1}^k \sum_{j=1}^{m_i} \frac{1}{m_ik^2}  } =    {  \max_{ \{ \{ m_i \}_{i \le k} , k  \} }  \sum_{i=1}^k \frac{1}{k^2}  }  =
	{   \max_{1\le   k   \le n}   \frac{1}{k}  }.  \label{Eqn_USO_initial_no_adv}
	\end{eqnarray}
	Since the minimum possible value of $k$ is 1, we have GC$^o$ is the only SO-partition in this case. \eop
	
	\section{Appendix C}
 \label{sec_appendixC_PEVA}
	
	\textbf{Proof of Lemma \ref{min_worth_sub_coal_general}: } 
Consider any partition $\P = \{S_1,\cdots,S_k\}$. From Theorem \ref{Thm_util_coal_general}, there exists a $M^\P \leq k$ such that only the coalitions $S_j$ for $j \leq M^\P$ obtains non-zero utility. As already defined, the worth of any sub-coalition, say $\nu_C^\P \text{ for }C \subset S_j \text{ with } 1 \leq j \leq M^\P$, is defined to be the minimum utility among all possible partitions (i.e., with any possible arrangement of players in $S_j \backslash C$). 

From Theorem \ref{Thm_util_coal_general}, we know 
\begin{equation}
	\sum_{i=1}^{M^\P} \frac{1}{\lambda_i} - \frac{M^\P-1}{\lambda_j} > 0 \text{ for } j = M^\P \text{ and } \sum_{i=1}^{M^\P+1} \frac{1}{\lambda_i} - \frac{M^\P}{\lambda_j} \leq 0 \text{ for } j = M^\P +1.  
	\label{Eqn_small_player_added}
\end{equation}

For simpler notations, we represent the coalitions with the highest influence factor among the members of coalitions and represent these factors directly by $\lambda$'s and not by $\um$'s.  Because of the monotonicity of $\lambda_j$, the first inequality in \eqref{Eqn_small_player_added} is true for all $j \leq M^\P$. Thus, the utility of any coalition $S_j; j \leq M^\P$ equals,
\begin{equation}
	\left( \frac{\sum_{i=1}^{M^\P} \frac{1}{\lambda_i}-\frac{M^\P-1}{\lambda_j}}{\sum_{i=1}^{M^\P} \frac{1}{\lambda_i}}\right)^2 = \left( \frac{s^{M^\P}-\frac{M^\P-1}{\lambda_j}}{s^{M^{\P}}}\right)^2 =  \left( 1-\frac{M^\P-1}{s^{M^\P}\lambda_j}\right)^2.
	\label{Eqn_Util_before}
\end{equation}
Next, we consider addition of a coalition with influence factor $\lambda'$, and let $\P'$ be the new partition. The new number of active coalitions is denoted by $M^{\P'}$. The utility of any active coalition $S_j; j \leq M^{\P'}$ is given by,
\begin{equation}
\left( \frac{\sum_{i=1}^{M^{\P'}} \frac{1}{\lambda_i}-\frac{M^{\P'}-1}{\lambda_j}}{\sum_{i=1}^{M^{\P'}} \frac{1}{\lambda_i}}\right)^2 = \left( \frac{s^{M^{\P'}}-\frac{M^{\P'}-1}{\lambda_j}}{s^{M^{\P'}}}\right)^2 =  \left( 1-\frac{M^{\P'}-1}{s^{M^{\P'}}\lambda_j}\right)^2.
	\label{Eqn_Util_after}
\end{equation}
From \eqref{Eqn_Util_before}, \eqref{Eqn_Util_after} and by definition of $M^\P$ and $M^{\P'}$ we know that
$$
\frac{M^\P-1}{s^{M^\P}\lambda_j} <1 \text{ and } \frac{M^{\P'}-1}{s^{M^{\P'}}\lambda_j} <1.
$$
Hence, to prove the result it suffices to show the following,
\begin{equation}
	\frac{M^\P-1}{s^{M^\P}\lambda_j} \le  \frac{M^{\P'}-1}{s^{M^{\P'}}\lambda_j} \mbox{, or equivalently prove that, } \frac{M^\P-1}{s^{M^\P}} \le  \frac{M^{\P'}-1}{s^{M^{\P'}}}. \label{Eqn_proof}
\end{equation}
Once this is proved it implies the utility of $C$ under   $\P$ is    more than   that under $\P'$ (one obtained after adding another coalition to $\P$). This equivalently implies coalition $C$ obtains least utility when all other members of its parent coalition are alone as in \eqref{Eqn_partition_min_gen}. Thus, it suffices to prove \eqref{Eqn_proof}.

\textbf{Case 1:} When $\lambda_{M^\P+1} \leq \lambda' \le  \lambda_{M^\P}$. 

In this case,  $\P' = \{ \lambda_1, \cdots, \lambda_{M^\P}, \lambda', 
 \lambda_{M^\P+1}, \cdots, \lambda_n\}$ after addition and rearrangement of partition such that the influence factors are in order. 
 After this addition, we have
\begin{eqnarray}
\label{Eqn_case1}
	\sum_{i=1}^{M^\P} \frac{1}{\lambda_i} + \frac{1}{\lambda'} - \frac{M^\P}{\lambda'}   = \sum_{i=1}^{M^\P} \frac{1}{\lambda_i} - \frac{M^\P-1}{\lambda_{M^\P}}  - (M^\P-1) \left ( \frac{1}{\lambda'}- \frac{1}{\lambda_{M^\P}} \right ), 
\end{eqnarray}
if the above is positive we will have that $M^{\P'} = M^\P+1 $ (by writing similar equation considering  $\lambda_{M^\P+1}$ also, one can easily observe that the equation {\footnote {$	\sum_{i=1}^{M^\P+1} \frac{1}{\lambda_i} + \frac{1}{\lambda'} - \frac{M^\P+1}{\lambda_{M^\P +1}} = \sum_{i=1}^{M^\P+1} \frac{1}{\lambda_i} - \frac{M^\P}{\lambda_{M^\P +1}} + \frac{1}{\lambda'}- \frac{1}{\lambda_{M^\P +1}} \leq 0 \text{ from \eqref{Eqn_small_player_added} and } \lambda_{M^\P+1} \leq \lambda'.$ \label{footnote_min}}} would be negative), else $M^{\P'} = M^\P.$ In the first case, by positivity of    \eqref{Eqn_case1}
$$
\frac{ s^{M^{\P'}}}{M^{\P'} -1 }  = \frac{ \sum_{i=1}^{M^\P} \frac{1}{\lambda_i} + \frac{1}{\lambda'} }{M^{\P}  }   =  \frac{ \sum_{i=1}^{M^\P} \frac{1}{\lambda_i}   }{M^{\P} -1 } 
+ \frac{   \frac{1}{\lambda'}  -  \frac{  \sum_{i=1}^{M^\P} \frac{1}{\lambda_i} } {M^{\P} -1 }  }{M^{\P}   }  <  \frac{ s^{M^\P}}{M^{\P} -1 } .
$$
For the second case, \eqref{Eqn_proof} holds with equality.

\textbf{Case 2:} When $ \lambda_{M^\P} \leq \lambda' \leq \lambda_{M^\P-1}$ 

In this case,  $\P' = \{ \lambda_1, \cdots,\lambda_{M^\P-1}, \lambda',\lambda_{M^\P},
, \cdots, \lambda_n\}$ after addition and rearrangement of partition such that the influence factors are in order. 
After this addition we have
\begin{eqnarray}
	\label{Eqn_case2}
	\sum_{i=1}^{M^\P} \frac{1}{\lambda_i} + \frac{1}{\lambda'} - \frac{M^\P}{\lambda_{M^\P}} = 	\sum_{i=1}^{M^\P} \frac{1}{\lambda_i} + \frac{1}{\lambda'} - \frac{M^\P-1}{\lambda_{M^\P}} - \frac{1}{\lambda_{M^\P}} = \sum_{i=1}^{M^\P} \frac{1}{\lambda_i} - \frac{M^\P-1}{\lambda_{M^\P}}+ \frac{1}{\lambda'}- \frac{1}{\lambda_{M^\P}},
\end{eqnarray}
if the above is positive we will have that $M^{\P'} = M^\P+1 $ (by replacing $M^\P+1$ with $M^\P$ in footnote \ref{footnote_min}), else $M^{\P'} = M^\P$ (by writing similar equation{\footnote{$\sum_{i=1}^{M^\P-1} \frac{1}{\lambda_i} + \frac{1}{\lambda'} - \frac{M^\P-1}{\lambda'} =  \sum_{i=1}^{M^\P} \frac{1}{\lambda_i}- \frac{M^\P-1}{\lambda_{M^\P}}+ (M^\P-2) \left[\frac{1}{\lambda_{M^\P}}-\frac{1}{\lambda'} \right] > 0 \text{ from \eqref{Eqn_small_player_added} and  as, } \lambda_{M^\P} \leq \lambda'.$}} considering $\lambda'$ also, one can easily observe that the equation would be positive). In the first case, by positivity of    \eqref{Eqn_case2} and from the monotonicity of $\lambda$, we have
$$
\frac{ s^{M^{\P'}}}{M^{\P'} -1 }  = \frac{ \sum_{i=1}^{M^\P} \frac{1}{\lambda_i} + \frac{1}{\lambda'} }{M^{\P}  }   =  \frac{ \sum_{i=1}^{M^\P} \frac{1}{\lambda_i}   }{M^{\P} -1 } 
+ \frac{   \frac{1}{\lambda'}  -  \frac{  \sum_{i=1}^{M^\P} \frac{1}{\lambda_i} } {M^{\P} -1 }  }{M^{\P}   }  <  \frac{ s^{M^\P}}{M^{\P} -1 } .
$$

For the second case, \eqref{Eqn_proof} follows as $\lambda' \geq \lambda_{M^\P}$ and hence
 $ s^{M^{\P'}} \le s^{M^\P}$.


\textbf{Case 3:} When $\lambda' > \lambda_{M^\P-1}$

In this case,  $\P' = \{ \lambda_1, \cdots,\lambda_l,\lambda',\lambda_{l+1},\cdots,\lambda_{M^\P}, \cdots, \lambda_n\}$ after addition and rearrangement of partition such that the influence factors are in order. 
After this addition, we have
\begin{eqnarray}
	\label{Eqn_case3}
	\sum_{i=1}^{M^\P} \frac{1}{\lambda_i} + \frac{1}{\lambda'} - \frac{M^\P}{\lambda_{M^\P}} = 	\sum_{i=1}^{M^\P} \frac{1}{\lambda_i} + \frac{1}{\lambda'} - \frac{M^\P-1}{\lambda_{M^\P}} - \frac{1}{\lambda_{M^\P}} = \sum_{i=1}^{M^\P} \frac{1}{\lambda_i} - \frac{M^\P-1}{\lambda_{M^\P}}+ \frac{1}{\lambda'}- \frac{1}{\lambda_{M^\P}},
\end{eqnarray}
if the above is positive we will have that $M^{\P'} = M^\P+1 $ (by footnote \ref{footnote_min}), else $M^{\P'} \leq M^\P.$ In the second case,  we first check  the possibility  of $M^{\P'}=M^\P$. Towards this consider,
\begin{eqnarray}
	\sum_{i=1}^{M^\P-1} \frac{1}{\lambda_i} + \frac{1}{\lambda'} - \frac{M^\P-1}{\lambda_{M^\P-1}} &=& 	\sum_{i=1}^{M^\P} \frac{1}{\lambda_i}-\frac{1}{\lambda_{M^\P}} + \frac{1}{\lambda'} - \frac{M^\P-1}{\lambda_{M^\P}}+ \frac{M^\P-1}{\lambda_{M^\P}}- \frac{M^\P-1}{\lambda_{M^\P-1}},  \nonumber \\
	& = & \sum_{i=1}^{M^\P} \frac{1}{\lambda_i}  - \frac{M^\P-1}{\lambda_{M^\P}}+(M^\P-1)\left [ \frac{1}{\lambda_{M^\P}} - \frac{1}{\lambda_{M^\P-1}}\right]-\left( \frac{1}{\lambda_{M^\P}} - \frac{1}{\lambda'} \right), \nonumber
\end{eqnarray}
and observe that 
if the above is positive  then  $M^{\P'} = M^\P$, else $M^{\P'} < M^\P.$ The remaining  two cases are similar to Case 2,  we are left with the case,  $M^{\P'} < M^\P$, i.e., $M^{\P'}+1 \leq M^\P$.

 	From the definition of $M^\P$ as in \eqref{Eqn_small_player_added}, we have
 		\begin{equation}
 		\frac{s^{M^\P}}{{M^\P}-1} > \frac{1}{\lambda_{M^\P}} \text{ and } \frac{s^{M^\P+1}}{{M^\P}} \leq \frac{1}{\lambda_{M^\P+1}}.
 		\label{Eqn_case3_cond}
 	\end{equation}
 From \eqref{Eqn_case3} and \eqref{Eqn_case3_cond}, we have
 \begin{eqnarray}
 		\frac{s^{M^\P}}{{M^\P}-1} > \frac{1}{\lambda_{M^\P}} \geq \frac{1}{\lambda_{M^{\P'}+1}} > \frac{s^{M^{\P'}+1}}{{M^{\P'}}} \stackrel{ (a) }{>} \frac{s^{M^{\P'}}}{{M^{\P'}-1}}.
 \end{eqnarray}
The inequality $(a)$ is obtained through following intermediate calculations:
\begin{eqnarray*}
	\frac{s^{M^{\P'}}}{{M^{\P'}-1}} & = & \frac{\sum_{i=1}^{M^{\P'}}\frac{1}{\lambda_i}+\frac{1}{\lambda_{M^{\P'}+1}}}{M^{\P'}-1} - \frac{1}{\lambda_{M^{\P'}+1}(M^{\P'}-1)} \\ 
	& = &   \left( \frac{s^{M^{\P'}+1}}{M^{\P'}-1} \right) \left( \frac{M^{\P'}-1+1}{M^{\P'}} \right) - \frac{1}{\lambda_{M^{\P'}+1}(M^{\P'}-1)} \\
	& = & \frac{s^{M^{\P'}+1}}{M^{\P'}}+ \frac{s^{M^{\P'}+1}}{M^{\P'}(M^{\P'}-1)}- \frac{1}{\lambda_{M^{\P'}+1}(M^{\P'}-1)} \\
 & = & \frac{s^{M^{\P'}+1}}{M^{\P'}}+\frac{1}{(M^{\P'}-1)} \left[\frac{s^{M^{\P'}+1}}{M^{\P'}}- \frac{1}{\lambda_{M^{\P'}+1}} \right] \\
		& < & \frac{s^{M^{\P'}+1}}{M^{\P'}} \text{ from the definition of } M^{\P'}.
	\end{eqnarray*}
Observe that when $\lambda' < \lambda_{M^\P+1}$, from \eqref{Eqn_small_player_added}, the number of coalitions obtaining strictly positive utility remains the same even on addition of a player. Hence, we have the result.
\eop 

\textbf{Proof of Lemma \ref{Lem_SV_A_player}:} Let the size of coalition $C_\beta$ be denoted by $m_\beta = \kb +1$. Using symmetry, one can estimate the number of sub-coalitions (with only S-players) of size $l < |\Cb|$. Thus from \eqref{Eqn_SV}, the Shapley value of A-player in coalition $\Cb$ (with $\kb \geq 1$) is given by,
	\begin{eqnarray}
\phi_{\beta}^\P& \hspace{-1mm} = \hspace{-1mm}&   \binom{\kb}{0}\frac{0!\kb!}{(\kb+1)!}\nu_{\{\beta\lambda\}}+ \sum_{l=1}^{\kb} \binom{\kb}{l} \frac{l!(\kb-l)!}{(\kb+1)!} (\nu_{\{\beta\lambda,1,\cdots,l\}} - \nu_{\{1,\cdots,l\}}), \nonumber \\
& \hspace{-1mm}  \hspace{-1mm}&  \text{ where } \binom{\kb}{l} \text{ denotes number of sub-coalitions of size $l$ with different S-players.} \nonumber \\
&\hspace{-1mm} = \hspace{-1mm}& \frac{1}{(\kb+1)} \left [ \nu_{\{\beta\lambda\}} +\sum_{l=1}^{\kb}   (\nu_{\{\beta\lambda,1,\cdots,l\}} - \nu_{\{1,\cdots,l\}}) \right] \nonumber \\ &\hspace{-1mm} = \hspace{-1mm}& \frac{1}{(\kb+1)} \left [\sum_{l=0}^{\kb-1}   (\nu_{\{\beta\lambda,1,\cdots,l\}} - \nu_{\{1,\cdots,l+1\}})+ \nu_{\Cb} \right] .  
\label{shap_val}
\end{eqnarray}
Using equation \eqref{Eqn_USm_asym} and Lemma \ref{min_worth_sub_coal_general}, the worth of any sub-coalition with A-player or with only S-player(s) is given by,
	 \begin{eqnarray*}
	\nu_{\{\beta\lambda,1,\cdots,l\}}
	&=& 
	\left( \frac{\frac{1}{\beta\lambda}+\frac{\kb-l}{\lambda}+\frac{k}{\lambda}-\frac{\kb-l+k}{\beta\lambda}}{\frac{1}{\beta\lambda}+\frac{\kb-l}{\lambda}+\frac{k}{\lambda}} \right)^2  \text{ for any } 0 \leq l \leq \kb \text{ and, } \\ \nu_{\{1,\cdots,l\}} &=& 	\left( \frac{\frac{1}{\beta\lambda}+\frac{\kb-l+1}{\lambda}+\frac{k}{\lambda}-\frac{\kb-l+k+1}{\lambda}}{\frac{1}{\beta\lambda}+\frac{\kb-l+1}{\lambda}+\frac{k}{\lambda}} \right)^2 \text{ for any } 0 < l \leq \kb.	  
\end{eqnarray*}

Let $C_1 = \frac{1}{\beta\lambda}+\frac{\kb-l+k}{\lambda}$ and $C_2 = \frac{\kb-l+k}{\lambda}$. 
Thus, the worths of the sub-coalitions can be re-written as  below:
\begin{eqnarray*}
	\nu_{\{\beta\lambda,1,\cdots,l\}}
	= 
	\left( \frac{C_1-\frac{C_2}{\beta}}{C_1} \right)^2    
\mbox{ and, }
	\nu_{\{1,\cdots, (l+1) \}}
=		\left( \frac{C_1-C_2}{C_1} \right)^2   \text{ for any } l\geq 0.   
	\end{eqnarray*}
Using the above equations, the terms in \eqref{shap_val}, can be simplified as,

\begin{eqnarray*}
	\nu_{\{\beta\lambda,1,\cdots,l\}} - \nu_{\{1,\cdots,l+1\}} & = & \frac{\left(C_1-\frac{C_2}{\beta}\right)^2 - (C_1-C_2)^2}{C_1^2} 
	 = \frac{\frac{C_2^2}{\beta^2}\left(1 -\beta^2\right)-2\frac{C_1C_2}{\beta} (1-\beta)}{C_1^2} \\ &=& \frac{\frac{C_2}{\beta^2} (1-\beta) \left[ C_2(1+\beta) -2C_1\beta\right]}{C_1^2} \\
	& = & \frac{\frac{C_2}{\beta^2} (1-\beta) \left[  \left( \frac{\kb-l+k}{\lambda} \right) (1+\beta) -2\beta\left( \frac{1}{\beta\lambda}+\frac{\kb-l+k}{\lambda} \right)\right]}{C_1^2} 
	\\
	& = & \frac{\frac{C_2}{\beta^2\lambda} (1-\beta) \left[ (\kb-l+k-2) -\beta(\kb-l+k) \right]}{\left( \frac{1}{\beta\lambda}+\frac{\kb-l+k}{\lambda} \right)^2} \\
	& = & \frac{{C_2 \lambda} (1-\beta) \left[ (\kb-l+k-2) -\beta(\kb-l+k) \right]}{\left( {1+(\kb-l+k)\beta} \right)^2}. 
\end{eqnarray*}
Substituting these values in \eqref{shap_val} we have,
\begin{eqnarray}
\phi_{\beta}^\P & = & \frac{1}{\kb+1} \Bigg[ \sum_{l=0}^{\kb-1}\frac{{(\kb-l+k)} (1-\beta) \Big[ (\kb-l+k-2) -\beta(\kb-l+k) \Big]}{\Big( {1+(\kb-l+k)\beta} \Big)^2} + \Bigg( \frac{1-k+k\beta}{1+k\beta} \Bigg)^2 \Bigg]. \nonumber
\end{eqnarray}
Using this and the symmetry of remaining players (all S-players), one can easily calculate the Shapley value of S-players in coalition $\Cb$ as given in \eqref{Eqn_SV_S_player}.
\eop

\textbf{Proof of Theorem \ref{Thm_stable_big_asym}: } 
We begin with few definitions.  Consider  partitions in which the coalitions with only S-players are singletons, as below, 
 \begin{eqnarray}
 \P=\{\{\beta\lambda,1,\cdots,\kb\}, \underbrace{ \{\kb+1\},\cdots, \{n\}}_{\mbox{$k$ coalitions}} \} \mbox{ with }  k :=  n-\kb,  \nonumber
 \label{Eqn_SS_partition}
 \end{eqnarray}and refer such  partitions as SS $(\kb, k)$ partitions.  Observe here that $k$ is the number of coalitions with only S-players. 
  We refer the rest of them as non-SS  $(\kb, k)$ partitions.  By convention we let $C_\beta :=\{\beta\lambda,1,\cdots,\kb\}$ represent the first coalition of the partition that contains A-player. 

The major steps of this proof are the following:

 (i) SS$(\kb, k)$-partitions with
	$\kb \geq 1$ and $k \geq 0$  are U-stable if and only if \eqref{Eqn_NE_partition_single} is satisfied.

(ii) SS$(\kb, k)$-partition   with $\kb =0$   is U-stable as it is  the ALC partition.

(iii)  Non-SS $(\kb, k)$ partitions with  $k \ge 3$   are not U-stable .

(iv)  Non-SS$(\kb, k)$-partitions    with $\kb \geq 0$ and $k=2$ are not U-stable.

(v) Non-SS$(\kb, k)$-partitions    with    $\kb \geq 0$ and $k=1$ are not U-stable.

Now, we begin with the step-wise details of the steps above. For the ease of notations, at places we drop the superscript $^\P$.

\textbf{Step 1:  Stability conditions for SS$(\kb, k)$-partitions   with  $\kb \ge1$  and $k \geq 0$:}  






%
Consider the natural strategy profile $\text{{\small $\underbar{x}$}} = \underbar{x}^\P$ leading uniquely to such a partition. 
Next we consider any player, say A-player in coalition $C_\beta = \{\beta\lambda, 1,\cdots,\kb\}$. Observe that this player can deviate to any one of the following possible strategies, that can lead to partitions different from the original:
$$
\{\beta\lambda\}, \{\beta\lambda,1\},\cdots,\{\beta\lambda,1,\cdots,\kb-1\}, \{\beta\lambda,\kb+1\},\cdots,\{\beta\lambda,n\},\cdots
$$
If the player deviates unilaterally to  any one of the above strategies, say to $\{\beta \lambda, 1, \cdots, l\}$ with $l > 0$ (similarly one can show for others),  
then the strategy profile after the unilateral deviation leads to the
following  two partitions (by definition of  natural strategy profile, $\text{{\small $\underbar{x}$}}^\P  $),
\begin{eqnarray}
   \mathcal{P}_1 = \{\{\beta\lambda,1,\cdots,l\},\{l +1,\cdots,\kb\},\{\kb+1\},\cdots, \{n\}\}, \text{ and } \nonumber \\ 
   \mathcal{P}_2 = \{\{\beta\lambda\},\{1,\cdots,\kb\},\{\kb+1\},\cdots, \{n\}\}. \nonumber 
\end{eqnarray}
Clearly the size of  both the partitions is the same, and hence from equation \eqref{Eqn_USm_asym} the  A-player gets bigger  utility in $\mathcal{P}_2$; by symmetry of S-players and equal sizes of the two partitions, the utility of coalition containing  $\beta\lambda$ player is the same, but in the first case it has to share its utility with other  S-players also. 
Observe this is also the utility derived by  A-player when it unilaterally deviates  to the strategy of being alone.  Hence, in all,  A-player 
does not  find it  advantageous to deviate unilaterally (to any of the available choices)   if and only if its Shapley value (SV) in the original partition is higher than the one derived in $\mathcal{P}_2$. 
 Similarly S-player  (without loss of generality, player 1) does not  find it  advantageous to deviate unilaterally,  if and only if its SV is higher than the utility it obtains  in 
 $ \left \{\{\beta\lambda, 2,\cdots,\kb\}, \{1\}, \{\kb+1\},\cdots, \{n\} \right \}.$

Recall SV of  A-player is given by  $\phi_\beta^\P$, utility of  coalition $\Cb$ is given by  $\varphi^*_{\Cb} (\P)$   and let the utilities of A-player and S-player after deviating to the strategy of being alone be denoted by $U_{\beta}$ and $U_{\lambda}$ respectively. Thus, for partition $\P$ to be stable, we require the following:
\begin{eqnarray}
	\label{Eqn_stable_cond}
\phi_\beta^\P - U_\beta  \ \geq  \   0 \text{, and }  \ \
\frac{1}{\kb} \left ( \varphi^*_{\Cb} (\P) - \phi_\beta^\P \right )-U_{\lambda} \  \geq  \ 0.
\end{eqnarray}
Hence, we have the result from equation \eqref{Eqn_USm_asym} of Theorem \ref{Thm_util_coal_general} and Lemma \ref{Lem_SV_A_player}.

\textbf{Step 3: Non-SS $(\kb,k)$ partitions with $k \geq 3$  are not U-stable:}  

In the third step of the proof, we show that  these partitions  are not U-stable partitions. Consider the n.s.p. (natural strategy profile) which leads to the partition of the following form:
$$
\mathcal{P} = \{\{\beta\lambda,1,2,\cdots,\kb\},\underbrace{\{\},\cdots, \{\}}_{\mbox{k coalitions}}\} \text{ with } k \geq 3 \text{ and } n-\kb \neq k.
$$
The conditions $k \geq 3$ and $n-\kb \neq k$ signifies that the partition contains at least three coalitions of symmetric players (other than one coalition with A-player) such that at least one of these $k$ coalitions have more than one player. From Lemma \ref{Lemma_Partition_weak} (here, coalition with A-player can be considered as an adamant player), such $\P$ is a weak partition (see Section \ref{sec_weak}) and hence, cannot be U-stable. 


\textbf{Step 4: Non-SS $(\kb, k)$ partitions    with $\kb \geq 0$ and $k=2$ are not U-stable.}


(i) With $\kb=0\text{ or } 1$, $n>5$ and $k=2$, at least one of the coalitions without A-player has more than 2 players and thus from Lemma \ref{Lemma_Partition_weak} (here, coalition with A-player can be considered as an adamant player), the partition is weak. Hence, such a partition is not a U-stable partition.


(ii) With $\kb=2$ and $k=2$, we show that the A-player finds it better to unilaterally deviate to the strategy of being alone 
 if it is a unique partition emerging from some strategy profile. Also, for a given value of $k$, the individual share of each player in $\Cb$  decreases with increase in $\kb$ (since $\nu(C_\beta)$ remains same for the given value of $k$). Thus, if a non-SS partition with $\kb=2$ and $k=2$ is not a U-stable partition, then any non-SS partition with $\kb >2$ and $k=2$ is also not U-stable. 
From Lemma \ref{Lem_SV_A_player}, Shapley value of A-player for $\kb=k =2$ is given by: 
\begin{eqnarray}
\phi_{\beta}^\P & = & \frac{1}{3} \left[ \frac{4(\beta-1)(4\beta-2)}{(4\beta+1)^2} + \frac{3(\beta-1)(3\beta-1)}{(3\beta+1)^2} + \left( \frac{2\beta-1}{2\beta+1} \right)^2 \right]. \quad \nonumber
\end{eqnarray}
From \eqref{Eqn_USm_asym}, utility of A-player after unilaterally deviating to the strategy of being alone equals,
\begin{eqnarray}
U_{\beta} & = &  \left( \frac{3\beta-2}{3\beta+1} \right)^2. \nonumber
\end{eqnarray}
To prove that the partition is not U-stable, it is enough to show that $\phi_\beta^\P < U_\beta$. Define $f(\beta) = U_{\beta}-\phi_{\beta}^\P$ and thus we have:
\begin{eqnarray}
f(\beta) & = &\left( \frac{3\beta-2}{3\beta+1} \right)^2- \frac{1}{3} \left[ \frac{4(\beta-1)(4\beta-2)}{(4\beta+1)^2} + \frac{3(\beta-1)(3\beta-1)}{(3\beta+1)^2} + \left( \frac{2\beta-1}{2\beta+1} \right)^2 \right]  \nonumber \\
& = & -\frac{1}{3} \left[\frac{4(\beta-1)(4\beta-2)}{(4\beta+1)^2}  + \left( \frac{2\beta-1}{2\beta+1} \right)^2 - 2\left( \frac{3\beta-2}{3\beta+1} \right)^2- \left( \frac{3\beta-2}{3\beta+1} \right)^2 + \frac{3(\beta-1)(3\beta-1)}{(3\beta+1)^2} \right]\nonumber \\
& = & -\frac{1}{3} \left[\frac{4(\beta-1)(4\beta-2)}{(4\beta+1)^2}  + \left( \frac{2\beta-1}{2\beta+1} \right)^2 - 2\left( \frac{3\beta-2}{3\beta+1} \right)^2- \frac{9\beta^2+4-12\beta-9\beta^2+12\beta-3}{(3\beta+1)^2}\right] \nonumber \\
& = & -\frac{1}{3} \left[\left( \frac{2\beta-1}{2\beta+1} \right)^2-\left( \frac{3\beta-2}{3\beta+1} \right)^2 - \left[ \left( \frac{3\beta-2}{3\beta+1} \right)^2 - \frac{4(\beta-1)(4\beta-2)}{(4\beta+1)^2} \right] - \frac{1}{(3\beta+1)^2}\right] \nonumber \\
& \stackrel{\rm (a) \& (b)}{=} & -\frac{1}{3} \left[\frac{(12\beta^2-2\beta-3)}{(2\beta+1)^2(3\beta+1)^2} - \frac{(33\beta^2-4\beta-4)}{(3\beta+1)^2(4\beta+1)^2}  - \frac{1}{(3\beta+1)^2}\right]  \nonumber \\ 
&=&-\frac{1}{3(3\beta+1)^2} \left[\frac{(12\beta^2-2\beta-3)}{(2\beta+1)^2} - \frac{(33\beta^2-4\beta-4)}{(4\beta+1)^2}  - 1\right]  \nonumber \\
&=&\frac{1}{3(3\beta+1)^2} \left[\frac{(49\beta^2+4\beta-3)}{(4\beta+1)^2}-\frac{(12\beta^2-2\beta-3)}{(2\beta+1)^2} \right]. \nonumber 
\end{eqnarray}
The justification of (a) and (b) is given at the end of the proof of this step. Thus to prove positivity of $f(\beta)$ it is sufficient to prove the same for 
\begin{eqnarray}
g(\beta) := 
	\frac{(49\beta^2+4\beta-3)}{(4\beta+1)^2}-\frac{(12\beta^2-2\beta-3)}{(2\beta+1)^2}.   \nonumber\end{eqnarray}
By differentiating $g(\beta)$ with respect to $\beta$, we have

\vspace{-4mm}
{\small \begin{eqnarray}
g'(\beta)& = & \frac{(4\beta+1)^2(98\beta+4)-8(49\beta^2+4\beta-3)(4\beta+1)}{(4\beta+1)^4}-\frac{(2\beta+1)^2(24\beta-2)-4(12\beta^2-2\beta-3)(2\beta+1)}{(2\beta+1)^4} \nonumber \\
& = & \frac{(4\beta+1)(98\beta+4)-8(49\beta^2+4\beta-3)}{(4\beta+1)^3}-\frac{(2\beta+1)(24\beta-2)-4(12\beta^2-2\beta-3)}{(2\beta+1)^3} \nonumber \\
& = & \frac{82\beta+28 }{(4\beta+1)^3}-\frac{28\beta+10}{(2\beta+1)^3} 
= \frac{(82\beta+28)(8\beta^3+1+12\beta^2+6\beta)-(28\beta+10)(64\beta^3+1+48\beta^2+12\beta)}{(2\beta+1)^3(4\beta+1)^3} \nonumber \\ 
& = & -2 \left( \frac{568\beta^4+388\beta^3-6\beta^2-51\beta-9}{(2\beta+1)^3(4\beta+1)^3} \right) \ < \ 0.\nonumber
\end{eqnarray}}
Observe that $g(1) > 0$, $g(\beta)$ is a strictly decreasing function of $\beta$ and as $\beta \to \infty$, $g(\beta) > 0$. Thus, $g(\beta)$ is strictly positive for all values of $\beta >1$ and hence, the player obtains higher utilities after deviating (for at least one strategy). This proves that the partition is not U-stable.

\textbf{Proof of (a) and (b):}
\begin{eqnarray}  \Bigg( \frac{2\beta-1}{2\beta+1} \Bigg)^2-\Bigg( \frac{3\beta-2}{3\beta+1} \Bigg)^2 & = & \frac{(2\beta-1)^2(3\beta+1)^2- (3\beta-2)^2(2\beta+1)^2}{(2\beta+1)^2(3\beta+1)^2} \nonumber \\
	& = & \frac{(4\beta^2-4\beta+1)(9\beta^2+6\beta+1)- (9\beta^2-12\beta+4)(4\beta^2+4\beta+1)}{(2\beta+1)^2(3\beta+1)^2} \nonumber \\
	& = & \frac{(12\beta^2-2\beta-3)}{(2\beta+1)^2(3\beta+1)^2}. \nonumber \\
	 \Bigg( \frac{3\beta-2}{3\beta+1} \Bigg)^2 - \frac{4(\beta-1)(4\beta-2)}{(4\beta+1)^2} & = & \frac{(3\beta-2)^2(4\beta+1)^2-4(\beta-1)(4\beta-2)(3\beta+1)^2}{(3\beta+1)^2(4\beta+1)^2} \nonumber \\
	& = & \frac{(9\beta^2-12\beta+4)(16\beta^2+8\beta+1)-(16\beta^2-24\beta+8)(9\beta^2+6\beta+1)}{(3\beta+1)^2(4\beta+1)^2} \nonumber \\
	& = & \frac{(33\beta^2-4\beta-4)}{(3\beta+1)^2(4\beta+1)^2}. \nonumber 
\end{eqnarray}


\textbf{Step 5: Non-SS$(\kb, k)$-partitions    with $\kb \geq 0$ and $k=1$ are not U-stable.}

(i) With $\kb=0 \text{ or } 1$, $n>5$ and $k=1$, coalition without A-player has atleast 5 players and thus from Lemma \ref{Lemma_Partition_weak} (here, coalition with A-player can be considered as an adamant player), the partition is weak. Hence, such a partition is not a U-stable partition.

(ii) With $\kb=2$, $n>5$ and $k=1$, coalition without A-player has 4 players. Assuming $\Cb$ to be adamant player, one can show that one of the players in coalition without A-player obtains strictly better utilities after unilaterally deviating to the strategy of being alone (see Table \ref{tab:my-table}). Thus, the partition is weak and hence it cannot be a U-stable partition.

(iii) With $\kb=3$ and $k=1$, we will show (below) that the A-player finds it better to unilaterally deviate to the strategy of being alone 
if it is a unique partition emerging from natural strategy profile. Prior to that, observe that,  for a given value of $k$, the individual share of each player in $\Cb$  decreases with increase in $\kb$ (since $\nu(C_\beta)$ remains same for the given value of $k$). Thus, if a non-SS partition with $\kb=3$ and $k=1$ is not a U-stable partition, then any non-SS partition with $\kb >3$ and $k=1$ is also not U-stable.
From Lemma \ref{Lem_SV_A_player}, Shapley value of A-player for $\kb=3, k=1$ is given by:  
\begin{eqnarray}
\phi_{\beta} & = & \frac{1}{4} \left[ \frac{4(\beta-1)(4\beta-2)}{(4\beta+1)^2} + \frac{3(\beta-1)(3\beta-1)}{(3\beta+1)^2} +\frac{4\beta(\beta-1)}{(2\beta+1)^2}+ \left( \frac{\beta}{\beta+1} \right)^2 \right]. \quad \nonumber
\end{eqnarray}

From \eqref{Eqn_USm_asym}, utility of A-player after unilaterally deviating to the strategy of being alone equals,
\begin{eqnarray}
U_{\beta} & = &  \left( \frac{2\beta-1}{2\beta+1} \right)^2. \nonumber
\end{eqnarray}
To prove that the partition is not U-stable, it is enough to show that $\phi_\beta^\P < U_\beta$. Define $f(\beta) = U_{\beta}-\phi_{\beta}^\P$ and thus we have:
\begin{eqnarray}
f(\beta) & = & \left( \frac{2\beta-1}{2\beta+1} \right)^2-\frac{1}{4} \left[ \frac{4(\beta-1)(4\beta-2)}{(4\beta+1)^2} + \frac{3(\beta-1)(3\beta-1)}{(3\beta+1)^2} +\frac{4\beta(\beta-1)}{(2\beta+1)^2}+ \left( \frac{\beta}{\beta+1} \right)^2 \right] \nonumber \\
& = & \frac{1}{4} \left[3\left( \frac{2\beta-1}{2\beta+1}\right)^2+\frac{1}{(2\beta+1)^2} - \frac{4(\beta-1)(4\beta-2)}{(4\beta+1)^2} - \frac{3(\beta-1)(3\beta-1)}{(3\beta+1)^2} - \left( \frac{\beta}{\beta+1} \right)^2 \right]   \nonumber \\ 
 & \stackrel{\rm (a), (b) \& (c)}{=} & \frac{1}{4} \left[\frac{(16\beta^2+2\beta-2)}{(2\beta+1)^2(3\beta+1)^2}+ \frac{(36\beta^2-4\beta-7)}{(2\beta+1)^2(4\beta+1)^2}+\frac{1}{(2\beta+1)^2}-\frac{(2\beta+4\beta^2-1)}{(2\beta+1)^2(\beta+1)^2} \right]  \nonumber \\ 
 & = & \frac{1}{4(2\beta+1)^2} \left[\frac{(16\beta^2+2\beta-2)}{(3\beta+1)^2}+ \frac{(52\beta^2+4\beta-6)}{(4\beta+1)^2}-\frac{(2\beta+4\beta^2-1)}{(\beta+1)^2} \right] \nonumber \\ 
 & \stackrel{\text{(d)}}{=} & \frac{1}{4(2\beta+1)^2} \left[\frac{(724\beta^4+508\beta^3+22\beta^2-46\beta-8)}{(3\beta+1)^2(4\beta+1)^2}-\frac{(2\beta+4\beta^2-1)}{(\beta+1)^2} \right]. \nonumber
\end{eqnarray}
Thus to prove positivity/negativity of $f(\beta)$ it is sufficient to prove the same for 
\begin{eqnarray}
	g(\beta) 
	& = & (724\beta^4+508\beta^3+22\beta^2-46\beta-8)(\beta+1)^2-(2\beta+4\beta^2-1)(3\beta+1)^2(4\beta+1)^2. \nonumber
\end{eqnarray}
By simplifying it further, we have

\vspace{-4mm}
{\small \begin{eqnarray}
g(\beta) & = & (724\beta^6+508\beta^5+22\beta^4-46\beta^3-8\beta^2)+(1448\beta^5+1016\beta^4+44\beta^3-92\beta^2-16\beta) \nonumber \\
& & +(724\beta^4+508\beta^3+22\beta^2-46\beta-8)-(2\beta+4\beta^2-1)(3\beta+1)^2(4\beta+1)^2, \nonumber \\
& = & (724\beta^6+1956\beta^5+1762\beta^4+506\beta^3-78\beta^2-62\beta-8 ) -(2\beta+4\beta^2-1)(3\beta+1)^2(4\beta+1)^2, \nonumber \\
& = & (724\beta^6+1956\beta^5+1762\beta^4+506\beta^3-78\beta^2-62\beta-8)  -(2\beta+4\beta^2-1)(9\beta^2+6\beta+1)(4\beta+1)^2, \nonumber \\
& = & (724\beta^6+1956\beta^5+1762\beta^4+506\beta^3-78\beta^2-62\beta-8) -(18\beta^3+36\beta^4-9\beta^2+12\beta^2+24\beta^3-6\beta \nonumber \\
& & +2\beta+4\beta^2-1)(4\beta+1)^2, \nonumber \\
& = & (724\beta^6+1956\beta^5+1762\beta^4+506\beta^3-78\beta^2-62\beta-8) -(36\beta^4+42\beta^3+7\beta^2-4\beta-1)(4\beta+1)^2, \nonumber \\
& = & 148\beta^6 + 996\beta^5 + 1278\beta^4 + 472\beta^3 - 37\beta^2 - 50\beta - 7 \ > \ 0. \nonumber
\end{eqnarray}}
 Thus, $g(\beta)$ is strictly positive for all values of $\beta >1$ and hence, the player obtains higher utilities after deviating (for at least one strategy). This proves that the partition is not U-stable.
 
\textbf{Proof of (a), (b) and (c):}
\begin{eqnarray}
\Bigg( \frac{2\beta-1}{2\beta+1}\Bigg)^2 - \Bigg( \frac{\beta}{\beta+1} \Bigg)^2 & = & \frac{(2\beta-1)^2(\beta+1)^2-\beta^2(2\beta+1)^2}{(2\beta+1)^2(\beta+1)^2} \nonumber \\
& = & \frac{(4\beta^2-4\beta+1)(\beta^2+1+2\beta)-\beta^2(4\beta^2+4\beta+1)}{(2\beta+1)^2(\beta+1)^2} \nonumber \\
& = & \frac{(-2\beta-4\beta^2+1)}{(2\beta+1)^2(\beta+1)^2}. \nonumber \\
\Bigg( \frac{2\beta-1}{2\beta+1}\Bigg)^2 - \frac{3(\beta-1)(3\beta-1)}{(3\beta+1)^2} & = &  \frac{(2\beta-1)^2(3\beta+1)^2-3(\beta-1)(3\beta-1)(2\beta+1)^2}{(2\beta+1)^2(3\beta+1)^2} \nonumber \\
&  = &  \frac{(4\beta^2-4\beta+1)(9\beta^2+6\beta+1)-(9\beta^2-12\beta+3)(4\beta^2+4\beta+1)}{(2\beta+1)^2(3\beta+1)^2} \nonumber \\
&  = &  \frac{(16\beta^2+2\beta-2)}{(2\beta+1)^2(3\beta+1)^2}. \nonumber \\ 
\Bigg( \frac{2\beta-1}{2\beta+1}\Bigg)^2 - \frac{4(\beta-1)(4\beta-2)}{(4\beta+1)^2} & = &  \frac{(2\beta-1)^2(4\beta+1)^2-4(\beta-1)(4\beta-2)(2\beta+1)^2}{(2\beta+1)^2(4\beta+1)^2} \nonumber \\
& = & \frac{(4\beta^2-4\beta+1)(16\beta^2+8\beta+1)-(16\beta^2-24\beta+8)(4\beta^2+4\beta+1)}{(2\beta+1)^2(4\beta+1)^2} \nonumber \\
&  = & \frac{(36\beta^2-4\beta-7)}{(2\beta+1)^2(4\beta+1)^2}. \nonumber 
\end{eqnarray}
\textbf{Proof of (d)}
\begin{eqnarray}
\frac{(16\beta^2+2\beta-2)}{(3\beta+1)^2}+ \frac{(52\beta^2+4\beta-6)}{(4\beta+1)^2} & = &  \frac{(16\beta^2+2\beta-2)(4\beta+1)^2+(52\beta^2+4\beta-6)(3\beta+1)^2}{(3\beta+1)^2(4\beta+1)^2} \nonumber \\
& = &  \frac{(16\beta^2+2\beta-2)(16\beta^2+8\beta+1)+(52\beta^2+4\beta-6)(9\beta^2+6\beta+1)}{(3\beta+1)^2(4\beta+1)^2} \nonumber \\
& = &  \frac{(724\beta^4+508\beta^3+22\beta^2-46\beta-8)}{(3\beta+1)^2(4\beta+1)^2}. \nonumber \mbox{ \eop }
\end{eqnarray}
%


\section{Appendix D}
\label{sec_AppendixD_PEVA}

 \textbf{Proof of Lemma \ref{Lem_SV_general}: } 
(a) 
Consider any partition $\P = \{S_1, \cdots, S_k\}$ that plays RSG with $1 \in S_1$ and $2 \in S_2$. By Lemma \ref{Lem_Insignificant}, under {\bf A}.1,
\begin{equation}
\varphi_{S_m}^*(\P) = 0 \mbox{ for any }  m \text{ such that } 1,2 \notin S_m.
\end{equation}
Thus by  \eqref{Eqn_SV}-\eqref{Eqn_partition_min_gen} and Lemma \ref{min_worth_sub_coal_general}, the  share $\phi_i^\P$ of player $i \in S_m$ equals 0 if $m > 2$. 

Consider any $i \in S_1$ with $i \ne 1$.  Again under {\bf A}.1, and by Lemmas \ref{Lem_Insignificant} and \ref{min_worth_sub_coal_general}  (see \eqref{Eqn_partition_min_gen}), one can show that $\nu_C^\P = 0$ for all $C \subset  S_1$ with $1 \notin C$, as $C$ does not even contain $2$.  It is easy to observe from  Theorem \ref{Thm_util_coal_general},  that $\nu_C^\P = \nu_{C \cup \{i\}}^\P = \varrho_2^2  $, when $C \subset  S_1$ with $1 \in C$.  Hence from  \eqref{Eqn_SV}, $\phi_i^\P = 0$. Similarly $\phi_i^\P = 0$ for all $i \in S_2$ with $i \ne 2.$  Since all players except player $1$ in coalition $S_1$ obtains zero utility, player $1$ obtains the coalitional utility as its share , i.e., $\phi_1^\P = \varrho_2^2 = \phi_1^{\P'}$ with $\phi_1^{\P'}$ being the utility of player $1$ after unilateral deviation.
Working in a similar way, one can show that $\phi_2^\P = (1-\varrho_2)^2 = \phi_2^{\P'}$ and no player finds it beneficial to deviate.

(b) Now consider any partition $\P = \{S_1, \cdots, S_k\}$, with $\{1,2, \cdots,l\} \subset S_1$, $l+1 \notin S_1$ where $2 \le l \le n$.
Again 
from Lemma \ref{Lem_Insignificant},  under {\bf A}.1, we have
$$
\varphi_{S_m}^*(\P)=0 \text{ for any } m > 2. 
$$
 Thus, on similar lines as in part (i),  the share $\phi_i^\P$ of any player $i > l+1$  equals zero and hence the  share of player $l+1$ equals $\left( 1- \varrho_{l+1} \right)^2$, the share of the coalition 
$\nu_{S_2}^\P$  
  (obtained by the utilities at NE of the appropriate RSG). 
 
 The coefficient in the Shapley Value expression can be interpreted as the probability that in any permutation of $N_C$, the members of $C$ are ahead of player $i$ (under consideration)
  and the members of $N_C - C -\{i\}$ are after $i$. 
This  kind  of a configuration can be chosen in $|C|!(n-|C|-1)!$ ways and that divided by $n!$ represents the probability of choosing a subset $C$ in such a manner. 

We first begin with the Shapley Value of player $1$. Define a class $\mathcal{F}_j$ as follows,
$$
\mathcal{F}_j = \left \{ C: 2,\cdots, j-1  \in C \text{ and } 1,j \notin C \right \},
$$and observe that the term,  $\nu_{C \cup \{1\}}^\P - \nu_C^\P =    \varrho_j^2-(1-\varrho_2)^2 $, remains the same for all $C \in \mathcal{F}_j $. Thus to estimate 
the SV of 1 coming from such coalitions, it suffices to find the joint probability of permutations that support class $\mathcal{F}_j $. 
%
Towards estimating this joint probability, 
by definition of $\mathcal{F}_j$, for any $C \in \mathcal{F}_j$  we require $2,\cdots,j-1$ (i.e., $j-2$) players out of $j$ players (including player $1$ and $j$)  to be ahead of player $1$ and player-$j$ after player-1,  while the arrangement of other players is immaterial. Thus, for any $j \le l$, the required joint probability is given by:
$$
\sum_{C \in \mathcal{F}_j} \frac{(j-2)!(j-(j-2)-1)!}{j!} = \frac{1}{j(j-1)}.
$$
Using the above arguments, (from \eqref{Eqn_SV}) the SV of player $1$ equals,
\begin{eqnarray}
\phi_1^{\P} & = &  \sum_{j=3}^l \sum_{C \in \mathcal{F}_j} \frac{(j-2)!(j-(j-2)-1)!}{j!} \left [ \varrho_j^2-(1-\varrho_2)^2\right] + \frac{0!(2-0-1)!}{2!} \varrho_2^2 \nonumber \\
& & + \frac{(l-1)!(l-(l-1)-1)!}{l!} \left[\varrho_{l+1}^2- \left (1-\varrho_2 \right  )^2 \right],\nonumber  \\
& = & \sum_{j=3}^l \frac{1}{j(j-1)} \left [ \varrho_j^2-(1-\varrho_2)^2\right] + \frac{1}{2} \varrho_2^2 + \frac{1}{l} \left[\varrho_{l+1}^2- \left (1-\varrho_2 \right  )^2 \right].\nonumber
\end{eqnarray}
Observe that the second and last terms in the first equation correspond to $j=2$  (all $C$ such that $1, 2 \notin C$) and $j=l+1$  (all $C$ such that $1 \notin C $
and $2, \cdots, l \in C$ and observe arrangement of  $j+1 \notin S_1$ is immaterial) respectively and can be handled similarly.
Similarly, one can write the equation for Shapley Value of player $2$. Next we write the SV of player $i$ such that $ 3 \leq i <l$.  Define a class $\mathcal{F}_{i,m}$ as follows,
$$
\mathcal{F}_{i,m} = \left \{C: 1,2,\cdots, i-1,i+1, \cdots,i+m-1  \in C \text{ and } i,i+m \notin C \right \}.
$$
Using the same procedure as explained above, the SV of player $3 \leq i < l$ equals,
{\small \begin{eqnarray}
\phi_i^\P & = & \sum_{m=1}^{l-i} \sum_{C \in \mathcal{F}_{i,m}} \frac{(i+m-2)!(i+m-(i+m-2)-1)!}{(i+m)!} \left[ \varrho_{i+m}^2 - \varrho_i^2 \right] + \frac{(l-1)!(l-(l-1)-1)!}{l!}  \left[ \varrho_{l+1}^2 - \varrho_i^2 \right], \nonumber \\
& = & \sum_{m=1}^{l-i} \frac{1}{(i+m)(i+m-1)} \left[ \varrho_{i+m}^2 - \varrho_i^2 \right] + \frac{1}{l}  \left[ \varrho_{l+1}^2 - \varrho_i^2 \right]. \nonumber 
\end{eqnarray}}
Observe again that the last term in the first equation corresponds to $m = l-i+1$.

Next, we write the SV of player $l$. It is easy to see that player $l$ obtains positive utilities when $C = \{1,2,\cdots,l-1\}$. Thus as explained above, the Shapley Value of player $l$ equals,
\begin{eqnarray}
\phi_l^\P & = & \frac{1}{l} \left[ \varrho_{l+1}^2 - \varrho_l^2 \right].
\end{eqnarray}
Hence, we have the result.
\eop

\textbf{Proof of Theorem \ref{Thm_general}: } From footnote \ref{Foot_note1}, it suffices to show the stability of all partitions against the unilateral deviation of being alone. 
 
(i) As a first step consider any partition $\P = \{S_1, \cdots, S_k\}$ that plays RSG with $1 \in S_1$ and $2 \in S_2$. By Lemma \ref{Lem_Insignificant}, under {\bf A}.1,
\begin{equation}
\varphi_{S_m}^*(\P) = 0 \mbox{ for any }  m \text{ such that } 1,2 \notin S_m.
\end{equation}
Thus by  \eqref{Eqn_SV}-\eqref{Eqn_partition_min_gen} and Lemma \ref{min_worth_sub_coal_general}, the  share $\phi_i^\P$ of player $i \in S_m$ equals 0 if $m > 2$. 

Consider any $i \in S_1$ with $i \ne 1$.  Again under {\bf A}.1, and by Lemmas \ref{Lem_Insignificant} and \ref{min_worth_sub_coal_general}  (see \eqref{Eqn_partition_min_gen}), one can show that $\nu_C^\P = 0$ for all $C \subset  S_1$ with $1 \notin C$, as $C$ does not even contain $2$.  It is easy to observe from  Theorem \ref{Thm_util_coal_general},  that $\nu_C^\P = \nu_{C \cup \{i\}}^\P = \varrho_2^2  $, when $C \subset  S_1$ with $1 \in C$.  Hence from  \eqref{Eqn_SV}, $\phi_i^\P = 0$. Similarly $\phi_i^\P = 0$ for all $i \in S_2$ with $i \ne 2.$ In all, none of the agents $i$, with $i >2$, have any incentive in changing their strategy. 

Since all players in $S_1$ other than player $1$ obtain zero utility, player 1 obtains whole coalitional utility as its share. 
The utility of player $1$ after unilateral deviation, which results in partition $\P' = \{ \{1\}, S_1-\{1\},  S_2, \cdots, S_k\}$, using exactly similar logic equals  $\phi_1^{\P'} = \varrho_2^2  = \phi_1^\P$. Thus  player $1$ does not find it beneficial to deviate. Working in a similar way, player 2 also does not find it beneficial to deviate and the partition is stable.

(ii) Now consider any partition $\P = \{S_1, \cdots, S_k\}$, with $2 \le t<n$ being the biggest number such that $1, 2, \cdots, t \in S_1$ and $t+1 \in S_2$.
Again 
from Lemma \ref{Lem_Insignificant},  under {\bf A}.1, we have
$$
\varphi_{S_m}^*(\P)=0 \text{ for any } m > 2. 
$$
 Thus, on similar lines as in part (i),  the share $\phi_i^\P$ of any player $i > t+1$  equals zero. 
For any coalition $C \subset S_1$, define $j_C := \min \{ m:  m \in S_1 - C, m \ne 1, 2 \} $.  Observe as before that for any $C$ with $1\notin C$ and $2 \in C$ we have following: 
$$ \left[\nu^\P_{C\cup \{1\}} -\nu^\P_C \right]  =  \varrho_{j_C}^2 -  \left(1-\varrho_2 \right)^2 \ge  \varrho_{3}^2 -  \left(1-\varrho_2 \right)^2, \mbox{ as } j_C \ge 3. $$
Thus SV of player $1$ is given by (see \eqref{Eqn_SV}),
 {\small \begin{eqnarray}
	\phi_1^\P & = & \sum_{C \subset S_1; 1,2 \notin C}  \frac{|C|! (|S_1|-|C|-1)!}{|S_1|!}  \left[\nu^\P_{C\cup \{1\}} -\nu^\P_C \right] + \sum_{C \subset S_1; 1 \notin C \& 2 \in C}
 \frac{ |C|! (|S_1|-|C|-1)!}{|S_1|!} 	
	 \left[\nu^\P_{C\cup \{1\}} -\nu^\P_C \right] , \nonumber \\
	& \geq &  \sum_{l=0}^{|S_1|-2} \binom{|S_1|-2}{l} \frac{l!(|S_1|-l-1)!}{|S_1|!} \varrho_2^2 + \sum_{l=0}^{|S_1|-2} \binom{|S_1|-2}{l}\frac{(l+1)!(|S_1|-l-2)!}{|S_1|!} 	
	 \left[ \varrho_3^2- \left( 1- \varrho_2 \right)^2 \right],\nonumber \\
	& = & \sum_{l=0}^{|S_1|-2} \frac{(|S_1|-l-1)}{|S_1|(|S_1|-1)} \varrho_2^2 + \sum_{l=0}^{|S_1|-2} \frac{l+1}{|S_1|(|S_1|-1)} \left[ \varrho_3^2- \left( 1- \varrho_2 \right)^2 \right], \nonumber \\
	& = &  \frac{1}{2}\varrho_2^2+\frac{1}{2} \left[ \varrho_3^2- \left( 1- \varrho_2 \right)^2 \right].
\end{eqnarray}}
The utility of player $1$ after unilateral deviation equals  $\phi_1^{\P'} = \varrho_2^2  \le \phi_1^\P$ as before by {\bf A}.2 with $j=2$ 
 and hence, the player $1$ does not find it beneficial to deviate. Similarly, one can write the SV shares and utility after unilateral deviation for player 2 which does not find it beneficial to deviate. 

Proceeding in a similar fashion (under {\bf A}.1) one can observe that the Shapley value of a player $2<j <t+1$ is derived only\footnote{It is easy to observe that $\nu_{C \cup \{j\} }^\P  - \nu_C^\P=0$  for the rest of the sub-coalitions $C$, as  
with $\varrho_l := \lambda_1 / (\lambda_l +\lambda_1)$:
\begin{eqnarray*}
\nu_{C \cup \{j\} }^\P  - \nu_C^\P = \left \{ 
\begin{array}{lllll}
\varrho_2^2 -  \varrho_2^2  & \mbox{ when }  1 \in C, \mbox{ but }  2 \notin C,   j \notin C \\ 
(1-\varrho_2)^2 - (1- \varrho_2)^2  & \mbox{ when }  2 \in C, \mbox{ but }  1 \notin C,     j \notin C \\
0  -  0  & \mbox{ when } 1, 2 \notin C,  \\
\varrho_{l}^2 -  \varrho_l^2   & \mbox{ when }  1, 2, \cdots, (l-1)  \in C,   l \notin C, \mbox{ with }  l-1 < j,     j \notin C.
\end{array} \right .
\end{eqnarray*}} using  the sub-coalitions $C$ such that $\{1,\cdots,j-1\} \subset C$ and $j \neq C$. Observe as before that for any such $C$ we have following: 
$$ \left[\nu^\P_{C\cup \{j\}} -\nu^\P_C \right]  =  \varrho_{j_C}^2 -  \varrho_{j}^2 \ge  \varrho_{j+1}^2 -   \varrho_{j}^2, \mbox{ as } j_C \ge j+1. $$
\vspace{-2mm}
and so, the SV of player $j$ equals,
\begin{eqnarray}
\label{Eqn_bound_phi_j_P}
\phi_j^\P 
& \geq & \sum_{C \subset S_1; 1,\cdots,j-1 \in C} \frac{|C|!(|S_1|-|C|-1)!}{|S_1|!} \left [ \varrho_{j+1}^2 - \varrho_{j}  ^2\right ] \nonumber \\
& = &  \sum_{l=0}^{|S_1|-j} \binom{|S_1|-j}{l}  \frac{(l+j-1)!(|S_1|-l-j)!}{|S_1|!} \left [ \varrho_{j+1}^2 - \varrho_{j}  ^2\right ], \nonumber \\
& = & \left [ \varrho_{j+1}^2 - \varrho_{j}  ^2\right ] \sum_{l=0}^{|S_1|-j} \binom{|S_1|-j }{l} \frac{(l+j-1)!(|S_1|-l-j)!}{|S_1|!} = \frac{1}{j}\left [ \varrho_{j+1}^2 - \varrho_{j}  ^2\right ], 
\end{eqnarray} 
where the last equality follows from Lemma \ref{Lem_SV_coefficient}.
The utility of player $j$ after unilateral deviation equals  $\phi_j^{\P'} = \left( 1-\varrho_j \right)^2  \le \phi_j^\P$ by {\bf A}.2 with $j >2$ and hence, the player $j$ does not find it beneficial to deviate.

Now consider the case when {\bf A}.2 is not satisfied for some $j$, then there exist  some  partitions  which are not U-stable because of the following: 
Consider any partition 
of the form $\P = \{\{1,2,\cdots,j\},\{j+1,\cdots\}, \{\}, \cdots \}$ (any $\P$ with $S_1 = \{1, 2, \cdots, j\}$),   for which the lower bound derived in \eqref{Eqn_bound_phi_j_P} is exact, and thus negation 
of {\bf A.}2 for that $j$ implies these partitions are not U-stable.


(iii) Next, we consider GC, denoted by $\P = \{S_1\}$ such that $S_1 = \{1,\cdots,n\}$.

Following the same procedure as in (ii) above, we get similar conditions as {\bf A}.2 under which GC is stable for all players (since there is atleast one extra term with $\lambda_{j_C} =0$) except player $n$. Observe that the Shapley value of player $n$ is only given by the sub-coalitions $C$ such that $\{1,\cdots,n-1\} \subset C$ and hence equals,
\begin{eqnarray*}
\phi_n^\P = \frac{1}{n} \left (\varrho_{n+1}^2 - \varrho_n^2 \right ) \geq \phi_n^{\P'} \text{ under {\bf A}.1 and {\bf A}.2.}
\end{eqnarray*}
Thus assuming {\bf A}.1 to be true, all partitions are stable if and only if {\bf A.2} is satisfied. \eop

\begin{lemma}
Under {\bf A}.1,  for any partition $\P = \{ S_1,\cdots,S_k\}$,   we have,
$
\varphi_{S_m}^*(\P) = 0 \text{ for all } S_m \in \P \text{ with } m \geq 3.
$
\label{Lem_Insignificant}
\end{lemma}
\textbf{Proof:} Consider any partition $\P = \{S_1,\cdots, S_k\}$ such that $k_1 \in S_2$ and $k_2 \in S_3$ (with $k_2 \ge k_1+1$) respectively are  the active players in coalitions $S_2$ and $S_3$. By {\bf A}.1 
\begin{eqnarray*}
 \frac{1}{\lambda_1}+\frac{1}{\lambda_{k_1}}- \frac{1}{\lambda_{k_2}}  &\le & \frac{1}{\lambda_1}+\frac{1}{\lambda_{k_1}}- \frac{1}{\lambda_{k_1+1}} \le 0. 
\end{eqnarray*}
Hence by Theorem \ref{Thm_util_coal_general}, $M^\P = 2$ and hence  any coalition $S_m$ with  $m>2$ obtains zero utility. \eop

\textbf{Proof of Theorem \ref{Thm_general_sym}:} From footnote \ref{Foot_note1}, it suffices to show the stability of all partitions against the unilateral deviation of being alone.

(i) As a first step consider any partition $\P = \{S_1, \cdots, S_k\}$   with $1 \in S_1$ and $2 \in S_2$. As in case (i) of Theorem \ref{Thm_general}  all such partitions are stable under {\bf A}.1$'$ 
 now using Lemma \ref{Lem_Insignificant_sym}. Basically,  with $1$ and $2$ in distinct coalitions, all other players including $3$ and $4$ derive zero shares under {\bf A}.1$'$, irrespective of their position in the partition.

(ii) Now consider any partition $\P = \{S_1, \cdots, S_k\}$, with $2 < t<n$ being the biggest number such that $1, 2, \cdots, t \in S_1$ and $t+1 \in S_2$.
Since the coalition $S_1$ contains atleast one of the players $3$ or $4$,  we have under {\bf A}.1$'$,
$$
\varphi_{S_m}^*(\P)=0 \text{ for any } m > 2. 
$$
 Thus, on similar lines as in part (i),  the share $\phi_i^\P$ of any player $i > t+1$  equals zero.  We now consider computing $\phi_1^\P$.
For any coalition $C \subset S_1$ such that $2 \in C$, define $j_C := \min \{ m:  m \in S_1 - C, m \ne 1, 2 \} $. Using {\bf A}.1$'$ and Lemma \ref{min_worth_sub_coal_general}, the worth of  any sub-coalition $C \subset S_1$ when player $1$ is added to it is given by,
 \begin{eqnarray}
\nu_{C \cup \{1\}}^\P
	= \left \{   \begin{array}{llll}
	  \left( \frac{2\lambda_1-\lambda_3}{2\lambda_1+\lambda_3} \right)^2   & \text{ when } 3\notin C \mbox{ and } 4 \notin C,   \\
	  \varrho_{j_C}^2  & \text{ else. }
	\end{array} \right.
\end{eqnarray}
 For computing the worth in the first line, from \eqref{Eqn_partition_min_gen} under {\bf A}.1$'$, the RSG eventually involves players $1$, $3$ and $4$, irrespective of
whether 3 or 4 is in $S_2$.
One can easily  check that
$
 (2\lambda_1-\lambda_3)^2/(2\lambda_1+\lambda_3)^2   < \varrho_3^2 \leq \varrho_{j_C}^2 \text{ for any } j_C >2.
$ Hence, as before, for any $C$ with $1\notin C$ and $2 \in C$ we have following: 
$$ \left[\nu^\P_{C\cup \{1\}} -\nu^\P_C \right]  
 \ge  \frac{(2\lambda_1-\lambda_3)^2}{(2\lambda_1+\lambda_3)^2} -  \left(1-\varrho_2 \right)^2. 
  $$
Thus SV of player $1$ is given by (see \eqref{Eqn_SV}),
 {\small \begin{eqnarray}
	\phi_1^\P & = & \sum_{C \subset S_1; 1,2 \notin C}  \frac{|C|! (|S_1|-|C|-1)!}{|S_1|!}  \left[\nu^\P_{C\cup \{1\}} -\nu^\P_C \right] + \sum_{C \subset S_1; 1 \notin C \& 2 \in C}
 \frac{ |C|! (|S_1|-|C|-1)!}{|S_1|!} 	
	 \left[\nu^\P_{C\cup \{1\}} -\nu^\P_C \right] , \nonumber \\
	& \geq &  \sum_{l=0}^{|S_1|-2} \binom{|S_1|-2}{l}\frac{l!(|S_1|-l-1)!}{|S_1|!} \varrho_2^2 + \sum_{l=0}^{|S_1|-2} \binom{|S_1|-2}{l}\frac{(l+1)!(|S_1|-l-2)!}{|S_1|!} 	
	 \left[  \left( \frac{2\lambda_1-\lambda_3}{2\lambda_1+\lambda_3} \right)^2 - \left( 1- \varrho_2 \right)^2 \right],\nonumber \\
	& = & \sum_{l=0}^{|S_1|-2} \frac{(|S_1|-l-1)}{|S_1|(|S_1|-1)} \varrho_2^2 + \sum_{l=0}^{|S_1|-2} \frac{l+1}{|S_1|(|S_1|-1)} \left[  \left( \frac{2\lambda_1-\lambda_3}{2\lambda_1+\lambda_3} \right)^2 - \left( 1- \varrho_2 \right)^2 \right], \nonumber \\
	& = &  \frac{1}{2}\varrho_2^2+\frac{1}{2} \left[  \left( \frac{2\lambda_1-\lambda_3}{2\lambda_1+\lambda_3} \right)^2 - \left( 1- \varrho_2 \right)^2 \right].
\end{eqnarray}}
The utility of player $1$ after unilateral deviation equals  $\phi_1^{\P'} = \varrho_2^2  \le \phi_1^\P$  by {\bf A}.2$'$ with $j=2$. Similarly, one can write the SV shares and utility after unilateral deviation for player $2$, which does not find it beneficial to deviate under {\bf A}.2$'$. 

We now consider player 3, and related some sub-cases. 

\textbf{Case 1:} When $4 \in S_2$

Observe that the player $3$ derives positive utility only with sub-coalitions $C$, with $1,2 \in C$.  The additional worth generated by the player when it adds to such a sub-coalition $C$ is given by,
$$
\nu_{C \cup \{3\}}^\P - \nu_C^\P
	=  \varrho_4^2 - \left( \frac{2\lambda_1-\lambda_3}{2\lambda_1+\lambda_3} \right)^2.
$$
Thus, the SV of player $3$ is given by,
\begin{eqnarray}
\phi_3^\P & = &  \sum_{C \subset S_1; 1, 2 \in C}
 \frac{ |C|! (|S_1|-|C|-1)!}{|S_1|!} 	
	 \left[\nu^\P_{C\cup \{3\}} -\nu^\P_C \right] \nonumber \\
  & = &  \sum_{l=0}^{|S_1|-3} \binom{|S_1|-3}{l} \frac{(l+2)!(|S_1|-l-3)!}{|S_1|!} \left [ \varrho_4^2 - \left( \frac{2\lambda_1-\lambda_3}{2\lambda_1+\lambda_3} \right)^2 \right] \nonumber \\
	 & = & \frac{1}{3} \left [ \varrho_4^2 - \left( \frac{2\lambda_1-\lambda_3}{2\lambda_1+\lambda_3} \right)^2 \right], \text{using Lemma \ref{Lem_SV_coefficient}}. \nonumber
\end{eqnarray}
From Theorem \ref{Thm_util_coal_general}, the utility of the same player after unilateral deviation to being alone equals $\left( \frac{\lambda_3}{2\lambda_1+\lambda_3}  \right)^2$ (since $M^\P =3$ and hence players $3$ and $4$ obtain strictly positive utility) and hence the partition is stable under {\bf A}.2$'$ for $j=3$.

\textbf{Case 2:} When $4 \in S_1$

Observe again that the player $3$ derives positive utility only with sub-coalitions $C$, with $1,2 \in C$.  The additional worth generated by the player when it adds to such a sub-coalition $C$ is given by,
 \begin{eqnarray}
\nu_{C \cup \{3\}}^\P - \nu_C^\P
	= \left \{   \begin{array}{llll}
	  \varrho_4^2 - \left( \frac{2\lambda_1-\lambda_3}{2\lambda_1+\lambda_3} \right)^2   & \text{ when } 4 \notin C,   \\
	  \varrho_5^2 - \varrho_4^2  & \text{ when  } 4 \in C.   \\
	\end{array} \right.
\end{eqnarray}
Then,
the Shapley Value of player $3$ is given by,
 \begin{eqnarray}
\phi_3^\P & = &  \sum_{C \subset S_1; 1, 2 \in C , 4 \in C}
 \frac{ |C|! (|S_1|-|C|-1)!}{|S_1|!} 	
	 \left[\nu^\P_{C\cup \{3\}} -\nu^\P_C \right] \nonumber \\
  && + \sum_{C \subset S_1; 1, 2 \in C , 4 \notin C}
 \frac{ |C|! (|S_1|-|C|-1)!}{|S_1|!} 	
	 \left[\nu^\P_{C\cup \{3\}} -\nu^\P_C \right], \nonumber \\
	  & \stackrel{ \text{ (a) }}{=} & \frac{3!(4-3-1)!}{4!} \left[ \varrho_5^2 - \varrho_4^2 \right] + \frac{2!(4-2-1)!}{4!} \left [  \varrho_4^2 - \left( \frac{2\lambda_1-\lambda_3}{2\lambda_1+\lambda_3} \right)^2 \right], \nonumber  \\
	  & = & \frac{1}{4} \left[ \varrho_5^2 - \varrho_4^2 \right] + \frac{1}{12} \left [  \varrho_4^2 - \left( \frac{2\lambda_1-\lambda_3}{2\lambda_1+\lambda_3} \right)^2 \right]. \nonumber 
\end{eqnarray} 
The equation (a) follows from the interpretation of coefficients in the Shapley Value, i.e., the probability that in any permutation, the members of $C$ are ahead of player $i$ (player $3$ in this case) The utility of the same player after unilateral deviation to being alone equals $\left( 1-\varrho_4 \right)^2$  and hence the partition is stable under {\bf A}.2$'$ for $j=4$.

Note that by symmetry, $\phi_3 = \phi_4$.  The SV of  $j >4$ (applicable only when $t\ge j$),   can be upper bounded with exactly the same term as in \eqref{Eqn_bound_phi_j_P} and hence  the agent  does  not find it beneficial to deviate by condition {\bf A}.2$'$ with $j$ which is the same as  that under {\bf A}.2.

(iii) Proceeding similarly, one can show that GC is stable under {\bf A}.2$'$.

(iv) This partition $\P = \{S_1,\cdots,S_k\}$ such that $1,2,4 \in S_1$ and $3 \notin S_1$ is same as the partition where $1,2,3 \in S_1$ and $4 \notin S_2$, which is already covered in (ii).

(v) The next partition is $\P = \{S_1,S_2,\cdots,S_k\},$ such that $1,2 \in S_1$, $ 3 \in S_2$ and $4 \in S_3$. One can observe from (ii) that the lower bound derived for SV of player $1$ and $2$ is exact in this case. Hence, under {\bf A}.2$'$ all partitions are stable.

(vi) The partition that we are left with is $\Q = \{S_1,S_2,\cdots,S_k\},$ such that $1,2 \in S_1$ and $ 3,4 \in S_2$. Again,  under {\bf A}.1$'$, we have
$$
\varphi_{S_m}^*(\P)=0 \text{ for any } m > 2. 
$$ 
The conditions derived under {\bf A}.2$'$ for $j=2$ ensures that no player in $S_1$ can obtain higher utilities on unilateral deviation. The SV of player $3$, which by symmetry equals that of   player $4$, is given by
\begin{eqnarray}
\phi_3^\P = \frac{1}{2} \varrho_4^2 \ \geq \ \left( \frac{\lambda_3}{2\lambda_1+\lambda_3} \right)^2 \ = \ \phi_3^{\P'} \text{ true under {\bf A}.2$'$ for  } j=3,
\end{eqnarray}
where $\phi_3^{\P'}$ is the utility of player $3$ after unilateral deviation. \eop

\begin{lemma}
Under {\bf A}.1$'$,  for any partition $\P = \{ S_1,\cdots,S_k\}$
\begin{enumerate}[label=(\roman*)]
\item With $3,4 \notin S_i \text{ for } i \geq 3$   we have,
$
\varphi_{S_m}^*(\P) = 0 \text{ for all } S_m \in \P \text{ with } m \geq 3.
$
\item  With $1,2 \in S_1$, $3 \in S_2$ and $4 \in S_3$, 
$
\varphi_{S_m}^*(\P) = 0 \text{ for all } S_m \in \P \text{ with } m \geq 4.
$
\end{enumerate}
\label{Lem_Insignificant_sym}
\end{lemma}
\textbf{Proof:} (i) The proof of this part is exactly same as Lemma \ref{Lem_Insignificant}.

(ii) Consider any partition $\P = \{S_1,\cdots, S_k\}$ such that $1,2 \in S_1$, $3 \in S_2$ and $4 \in S_3$. This means that player $3$ and $4$ are  the active players in coalitions $S_2$ and $S_3$. Define $j_C := \min \{ m :  m \in S_4 , \lambda_m \ge \lambda_i \text{ for } i \in S_4 \} $. By {\bf A}.1$'$ we have,
\begin{eqnarray*}
 \frac{1}{\lambda_1}+\frac{2}{\lambda_{4}}- \frac{2}{\lambda_{j_C}}  \le  \frac{1}{\lambda_1}+\frac{2}{\lambda_{4}}- \frac{2}{\lambda_{5}}  =  2 \left( \frac{1}{\lambda_1}+\frac{1}{\lambda_{4}}- \frac{1}{\lambda_{5}} \right) - \frac{1}{\lambda_1} < 0, \text{ note } j_C \geq 5. 
\end{eqnarray*}
Hence by Theorem \ref{Thm_util_coal_general}, $M^\P = 2 $ in first case and $3$ in second  case. Hence  any coalition $S_m$ with  $m>M^\P$ obtains zero utility. \eop

\begin{lemma}
Consider a partition $\P = \{S_1,S_2,\cdots,S_k\}$. Then  for any $j \le |S_1|$, we have
$$
\sum_{l=0}^{|S_1|-j} \binom{|S_1|-j}{l} \frac{(l+j-1)!(|S_1|-l-j)!}{|S_1|!} = \frac{1}{j}.
$$
\label{Lem_SV_coefficient}
\end{lemma}
\vspace{-10mm}
\textbf{Proof:} The above term can be simplified as following:
\vspace{3mm}
{\small \begin{eqnarray}
 \sum_{l=0}^{|S_1|-j} \binom{|S_1|-j}{l} \frac{(l+j-1)!(|S_1|-l-j)!}{|S_1|!} =  \sum_{l=0}^{|S_1|-j} \frac{(|S_1|-j)!}{l!} \frac{(l+j-1)!}{|S_1|!}= \frac{(|S_1|-j)!}{|S_1|!} \sum_{l=0}^{|S_1|-j} \frac{(l+j-1)!}{l!}. \nonumber
\end{eqnarray}}

For ease of notations, let $m = |S_1|$. One can check that our conjecture is true for $m=3$ and we assume,
\begin{equation}
a_m = \frac{(m-j)!}{m!} \sum_{l=0}^{m-j} \frac{(l+j-1)!}{l!} = \frac{1}{j} \text{ for } 2 <j<n \text{ to be true}.
\end{equation}
We use induction to show that $a_{m+1}=1/j$.
\begin{eqnarray}
a_{m+1} & = & \frac{(m+1-j)!}{(m+1)!} \sum_{l=0}^{m+1-j} \frac{(l+j-1)!}{l!} \ = \ \left( \frac{m+1-j}{m+1} \right) \frac{(m-j)!}{m!} \left[ \sum_{l=0}^{m-j} \frac{(l+j-1)!}{l!} + \frac{m!}{(m+1-j)!} \right], \nonumber \\
& = & \left( \frac{m+1-j}{m+1} \right)a_m + \frac{(m+1-j)!}{(m+1)!}\frac{m!}{(m+1-j)!} \ = \  \frac{m+1-j}{j(m+1)}  + \frac{m!}{(m+1)!} \nonumber \\
& = & \frac{m+1-j}{j(m+1)}  + \frac{1}{(m+1)} \ = \  \frac{1}{j}. \nonumber 
\end{eqnarray} \eop

\textbf{Proof of Theorem \ref{Thm_monotonicity}:} Consider any $\P = \text{SS}(C)$ partition with $C = \{j, k\}$.  Then from SVs given by  \eqref{Eqn_SV} as applied to coalition $C \in \P$ and by Lemma \ref{min_worth_sub_coal_general} and Theorem \ref{Thm_util_coal_general}, we have that 
\begin{eqnarray}
\phi_j^\P - \phi_j^{ALC}  &=& \left   \{   \frac{1}{2} \left[ \left( \frac{\bar{w} - w_k - (n-2)w_j}{\bar{w} - w_k} \right)^2 - \left( \frac{\bar{w}  - (n-1)w_k}{\bar{w}} \right)^2 \right]+ \frac{1}{2} \left( \frac{\bar{w}  - (n-1)w_j}{\bar{w}} \right)^2 \right  \} \nonumber \\
&& -  \left( \frac{\bar{w}  - (n-1)w_j}{\bar{w}} \right)^2 \nonumber \\
 &=&  \frac{1}{2}    \left \{ \left( \frac{\bar{w} - w_k - (n-2)w_j}{\bar{w} - w_k} \right)^2 - \left( \frac{\bar{w}  - (n-1)w_k}{\bar{w}} \right)^2  -  \left( \frac{\bar{w}  - (n-1)w_j}{\bar{w}} \right)^2 
 \right \}. \label{Eqn_jk_j_k}
\end{eqnarray}
It is easy to observe that $\phi_k^\P - \phi_k^{ALC}  $ is also given by the same expression, and thus partition $\P$ would be stable  if and only if the above is positive;  observe any possible unilateral deviation (by either $j$ or $k$ and from such an SS partition) leads to ALC. 

To prove the said result, it suffices to show that the value in RHS of \eqref{Eqn_jk_j_k}  increases (or remains the same)  when one replaces $j$ with a $j' < j$, or equivalently when   $w_j $  is replaced with $w_{j'}$ without changing other terms.  We achieve to show this by embedding the RHS into the following  continuous function of $w$ (without perturbing ${\bar w}$) and then by  showing    the resulting function to be non-decreasing of 
$w$  on interval $[w_1, w_j]$   (observe second term of RHS of \eqref{Eqn_jk_j_k}  does not change with this $w$): 
 \begin{eqnarray}
 g(w) & := &
  \left( \frac{\bar{w} - w_k - (n-2)w}{\bar{w} - w_k} \right)^2 - \left( \frac{\bar{w}  - (n-1)w}{\bar{w}} \right)^2  \  =  \  \left( 1-\frac{(n-2)w}{\bar{w} - w_k} \right)^2 - \left( 1-\frac{ (n-1)w}{\bar{w}} \right)^2 \nonumber \\
   & = & \left( \frac{ (n-1)w}{\bar{w}} - \frac{(n-2)w}{\bar{w} - w_k} \right) \left( 2 - \frac{(n-2)w}{\bar{w} - w_k} - \frac{ (n-1)w}{\bar{w}}\right) \nonumber \\
  & =  &  c_1w \left( \frac{2\bar{w}(\bar{w}-w_k)-(n-2)w\bar{w}-(n-1)w(\bar{w}-w_k)}{\bar{w}(\bar{w}-w_k)}\right) \nonumber \\
& = &  c_1w \left( \frac{2\bar{w}[\bar{w}-(n-1)w]-w_k[\bar{w}-(n-1)w]+\bar{w}(w-w_k)}{\bar{w}(\bar{w}-w_k)}\right) \nonumber  \\
& =  &  c_1w \left( \frac{(2\bar{w}-w_k)[\bar{w}-(n-1)w]+\bar{w}(w-w_k)}{\bar{w}(\bar{w}-w_k)}\right) ,\nonumber 
\end{eqnarray}
where constant, $c_1 := (n-1)/{\bar w} - (n-2)/({\bar w}-w_k)$.
Differentiating it with respect to $w$ we obtain,
\begin{eqnarray}
\frac{d g}{dw}  & = & (2\bar{w}-w_k)[\bar{w}-2(n-1)w]+\bar{w}(2w-w_k) \nonumber \\
& = & (2\bar{w}-w_k)[\bar{w}-2(n-2)w]-2w(2\bar{w}-w_k)+\bar{w}(2w-w_k) \nonumber \\
 & = & (2\bar{w}-w_k)[\bar{w}-2(n-2)w]-2w\bar{w} + 2ww_k - \bar{w}w_k \nonumber \\
  & = & (2\bar{w}-w_k)[\bar{w}-2(n-2)w] + 2w(w_k -\bar{w}) - \bar{w}w_k \ < \ 0, \nonumber
\end{eqnarray}for all $w \ge w_1$, 
if $\bar{w}-2(n-2)w_1 \le 0$. And hence the result. 

 Next, we show that $\bar{w}-2(n-2)w_1 \le 0$ implies $\Delta_w < w_1$. Observe that  $\bar{w}-2(n-2)w_1 \le 0$ implies $w_1 \geq \frac{\bar{w}}{2(n-2)}$. From the definition of $\Delta_w$ as in \eqref{Eqn_def_w_delta}, we know that $\Delta_w = w_n-w_1$. Since all players are significant, we have $\bar{w} - (n-1)w_n >0$ which implies $w_n < \frac{\bar{w}}{n-1}$. Thus, we have
 \begin{eqnarray}
 \Delta_w < \frac{\bar{w}}{n-1} - \frac{\bar{w}}{2(n-2)} <  \frac{\bar{w}}{n-2} - \frac{\bar{w}}{2(n-2)} = \frac{\bar{w}}{2(n-2)} \leq w_1. \mbox{ \eop }\nonumber
 \end{eqnarray}

\textbf{Proof of Theorem \ref{Thm_monotonicity_new}:} As in Theorem \ref{Thm_monotonicity}, one needs to show the positivity of the following function for any $w_{k'}$ with $k'>k$ (recall the coalition is of two members and $w_j \le w_k$),
 \begin{eqnarray}
 \phi_{k'}^\P - \phi_{k'}^{ALC}
 &=&  \frac{1}{2}    \left \{ \left( \frac{\bar{w} - w_{k'} - (n-2)w_j}{\bar{w} - w_{k'}} \right)^2 - \left( \frac{\bar{w}  - (n-1)w_{k'}}{\bar{w}} \right)^2  -  \left( \frac{\bar{w}  - (n-1)w_j}{\bar{w}} \right)^2 
 \right \}, \nonumber \end{eqnarray}
 to show both members of $\{j,k'\}$ are stable against unilateral deviations.
 Following the steps of Theorem  \ref{Thm_Sym_to_Asym} which are used to define $f(\delta)$, we have
 {\small \begin{eqnarray}
  \frac{2}{\epsilon} \left [\phi_{k'}^\P - \phi_{k'}^{ALC} \right] & = & \frac{ w_j}{\left[ \bar{w}(\bar{w}-w_{k'}) \right]^2} \left [ 2\bar{w} c-\epsilon w_j \right] - \frac{\epsilon}{\bar{w}^2} \text{ with } c := [\bar{w}-w_{k'}-(n-2)w_j ] \text{ and } \epsilon := {\bar w}-(n-1)w_{k'}, \nonumber \\
 & = & \frac{1}{\bar{w}^2} \left[ \frac{ w_j}{(\bar{w}-w_{k'})^2} \left [ 2\bar{w} c-\epsilon w_j \right] - \epsilon \right].  \nonumber
 \end{eqnarray}}
 It suffices to show the positivity of the term on RHS after leaving $1/\bar{w}^2$. Towards this, we show the positivity of following function with $x:= w_{k'}$ (observe that $\bar{w}-w_{k'} = \sum_{i \neq k'} w_i := \tilde{w}$),
 \begin{eqnarray}
 g(x) & = & \frac{w_j}{\tilde{w}^2} \left[2(\tilde{w}+x) c-\left[\tilde{w}-(n-2)x\right] w_j \right]- \left[\tilde{w}-(n-2)x\right] \text{ with } c:= \tilde{w}-(n-1)w_j. \nonumber 
 \end{eqnarray}
 Clearly, $g(x)$ is a linear function of $x$ and its slope is given by, 
 \begin{eqnarray}
 g'(x) & = & \frac{2cw_j}{\tilde{w}^2} + (n-2)\left [\left(\frac{w_j}{\tilde{w}}\right)^2+1\right] \ > \ 0,\nonumber 
 \end{eqnarray}

which implies that $g(x)>0$ and hence $\phi_{k'}^\P - \phi_{k'}^{ALC}>0$ for larger value of $w_{k'}$ once SS-$(\{j,k\})$ is stable; and thus, we have that SS-$(\{j,k+1\})$ is stable.\eop 
 
 \textbf{Proof of Theorem \ref{Thm_Sym_to_Asym}:}  Let us begin with a system in which SS($C$) with $C = \{1, n\}$ is not U-stable.  All players are significant at ALC, and hence
 ${\bar w} - (n-1) w_j > 0$ for all $j$ including $n$.
Consider a sequence of systems one for each $\delta$ in which only  influence factors of players 1, $n$ change to  $w_1' = w_1 - \delta$ and $w_n' = w_n + \delta $. 
  As in \eqref{Eqn_jk_j_k} of proof of Theorem \ref{Thm_monotonicity}, SS($C$) partition in  any $\delta$-system becomes U-stable if the following function becomes positive (with $\epsilon := {\bar w}-(n-1)w_n'$):
   \begin{eqnarray}
   f(\delta)   & := &\frac{1}{\epsilon} \left ( \left( 1-\frac{(n-2)w_1'}{\bar{w}-w_n'} \right)^2 - \left( 1-\frac{(n-1)w_1'}{\bar{w}} \right)^2- \left( 1-\frac{(n-1)w_n'}{\bar{w}} \right)^2 \right ) \nonumber \\
 & = &\frac{1}{\epsilon} \left (  w_1' \left[ \frac{n-1}{\bar{w}}-\frac{n-2}{\bar{w}-w_n'} \right] \left[ 2-\frac{(n-1)w_1'}{\bar{w}}-\frac{(n-2)w_1'}{\bar{w}-w_n'} \right] - \frac{\epsilon^2}{\bar{w}^2}  \right ) \nonumber \\
& = &  \frac{ w_1'}{\left[\bar{w}(\bar{w}-w_n')\right]^2} \left [ 2\bar{w} c-\epsilon w_1' \right] - \frac{\epsilon}{\bar{w}^2}, \nonumber \end{eqnarray}with $c := [\bar{w}-w_n'-(n-2)w_1' ]$, because of the following:
 \begin{eqnarray}
2-\frac{(n-1)w_1'}{\bar{w}}-\frac{(n-2)w_1'}{\bar{w}-w_n'} & = & \frac{2\bar{w}(\bar{w}-w_n')-(n-2)w_1'\bar{w}-(n-1)w_1'(\bar{w}-w_n')}{\bar{w}(\bar{w}-w_n')}, \nonumber \\
&   & \hspace{-25mm} = \  \frac{2\bar{w} [\bar{w}-w_n'-(n-2)w_1' ]+ (n-2)w_1'\bar{w}-(n-1)w_1'(\bar{w}-w_n')}{\bar{w}(\bar{w}-w_n')} \nonumber \\
& =  & \frac{2\bar{w} [\bar{w}-w_n'-(n-2)w_1' ]-\epsilon w_1'}{\bar{w}(\bar{w}-w_n')} \nonumber \\
&   & \hspace{-25mm} = \ \frac{2\bar{w} c-\epsilon w_1'}{\bar{w}(\bar{w}-w_n')} \mbox{, and  because, }  \frac{n-1}{\bar{w}}-\frac{n-2}{\bar{w}-w_n'} =  \frac{ \epsilon }{\bar{w}(\bar{w}-w_n')} .
 \nonumber 
\end{eqnarray}
 We consider only systems in which $n$-th player is significant at ALC, i.e., such that $w_n' = w_n + \delta <  {\bar w} / (n-1) $.  When one considers the limit 
 $w_n'  \uparrow  {\bar w} / (n-1) $ or equivalently $\epsilon \to 0$, we have (recall ${\bar w}-(n-1) w_n > 0$):
 \begin{eqnarray}
  c \to c^*,  \mbox{ with, } \   \   c^* & = & \bar{w}-\frac{\bar{w}}{n-1}-(n-2)\left(w_1+w_n-\frac{\bar{w}}{n-1} \right)  \nonumber \\
 & = & (n-2) \left [ \frac{\bar{w}}{n-1} -\left(w_1+w_n-\frac{\bar{w}}{n-1} \right ) \right] \nonumber \\
&  = &   (n-2) \left [ \frac{2\bar{w}}{n-1} -\left(w_1+w_n\right ) \right] \ > \ 0.
 \end{eqnarray}
 By continuity this implies $f(\delta_o ) > 0$ for $\delta_o := {\bar w}/(n-1) - w_n$, which in turn implies the existence of a threshold ${\bar \delta}$ such that 
 $f(\delta) > 0$ and hence  SS($C$)  is U-stable,  when  ${\bar \delta} \le   \delta  <    \delta_o $ and hence the theorem. 
 \eop
 
  \textbf{Proof of Theorem \ref{Thm_no_identical}: }
Consider a partition $\mathcal{P} = \{S_1,S_2,\cdots,S_k\}$ with $k \geq 3$ significant coalitions (i.e., obtains non-zero utility). Let $\bar{w} := \sum_{i=1}^k 1/\um^\P_i$.
W.l.o.g., let $S_1$ be the coalition with atleast two identical players (having maximum influence factor). For $k$ coalitions to be significant, one requires $\bar{w}-(k-1)/\um^\P_k > 0$. 
 Then the utility of $\lambda$ player in $S_1$ (from Theorem 6.1) is upper bounded by (and equality only if $S_1 = \{\lambda, \lambda\}$),
\begin{equation}
\frac{1}{2} \left( \frac{\bar{w}-\frac{k-1}{\lambda}}{\bar{w}} \right) ^2.
\end{equation}
Consider $\lambda$ player unilaterally deviates to being alone. Then by \eqref{Eqn_Counter}, we have $\bar{w}+1/\lambda-k/\um^\P_k > 0$ 
and hence the utility of the same player after unilateral deviation is given by (from footnote \ref{Foot_note1}, it suffices to show the stability of all partitions against the unilateral deviation of being alone),
\begin{equation}
 \left( \frac{\bar{w}+\frac{1}{\lambda}-\frac{k}{\lambda}}{\bar{w}+\frac{1}{\lambda}} \right) ^2.
\end{equation}
 Numerators are the same while the denominator of the first term is larger since,
\begin{eqnarray*}
\sqrt{2} \bar{w} = \bar{w} + 0.414 \bar{w} \stackrel{\mbox{(a)}}{\ge} \bar{w} + 0.414 \left(\frac{3}{\lambda} \right) > \bar{w} + \frac{1}{\lambda}.
\end{eqnarray*}
and hence the utility after deviation is larger. The inequality (a) follows from the definition of $\bar{w}$, size of partition being atleast 3 and $\lambda \geq \lambda_i \text{ for all } i$.
 Thus, the partition is not stable. \eop 

 \textbf{Proof of Lemma \ref{Lem_worst_NE_util}:} 
 Recall   ALC is always U-stable.  Thus,  it is sufficient to show that the sum of the utilities of the coalitions in the partition $\P'$ obtained by addition of a player with influence factor $\lambda$  to any partition  $\P$ is less than or equal to the sum of the utilities of the coalitions under  $\P$. For example, $\P'$ can result when a coalition 
 in $\P$ splits and ALC is obtained when no further split is possible. 
 
 Consider any partition $\P= \{S_1,\cdots,S_k\}$, with influence factors (of active players)  represented by  $\lambda_1, \cdots, \lambda_k$, with $w_k \geq \cdots \geq w_1$ and let $\P' = \P \cup \{ S_\lambda\}$.   
 We prove the above results by considering different cases. 
  Towards this, we   first consider the case when all players are significant (before and after addition of a new player). This study is divided into two cases.

\textbf{Case 1:  When $w \geq w_k$ with $w: = 1/\lambda$ and when  $\sum_{m=1}^{k}w_m+w-kw > 0$.}
Let $f$ denote the function representing the sum of utilities of coalitions in partition $\P'$ (by Theorem \ref{Thm_util_coal_general}) and also define another function $U_{SO}$ as below: 
\begin{eqnarray}
f(w) & := &  U_{SO} (w_1, \cdots, w_k, w) =  \sum_{m=1}^{k} \left ( \frac{\sum_{i=1}^{k} w_i+w-kw_m}{\sum_{i=1}^{k} w_i+w} \right)^2 +  \left ( \frac{\sum_{i=1}^{k} w_i +(1-k)w}{\sum_{i=1}^{k} w_i +w} \right)^2. \nonumber
\end{eqnarray}
Observe that the same sum under $\P$ is given by $ U_{SO} (w_1, \cdots, w_k) $. 
By differentiating $f(w)$ with respect to $w$, we have

\begin{eqnarray}
f'(w) & = &  \sum_{m=1}^{k} \left [ \frac{2\left(\sum_{i=1}^{k} w_i + w-kw_m \right)\left(\sum_{i=1}^{k} w_i+w \right)^2 -2\left(\sum_{i=1}^{k} w_i+w-kw_m \right)^2\left(\sum_{i=1}^{k} w_i + w\right) }{\left(\sum_{i=1}^{k} w_i + w \right)^4} \right]\nonumber \\
& &  +   \left [ \frac{2\left(\sum_{i=1}^{k} w_i+(1-k)w \right)\left(\sum_{i=1}^{k} w_i + w \right)^2\left( 1-k \right) -2\left(\sum_{i=1}^{k} w_i+(1-k)w \right)^2\left(\sum_{i=1}^{k} w_i + w\right) }{\left(\sum_{i=1}^{k} w_i + w\right)^4} \right], \nonumber \\
& = & \sum_{m=1}^{k} \left [ \frac{2kw_m\left(\sum_{i=1}^{k} w_i+w-kw_m \right) }{\left(\sum_{i=1}^{k} w_i + w\right)^3} \right]  +   \left [ \frac{2\left(\sum_{i=1}^{k} w_i+(1-k)w \right)\left(-k\sum_{i=1}^{k} w_i  \right) }{\left(\sum_{i=1}^{k} w_i + w \right)^3} \right], \nonumber \\
& = & \frac{2k}{\left(\sum_{i=1}^{k} w_i + w\right)^3} \left[ \sum_{m=1}^{k} \left [ w_m\left(\sum_{i=1}^{k} w_i + w-kw_m \right) \right]    -\left(\sum_{i=1}^{k} w_i+(1-k)w \right)\left(\sum_{i=1}^{k} w_i \right)   \right], \nonumber
\end{eqnarray}
\begin{eqnarray}
 & = & \frac{2k}{\left(\sum_{i=1}^{k} w_i + w \right)^3} \left[ -k\sum_{i=1}^k w_i^2 +kw\left(\sum_{i=1}^{k} w_i \right)\right] \nonumber \\
  & = & \frac{2k^2}{\left(\sum_{i=1}^{k} w_i + w \right)^3} \left[ -\sum_{i=1}^k w_i^2 +w \left(\sum_{i=1}^{k} w_i \right)\right] > 0 \text{ since $w \geq w_i$ for all $i \leq k$.} \nonumber
\end{eqnarray}
Thus  $f$ is an increasing function of $w$ and for significance we required $w < \frac{ \sum_{m=1}^k w_m}{k-1}$. Thus it is sufficient to show  $f  \left (  \frac{ \sum_{m=1}^k w_m}{k-1} \right )$ is smaller than
$U_{SO} (w_1, \cdots, w_k)$ and this is true because:
\begin{eqnarray}
f\left (  \frac{ \sum_{m=1}^k w_m}{k-1} \right )& = & \sum_{m=1}^{k} \left ( \frac{\sum_{i=1}^{k} w_i +\frac{ \sum_{m=1}^k w_m}{k-1}-kw_m}{\sum_{i=1}^{k} w_i + w} \right)^2 +  \left ( \frac{\sum_{i=1}^{k} w_i +(1-k)\frac{ \sum_{m=1}^k w_m}{k-1}}{\sum_{i=1}^{k} w_i + w} \right)^2, \nonumber \\
& = &  \sum_{m=1}^{k} \left ( \frac{\frac{ k}{k-1}\sum_{i=1}^{k} w_i -kw_m}{\frac{ k}{k-1}\sum_{i=1}^{k} w_i} \right)^2 \ = \ \sum_{m=1}^{k} \left ( \frac{\sum_{i=1}^{k} w_i -(k-1)w_m}{\sum_{i=1}^{k} w_i} \right)^2 \ = \ U_{SO} (w_1, \cdots, w_k). \nonumber 
\end{eqnarray}
Thus, when all the players are significant, the sum of utilities of the coalitions in $\P$ is greater than the sum of utilities of the coalitions in $\P'$.

\textbf{Case 2:  When  $w_{j} \leq w \leq w_{j+1}$ for some $j < k$ and  $\sum_{m=1}^{k}w_m+w-kw_k > 0$.}
From Case 1,   $f(w)$ increases with $w$ and the sum utility (also represented by $U_{SO}$) decreases when a member is added to the end.  Using this progressively 
$U_{SO}$ increases after removing the tail ($\{w_{j+2}, \cdots, w_k\}$) as below, 
$$
U_{SO} ( w_1,\cdots,w_j,w,w_{j+1},\cdots,w_k) \leq  U_{SO} ( w_1,\cdots,w_j,w) \stackrel{\text{(a)}}{\leq} U_{SO} ( w_1,\cdots,w_j,w_{j+1}),
$$
and inequality $(a)$ is once again by Case 1, now by replacing $w$ with a bigger $w_{j+1}$ (without affecting the number playing RSG).

{\bf Case 3: Some coalitions become insignificant:}
Next, we consider the case when some coalitions become insignificant on addition of a coalition/player.  If some coalitions were insignificant  in $\P$ and if $w$ is added after those coalitions, then sum utilities are the same,  with $w$ also remaining insignificant in $\P'$ (see definition of $M^\P$ in Theorem \ref{Thm_util_coal_general}).  We are left to consider when $w$ is added in between significant coalitions in $\P$, or just after the last significant coalition.  

Thus  without loss of generality  assume $S_k$ is significant under $\P$. 
If  $w \geq w_k$, and if   $w$ becomes insignificant, then clearly  the  sum   utilities  under   both the partitions ($\P$ and $\P'$) remain the same.

 Remains to consider the case when $w_{j} \leq w \leq w_{j+1}$ for some $j < k$. 
Since    $w_1,\cdots,w_{j+1}$ were significant  in $\P$, 
  $w$ remains significant after addition  (as $w \le w_{j+1}$).  
If $S_{j+1}$ becomes insignificant (for this $w < w_{j+1}$)  after addition of $w$, then $U^*_{SO} $ under $\P$ is more than that under $\P'$  by using monotonicity of Case 1 ($w$ replacing $w_{j+1}$).
If other wise, 
  $U^*_{SO} $ under $\P$ is  still more than or equal to that under $\P'$ using the same steps as in Case 2 (progressively till the coalition that becomes insignificant after adding $w$).

 Thus, ALC is the worst NE-partition. Thus the lemma. \eop


\textbf{Proof of Theorem \ref{Thm_GC_CS}: }
We first  prove  that GC is stable against coalitional deviations. 
Observe that all players obtain non-zero utility under GC.  Also by Theorem \ref{Thm_general}    GC is stable under unilateral deviation, i.e., a deviation of  singleton coalition, say $C' =\{j\}$, is not  beneficial to the player  $j$.

(i)  Consider a coalition $C'$ such that $1,2 \notin C'$, which attempts to deviate from GC. Observe that all players obtain non-zero utility under GC. Then from Lemma \ref{Lem_SV_general} part (ii) we know that only the active player of $C'$, i.e., the player with the maximum influence factor in $C'$ obtains non-zero utility and thus, such a coalitional deviation is not beneficial to all the players in $C'$.

(ii) Consider a coalition $C'$ that attempts to deviate from GC,  such that exactly one of $1,2 \in C'$.  Then from Lemma \ref{Lem_SV_general} part (i) we know that only players $1$ and $2$ obtains strictly positive utility and thus, such a coalitional deviation is not beneficial to all the players in $C'$.

(iii)  Consider a coalition $C'$ such that $1,2 \in C'$, deviates from GC.   Towards this, we compare the utility of player $1$ between two partitions $\P_1$ and $\P_2$ such that $\{1,2,\cdots,l\} \subseteq S_1 \in \P_1 \text{ and } l+1 \in S_2 \in \P_1$ and $\{1,2,\cdots,l+1\} \subseteq S'_1 \in \P_2 \text{ and } l+2 \in S'_2 \in \P_2$. Then from Lemma \ref{Lem_SV_general} part (ii) the difference in SV shares of player $1$ in the two partition equals,
\begin{eqnarray}
\phi_1^{\P_2} - \phi_1^{\P_1} & = &  \frac{1}{l(l+1)} \left( \varrho_{l+1}^2-(1-\varrho_2)^2 \right) + \frac{1}{l+1}  \left( \varrho_{l+2}^2-(1-\varrho_2)^2 \right) - \frac{1}{l}  \left( \varrho_{l+1}^2-(1-\varrho_2)^2 \right), \nonumber \\
& = & \left[ \frac{1}{l(l+1)} \left( \varrho_{l+1}^2-(1-\varrho_2)^2 \right)  - \frac{1}{l}  \left( \varrho_{l+1}^2-(1-\varrho_2)^2 \right) \right] + \frac{1}{l+1}  \left( \varrho_{l+2}^2-(1-\varrho_2)^2 \right), \nonumber \\
& = & - \frac{1}{l+1}\left( \varrho_{l+1}^2-(1-\varrho_2)^2 \right) + \frac{1}{l+1}  \left( \varrho_{l+2}^2-(1-\varrho_2)^2 \right), \nonumber \\
& = & \frac{1}{l+1} \left( \varrho_{l+2}^2- \varrho_{l+1}^2 \right) > 0 \text{ since } \varrho_{l+2}^2 > \varrho_1^2. \nonumber 
\end{eqnarray} 
Proceeding in a similar fashion (by induction), one can show that player $1$ obtains strictly better utilities in GC and thus, GC is stable against any coalitional deviations.

It remains to prove that none of the coalitions other than GC are stable under   additional   assumption {\bf A.2}.  
Consider any partition $\P_1$ with $S_1$ containing first $l$ players and with $(l+1) \in S_2$ for some $l < n$. Once again let $\P_2$ be the partition after coalitional deviation by  coalition $C' = \{1, 2, \cdots, l+1\}$.  We will show that each and every player in $C'$ derives better after deviation, i.e., under $\P_2$. Basically $\P_2$ has first $(l+1)$ players together in $S_1'$, i.e., $(l+1)$ is added to $S_1'$ and the rest of the players are in the same coalitions as in the original partition $\P_1.$
 
 As in case (iii) of previous step, one can show that Shapley value of every player (except for  $(l+1)$) in $C'$ under $\P_2$ is strictly better than that  under  $\P_1$. For last player $(l+1)$ by  {\bf A}.2 for $j=l+1$
 and {\bf A}.1 
 and proceeding as in proof of Theorem \ref{Thm_general} the player derives better after deviation. 
 
 Observe that the  above estimates are independent of the configuration of the rest of the players $j \ge (l+2)$,  because these players derive 0 utility irrespective of their placement,  from Lemma \ref{Lem_SV_general}. 
 
 Next  consider the  partition in which player $1$ and $2$ are in different coalitions.  Now using  {\bf A}.2 with  $j=2$, the  SVs  of players 1 and 2 in GC is bigger than that  in original partition and thus the partition is blocked by $C' = GC$.  
\eop

  \textbf{Proof of Theorem \ref{Thm_CS_None}:} From Corollary \ref{corollary_ALC_NE}, we know that all partitions other than ALC are not stable against unilateral deviations. Thus, it suffices to show that ALC is not stable against coalitional deviations. 
Towards this, consider the players deviate together to form GC. Then, the share of each player equals $1/n$ (from \eqref{Eqn_USm_player}). From Theorem \ref{thm:Thm1}, the utility of each player in ALC equals $1/n^2$ which is smaller than the utility derived under GC.

Hence, ALC is also not C-stable. \eop

\section{Appendix E}
\label{sec_appendixE_PEVA}

We only give the computations for $n=3$ with adamant player as one can derive similarly for others.
		Consider the case when player 2  and 3 chooses $GC = \{1,2,3\}$. Player 1 could choose strategy GC, $\{1,2\}$ (or equivalently $\{1,3\}$) or $\{1\}$. From
		\eqref{Eqn_USm_player} the utility of player 1  if  it
		chooses  $GC = \{1,2, 3\}$ (partition GC is formed) 
		is 
		\begin{eqnarray}
		\label{util_gc_3}
		\frac{1}{3}\left(\frac{\lambda}{\lambda+\lambda_0}\right)^2 =   \frac{1}{3}\left(\frac{1}{1+\eta}\right)^2.
		\end{eqnarray}
		From \eqref{Eqn_USm_player} and \eqref{min_util_mult_partition}, if player
		chooses  $ \{1,2\}$ (multiple partitions are formed), it obtains inferior utility than when it chooses $\{1\}$ ($\mathcal{P}_2$ type partition is formed). The utility of player $1$ when it chooses the latter strategy equals, 
		\begin{eqnarray}
		\label{util_alone_3}
		\left(\frac{\lambda}{\lambda+2\lambda_0}\right)^2 =   \left(\frac{1}{1+2\eta}\right)^2.
		\end{eqnarray}
		Thus, comparing \eqref{util_gc_3} and \eqref{util_alone_3}, $\{1\}$ lies in best response of player 1 against $\{1,2,3\}$ strategy of player 2 and 3, if $\eta \leq 2.732$ else, GC is a NE-partition (also when $\eta = 2.732$).
One can check that adversary is significant in all possible partitions in this range. ${\mathcal P}_2$ can also be a NE-partition in this range for some strategy profiles. For example, $x_1 = \{1\}$, $x_2= \{2,3\}$ and $x_3= \{2,3\}$.
		
		When player 1 and 3 chooses $\{1\}$ and $\{1,2,3\}$ respectively, player 2 could either form coalition with player 3 or remain alone. From \eqref{Eqn_USm_player} the utility of player 2  if  it chooses  $\{1,2,3\}$ (or equivalently $\{2,3\}$) is given by,
		\begin{eqnarray}
		\label{util_mult_3}
		\frac{1}{2}\left(\frac{\lambda}{\lambda+2\lambda_0}\right)^2 =   \frac{1}{2}\left(\frac{1}{1+2\eta}\right)^2.
		\end{eqnarray}
		Similarly, the utility of player 2  if  it chooses  $\{2\}$ (or equivalently $\{1,2\}$) is
		\begin{eqnarray}
		\label{util_alc_3}
		\left(\frac{\lambda}{\lambda+3\lambda_0}\right)^2 =   \left(\frac{1}{1+3\eta}\right)^2.
		\end{eqnarray}
		Comparing \eqref{util_mult_3} and \eqref{util_alc_3}, ALC is formed ,i.e., $\{2\}$ lies in best response of player 2 against $\{1\}$ and $\{1,2,3\}$ strategy of player 1 and 3, if $\eta \leq 2.414$, until the adversary becomes insignificant , i.e., $0.57 < \eta  \leq 2.414$ (from \eqref{Eqn_USm}).
Thus, $\mathcal{P}_2$ type partitions are NE-partitions when $2.414 < \eta  \leq 2.732$. Adversary is significant for all possible partitions in this range as well.
		
		Similarly, when $0.5 < \eta  \leq0.57$ adversary becomes insignificant in ALC partition. Hence, we need to check player 2's best response. One can easily observe that $\mathcal{P}_2$ and ALC$^o$ are the NE-partitions in this range.
		
		When $\eta \leq 0.5$, we have GC as the only partition where adversary is significant.
		Now consider best response of player 1 against strategy $GC = \{1,2,3\}$ of player 2 and 3.
		
		Utility of player 1 when he chooses strategy $GC = \{1,2,3\}$ is given by \eqref{util_gc_3}. From \eqref{min_util_mult_partition} utility of player 1 if it chooses $\{1,2\}$ (or equivalently $\{1,3\}$) is:
		\begin{eqnarray}
		\frac{1}{2}\frac{1}{2^2} &=& \frac{1}{8}.
		\label{util_p2_3}
		\end{eqnarray}
		Similarly, utility of player 1 if it chooses $\{1\}$ (or equivalently $\{1,3\}$) is:
		\begin{eqnarray}
		\label{util_p3_3}
		\frac{1}{2^2} &=& \frac{1}{4}.
		\end{eqnarray}
		From \eqref{util_p2_3} and \eqref{util_p3_3}, player 1 gets strictly better utilities when he chooses $\{1\}$. Thus, comparing \eqref{util_gc_3} and \eqref{util_p3_3}, GC is formed ,i.e., $\{1,2,3\}$ lies in best response of player 1 against $\{1,2,3\}$ strategy of player 2 and 3, if $\eta \le 0.15$. Hence, GC and $\mathcal{P}_2^o$  is the NE-partition when $\eta \le 0.15$ and $0.15 < \eta \leq 0.5$ respectively. The SO-partitions can be calculated from Lemma \ref{Lemma_SO_Partition} as before.
	\chapter{Proofs Related to Chapter~\ref{chap_OR}}
\label{chap_appendixII}

\section{Characteristic Form games }
\label{appendix_A}

A game in characteristic form~\cite{aumann1961} can be defined using  the tuple,~$(\mathcal{N},  \nu,\mathcal{H})$, 
where: 
(a)~$\mathcal{N}$ denotes the set of~$n$ agents; 
(b)~$\nu$ is called a characteristic function and  for any~$C \subseteq \mathcal{N}$,~$\nu(C)$  denotes the set  of all possible
 payoff vectors of dimension~$n$ that agents in~$C$ can jointly achieve;  
 and
(c)~$\mathcal{H}$ is the set of all possible payoff vectors of dimension~$n$ (such vectors are also referred to as allocation vectors in literature), which are achievable. We say~$(\mathcal{N},  \nu,\mathcal{H})$ is an \textit{ordinary game} (see~\cite{aumann1961}) if:~$\x \in \nu(\mathcal{N})$ if and only if there is a~$\textbf{y} \in \mathcal{H}$ such that~$x_i \leq y_i  \text{ for all }  i.$ 

In this appendix, we provide the details of how our problem can be recast as a characteristic game.  
Let~${\cal F} (\P) $ be the set of all feasible payoff vectors
under partition~$\P$,  these are the vectors that satisfy the following:     the sum of payoffs of all  agents in any coalition~$S$ is less than or equal to that obtained by~$S$ under partition~$\mathcal{P}$ at WE,~$\lambda_S^\mathcal{P}$. Hence  
\begin{equation}
{\cal F} (\P) := \left \{\x = [\xs_i] : \sum_{i \in S} \xs_i \leq \lambda_S^\mathcal{P} \text{ for all }  S \in \P   \right \}.
\end{equation}
Thus $\mathcal{H}$, the set of all achievable/feasible payoff vectors is~$
\mathcal{H} = \cup_{\mathcal{P}} {\cal F} (\P) .
$
Observe that for grand coalition,~$\mathcal{F}(\mathcal{N}) = \mathcal{H}$ and hence is convex. We are now left to define the characteristic function~$\nu$.

{\bf  Characteristic function using pessimal rule:}
The  characteristic function precisely describes the set of all possible divisions of the anticipated worth  of any coalition. One can define such a function for partition form games using an appropriate anticipation rule~\cite{pessimistic}. There are many known anticipatory rules to define characteristic function, also described in Section~\ref{sec:classical}. 

According to the most widely used \textit{pessimistic anticipation rule}~\cite{pessimistic}, the agents in deviating coalition~$C$ assume that the outside agents arrange themselves to hurt the agents in~$C$ the most. 
Further, the minimum utility that coalition~$C$ can achieve irrespective of the arrangement of the agents outside this coalition is given  by~$
{\underline \nu}_C   := \min_{ \P : C \in \P } \max_{\x \in {\cal F} (\P)}  \sum_{i \in C}  \xs_i  
$
(observe in our case,~${\underline \nu}_C = \ulam_C$).
Thus, the characteristic function~$\{\nu(C); \text{ for all } C\}$ under pessimal rule is given by the following: for any coalition~$C$,~$
\nu(C) = \left \{\x : \sum_{i \in C}  \x_i \leq \underline{\nu}_C \right \},
$ is  the set of possible payoff vectors that agents in~$C$ can jointly achieve independent of the arrangement of outside agents.
From the above definition, it is clear that our game is an ordinary game. 

\textbf{Stability:} To study the stability aspects, one needs to understand if a certain coalition can `block' any payoff vector.
Blocking by a coalition implies that coalition is working as an independent unit and has an anticipation of the value it can achieve (e.g., irrespective of arrangements of others under pessimal rule). 
If the division of this anticipated value among the members of the coalition, under any given allocation rule,  renders the members to achieve more than that in the current payoff vector then the coalition has tendency to oppose the current arrangement or the payoff vector. 

\textit{Blocking:} 
A payoff vector~$\x \in \mathcal{H}$ is blocked by a coalition~$C$ if there exist a payoff vector~$\textbf{y} \in \nu(C)$ such that~$
y_i > \xs_i \text{ for all } i \in C.
$

With these definitions in place, we now give define a related solution concept called R-core, which is an extension of the classical definition of \textit{core}, for transferable utility games (in non-partition form games). 

\textbf{R-core~\cite[Section 3]{aumann1961}:} 
We define R-core~$\mathscr{C}(\mathcal{H})$ to be the set of vectors in~$\mathcal{H}$ which cannot be blocked by any other member of~$\mathcal{H}$. 

The authors in~\cite{hafalir} studied the properties of this core under the name \textit{c-core} (which is also popular by the name $\alpha$-core in literature). In~\cite[Corollary 2]{hafalir}, they showed that a convex partition form game necessarily has a non-empty core. However, one can easily check that our game is not convex as in~\cite{hafalir} and hence, it is not clear if core is non-empty or not. In fact, in Theorem~\ref{thm_impossible}, we showed that the R-core is empty for our game. We hence introduce more generalised and relevant notions of stability in this paper.

\section{Proof of Theorem~\ref{Thm_WE}}
\label{appendix_B}

\textbf{ Proof of  Existence and Uniqueness:}
Let the size of a partition be denoted by~$p$.
The first step of this proof is to show the existence and uniqueness of WE for the case when~$p=2$.  In the next step,  using induction we prove the existence for any general~$p=m>2$ using the corresponding results for~$m-1$. In the third step we show the continuity of the WE, to be precise the arrival rates at WE for~$m$. The last step attributes to the uniqueness of our solution.

\textit{Step 1: Existence and Uniqueness of WE for~$p=2$} 

To obtain WE, the following equation needs to be solved:~$
B_{C_1}^\P(N_{C_1},a_{C_1}^\P) = B_{C_2}^\P(N_{C_2},a_{C_2}^\P).
$
Define a function~$f:= B_{C_1}^\P(N_{C_1},a_{C_1}^\P) - B_{C_2}^\P(N_{C_2},a_{C_2}^\P)$. Then,~$f$ is a function of~$\lambda_{C_1}^\mathcal{P} \in [0,\Lambda]$ since~$\lambda_{C_2}^\mathcal{P}= \Lambda-\lambda_{C_1}^\mathcal{P}$. 
\begin{itemize}
	\item 
	At~$\lambda_{C_1}^\mathcal{P}=0$ we have~$B_{C_1}^\P(N_{C_1},a_{C_1}^\P)=0$ and~$B_{C_2}^\P(N_{C_2},a_{C_2}^\P) >0$, thus~$f(0) <0$.
	
	\item At~$\lambda_{C_1}^\mathcal{P}=\Lambda$ we have~$B_{C_1}^\P(N_{C_1},a_{C_1}^\P)>0$ and~$B_{C_2}^\P(N_{C_2},a_{C_2}^\P) =0$, thus~$f(\Lambda) >0$. 
\end{itemize}
Then,~$B_{C_1}^\P(N_{C_1},a_{C_1}^\P)$ and~$B_{C_2}^\P(N_{C_2},a_{C_2}^\P)$ are polynomial functions with denominator~$> 1$ and hence are continuous functions. This implies that~$f$ is a continuous function.

Thus,~$f$ satisfies the hypothesis of Intermediate Value Theorem (IVT). Using IVT, there exists a value of~$\lambda_{C_1}^\mathcal{P} = \lambda^* \in (0,\Lambda)$ such that~$f(\lambda^*)=0$.
The uniqueness of~$\lambda^*$ follows since~$B_{C_1}^\P(N_{C_1},a_{C_1}^\P)$ and~$B_{C_2}^\P(N_{C_2},a_{C_2}^\P)$ are strict increasing functions of~$\lambda_{C_1}^\mathcal{P}$ and~$\lambda_{C_2}^\mathcal{P}$ respectively. 

 \textit{Step 2: Existence for general~$p = m > 2$}
 
 To prove the existence for any general~$m>2$, we assume that a unique WE exists for~$p=m-1$, i.e.,~$\lambda_{C_1}^{\P}, \cdots, \lambda_{C_{m-1}}^{\P}$ with corresponding common blocking probability~$B^*$.
 
 With~$m$ units we can initially fix~$\lambda_{C_{m}}^{\P} = 0$ and obtain WE corresponding to the remaining units, which we have assumed to exist. With increase in~$\lambda_{C_m}^{\P}$, $\Lambda - \lambda_{C_m}^{\P}$ which is the total share of remaining agents, decreases. From part $(i)$ of this theorem  applied to the case with~$m-1$, we know that the corresponding WE solution for these agents also decreases. This implies that the common blocking probability for~$C_1, \cdots, C_{m-1}$ reduces while blocking probability of~$C_m$ increases (see~\eqref{Eqn_PB}). Using similar arguments as above and treating~$C_1, \cdots, C_{m-1}$ as one while defining function for IVT (continuity is obtained from Step $3$, with~$m-1$), one can show that WE exists.

\textit{Step 3: Continuity of Optimisers, i.e., WE:} 
Consider the following function~$g$ for~$m$ coalitions in partition~$\P$:~$
g(\Lambda,{\bm \lambda}) := \sum_{C_j \in \mathcal{P}; 1<j\leq m} (B_{C_1}^\P-B_{C_j}^\P)^2,
$
where~${\bm \lambda}$ is the vector of arrival rates for all~$C_j \in \P$. Then, we define~$g^*(\Lambda,{\bm \lambda}^*) = \min_{\{{\bm \lambda}: \sum_j \lambda_j = \Lambda \} }g(\Lambda,{\bm \lambda}) $.
Observe that the (unique) minimizer~${\bm \lambda}^*$ of the function~$g$ is the (unique) WE for our queueing model, and that the function~$g$ is jointly continuous. Thus, using Maximum Theorem we have that~$g^*$ and~${\bm \lambda}^*$ is continuous in~$\Lambda$.

\textit{Step 4: Uniqueness of WE }
To prove the uniqueness of the WE, we assume the contradiction,  i.e., say~$(\lambda_1,\cdots,\lambda_m)$ and~$(\lambda_1',\cdots,\lambda_m')$ are two distinct WEs. One can have the following cases: 

 \textit{Case 1: There exist multiple WEs with same common blocking probability~$B^*$} 
  This implies that some of the units in partition are obtaining different arrival rates in the multiple WEs such that they have common~$B^*$, i.e., say $\lambda_i' \ne \lambda_i$. However, this is not possible since blocking probability is a strictly increasing function of arrival rate. 
 
 \textit{Case 2: There exist multiple WEs with different common blocking probability~$B^*$ and~${\hat B}^*$} 
 
 Without loss of generality, we can assume that~$B^* < \hat{B}^*$. This implies that the arrival rates to the units with common blocking probability~$\hat{B}^*$ is more (since blocking probability is an increasing function of arrival rate). However, the total arrival rate is fixed at~$\Lambda$ which implies that one of the WE does not satisfy~$\sum_{C_j \in \P} \lambda_{C_j}^{\P} = \Lambda$.

\textbf{Proof of  All units used}  
For contradiction, let us assume that the customers split themselves amongst some strict subset of units of partition~$\P$. Then, each unit with zero arrivals have a zero blocking probability while units with non-zero arrivals have some strict positive blocking probability. However, this contradicts the fact that the coalitions having zero arrivals should have a higher blocking probability than others at WE.

Hence at WE, each of the units in partition~$\P$ obtain non-zero arrival rates.

\textbf{Proof of part (i)}  Let~$\lambda_{C_1}^\P, \cdots, \lambda_{C_k}^\P$ be the individual arrival rates corresponding to partition~$\P$ at WE  (satisfies~\eqref{Eqn_WE_properties}) for the coalitions~$C_1, \cdots, C_k$ respectively with the total arrival rate~$\Lambda>0$. Let the   corresponding  common  blocking probability be~$B^*$. When the total arrival rate is increased to~$\Lambda'$, the individual arrival rates to the providers at WE are changed to~$\lambda_{C_1}^{'\P},\cdots, \lambda_{C_k}^{'\P}$ and the   corresponding  common  blocking probability is changed to ~$\hat{B}^*$. Note that these splits to the individual operating units must satisfy:
\begin{equation}
\sum_{i=1}^k \lambda_{C_i}^{'\P} = \Lambda' > \Lambda =  \sum_{i=1}^k \lambda_{C_i}^{\P}  \text{ and }  \text{ for any partition } \P.
\label{total_conserved}
\end{equation} 

\noindent
Next we will show that~$\lambda_{C_j}^{'\P} \leq \lambda_{C_j}^{\P}$ is not possible for any~$C_j \in \P$. Using~\eqref{total_conserved}, we know that at least one of the units have higher individual arrival rates at new WE, i.e,~$
\lambda_{C_j}^{'\P} > \lambda_{C_j}^{\P} \text{ for at least one } C_j \in \P.
$
This means that the common blocking probability at new WE is increased, i.e.,~$\hat{B}^* > B^*$. Now since blocking probability is a strictly increasing function of arrival rates, we have that arrival rate to each coalition is increased at new WE for~$\Lambda'$, i.e.,~$\lambda_{C_j}^{'\P} > \lambda_{C_j}^{\P}$ for all~$C_j \in \P$. 

Hence, WE is an increasing function of  $\Lambda$.

\textbf{Proof of part (ii)} Let~$\lambda_{C_1}^\P, \cdots, \lambda_{C_k}^\P$ be the individual arrival rates corresponding to partition~$\P$ at WE for the coalitions~$C_1, \cdots, C_k$ respectively. 
 Let the   corresponding  common  blocking probability be~$B^*$. Observe that the blocking probability of~$C_i$ and~$C_j$ units also equals~$B^*$, and hence the merger~$M = C_i \cup C_j \neq \mathcal{N}$ has strictly smaller  blocking probability, i.e.,~$B_M < B^*$,  if the joint arrival rate was~$\lambda_{C_i}^\P + \lambda_{C_j}^\P$. 
From~\eqref{Eqn_PB} the  blocking probability is a strictly increasing function of arrival rate. Thus the new WE after merger is formed with a (strict) bigger arrival rate to the merger, as again at the new WE the new blocking probabilities of all coalitions~$C \in {\P}'$ should be equal by~\eqref{Eqn_WE_properties}.

\textbf{Proof of  part  (iii)} 
 Consider a system with identical servers. We know that when any number of identical servers combine with their arrival rates, the combined blocking probability reduces. This reduction is more when the number of servers combining are more, i.e.,
\begin{equation}
B(N,a) > B(LN,La) > B(MN,Ma).
\label{Eqn_common_result}
\end{equation}
where~$0<L< M$ are constants,~$B$ is the blocking probability,~$N$ is the number of servers and~$a$ is the offered load.
Now if we consider that the coalition with~$N_{C_1}$ and~$N_{C_2}$ servers gets exactly~$N_{C_1}/N$ and~$N_{C_2}/N$ share of total arrival rate~$\Lambda$ at WE respectively.
Using~\eqref{Eqn_common_result}, we have that coalition with~$N_{C_1}$ servers has strictly smaller blocking probability. From~\eqref{Eqn_WE_properties}, the blocking probability of each unit at WE is same. So, the arrival rate to coalition with~$N_{C_1}$ and~$N_{C_2}$ servers need to be increased and reduced respectively to achieve the WE. 

Hence, coalition with~$N_{C_1}$ and~$N_{C_2}$ servers satisfy~$ 
\frac{\lambda_{C_1}^{\P}}{N_{C_1}} > \frac{\Lambda}{N} > \frac{\lambda_{C_2}^{\P}}{N_{C_2}}.
$

\section{Rest of the proofs}
\label{appendix_C}

\textbf{Proof of Theorem~\ref{thm_impossible}:}
 Consider any  configuration, say~$(\P,\Phi)$. 
From~\eqref{eq:blocking_PA}, the configuration is stable if and only if
\begin{equation}
\label{Eqn_reqd_conditions}
\sum_{i \in C} \phi_i \ge \underline{\lambda}_C \text{ for all } C \notin \P \text{ and } C \subset \mathcal{N}. 
\end{equation} 

\textit{Case 1: All players are alone in~$\P$}

In such a case, for some player~$j$, consider the merger coalition~$M = \mathcal{N}-\{j\}$. Then from Theorem~\ref{Thm_WE}.$(ii)$,
\begin{equation*}
 \ulam_M  > \sum_{C_l \in M} \lambda_{C_l}^ \mathcal{P} = \sum_{i\ne j} \phi_i,
\end{equation*}
which implies that~$M$ blocks the prevalent configuration under the GB-PA rule.

\textit{Case 2: There exists at least one coalition~$C \in \P$ such that~$|C| \ge 2$}

This implies that~$S_a = \mathcal{N}-\{a\} \notin \P$ for all~$a \in C$. We will show that some~$a \in C,$ either~$S_a$ or~$\{a\}$ will block the prevailing configuration. 

\noindent
\textit{Case 2(a): The configuration is blocked by~$S_a$ for some~$a \in C.$} In this case, the instability of the coalition follows immediately.

\noindent
\textit{Case 2(b): The configuration is not blocked by~$S_a$ for any~$a \in C.$} In this case,
$$
\sum_{i \in S_a} \phi_i \ge \underline{\lambda}_{S_a} = \Lambda - \ulam_{\{a\}}
$$ for all~$a \in C.$ This is equivalent to the statement~$\phi_a \leq \ulam_{\{a\}}$ for all~$a \in C.$
However, there exists a ${\hat a} \in C$ such that~$\phi_{\hat a} < \ulam_{\{{\hat a}\}}$ since~$\sum_{q \in C} \phi_i = \sum_{q \in C} \ulam_{\{q\}} \le \sum_{q \in C} \lambda_{\{q\}}^{\P'}  < \lambda_C^\P$. Thus, the configuration $(\P, \Phi)$  is blocked by~$\{{\hat a}\}.$  
\eop


\textbf{Proof of Theorem~\ref{Thm_duo_mono}:}
Consider a partition~$\mathcal{P} = \{C_1,C_2,\cdots, C_k\}$ with cardinality greater than~$2$. Let~$M$ be the merger coalition containing all coalitions of~$\P$ except one, i.e.,~$
M = \cup_{i=2}^k C_i \text{ and } \P' = \{C_1,M\}.
$ Then from Theorem~\ref{Thm_WE}.$(ii)$,~$
 \lambda_{M}^{\mathcal{P}'} = \ulam_M^{\P'}  > \sum_{C_i \in M} \lambda_{C_i}^ \mathcal{P},
$
which is same as the condition required for blocking by mergers under RB-IA rule. 

 Hence, there exists a configuration/payoff vector such that each of the members in $M$ obtain strictly better and thus, such a partition is not stable.
 \eop

\textbf{Proof of Theorem~\ref{Thm_GC}:} There can be no merger from~$\P_G$,  and we only need to check if an appropriate split can block a configuration~$(\P_G, \Phi)$, under consideration.

\begin{enumerate}[label=(\roman*)]
    \item \textit{When~$N_1 < \sum_{i \in \mathcal{N}; i \neq 1}N_i$}

\begin{enumerate}[label=(\alph*)]
\item We first consider payoff vectors~$\Phi$ that satisfy 
\begin{equation}
   \sum_{i=2}^n \phi_i < \left (   1-\frac{N_1}{N} \right ) \Lambda.
   \label{Eqn_payoff1}
\end{equation} 
Let~$S :=\{2, 3, \cdots, n\}$  be the coalition  made of all agents except agent $1$. We will prove that this  coalition will block the configuration of the form stated above.

 Since coalition~$S$ has more than $N/2$ servers, it must satisfy  the following (from Theorem~\ref{Thm_WE}.$(iii)$):~$
\lambda_S^{\P'} =  \underline{\lambda}_S > \Lambda \left (  1-\frac{ N_1}{N} \right )    ,  \mbox{ where }   {\P'}  := \{S, \{1\} \}, 
$
which is same as~\eqref{Eqn_condition_S}.
Further, from~\eqref{Eqn_payoff1},~$
\lambda_S^{\P'} > \Lambda \left (  1-\frac{ N_1}{N} \right ) > \sum_{i=2}^n \phi_i, 
$
which implies that~\eqref{Eqn_condition_S_pt2} is also satisfied by coalition~$S$.

 Hence,~$(\P_G,\Phi)$ is blocked by coalition~$S$.

\item Next, we consider payoff vectors that satisfy \begin{eqnarray}
\sum_{i=2}^n \phi_i \geq  \left (  1-\frac{N_1}{N}  \right ) \Lambda. \label{Eqn_condition1_new}
\end{eqnarray}

Suppose, for the sake of obtaining a contradiction, that,~$\{\P_G,\Phi\}$ is stable. Since~$N_1$ is the agent with  maximum number of servers,~$S_k := S \backslash \{k\} \cup \{1\}$ has~$N_{S_k}  > N/2$ for any~$k \ge 2$. By Theorem~\ref{Thm_WE}.$(iii)$ such coalitions satisfy~\eqref{Eqn_condition_S}. Thus, the stability of~$\{\P_G,\Phi\}$ implies that~\eqref{Eqn_condition_S_pt2} must be violated for the same coalitions. That is, we have~$
\sum_{i \in S_k} \phi_i \ge \ulam_{S_k}  >  \frac{\sum_{i \in S_k} N_i}{N}\Lambda,  \mbox{ for each }  k > 1, 
$ in view of Theorem~\ref{Thm_WE}.$(iii)$.
By adding all the above inequalities with~$k=2, \cdots, n$, we have:
\begin{eqnarray*}
(n-1)\phi_1+(n-2) \sum_{i=2}^n \phi_i   >   \frac{(n-1)N_1+(n-2)\sum_{i=2}^n N_i}{N} \Lambda,
\end{eqnarray*}
which implies,
 \begin{eqnarray*}
 \phi_1 + (n-2)\Lambda > \Bigg(\frac{N_1+(n-2)N }{N} \Bigg)\Lambda = \frac{N_1}{N}\Lambda + (n-2)\Lambda, 
\end{eqnarray*}
since~$\sum_{i=1}^n \phi_i = \Lambda,   \sum_{i=1}^n N_i = N.$
Thus we have,~$
\phi_1    >  \frac{N_1}{N}\Lambda $ which contradicts~\eqref{Eqn_condition1_new}. Thus,~$(\P_G,\Phi)$ is unstable under RB-IA rule.
\end{enumerate}

\item \textit{When~$N_1 \ge \sum_{i \in \mathcal{N}; i \neq 1}N_i$}

In this case, the coalitions that satisfy condition~\eqref{Eqn_condition_S} for blocking under RB-IA are exactly those coalitions that contain player~$1$ (from Theorem \ref{Thm_WE}.$(iii)$). However, for any such coalition, the condition~\eqref{Eqn_condition_S_pt2} for blocking under RB-IA gets violated so long as~$
\phi_1 \ge \max_{C} \ulam_C \text{ for all } C \subset \mathcal{N} \text{ containing agent 1.}
$
Thus, any allocation~$\Phi$ satisfying the above bound on~$\phi_1$ is guaranteed to be stable under RB-IA. \eop
\end{enumerate}

\textbf{Proof of Theorem \ref{Thm_two_partition}:}
Part~$(i)$ follows from part~$(ii)$, proved below,  as~$k^*$ exists.

$(ii)$ Any~$2$-partition~$\P = \{C_1,C_2\}$ cannot be blocked by mergers since merger lead to~$\P_G$ and~\eqref{Eqn_condition_M} is not satisfied. Next we look at splits. Say~$C_1 \in \mathbb{C}^*$. Then it follows from the definition of~$\mathbb{C}^*$ that there exists no coalition~$C \subset C_1$ such that it satisfies~\eqref{Eqn_condition_S}. Further, coalition~$C_2$ cannot do better by splitting.
Hence, any partition with one of the coalitions belonging to~$\mathbb{C}^*$ is a stable partition under RB-IA rule.

$(iii)$ Once again, it is easy to verify that a merger cannot block any~$2$-partition~$\P.$ Next, we check for splits.
    Any split leads to a coalition with a number of servers less than~$N/2$, and hence from Theorem~\ref{Thm_WE}.$(iii)$,~\eqref{Eqn_condition_S} is not satisfied and hence, no split is feasible. Thus,~$\P$ is stable under RB-IA rule. 
    \eop

\textbf{Proof of Theorem \ref{Thm_R_rule_stability}: } 
$(i)$ Consider any configuration~$(\P_G, \Phi )$  with GC.
The proof of this part can be split into two cases:

\textit{Case 1: When~$N_1 < \sum_{i \in S; i \neq 1}N_i \text{ for some } S \subset \mathcal{N}$}

Under RB-PA rule for the configuration to be stable, we need to ensure that the  following system of equations are satisfied simultaneously.
\begin{eqnarray}
\sum_{i \in C} \phi_i \geq \ulam_C \text{ for all } C \subset \mathcal{N} \text{ and, }
\sum_{i \in \mathcal{N}} \phi_i  = \Lambda.  \label{Eqn_conditions}
\end{eqnarray} 
However, a subset of these equations itself admit no feasible solution (as proved in Theorem~\ref{Thm_GC}).   Thus, such a system of equations does not have a solution and  hence~$(\P_G, \Phi )$ is unstable for any payoff vector $\Phi$.

\textit{Case 2: When~$N_1 \ge \sum_{i \in \mathcal{N}; i \neq 1}N_i$}

Once again we need to satisfy~\eqref{Eqn_conditions} to prove that~$(\P_G, \Phi)$ is stable. In particular those equations will also have to be satisfied  for subsets~$S$ such that~$|S| = n-1$.
If there exists a payoff vector~$\Phi$  that satisfies all such conditions,  consider one such~$S$ and say~$j \notin S$.  Then  from~\eqref{Eqn_conditions},~$\phi_j = \Lambda - \sum_{i \in S} \phi_i  \leq  \Lambda -\ulam_S 
= \ulam_{\{j\} }.$
If~$\phi_j < \ulam_{\{j\}}$ for some~$j$ then configuration~$(\P_G, \Phi)$  is blocked by~$\{j\}$ under RB-PA rule. Otherwise if~$\phi_j = \ulam_{\{j\}}$ for all~$j \in \mathcal{N}$ then~$\sum_{i \in \mathcal{N}} \phi_i = \sum_{i \in \mathcal{N}} \ulam_{\{j\}} < \Lambda$ and thus~\eqref{Eqn_conditions} is not satisfied.
Hence~$(\P_G, \Phi)$ is unstable for any payoff~$\Phi$. 

$(ii)$ Since the condition required for a merger to be successful under RB-PA rule is same as under RB-IA rule, the result follows from Theorem~\ref{Thm_duo_mono}.

$(iii)$ 
%
When the  payoff vector is given by equation~\eqref{Eqn_PSA}, the RB-PA and RB-IA rules 
are equivalent to each other.  
Thus, the result follows from Theorem~\ref{Thm_two_partition}.


 Moreover because of the continuity of~$\Phi$, we have the next result. 
  \eop
	
 \textbf{Proof of Theorem \ref{Thm_SV_small}: }Consider any~$2$-partition~$\P = \{C_1,C_2\}$.
	 
$(i)$ W.l.o.g., say coalition~$C_1 = \{i,j\} $.
	 From~\eqref{Eqn_SV} and~\eqref{eq_subcoalition_worth}, the share of player~$i$ is given by:
	 \begin{eqnarray}
 \phi_i  & = & \frac{1}{2} \left( \lambda_{C_1}^\P - \lambda_{\{j\}}^{\P'} \right) + \frac{1}{2} \lambda_{\{i\}}^{\P'} \ > \lambda_{\{i\}}^{\P'} > \ulam_{\{i\}}, \text{ where } \P' = \{\{i\},\{j\},C_2\}.\nonumber 
  \end{eqnarray}
   The first inequality holds since~$\lambda_{C_1}^\P > \lambda_{\{i\}}^{\P'}+\lambda_{\{j\}}^{\P'},$ and the second follows from Theorem~\ref{Thm_WE}.$(ii)$. Thus, a split of~$C_1$ does not block the configuration~$(\P, \Phi_s^\P)$. Further, a merger cannot block the configuration due to the constant sum nature of the game.
  
An identical argument also applies for part $(ii)$. 
	 \eop

\textbf{Proof of Theorem \ref{Thm_heavy}: }Consider any~$2$-partition~$\P = \{C_1,C_2\}$ with~$k := N_{C_1} \ge N_{C_2}$ It is easy to see that the~$2$-partition cannot be blocked by a merger under RB-IA/RB-PA rules. It therefore suffices to check for stability against splits.
\begin{enumerate}[label=(\roman*)]
    \item Be relaxing $k$ to be a real number, we show that $\Psi$ is increasing for all $k$ in Lemma~\ref{lem_psi_inc}. Using Lemma~\ref{lem_psi_inc},  no split satisfies~\eqref{Eqn_condition_S} and hence~$\P$ is stable.
    \item Under the proportional payoff vector~$\phi_p^\P$,~\eqref{eq_def_RBPA} is equivalent to~\eqref{Eqn_condition_S}, and hence the result follows.
\end{enumerate}
Below, we prove Lemma~\ref{lem_psi_inc}.
\eop

\begin{lemma}
\label{lem_psi_inc}
Consider any $\epsilon>0$.  Then there exists a $\bar{\Lambda}$ such that for all $\Lambda \ge {\bar \Lambda},$ $\Psi:=\lambda_1/k$ is strictly increasing in $k$ over $N/2 \le k \le N-\epsilon$. 
\end{lemma}
\textbf{Proof:}
To prove this result, we work with the analytical extension of the Erlang-B formula (see~\cite{jagerman}) so that~$k$ may be treated as a real number. Under this extension, it is easy to see that the Wardrop splits are uniquely defined for real-valued service capacities. For any 2-partition, differentiating~$\Psi$ with respect to~$k$, we have
$
    \frac{d \Psi}{d k}  =  \frac{d}{dk} \left(\frac{\lambda_1}{k} \right) \ = \ \frac{1}{k} \left(\frac{d \lambda_1}{d k} - \frac{\lambda_1}{ k} \right). \nonumber
$
Thus, to prove the theorem, it suffices to show that given~$\epsilon > 0$, there exists a~$\bar{\Lambda}$ such that for any~$\Lambda \geq \bar{\Lambda}$,
\begin{equation}
    \frac{d \lambda_1}{d k} - \frac{\lambda_1}{ k} > 0 \text{ for all } k \in \left[\frac{N}{2}+\epsilon,N-\epsilon \right]. \nonumber
\end{equation}
Towards this, we know that the arrival rates at WE~($\lambda_1$) are obtained by equating the blocking probabilities of the two coalitions. The reciprocal of the blocking probability of a coalition with~$k$ servers and offered load~$a$ admits the following integral representation (see~\cite{jagerman}):
\begin{equation}
    R(k,a) = a\int_0^\infty  h(t; a, k) dt
    \mbox{ where }
   h(t; a, k) =     (1+t)^k e^{-a t} . \nonumber
\end{equation}
Thus, the WE satisfies~$
    R(k,\lambda_1) - R(N-k,\Lambda-\lambda_1) = 0, \nonumber 
$
    which is equivalent to
   \begin{eqnarray}
    \lambda_1 \int_0^\infty h(t;\lambda_1,k) dt - (\Lambda-\lambda_1)\int_0^\infty h(t;\Lambda-\lambda_1,N-k) dt = 0.  
    \label{eq_integral_pb}
\end{eqnarray}
 Differentiating both sides of the above  with respect to~$k$ using Lemma~\ref{lem_interchange}, we have
 \begin{eqnarray}
    \frac{d \lambda_1}{dk} \left[\int_0^\infty(1+t)^k e^{-\lambda_1 t} dt \right] + \lambda_1 \left[ \int_0^\infty(1+t)^k ln(1+t)e^{-\lambda_1 t} dt - \int_0^\infty(1+t)^k e^{-\lambda_1 t} t \left(\frac{d \lambda_1}{dk} \right) dt\right]  \hspace{-7mm} & \nonumber \\
& \hspace{-140mm} + \frac{d \lambda_1}{dk} \left[\int_0^\infty(1+t)^{N-k} e^{-(\Lambda-\lambda_1) t} dt \right] - (\Lambda-\lambda_1) \left[-\int_0^\infty(1+t)^{N-k} ln(1+t)e^{-(\Lambda-\lambda_1) t} dt \right. \nonumber \\
& \hspace{-200mm} \left. + \int_0^\infty(1+t)^{N-k} e^{-(\Lambda-\lambda_1) t} t \left(\frac{d \lambda_1}{dk} \right) dt \right] = 0. \nonumber 
\label{Eqn_Integral}
\end{eqnarray}
Rearranging the above terms we obtain,

\vspace{-8mm}
{\small \begin{eqnarray}
    \frac{d \lambda_1}{d k}  =  \frac{-\lambda_1 \int_0^\infty h(t;\lambda_1,k) \ln(1+t)  dt - (\Lambda-\lambda_1 )\int_0^\infty h(t;\Lambda-\lambda_1,N-k) \ln(1+t) dt}{ \int_0^\infty h(t;\lambda_1,k) dt + \int_0^\infty h(t;\Lambda-\lambda_1,N-k) dt-\lambda_1 \int_0^\infty h(t;\lambda_1,k)t dt - (\Lambda-\lambda_1 )\int_0^\infty h(t;\Lambda-\lambda_1,N-k)t dt}.\nonumber \label{Eqn_derivlam1}
\end{eqnarray}}
Observe that each of the integrals in the above expression is of the form~$\int_0^\infty f(t;k)e^{-\lambda_1t} dt $ or $\int_0^\infty f(t;k)e^{-(\Lambda-\lambda_1)t} dt.$
In heavy traffic, since~$\lambda_1$  and~$\Lambda-\lambda_1$ tend to infinity (see Lemma~\ref{lem_accuracy_heavy} below) the value of these integrals is dominated by the behavior of the integrand around zero. Accordingly, one can approximate these integrals using a Taylor expansion of~$f(t)$ around~$t=0.$ Formally, using Lemma~\ref{lem_error_taylor} below (it is easy to show that all the integrals above satisfy the hypotheses of Lemma~\ref{lem_error_taylor}), we have~$\frac{d \lambda_1}{dk} =- \frac{T_1}{T_2}, \text{ where }$ 

\vspace{-9mm}
{\small \begin{align*}
T_1 &= \lambda_1\left(\frac{1}{\lambda_1^2}+\frac{2k-1}{\lambda_1^3}+\frac{3k^2-6k+2}{\lambda_1^4}\right)+(\Lambda-\lambda_1)\left(\frac{1}{(\Lambda-\lambda_1)^2}+\frac{2(N-k)-1}{(\Lambda-\lambda_1)^3}+\frac{3(N-k)^2-6(N-k)+2}{(\Lambda-\lambda_1)^4}\right)+o\left(\frac{1}{\Lambda^2}\right),\\
T_2 &= \frac{1}{\lambda_1}+\frac{k}{\lambda_1^2}+\frac{k(k-1)}{\lambda_1^3}+\frac{k(k-1)(k-2)}{\lambda_1^4}+ \frac{1}{\Lambda-\lambda_1}+\frac{N-k}{(\Lambda-\lambda_1)^2}+\frac{(N-k)(N-k-1)}{(\Lambda-\lambda_1)^3} \nonumber \\
& +\frac{(N-k)(N-k-1)(N-k-2)}{(\Lambda-\lambda_1)^4} -\lambda_1 \left(\frac{1}{\lambda_1^2}+\frac{2k}{\lambda_1^3}+\frac{3k(k-1)}{\lambda_1^4}+\frac{4k(k-1)(k-2)}{\lambda_1^4}\right)    - (\Lambda-\lambda_1)\left(\frac{1}{(\Lambda-\lambda_1)^2} \right. \nonumber \\
& \left. +\frac{2(N-k)}{(\Lambda-\lambda_1)^3}+\frac{3(N-k)(N-k-1)}{(\Lambda-\lambda_1)^4}+ \frac{4(N-k)(N-k-1)(N-k-2)}{(\Lambda-\lambda_1)^4}\right) +o\left(\frac{1}{\Lambda^3}\right). \nonumber 
\end{align*}}
Simplifying the above expression we get,
    \begin{eqnarray}
   \frac{d \lambda_1}{dk} =  \frac{-\left(\frac{1}{\lambda_1}+\frac{2k-1}{\lambda_1^2}+\frac{3k^2-6k+2}{\lambda_1^3}+\frac{1}{\Lambda-\lambda_1}+\frac{2(N-k)-1}{(\Lambda-\lambda_1)^2}+\frac{3(N-k)^2-6(N-k)+2}{(\Lambda-\lambda_1)^3}+o\left(\frac{1}{\Lambda^2}\right)\right)}{-\frac{k}{\lambda_1^2}-\frac{2k(k-1)}{\lambda_1^3}-\frac{3k(k-1)(k-2)}{\lambda_1^4}-\frac{N-k}{(\Lambda-\lambda_1)^2}-\frac{2(N-k)(N-k-1)}{(\Lambda-\lambda_1)^3}-\frac{3(N-k)(N-k-1)(N-k-2)}{(\Lambda-\lambda_1)^4}+o\left(\frac{1}{\Lambda^3}\right)}. \nonumber
\end{eqnarray}
Subtracting~$\lambda_1/k$ and by some simplification~$ \frac{d \lambda_1}{dk} - \frac{\lambda_1}{k}$ equals, 
 \begin{align*}
= & \frac{ \frac{k}{\Lambda-\lambda_1}-\frac{(N-k)\lambda_1}{(\Lambda-\lambda_1)^2}+T_3+o\left( \frac{1}{\Lambda^2} \right)
    }{k\left(\frac{k}{\lambda_1^2}+\frac{2k(k-1)}{\lambda_1^3}+\frac{3k(k-1)(k-2)}{\lambda_1^4}+\frac{N-k}{(\Lambda-\lambda_1)^2}+\frac{2(N-k)(N-k-1)}{(\Lambda-\lambda_1)^3}+\frac{3(N-k)(N-k-1)(N-k-2)}{(\Lambda-\lambda_1)^4}+o\left(\frac{1}{\Lambda^3}\right)\right)} \nonumber \\
    = & \frac{\frac{k(N-k)}{(\Lambda-\lambda_1)^2} \left(\frac{\Lambda-\lambda_1}{N-k}-\frac{\lambda_1}{k} \right) +T_3+o\left( \frac{1}{\Lambda^2} \right)
    }{k\left(\frac{k}{\lambda_1^2}+\frac{2k(k-1)}{\lambda_1^3}+\frac{3k(k-1)(k-2)}{\lambda_1^4}+\frac{N-k}{(\Lambda-\lambda_1)^2}+\frac{2(N-k)(N-k-1)}{(\Lambda-\lambda_1)^3}+\frac{3(N-k)(N-k-1)(N-k-2)}{(\Lambda-\lambda_1)^4}+o\left(\frac{1}{\Lambda^3}\right)\right)}. 
    \end{align*}
    where~$T_3 := \frac{k}{\lambda_1^2}+\frac{k\left[2(N-k)-1\right]}{(\Lambda-\lambda_1)^2}-\frac{2(N-k)(N-k-1)\lambda_1}{(\Lambda-\lambda_1)^3}+\frac{3k^2-4k}{\lambda_1^3}+\frac{k(3(N-k)^2-6(N-k)+2}{(\Lambda-\lambda_1)^3}-\frac{3(N-k)(N-k-1)(N-k-2)\lambda_1}{(\Lambda-\lambda_1)^4}.$
From Lemma~\ref{lem_accuracy_heavy}, it follows that~$\frac{\lambda_1}{\Lambda} = \frac{k}{N} + o(1)$ and~$\frac{\Lambda-\lambda_1}{\Lambda} = \frac{N-k}{N} + o(1)$ as~$\Lambda \to \infty,$ with the~$o(1)$ terms being uniform over~$k \in \left[N/2 ,N-\epsilon\right].$ Additionally, from Lemma~\ref{lem_accuracy_heavy_tight},~$\left(\frac{\lambda_1}{k} - \frac{\Lambda-\lambda_1}{N-k}\right) = \left(\frac{1}{N-k} - \frac{1}{k}\right) + o(1),$ with the~$o(1)$ term again being uniform over~$k \in \left[N/2,N-\epsilon\right].$ Now, multiplying  by~$\Lambda^2$ in the numerator and denominator above and applying these results, we obtain
    \begin{eqnarray}
   \lim_{\Lambda \to \infty} \left( \frac{d \lambda_1}{dk} - \frac{\lambda_1}{k}  \right) = \frac{\frac{N^2}{k}+\frac{kN^2}{(N-k)^2}-\frac{kN^2}{N-k}\left(\frac{1}{N-k}-\frac{1}{k} \right)}{kN^2\left(\frac{1}{k}+\frac{1}{N-k}\right)} = \frac{1}{k}  >  0. \nonumber 
\end{eqnarray}
Observe that the above limit is uniform over~$k \in \left[ \frac{N}{2},N-\epsilon \right]$.
\eop

\begin{lemma}
\label{lem_accuracy_heavy}
     $\frac{\lambda_1}{\Lambda} \to \frac{k}{N} \text{ and } \frac{\Lambda-\lambda_1}{\Lambda} \to \frac{N-k}{N}$   
    $\text{ uniformly over } k \in \left[ \frac{N}{2}, N \right] \text{ as } \Lambda \to \infty.$
\end{lemma}
\textbf{Proof:} 
We know that the blocking probability of a coalition with~$s$ servers and offered load~$a$, when~$\Lambda \to \infty$ is bounded as (see~\cite{harel}):~$
 1 - \frac{1}{\rho} < B(s,a) <  \frac{\rho}{1+\rho} \text{ where } \rho = a/s > 0. \nonumber
$
Using the upper bound for the larger coalition and the lower bound for the smaller coalition, the arrival rate~$\lambda_1$ at WE can be lower bounded by~$\hat{\lambda}_1,$ which satisfies:~$
    \frac{\frac{\hat{\lambda}_1}{k}}{1+\frac{\hat{\lambda}_1}{k}} = 1 - \frac{1}{\frac{\Lambda-\hat{\lambda}_1}{N-k}} 
  \implies  \hat{\lambda}_1 = k\left(\frac{\Lambda-N+k}{N}\right). $
Next, using the upper bound for the smaller coalition and the lower bound for the larger coalition, we obtain an upper bound~$\tilde{\lambda}_1$ of~$\lambda_1$ as follows:~$
     \nicefrac{\frac{\Lambda-\tilde{\lambda}_1}{N-k}}{\left (1+\frac{\Lambda-\tilde{\lambda}_1}{N-k} \right )} = 1 - \nicefrac{1}{\left(\nicefrac{\tilde{\lambda}_1}{k}\right)} 
  \implies   \tilde{\lambda}_1 = k \left(\frac{\Lambda+N-k}{N}\right). $
  
From the above, we obtain the following bounds on $\lambda_1$,~$
   k\left(\frac{\Lambda-N+k}{N}\right) \le \lambda_1 \le k\left(\frac{\Lambda+N-k}{N}\right). 
$
  It now follows that ~$
   \left | \frac{\lambda_1}{\Lambda} - \frac{k}{N} \right|  \stackrel{(a)}{\le} k\left(\frac{N-k}{N\Lambda} \right) \ \stackrel{(b)}{\le} \ \frac{N}{\Lambda}.
$
(the above inequalities lead to inequality $(a)$, while the  bound $(b)$ is obvious), which implies that~$
\lim_{\Lambda \to \infty} \left | \frac{\lambda_1}{\Lambda} - \frac{k}{N} \right| = 0 \text{ uniformly over } k \in \left[ \frac{N}{2},N \right].  
$
This implies the result.  
\eop

\begin{lemma}
\label{lem_error_taylor}
 Suppose~$f$ is~$m$-times differentiable on~$[0,\infty)$, such that~$f^{(m)}(t;k)$ is non-negative, monotonically increasing, and~$
    f^{(m)}(t;k) \le c_1 +c_2t^N  \text{ for all } t \ge 0$ and $k \in \left[ \frac{N}{2}, N - \epsilon \right]$,
 for some positive scalars~$c_1 \text{ and } c_2$. Further, $\nicefrac{\Lambda}{2} \le \lambda_1(k) \le \Lambda$ for all $k$.
Then
 \begin{eqnarray}
    \int_0^\infty f(t;k)e^{-\lambda_1(k) t} dt = \int_0^\infty \sum_{j=0}^{m-1} \left( f^{(j)}(0;k)\frac{t^j}{j!}\right)   e^{-\lambda_1(k) t} dt + o\left(\frac{1}{\Lambda^{m-1}}\right), \text{ with } f^{(0)}(\cdot;k) = f(\cdot;k), \nonumber 
\end{eqnarray}
 as~$\Lambda \to \infty.$ Here, the~$o\left(\frac{1}{\Lambda^{m-1}}\right) $  error is uniform over~$k \in \left[ \frac{N}{2}, N-\epsilon \right]$ for~$\epsilon >0$.
\end{lemma}

\textbf{Proof:}
Using the Taylor expansion of~$f(t;k)$ (for any~$t$) around~$0$~\cite{rudin}, we have
 \begin{eqnarray}
    \int_0^\infty f(t;k)e^{-\lambda_1(k)t} dt = \int_0^\infty \sum_{j=0}^{m-1} \left( f^{(j)}(0;k)\frac{t^j}{j!}\right)e^{-\lambda_1(k) t} dt + \int_0^\infty   \frac{f^{(m)}(c(t);k) t^m}{m!}   e^{-\lambda_1(k) t} dt. \nonumber
\end{eqnarray}  for some~$c(t)$ strictly between~$0$ and~$t$. Observe that the residue term above can be upper bounded  as~$
    \int_0^\infty \frac{f^{(m)}(c(t);k) t^m}{m!} e^{-\lambda_1(k) t} dt \le \int_0^\infty \frac{f^{(m)}(t;k) t^m}{m!} e^{-\lambda_1(k)t } dt,
$
 since the~$m^{th}$ derivative of~$f(\cdot)$ is strictly monotonically increasing in~$t$ and as~$c(t) \le t$. Under the hypothesis of this lemma, we have an upper bound independent of $k \in \left[ \frac{N}{2}, N- \epsilon \right]$, which further can be upper bounded:
 \begin{eqnarray}
     \int_0^\infty \frac{f^{(m)}(t;k) t^m}{m!} e^{-\frac{\Lambda}{2} t} dt  \le  \int_0^\infty \frac{\left(c_1+c_2t^N\right) t^m}{m!} e^{-\frac{\Lambda}{2} t}  dt \stackrel{(a)}{=}   \ o\left( \frac{1}{\Lambda^{m-1}} \right), \nonumber
 \end{eqnarray}
 where equality~$(a)$ follows from simple calculations (involving the gamma function).
\eop

 \begin{lemma}
\label{lem_accuracy_heavy_tight}
 For any $\epsilon>0$,~~$
     \frac{\lambda_1}{k}-\frac{\Lambda-\lambda_1}{N-k} \to \left(\frac{1}{N-k}-\frac{1}{k} \right) 
    \text{  uniformly over } k \in \left[ \frac{N}{2}, N-\epsilon \right] \text{ {\it as }} \Lambda \to \infty. $
\end{lemma}
\textbf{Proof:}
Observe that~\eqref{eq_integral_pb} coincides with the WE equation for  integral values of~$k$ and~$N-k$. Relaxing~$k$ to be a real-valued number such that~$k \in \left[ \frac{N}{2},N-\epsilon\right]$, observe that each integral is of the form~$
\int_0^\infty f(t;k)e^{-\lambda_1t} dt\text{ or } \int_0^\infty f(t;k)e^{-(\Lambda-\lambda_1)t} dt.
$
In heavy traffic, since~$\lambda_1$ and~$\Lambda-\lambda_1$ tend to infinity (see Lemma~\ref{lem_accuracy_heavy}) the value of these integrals is dominated by the behavior of the integrand around zero. Accordingly, one can approximate these integrals using a Taylor expansion of~$f(t;k)$ around~$t=0.$ Formally, using Lemma~\ref{lem_error_taylor} (it is easy to show that all the above integrals satisfy the hypotheses of Lemma~\ref{lem_error_taylor}) and solving the non-negligible integrals, equation~\eqref{eq_integral_pb} can be re-written as~$
 1+ \frac{k}{\lambda_1}+\frac{k(k-1)}{\lambda_1^2} = 1+ \frac{N-k}{\Lambda-\lambda_1}+\frac{(N-k)(N-k-1)}{(\Lambda-\lambda_1)^2} + o\left(\frac{1}{\Lambda^2}\right), 
$ with~$o\left( \frac{1}{\Lambda^2} \right)$ being uniform over $k \in \left[ \frac{N}{2}, N-\epsilon \right]$.
 Simplifying the above using Lemma~\ref{lem_accuracy_heavy} (e.g., $o\left( \nicefrac{1}{\Lambda-\lambda_1}\right) = o\left( \nicefrac{1}{\Lambda}\right)$), and using $\Lambda o\left( \nicefrac{1}{\Lambda^2} \right) = o(\nicefrac{1}{\Lambda})$,  
 \begin{eqnarray}
\frac{k}{\lambda_1}\left( 1 + \frac{k-1}{\lambda_1} \right) & = & \frac{N-k}{\Lambda-\lambda_1}\left(1+\frac{N-k-1}{\Lambda-\lambda_1} + o\left( \frac{1}{\Lambda} \right)  \right) \implies
\frac{\lambda_1}{k} =
\frac{\Lambda-\lambda_1}{N-k} \left[\frac{\left( 1 + \frac{k-1}{\lambda_1} \right)}{\left(1+\frac{N-k-1}{\Lambda-\lambda_1} \right) + o\left( \frac{1}{\Lambda} \right)} \right].\nonumber
\end{eqnarray}
Subtracting~$\frac{\Lambda-\lambda_1}{N-k}$ from both sides of the above equation, we have
\begin{eqnarray}
\frac{\lambda_1}{k}-\frac{\Lambda-\lambda_1}{N-k} 
= \frac{\Lambda-\lambda_1}{N-k}\left[\frac{\left(  \frac{k-1}{\lambda_1} \right)-\left(\frac{N-k-1}{\Lambda-\lambda_1}\right)-o\left(\frac{1}{\Lambda}\right)}{\left(1+\frac{N-k-1}{\Lambda-\lambda_1} \right) + o\left( \frac{1}{\Lambda} \right)}\right]. \nonumber
\end{eqnarray}
Note that as~$\Lambda \to \infty$, the denominator of the above expression goes to~$1$. Further, multiplying and dividing by~$\Lambda$ and using Lemma~\ref{lem_accuracy_heavy}, we have (observe all errors converge uniformly in~$k$)
 \begin{eqnarray}
\lim_{\Lambda \to \infty} \left(\frac{\lambda_1}{k}-\frac{\Lambda-\lambda_1}{N-k} \right) = \frac{1}{N} \left[\left( \frac{k-1}{k}\right) N - \left( \frac{N-k-1}{N-k}\right)N\right] = \frac{1}{N-k} - \frac{1}{k}.  \nonumber \mbox{\eop}
\end{eqnarray}

\begin{lemma}
While differentiating \eqref{eq_integral_pb}, the limits (derivative) and the integral can be interchanged.
\label{lem_interchange}
\end{lemma}
\textbf{Proof:}
Since the blocking probability of any coalition increases with increase in arrival rate, the derivative of the left hand side of~\eqref{eq_integral_pb} with respect to~$\lambda_1(k)$ is not zero. Thus using Implicit Function Theorem, we obtain~$\lambda_1(k)$ to be a continuously differentiable function of~$k$ and hence,~$T_4 := \sup_{ k' \in [k, k+{\bar h}]} \frac{d \lambda_1(k')}{dk}$ is  finite for some~${\bar h}>0$.

It is sufficient to consider the limit of the form~$
    \lim_{h \to 0} \int_0^\infty \left( \frac{(1+t)^{k+h}e^{-\lambda_1(k+h) t}-(1+t)^ke^{-\lambda_1(k) t}}{h} \right)  dt.
$
By differentiability for all~$t$,

\vspace{-8mm}
{\small $$
 y_h(t) := \left( \frac{(1+t)^{k+h}e^{-\lambda_1(k+h) t}-(1+t)^ke^{-\lambda_1(k) t}}{h} \right)  \to \bigg((1+t)^k\ln(1+t) - (1+t)^kt(d\lambda_1(k)/dk)\bigg)e^{-\lambda_1(k) t}.
$$}
Consider any~$h \in (0, {\bar h} ]$. By Mean Value Theorem, there exists a~$k' \in (k,k+h)$ such that

\vspace{-9mm}
 {\small \begin{eqnarray}
 y_h(t) \le \left((1+t)^{k'}\ln(1+t)-(1+t)^{k'}t\left(\frac{d \lambda_1(k')}{dk}\right)\right)e^{-\lambda_1(k')t} \le \left((1+t)^{k+{\bar h}}\ln(1+t)+(1+t)^{k+{\bar h}}t T_4\right)e^{-\lambda_1(k)t}. \nonumber
\end{eqnarray}}
The upper bound is integrable and hence the result follows by Lebesgue's Dominated Convergence Theorem. 
 \eop

 \textbf{Proof of Theorem \ref{low_traffic}:} 
 $(i)$ Consider any~$2$-partition~$\P = \{C_1,C_2\} \in \mathscr{P}$ and~$\Phi \in {\bm \Phi}^\P$. From Theorem~\ref{Thm_two_partition}.$(iii)$,~$\P$ is stable under RB-IA rule.
  
Now, consider a~$2$-partition~$\P = \{C_1,C_2\} \notin \mathscr{P}$. This implies there exists a~$C \subset C_1$ such that~$N_{C_1} > N_{C} > N/2.$ We will show that coalition~$C$ blocks the configuration~$(\P,\Phi^\P_p)$. 
 From Lemma~\ref{lem_light_new}, we have~$
	\nicefrac{\lambda_{C_1}^\P}{\Lambda N_{C_1}} \to \nicefrac{1}{N_{C_1}} \text{ and } \nicefrac{\underline{\lambda}_C}{(\Lambda N_{C})} \to \nicefrac{1}{N_{C}} > \nicefrac{1}{N_{C_1}}  \mbox{ as } \Lambda \to 0.
$ Thus there exists a~$\underline{\Lambda}>0$ such that for any~$\Lambda \le \underline{\Lambda}$,~$
\nicefrac{\underline{\lambda}_C}{N_C} > \nicefrac{\lambda_{C_1}^\P}{N_{C_1}}.
 $
 It now follows that coalition~$C$ satisfies condition~\eqref{Eqn_condition_S} for blocking.  Moreover, under the proportional payoff vector~$\Phi_p^\P$,~\eqref{Eqn_condition_S} implies~\eqref{Eqn_condition_S_pt2}. This means that~$C$ blocks the configuration~$(\P,\Phi^\P_p)$, which in turn implies that~$\P$ is not a stable partition under RB-IA rule.
   
    $(ii)$ Under proportional payoff vector~$\Phi_p^\P$,~\eqref{eq_def_RBPA} is equivalent to~\eqref{Eqn_condition_S}, and hence the result under RB-PA follows along similar lines.
\eop

\begin{lemma}
\label{lem_light_new}
Consider a coalition $C$ such that $N_C>N/2$, then~$
\frac{\underline{\lambda}_{C}}{\Lambda}
\to 1 \text{ as } \Lambda \to 0.
$
Consequently, for any~$2$-partition~$\P = \{C_1,C_2\}$ where~$N_{C_1} > N_{C_2}$,~$
\frac{\lambda_{C_1}^\P}{\Lambda} = \frac{\underline{\lambda}_{C_1}}{\Lambda}
\to 1 \text{ as } \Lambda \to 0.
$
\end{lemma}

\textbf{Proof:}
Let~$\lambda_1 =  \underline{\lambda}_{C}.$  It is sufficient to show that~$
\frac{\lambda_1}{\Lambda-\lambda_1} \to \infty \text{ as } \Lambda \to 0.
$
In light traffic, the reciprocal of the blocking probabilities of the two coalitions satisfy
$$
R(k,\lambda_1) \sim \frac{k!}{\lambda_1^k} \text{ and } R(N-k,\Lambda-\lambda_1) \sim \frac{(N-k)!}{(\Lambda-\lambda_1)^{N-k}},
$$
where $f(\Lambda)\sim g(\Lambda)$ means $\lim_{\Lambda \to 0} \frac{f(\Lambda)}{g(\Lambda)} = 1.$ We therefore obtain,
\begin{eqnarray}
    \frac{k!}{\lambda_1^k}  \sim  \frac{(N-k)!}{(\Lambda-\lambda_1)^{N-k}} 
   \Rightarrow \left(\frac{\lambda_1}{\Lambda-\lambda_1}\right)^{N-k} \lambda_1^{2k-N}  \sim \frac{k!}{(N-k)!}. \nonumber
\end{eqnarray}
With~$\Lambda \to 0$,~$\lambda_1^{2k-N} \to 0$ and R.H.S. is a finite constant, this implies,~$
\lim_{\Lambda \to 0} \left(\frac{\lambda_1}{\Lambda-\lambda_1}\right) = \infty.
$
Now observe that for~$2$-partition~$\P = \{C_1,C_2\}$ with~$N_{C_1} > N_{C_2}$, we have~$N_{C_1} > N/2$ and hence the result follows.
\eop

\textbf{Proof of Theorem~\ref{dynamic_conv}:} We first show that
 starting from any $k$-partition $\P$ with $k > 2$, the dynamics hits a 2-partition with probability one: 
  i) from any such $\P$, there exists at least one direct path to  a 2-partition with probability strictly greater than zero, as given in the proof of Theorem \ref{Thm_duo_mono}; ii) thus  there exists a non-zero uniform lower bound  ${\underline p} >0$ on  the probability of hitting a 2-partition,  irrespective of  the starting $k$-partition,  because of finitely many such  partitions; and  iii) thus by independence,  the dynamics hits a 2-partition with probability one in finite number of steps (uniformly upper bounded by a geometric random variable with parameter  ${\underline p}$). 
  
  Similarly, starting from   the grand coalition, the system either evolves to a 2-partition or stops. 

If the dynamics hits one of the stable  partitions (among $2$-partitions), we are done. If not, by {\bf A.}1,  the 2-partition (say $\P = \{C, {\cal N}   \backslash  C\}$)  is such that (without loss of generality)   $N_C> k^*$ and $C$ contains a $C^* \in {\cal C}^*$.   The movement from $\P$ to $\P_1  := \{C^*,  C  \backslash  C^*,    {\cal N}  \backslash  C\}$ is possible by \eqref{Eqn_condition_S}  because clearly  by definition of $k^*$ and $C^*$
$$
\frac{ \ulam_{C^*} }{k^*}  >   \frac{ \ulam_C}{N_C}  =  \frac{ \lambda_C^\P } {N_C} \mbox{ which implies }    \ulam_{C^*} >   \lambda_C^\P  \frac{k^*} {N_C}.
$$
From $\P_1$ merger of $ C  \backslash  C^*$ and   $  {\cal N}  \backslash  C$ to $\P_2  := \{C^*, {\cal N} \backslash C^* \}$  is possible  by \eqref{Eqn_condition_M}, as clearly 
$$
\ulam_{ {\cal N}  \backslash  C^*}  >  \lambda^{\P_1}_{C  \backslash  C^*} +   \lambda^{\P_1}_{{\cal N}  \backslash C^*},
$$as in the proof of Theorem \ref{Thm_duo_mono}. The succession of these two events occur with  probability that can be lower bounded by a strictly positive number ${\underline p}'$, uniformly across all such starting 2-partitions.  As in the previous paragraph, any upward movement will return to a 2-partition with probability one and in each of these returns there is  uniform lower bound ${\underline p}'$ on the probability of return to the stable 2-partition with a $C^*$. Hence  the theorem.  \eop


 \chapter{Proofs Related to Chapter~\ref{chap_wireless}}

\label{chap_appendixIII}

\section{Appendix A}
\label{sec_AppendixA_CN}

We derive the expression for various gradients, and provide their simplifications, in  this section.
We begin with the stationary distribution, $\bmu$.

 Using \eqref{eqn_stationary_orig}, one can work with a smaller vector $\tmu$, which is a $(\ns-1)$-dimensional row vector. We consider it to be defined over a reduced state space $\Sc'$, more precisely   the components of $\bmu$ corresponding to the first $(\ns-1)$ states as below,
\begin{eqnarray}
    \tmu := \left[\mu_{\bd}(1), \ \cdots, \ \mu_{\bd}(\ns-1) \right]. \nonumber
\end{eqnarray}
 One can re-write the fixed point equation in  \eqref{eqn_stationary_orig} in terms of $\tmu$ (as given below) which can then be calculated by solving $\tmu = {\bf b}_\bd^T(I-\Pb_\bd^{-1})$.
 
\begin{equation}
\label{eq_reduced_sd}
\begin{aligned}
 \tmu = \tmu \tP + {\bf b}^T_\bd  \text{ where }   {\bf b}_\bd = \begin{bmatrix}
\Pb_\bd(\ns,1) \\
\vdots \\
\Pb_\bd(\ns,\ns-1)
\end{bmatrix}\text{ and } \\
\tP(\rS,\rXn) = \Pb_\bd(\rS,\rXn)-\Pb_\bd(\ns,\rXn) \text{ for all } \rS,\rXn \in \Sc'. 
\end{aligned} 
\end{equation}

Then $\bmu$ is as given below,
\begin{equation}
\label{eq_bmu_from_tmu}
    \bmu = \left[\mu_{\bd}(1), \ \cdots, \ \mu_{\bd}(\ns-1), \ 1-\sum_{j=1}^{\ns-1} \mu_{\bd}(j) \right]. 
\end{equation}
Using \eqref{Eqn_optim},\eqref{eqn_rd}, and \eqref{eq_bmu_from_tmu} one can estimate the average utilities accumulated by each user, $\rbU_\bd$. One can again re-write \eqref{Eqn_optim} in terms of $\tmu$ as below,
\begin{equation}
 \rbU_{\bd} = \tmu\tr^T + {\bf z}_\bd \text{ with } 
 {\tilde r}^{(n)}_{\bd}(\rS) := r^{(n)}_{\bd}(\rS) - r^{(n)}_{\bd}(\ns)
\text{ and }
z_\bd(n) := r^{(n)}_{\bd}(\ns) \text{ for all } \rS \in \mathcal{S}', 
\label{eq_reduced_ubar}   
\end{equation}
where $\tr$ is the $N \times \Sc'$-dimensional matrix and ${\bf z}_\bd$ is a $N$-dimensional column vector.

Our aim here is to optimal policy $\bd$ which optimises the $\alpha$-fair function of average utilities in \eqref{Eqn_alpha_fair}. Towards this, \eqref{Eqn_d_term_c} defines $\bd$ in terms of $\bc$ and hence, $\bc$ can be obtained by using the following update equation,
\begin{equation}
    \bc \leftarrow \bc + \left(\nabla_{\rbU_\bd}^{\Gamma_\alpha} \right)\left(\nabla^{\rbU_\bd}_{\bc}\right)
\end{equation}
where $\nabla_{\rbU_\bd}^{\Gamma_\alpha}$ represents the derivative of $\Gamma_\alpha$ with respect to $\rbU_\bd$ while the remaining notations are as in Table \ref{tab_nabla_notations}. The expression for $\nabla_{\rbU_\bd}^{\Gamma_\alpha}$ is provided in \eqref{eqn_c_update}. Thus, we are now left to estimate $\nabla^{\rbU_\bd}_{\bc}$ as in \eqref{eq_grad_u_wrt_c}.

Differentiating \eqref{eq_reduced_ubar}  with respect to the policy $\bc$ and using simple algebra, we have
\begin{eqnarray}
\nabla^{\rbU_\bd}_{\bc} & = & \left(\nabla^{\rbU_\bd}_{\rd}\right)\left( \nabla^{\rd}_{\bc}\right) + \left(\nabla^{\rbU_\bd}_{\tmu}\right)\left(\nabla^{\tmu}_{\bc} \right), \\
& = & \bmu\left( \nabla^{\rd}_{\bc}\right) + \tr\left(\nabla^{\tmu}_{\bc} \right) \text{ where } 
\frac{\partial r^{(n)}_{\bd}(j)}{\partial \bc(k)} = \sum_a \left( r^{(n)}(j,a) \frac{\partial d(j,a)}{\partial \bc(k)} \right). \nonumber
\end{eqnarray}
Differentiating \eqref{eq_reduced_sd} with respect to $\bc$, we obtain 
\begin{eqnarray}
\label{eqn_deriv_tmu_c}
\frac{\partial \tmud(j)}{\partial \bc(k)}  & = &  \sum_{i \in \Sc'} \left[\frac{\partial {\tilde \mu}_\bd(i)}{\partial \bc(k)}\tP(i,j) + {\tilde \mu}_\bd(i)\frac{\partial \tP(i,j)}{\partial \bc(k)}\right] + \frac{\partial b_\bd^T(j)}{\partial \bc(k)}  \text{ for all } j \in \Sc', k \in \Sc, 
\end{eqnarray}
where the other partial derivatives are given by:
\begin{eqnarray}
 && \frac{\partial b_\bd^T(j)}{\partial \bc(k)} =
\sum_{a} \left[ p(j|\ns,a)\frac{\partial d(\ns,a)}{\partial {\bf c}(k)} \right] = \sum_{a} \left[ p(j|\ns,a)\frac{\partial d(\ns,a)}{\partial {\bf c}(\ns)} \right]\ind_{\{k=\ns\}}
 \text{, with } \nonumber \\
&& \frac{\partial d(i,a)}{\partial \bc(k)} 
= \left [ 
 \frac{\partial d(i,a)}{\partial  c(k,1)},      \cdots,
    \frac{\partial d(i,a)}{\partial c(k,\na)}
\right ] \nonumber \\
&& \frac{\partial \tP(i,j)}{\partial \bc(k)} = \sum_{a \in \Ac} \left[\frac{ \partial d(i,a)}{\partial \bc(k) }p(j|i,a)-\frac{\partial d(\ns,a)}{\partial \bc(k)}p(j|\ns,a) \right]  \mbox{, and, } \nonumber \\
&& \frac{\partial d(i,a)}{\partial c(k,a'')} = \frac{\partial }{\partial c(k,a'')} \left[ \frac{c(i,a)}{\sum_{a'} c(i,a')} \right] 
= \left( \frac{\sum_{a'} c(i,a')\ind_{\{a=a''\}}}{(\sum_{a'} c(i,a'))^2} -  \frac{c(i,a)}{(\sum_{a' \in \Ac} c(i,a'))^2}\right) \ind_{\{i=k\}}. \nonumber
\end{eqnarray}
Simplifying after removing zero terms (like the ones when $i\ne k$ in $\frac{\partial d(i,a)}{\partial c(k,a'')}$) we have 
$$
\frac{\partial b_\bd^T(j)}{\partial \bc(k)}  = 
 \sum_{a} \left[ p(j|\ns,a)\frac{\partial d(\ns,a)}{\partial {\bf c}(\ns)} \right]\ind_{\{k=\ns\}} = \Cv (j,L) \ind_{\{k=\ns\}} \mbox{ with }  \Cv(j,k) := \sum_{a \in \Ac} \left\{\frac{ \partial d(k,a)}{\partial \bc(k) }p(j|k,a)\right\}
$$
and then 
$ \text{ for all } j \in \Sc', k \in \Sc, $:
\begin{eqnarray*}
\frac{\partial \tmud(j)}{\partial \bc(k)}  
&   = & \sum_{i \in \Sc'} \frac{\partial {\tilde \mu}_\bd(i)}{\partial \bc(k)}\tP(i,j) + \sum_{i \in \Sc'} {\tilde \mu}_\bd(i)\sum_{a \in \Ac} \left\{\frac{ \partial d(i,a)}{\partial \bc(k) }p(j|i,a)\right\} \nonumber \\
&& - \sum_{i \in \Sc'}{\tilde \mu}_\bd(i) \sum_{a}  \frac{\partial d(\ns,a)}{\partial \bc(k)}p(j|\ns,a)+ \frac{\partial b_\bd^T(j)}{\partial \bc(k)} \nonumber  \\
&   = &  \sum_{i \in \Sc'} \frac{\partial {\tilde \mu}_\bd(i)}{\partial \bc(k)}\tP(i,j) +  {\tilde \mu}_\bd(k)\ind_{\{k<\ns\}}\sum_{a \in \Ac} \left\{\frac{ \partial d(k,a)}{\partial \bc(k) }p(j|k,a)\right\} \nonumber \\
&& - \ind_{\{k=\ns\}}\sum_{i \in \Sc'}{\tilde \mu}_\bd(i)\sum_{a}  \frac{\partial d(\ns,a)}{\partial \bc(L)}p(j|\ns,a)\ + \frac{\partial b_\bd^T(j)}{\partial \bc(k)} \nonumber  \\
&   = &  \sum_{i \in \Sc'} \frac{\partial {\tilde \mu}_\bd(i)}{\partial \bc(k)}\tP(i,j) +  {\tilde \mu}_\bd(k)\ind_{\{k<\ns\}}  \Cv (j,k)  - \ind_{\{k=\ns\}}\sum_{i \in \Sc'}{\tilde \mu}_\bd(i) \Cv (j,L)  + \frac{\partial b_\bd^T(j)}{\partial \bc(k)} \nonumber  \\
&   = &  \sum_{i \in \Sc'} \frac{\partial {\tilde \mu}_\bd(i)}{\partial \bc(k)}\tP(i,j) +  {\tilde \mu}_\bd(k)\ind_{\{k<\ns\}}  \Cv (j,k)  - \ind_{\{k=\ns\}}\sum_{i \in \Sc'}{\tilde \mu}_\bd(i) \Cv (j,L)  + \Cv (j,L) \ind_{\{k=\ns\}} \nonumber  \\
&   = &  \sum_{i \in \Sc'} \frac{\partial {\tilde \mu}_\bd(i)}{\partial \bc(k)}\tP(i,j) +  {\mu}_\bd(k) \ind_{\{k<\ns\}} \Cv (j,k) +  {\mu}_\bd(\ns) \Cv (j,L) \ind_{\{k=\ns\}} \nonumber   \\
&   = &  \sum_{i \in \Sc'} \frac{\partial {\tilde \mu}_\bd(i)}{\partial \bc(k)}\tP(i,j)  +  {\mu}_\bd(k) \Cv (j,k).  \nonumber 
\end{eqnarray*}

\section{Appendix B}
\label{sec_AppendixB_CN}

\textbf{Proof of Theorem \ref{lem_multichain}:} 
Take any SMR policy $\bd$ and initial condition $\rS_0$. Because of  finite state space, there are finitely many disjoint closed communicating classes $\{C_i\}_{i \le k}$ -- for each $i$ there exists a unique stationary distribution $\bm \mu_{\bd,C_i}$ with support over $C_i$ (see \cite{hoel}). Further for each initial condition, there exists a unique limiting  distribution (see \cite{hoel}) given by   the convex combination, 
$\sum_{i\le k}\sigma_{\bd,C_i}(\rS_0) \bm \mu_{\bd,C_i} $ where~$\sigma_{\bd,C_i}(\rS_0)$ represents the absorption probability to class~$C_i$ when Markov chain starts in $\rS_0$. Let  $\rbU_{\bm \mu_{\bd,C_i}}$ be the expected utility under distribution $\bm \mu_{\bd,C_i}$.
Then the corresponding time-average utilities in \eqref{eqn_time_avg_util} 
are given by (\cite{meyn}),
\begin{equation}
\label{eqn_util_mult}
 \rbU_{\bd}(\rS_0) = \sum_{ i \le k} \sigma_{\bd,C_i}(\rS_0)\rbU_{\bm \mu_{\bd,C_i}} \mbox{ with vector of utilities,  } \rbU_{\bd}(\rS_0) := \left [\rsbU^{(1)}_{\infty}(\rS_0), \cdots, \rsbU^{(N)}_{\infty}(\rS_0) \right ],   
\end{equation}

Now fix $\rS_0$ and
say $\bd^*$ is the optimal policy in the LHS of \eqref{eqn_sup_obj}. The result is straightforward when~$\bd^*$ is uSMR.
Say 
 there exist more than one   closed communicating classes under $\bd^*$. Let  $C^*$  be the class that maximizes the following:
 $$
 \max_{i \le k} f(\rbU_{\bm \mu_{\bd^*,C_i}}) = f(\rbU_{\bm \mu_{\bd^*,C^*}}).
 $$  
We now claim the existence of an uSMR policy $\bd'$ such that $C^*$ is the unique closed communicating class, i.e., such that  $\sigma_{\bd',C^*}(\rS'_0) = 1$ and $\bm \mu_{\bd', \rS'_0} = \bm \mu_{\bd^*,C^*}$,  for all $\rS'_0$  (also true for $\rS_0$). From convex combination in \eqref{eqn_util_mult} such an existence also establishes that  either $\bd^*$ itself is uSMR  (i.e., $k = 1$) or has $k > 1$ with  $\rbU_{\bm \mu_{\bd^*,C_i}}=\rbU_{\bm \mu_{\bd^*,C^*}}$ for each $i$. In the latter case, uSMR policy $\bd'$ achieves the supremum in RHS of \eqref{eqn_sup_obj}, once we establish the claim, which is done in the immediate following.

\medskip

\noindent
{\bf Proof of claim:}
By given hypothesis and from \eqref{Eqn_trans_prob},  $C^*=\mathcal{G}^* \times \mathcal{H}$, where $\mathcal{G}^* \subset \mathcal{G}$ denotes the set of age components of states in  $C^*$, and hence it is sufficient to consider reachability to $\mathcal{G}^*$. Again from \eqref{Eqn_trans_prob},  the transitions do not depend on data scheduler/decisions $a_c$, and  hence to construct the required $\bd'$ it is sufficient to define  age decisions, $\bd'(\rS, a_g)$ for all $a_g \in {\cal N}$ and $\rS \notin C^*$.  
In particular, define $\bd'$ as below:
\begin{equation}
   \bd'(\rS,\rA) =  \bd^*(\rS, \rA)  \mbox{ for any } \rS \in C^*\mbox{ and }  \bd'(\rS, a_g) = \frac{1}{N}   \mbox{ for any } \rS \notin C^* \mbox{ and for all }  a_g \in {\cal N}. 
   \label{eqn_d'}
\end{equation}

As new policy $\bd'$ matches with optimal $\bd^*$ on $C^*$, $C^*$ remains closed even under $\bd'$; by virtue of construction in \eqref{eqn_d'}, we will further show  that $\rS_0  \to  \rS' $  (leads to) for any  $\rS_0 \notin C^*$ and   $\rS' \in C^*$ and hence   all states in $(C^*)^c$ are transient  in the immediate  following. 

 Without loss of generality, consider   state $\rS' \in C^*$ such that  the corresponding age components satisfy, $\bar g \ge g_1' \ge g_2' \ge \cdots \ge g_{N-1}'>g_N'=1$, where   equality  $g_k'= g_l'$ (with $l \ne k$)
is possible only when $g_l'=g_k' = \bar g$. All states have such structure (w.l.g.) since only one user's age is updated at any time,  $\bar g >1$, and any user's age cannot exceed $\bar g$  (see \eqref{eqn_age_evolution}). 

To begin with, say $N=\bar g$ and say $\rS' = (\rG', \rH') \in C^*$ with $\rG' = (\bar g, \bar g -1,\cdots, 3,  2, 1) $ and say $\rS_\tau = \rS_0 $ for some $ \rS_0  \notin C^*$. 
Then the probability of reaching $C^*$  within $N$-steps,
$$
\mathbb{P}\bigg ( \S_t \in  C^* \mbox{ for some }  t \le \tau + N | \S_\tau = \rS_0  \bigg)  \ge \left ( \frac{1}{N} \right )^N (\mbox{and hence } \rS_0 \to \rS'),
$$
as i) either Markov chain gets absorbed into $C^*$ within $\tau+N$ steps, or, ii) by construction of $\bd'$ as in \eqref{eqn_d'}, the probability 
$$ \mathbb{P} ( A_{g,\tau+1}=1,A_{g,\tau+2}=2,\cdots,A_{g,\tau+N-1}=N-1,A_{g,\tau+N}=N |  \S_\tau = \rS_0, \S_{\tau+k} \notin C^* \mbox{ for all } k < N )  =  \left(\nicefrac{1}{N}\right)^N, $$  irrespective of $\rS_0 \notin C^*$ because of Markov property.

%
Now consider any $\rS' \in C^*$. As before such an $\rS'$ has the following structure of age components (after re-ordering the users required),  $\rG' = (\bar g, \bar g, \cdots, \bar g,g_{k+1}' = i_1, \cdots, g_{N-1}' = i_k, g_N' = i_{k+1})$ with $\bar g > i_1  > i_2 \cdots,  > i_{k} > i_{k+1} = 1$ and some $k \ge 0$.  Then using similar logic as before, the probability of reaching $C^*$  within $2N\bar g$-steps,
\begin{eqnarray}
\mathbb{P}\bigg ( \S_t \in  C^* \mbox{ for some }  t \le \tau + 2N\bar g | \S_\tau = \rS_0  \bigg)  & \ge & \left ( \frac{1}{N} \right )^{k \bar g}\left ( \frac{1}{N} \right )^{i_1-\sum_{j=2}^{k+1}i_j}\cdots\left ( \frac{1}{N} \right )^{i_k-i_{k+1}}\left ( \frac{1}{N} \right ) \nonumber \\
& \ge & \left ( \frac{1}{N} \right )^{(k+(N-1))\bar g} \ge \left ( \frac{1}{N} \right )^{(2N-1)\bar g} > 0. \nonumber   
\end{eqnarray}
Basically, such a path is constructed by updating the age of user $N$,  $(i_k-i_{k+1}) = i_{k}-1$ number of times at  end (of the sequence), age of user $N-1$, $(i_{k-1}-(i_k+i_{k+1}))$ number of times before updating that of user $N$, and so on till user $k+1$, and by updating the age of the remaining users at most by  $\bar g$ times.
%
%
Thus irrespective of the initial condition $\rS_0$, one of the states in $C^*$ is reached under SMR policy $\bd'$. 
This completes the proof of claim.
\eop

\noindent
\textbf{Proof of Theorem \ref{thm_existence}: }Consider a a special class of SMR policies $\bd_\del$ where channel is allocated to user $2$ with probability $\del$, irrespective of the state. 
 It is easy to observe that $\rsbU_{\bd_\del}^{(2)} - \theta\rsbU^{(1)}_{\bd_\del} > 0$   under a policy $\bd_\del$ with $\del=1$, and
 is strictly less than $0$  under another $\bd_\del$ with $\del = 0$. Further, $\rsbU_{\bd_\del}^{(2)} - \theta\rsbU^{(1)}_{\bd_\del}$ is continuous in $\del$ and hence the existence of the solution  follows from Intermediate Value Theorem.
\eop

\noindent
\textbf{Proof of Theorem \ref{thm_common_eps_del}:} For convenience, we represent realisations of channel conditions of users $1$ and $2$ by $h_1$ and $h_2$, in this proof. Say there exist two states  $\rS$ and $\rXn$ such that the corresponding $\rH$ components of the states satisfy
\begin{equation}
 \frac{\rsH_1-\rsH_2}{\theta\rsH_1+\rsH_2} > \frac{\rsH_1'-\rsH_2'}{\theta\rsH_1'+\rsH_2'}.  
 \label{eq_condition_eps}
\end{equation}
Let $\bd$ be an optimal decision and if possible say $\bd(\rS,a_c = 1) < 1 \text{ and } \bd(\rS',a_c = 1) > 0$.
Now consider a new decision rule $\bd'$ which matches with $\bd$ except for the following data decisions,
 $$
 \bd'(\rS,a_c = 1) = \bd(\rS,a_c = 1) + \Delta, \text{ and } \bd'(\rXn,a_c = 1) = \bd(\rXn,a_c = 1) - \Delta'.
 $$
From transition probability matrix given in \eqref{Eqn_trans_prob}, the data decisions do not alter stationary distribution, so we have $\bmu = {\bm \mu}_{\bd'}$, and hence the constraint in \eqref{eqn_general_theta_fair} modifies to the following under new rule $\bd'$
\begin{eqnarray}
 \rsbU^{(2)}_{\bd'}   \ = \ \rsbU^{(2)}_\bd - \Delta p(\rH)\rsH_2
+ \Delta' p(\rH')\rsH_2' \ = \ \theta\left(\rsbU^{(1)}_\bd + \Delta p(\rH)\rsH_1- \Delta' p(\rH')\rsH_1'\right) \ = \ \theta \rsbU^{(1)}_{\bd'}\nonumber
\end{eqnarray}
where $p(\rH) = {\bm \mu}_\bd(\H = \rH) =  {\bm \mu}_{\bd'}(\H =\rH)$ is the probability of channel estimates being $\rH$; the  equality ${\bm \mu}_\bd(\H = \rH) =  {\bm \mu}_{\bd'}(\H =\rH)$ again follows from \eqref{Eqn_trans_prob}.
Since $\bd$ satisfies the constraint, we have $\rsbU^{(2)}_\bd = \theta \rsbU^{(1)}_\bd $ and hence the above simplifies,
\begin{eqnarray}
     \Delta p(\rH)\left(\theta\rsH_1+\rsH_2\right) & = &  \Delta' p(\rH')\left(\theta \rsH_1'+\rsH_2'\right)  
     \label{eq_change_con}
\end{eqnarray}
The change in the objective function under the new decision rule $\bd'$ is as below,
\begin{eqnarray}
    \left({\rsbU}_{\bd'}^{(1)}+{\rsbU}_{\bd'}^{(2)} \right)  - \left({\rsbU}_{\bd}^{(1)}+{\rsbU}_{\bd}^{(2)} \right) =  \Delta p(\rH)\left( \rsH_1-\rsH_2\right)+ \Delta' p(\rH')\left( \rsH_2'-\rsH_1'\right). \nonumber
\end{eqnarray}
Using \eqref{eq_change_con} and \eqref{eq_condition_eps} specific to the two states, the above implies an improvement in the objective function under $\bd'$ as,
\begin{eqnarray}
    \frac{\Delta'p(\rH')}{\theta \rsH_1+\rsH_2} \left[\left( \rsH_1-\rsH_2\right) \left( \theta \rsH_1'+\rsH_2' \right) + \left( \theta \rsH_1+\rsH_2 \right)\left( \rsH_2'-\rsH_1'\right)\right] > 0.
\end{eqnarray}
This contradicts the optimality of $\bd$. Hence for any given $\theta$, the optimal data decisions   are ordered according to state-metric, $\nicefrac{h_1-h_2}{(\theta h_1 + h_2)}$,  and thus the optimal scheduler is as in \eqref{eqn_ep_del_h}.  Similar argument follows for $\del.$
\eop

\ignore{\medskip

\noindent
\textbf{Proof of Theorem \ref{thm_SD}:} We begin with showing the equality in \eqref{eqn_gam_last}. While solving  \eqref{eqn_stationary_orig} with transitions as in Figure \ref{fig:TD_MC}, we obtain $\mud(2_t) = \mud(2_o)$. Further the stationary distribution (pmf) satisfies the following set of equations:
\begin{align}
   & 2\mud(2_t) + \sum_{g=3}^{j} \mud(g_t) + \sum_{g=3}^{i} \mud(g_o) = 1, \nonumber \\
\mud(g_o) &= \left(\prod_{l=1}^{g-2}\gam_l\right)\mud(2_t)\text{ for any } g < i, \text{ and } \mud(i_o) = \left(\frac{\prod_{l=1}^{i-2}\gam_l}{1-\gam_{i-1}}\right)\mud(2_t), \text{ and } \nonumber \\
\mud(g_t) &= \left(\prod_{l=1}^{g-2}\bet_l\right)\mud(2_t) \text{ for any } g < j, \text{ and } \mud(j_t) = \left(\frac{\prod_{l=1}^{j-2}\bet_l}{1-\bet_{j-1}}\right)\mud(2_t). \nonumber
\end{align}
Thus we have the following,
\begin{align}
\label{eqn_sd}
    \small{\mud(2_t) = \frac{1}{ \sum_{g=2}^j  C_g + \sum_{g=2}^i D_g}, \text{ where $C_g = \prod_{l=1}^{g-2}\left[\bet_l\left(\mathds{1}_{\{g<j\}}+ \frac{1}{1-\bet_{j-1}}\ind_{\{g=j\}} \right)\right],$ and $D_g = \prod_{l=1}^{g-2}\left[\gam_l\left(\ind_{\{g<i\}}+\frac{1}{1-\gam_{i-1}}\ind_{\{g=i\}}\right)  \right].$}}
\end{align}
One can substitute the above stationary distribution into the constraint of the optimization problem  given in \eqref{eqn_general_theta_fair} and can derive an equation between two variables (say $\gamma_{i-2}$, $\gamma_{i-1}$) such that the corresponding scheduler satisfies \eqref{eqn_general_theta_fair}: this relation is defined using $\gamma_{ij}^*$ defined in theorem hypothesis as below:
\begin{align*}
    \gs := \frac{\gam_{i-2}}{1-\gam_{i-1}} = -\frac{ \sum_{g=2}^{j} \varpi_{g_t}C_g+  \sum_{g=2}^{i-1} \varpi_{g_o}D_g}{\prod_{l=1}^{i-3}\gam_l} \text{ where } \varpi_{g_t} =  \frac{\ut_{g_t}-\theta\uo_{g_t}}{\ut_{i_o}-\theta\uo_{i_o}} \text{ and } \varpi_{g_o} =  \frac{\ut_{g_o}-\theta\uo_{g_o}}{\ut_{i_o}-\theta\uo_{i_o}}.
\end{align*}
Observe that $\rsbU_{g_t}^{(n)}$ and $\rsbU_{g_t}^{(n)}$ are constants for any $g$  and $n$, given the system parameters.
Now, using \eqref{eqn_sd} and the expression of $\gs$ from above, the objective function of optimisation problem \eqref{eqn_general_theta_fair} can be written as:
\begin{align}
& = \frac{ \sum_{g=2}^{j} {\bar u}_{g_t}C_g + \sum_{g=2}^{i-1} {\bar u}_{g_o}D_g - {\bar u}_{i_o}\left(\prod_{l=1}^{i-3}\gam_l\right)\left(\frac{ \sum_{g=2}^{j} \varpi_{g_t}C_g+  \sum_{g=2}^{i-1} \varpi_{g_o}D_g}{\prod_{l=1}^{i-3}\gam_l}\right) }{\sum_{g=2}^j  C_g + \sum_{g=2}^{i-1} D_g -\left(\prod_{l=1}^{i-3}\gam_l\right)\left(\frac{ \sum_{g=2}^{j} \varpi_{g_t}C_g+  \sum_{g=2}^{i-1} \varpi_{g_o}D_g}{\prod_{l=1}^{i-3}\gam_l}  \right)}, \nonumber \\
& = \frac{\sum_{g=2}^{j} \left({\bar u}_{g_t}- {\bar u}_{i_o}\varpi_{g_t}\right)C_g + \sum_{g=2}^{i-1} \left({\bar u}_{g_o}- {\bar u}_{i_o}\varpi_{g_o}\right)D_g }{ \sum_{g=2}^j  \left(1 - \varpi_{g_t}\right)C_g + \sum_{g=2}^{i-1} \left(1- \varpi_{g_o}\right)D_g }. 
\nonumber
\end{align}
Now consider any further sub-class of schedulers where (wlog) say all $\bet_l$ for any $1 < l \le j-2$ and $\gam_l$ for any $1 \le l \le i-3$ are fixed except for $\beta_1$.  When one considers optimizing among this class,  
from the above   the objective function is of the form $\frac{Qx+R}{Tx+U}$ with $x = \beta_1$ (for appropriate $Q,R,T$ and $U$). Using elementary (derivative-based) arguments this function is either increasing or is decreasing in $x$. Thus $\beta_1 = 0$ or $\beta_1 = 1$ is the optimizer in the considered sub-class.  
Similar arguments follow for other $\bet_l$ and $\gam_l$.

Since, we consider optimisation over uSMR policies, it is not possible to have
 $\bet_{\bar  g -1} = 1$ when some $\beta_g = 0$ for some $g < {\bar  g -1},$ (as otherwise we have two stationary distributions).

Substituting it in \eqref{eqn_sd}, we obtain the stationary distribution in \eqref{eqn_optimal_sd}.

\eop

\ignore{\noindent
\textbf{Proof of Theorem \ref{thm_i_eps_del_scheduler}:} Say the solution exists for some $j=z.$ Define $\Ps_{z-2} := \frac{\gam_{j-2}}{1-\gam_{j-1}}$ and then using \eqref{eqn_optimal_sd}, the objective function of optimisation problem \eqref{eqn_2users_theta_fair} can be written as below:
\begin{align}
    O_z & =    \frac{\left(\sum_{g=2}^{i} {\bar u}_{g_t}+ \sum_{g=2}^{z-1} {\bar u}_{g_o} \right)  + {\bar u}_{z_o}\Ps_{z-2}}{i+z-3+\Ps_{z-2}} \text{ where } \Ps_{z-2}  = -\frac{ \sum_{g=2}^{i} \left(\ut_{g_t}-\theta\uo_{g_t} \right)+  \sum_{g=2}^{z-1} \left(\ut_{g_o}-\theta\uo_{g_o} \right)}{\left(\ut_{z_o}-\theta\uo_{z_o} \right)}.
\end{align}
Also, assume the solution exists for some other $z'>z$. Without loss of generality, let $z' = z+1$. Then, the difference of the objective functions at $z$ and $z+1$ equals,
\begin{align}
    O_z - O_{z+1} & = \frac{\left(\sum_{g=2}^{i} {\bar u}_{g_t}+ \sum_{g=2}^{z-1} {\bar u}_{g_o} \right)  + {\bar u}_{z_o}\Ps_{z-2}}{i+z-3+\Ps_{z-2}}-\frac{\left(\sum_{g=2}^{i} {\bar u}_{g_t}+ \sum_{g=2}^{z} {\bar u}_{g_o} \right)  + {\bar u}_{(z+1)_o}\Ps_{z-1}}{i+z-2+\Ps_{z-1}} \nonumber \\
    & = \frac{(1+\Ps_{z-1}-\Ps_{z-2})\left(\sum_{g=2}^{i} {\bar u}_{g_t}+ \sum_{g=2}^{z-1} {\bar u}_{g_o} \right) + (i+z-2+\Ps_{z-1}){\bar u}_{z_o}\Ps_{z-2}-(i+z-3+\Ps_{z-2})({\bar u}_{z_o}+\Ps_{z-1}{\bar u}_{(z+1)_o})}{(i+z-3+\Ps_{z-2})(i+z-2+\Ps_{z-1})}\nonumber \\
    & > \frac{(1+\Ps_{z-1}-\Ps_{z-2})\left(\sum_{g=2}^{i} {\bar u}_{g_t}+ \sum_{g=2}^{z-1} {\bar u}_{g_o} \right) + (i+z-2){\bar u}_{z_o}\Ps_{z-2}-(i+z-3)\left({\bar u}_{z_o}+\Ps_{z-1}{\bar u}_{(z+1)_o}\right)-\Ps_{z-2}{\bar u}_{z_o}}{(i+z-3+\Ps_{z-2})(i+z-2+\Ps_{z-1})}\nonumber \\
    & =  \frac{(1+\Ps_{z-1}-\Ps_{z-2})\left(\sum_{g=2}^{i} {\bar u}_{g_t}+ \sum_{g=2}^{z-1} {\bar u}_{g_o} \right) + (i+z-3){\bar u}_{z_o}\Ps_{z-2}-(i+z-3)\left({\bar u}_{z_o}+\Ps_{z-1}{\bar u}_{(z+1)_o}\right)}{(i+z-3+\Ps_{z-2})(i+z-2+\Ps_{z-1})}\nonumber \\
    & =  \frac{(1+\Ps_{z-1}-\Ps_{z-2})\left(\sum_{g=2}^{i} {\bar u}_{g_t}+ \sum_{g=2}^{z-1} {\bar u}_{g_o} \right) + (i+z-3)(\Ps_{z-2}-1){\bar u}_{z_o}-(i+z-3)\Ps_{z-1}{\bar u}_{(z+1)_o}}{(i+z-3+\Ps_{z-2})(i+z-2+\Ps_{z-1})}\nonumber \\
    & =  \frac{\Ps_{z-1}\left(\sum_{g=2}^{i} {\bar u}_{g_t}+ \sum_{g=2}^{z-1} {\bar u}_{g_o}-(i+z-3){\bar u}_{(z+1)_o}\right) + \left(1-\Ps_{z-2}\right)\left(\sum_{g=2}^{i} {\bar u}_{g_t}+ \sum_{g=2}^{z-1} {\bar u}_{g_o}-(i+z-3){\bar u}_{z_o}\right)}{(i+z-3+\Ps_{z-2})(i+z-2+\Ps_{z-1})}\nonumber \\
    & > \frac{\Ps_{z-1}\left((i+z-3){\bar u}_{z_o}-(i+z-3){\bar u}_{(z+1)_o}\right) + \left(1-\Ps_{z-2}\right)\left((i+z-3){\bar u}_{z_o}-(i+z-3){\bar u}_{z_o}\right)}{(i+z-3+\Ps_{z-2})(i+z-2+\Ps_{z-1})} \nonumber \\
    & =  \frac{\Ps_{z-1}(i+z-3)\left({\bar u}_{z_o}-{\bar u}_{(z+1)_o}\right)}{(i+z-3+\Ps_{z-2})(i+z-2+\Ps_{z-1})} \ > \ 0.\nonumber
\end{align}
Thus, the objective function is higher at the minimum value of $j$.
\eop}}

\medskip

\noindent
 \textbf{Proof of Theorem \ref{thm_efficient}:} The proof follows in three steps. Consider $\Omega_{j}$ for $2 \le g \le \bar g$ as defined in \eqref{eqn_ubar_avg_eff}.

\medskip
 \textbf{Step 1:} To show  that, $\mathbf{{\bar u}_{w_t}>{\bar u}_{3_o}}$ for all $2 \le w \le \et$.
 
 \medskip

By hypothesis,
we have ${\bar u}_{2_t} \ge {\bar u}_{3_t} > {\bar u}_{3_o}$. 
Hence when $\eta = 2$ or $3$, Step 1 is immediately true.   
 Now consider   any $\eta \ge 4$. It is already proved that 
${\bar u}_{w_t}>{\bar u}_{3_o}$  for $w \le 3$. We will prove using induction that ${\bar u}_{w_t}>{\bar u}_{3_o}$ for all $w \le \eta$ which eventually proves Step 1. Assume 
${\bar u}_{l_t}>{\bar u}_{3_o}$ for all $l \le w$   and consider $w+1$. 
%
By optimality of $\eta$, we have $\Omega_\eta \ge \Omega_{w}$ and hence 
one can't have $\Omega_{w+1} <  \Omega_{w}$, as otherwise it would contradict  Lemma \ref{lem_monotonicity_ue}. Thus we have $\Omega_{w+1} \ge \Omega_{w}$  and hence:
$$
\frac{2\Omega_{2}+\sum_{g=3}^{w+1}{\bar u}_{g_t}}{w+1} = \Omega_{w+1} \ge \Omega_{w} = \frac{2\Omega_{2}+\sum_{g=3}^w {\bar u}_{g_t}}{w} \stackrel{(a)}{>} \frac{2\Omega_{2}+\sum_{g=3}^{w}{\bar u}_{g_t}+{\bar u}_{3_o}}{w+1},
$$
where  inequality $(a)$ follows by  induction hypothesis and as $2 \Omega_2 = {\bar u}_{2_t}+ {\bar u}_{2_0}> 2 {\bar u}_{3_0}$. This implies ${\bar u}_{{(w+1)}_t}>{\bar u}_{3_o}$.

\medskip
\textbf{Step 2:} To show ${\bar u}_{{(\et+1)}_t}\le\Omega_{\et}$ and $ {\bar u}_{3_o}<\Omega_{\et}\le {\bar u}_{\et_t}$

\medskip
 By optimality of $\et$, we have $\Omega_{\et+1}\le \Omega_{\et}$ which on simple computations give ${\bar u}_{{(\et+1)}_t}\le \Omega_{\et}$. By definition of $\Omega_\et$ and Step 1, 
 we have $\Omega_{\et} > {\bar u}_{3_o}$. 
 Now we are left to show  ${\bar u}_{\et_t} \ge\Omega_{\et}$. Towards this, by the definition of $\eta$, we again have
 $\Omega_{\et-1} \le \Omega_{\et}$ which on simplification gives $
 \Omega_{\et-1} \le  {\bar u}_{\et_t}
 $ and thus,
 $$\Omega_{\et} = \frac{(\et-1)\Omega_{\et-1}+{\bar u}_{\et_t}}{\et} \le {\bar u}_{\et_t}.$$

\medskip
\textbf{Step 3:} To show that the optimal value of \eqref{Eqn_alpha_fair} (under dual  schedulers) with $\alpha  = 0$  equals $ \Omega_{\et}$

\medskip
From Step 2 and hypothesis,   ${\bar u}_{g_o} \le {\bar u}_{3_o} < \Omega_{\et}$ for all $g \ge 3$ and $ {\bar u}_{w_t} \le {\bar u}_{{(\et+1)}_t} \le \Omega_{\et}$  for all $w \ge \et+1$. Thus for any $\bd$, the objective function $O(\bd) = \sum_{n \in \Nc} {\bar u}^{(n)}_\bd$ can be upper bounded as below  (While solving  \eqref{eqn_stationary_orig} with transitions as in Figure \ref{fig:TD_MC}, we obtain $\mud(2_t) = \mud(2_o)$ for any $\bd$),  
\begin{eqnarray}
   O(\bd) &=&  \sum_{g=2}^{\bar g}  \mu_\bd (g_t) {\bar u}_{g_t} + \sum_{g=2}^{\bar g}  \mu_\bd (g_o) {\bar u}_{g_o}, \nonumber  \\
    & \le & 
    2\mud(2_t)\Omega_{2} + \sum_{g=3}^\et\mud(g_t)\bar{u}_{g_t} + \Omega_{\et} \left[ 1 - \left(2\mud(2_t) + \sum_{g=3}^\et\mud(g_t)\right)\right],  
    \nonumber \\
    & = & \left(\frac{2\mud(2_t)\Omega_{2} + \sum_{g=3}^\et\mud(g_t)\bar{u}_{g_t}}{2\mud(2_t)+\sum_{g=3}^\et\mud(g_t)} \right)\left(2\mud(2_t)+\sum_{g=3}^\et\mud(g_t)\right) + \Omega_{\et} \left[ 1 - \left(2\mud(2_t) + \sum_{g=3}^\et\mud(g_t)\right)\right], \nonumber \\
& \le & \Omega_\eta \left(\frac{2\mud(2_t)  + \sum_{g=3}^\et\mud(g_t) }{2\mud(2_t)+\sum_{g=3}^\et\mud(g_t)} \right)\left(2\mud(2_t)+\sum_{g=3}^\et\mud(g_t)\right) + \Omega_{\et} \left[ 1 - \left(2\mud(2_t) + \sum_{g=3}^\et\mud(g_t)\right)\right], \nonumber \\
%
 %
    & = & \Omega_{\et} \left(2\mud(2_t)+\sum_{g=3}^\et\mud(g_t)\right) + \Omega_{\et} \left[ 1 - \left(2\mud(2_t) + \sum_{g=3}^\et\mud(g_t)\right)\right]    \  = \  \Omega_{\et}. \nonumber
\end{eqnarray}
Further, $\Omega_{\et} = O(\bd_e^*)$ with  $\bd_e^*$ as defined in  hypothesis.

$(ii)$ From Steps 2 and 3 of part $(i)$,  for any $\bd$, $O(\bd)  \le \Omega_{\et} \le {\bar u}_{\et_t}$ and now $\et = \bar g$.  \eop
\ignore{Thus, we can move mass from any one of $\mud (g_o)$ with $g \ge 3$ to one of $\mud (g_t)$ with $2 \le g \le \bar g$ (since $\et = \bar g$) with improved objective function and hence we consider (without loss) optimization across $\bd$ such that $\mud (g_o)= 0 $ for all $g \ge 3$.  For such $\bd$, the $O(\bmu)$ has the following form
\begin{eqnarray*}
  O(\bmu) & = & 2\mud(2_t)\Omega_{2}+\sum_{g=3}^{\et} \bigg( \mud(g_t)\bar{u}_{g_t} \bigg). 
  \end{eqnarray*}

Further, we  know that the stationary distribution $\bmu$ is from the following feasible region. 
$$ \Fc_{2,\et} = \left \{ \mud(g_t) \le \mud((g-1)_t) \text{ for any } 3 \le g < \et, 0 \le \mud(\et_t) \le 1  \text{ and } 2\mud(2_t) + \sum_{g=3}^{\et} \mud(g_t) = 1 \right\}.$$
Next, we show that $\mud(2_t) = 0$ which implies that $\mud(g_t) = 0$ for any $g < \et$ and $\mud(\et_t) = 1.$ Say $\mud(2_t) > 0$. Then, from Step 2 and using the definition of $\eta$, we have ${\bar u}_{\et_t} > \Omega_{n} > \Omega_{2}.$ In view of this, one can again move mass from $\mud(2_t)$ to $\mud(\et_t)$, and thus $\mud(2_t) = 0.$ }

\begin{lemma}
 \label{lem_monotonicity_ue}
 For any fixed $j$, if $\Omega_{j} > \Omega_{j+1}$  then, $\Omega_{m} < \Omega_{j}$ for all $m \ge j+2$. 
 \end{lemma}
 \noindent
 \textbf{Proof: }Since $\bar{u}_{m_t} \le \bar{u}_{{(j+1)}_t}$ for all $m \ge j+2$, from \eqref{eqn_ubar_avg_eff} we have 
 \begin{eqnarray}
  \Omega_{m} \le \frac{\sum_{g=2}^j\bar{u}_{g_t}+\bar{u}_{2_o}+(m-j)\bar{u}_{{(j+1)}_t}}{m} & \stackrel{(a)}{<} & \frac{\sum_{g=2}^m\bar{u}_{g_t}+ \bar{u}_{2_o}+(m-j)\left(\frac{\sum_{g=2}^j\bar{u}_{g_t}+ \bar{u}_{2_o}}{j} \right)}{m} \nonumber \\
  & = & \frac{\sum_{g=2}^j\bar{u}_{g_t}+ \bar{u}_{2_o}}{j} \ = \ \Omega_{j}, \nonumber  
 \end{eqnarray}
 where inequality $(a)$ follows from $ \Omega_{j+1} < \Omega_{j}$.  \eop

\ignore{\begin{lemma}
Define $\Bc_{i,j}$ to be the bag of stationary distributions when the markov chain visits states $g_t$ for any $2 \le g \le j$, and $g_o$ for any $2 \le g \le i$.     If ${\bar u}_{j_t} > \Omega_{2}$ for any $j<\bar g$  then,
    $$
    \sup_{\bmu \in \mathcal{B}_{2,j}} O(\bmu)  = \Omega_{j} \text{ with } \mud^*(g_t) = \mud^*(2_o) = \frac{1}{j} \text{ for $2 \le g \le j$ and $0$ for other states.} \mbox{ \eop}
    $$
\label{lem_efficient_restricted_sd}
\end{lemma}
 \noindent
\textbf{Proof of Lemma \ref{lem_efficient_restricted_sd}:}
From the proof of Theorem \ref{thm_SD} we know $\mud(2_t) = \mud(2_o)$, and thus the required optimization  problem  is equivalent to the following:
\begin{eqnarray}
 \sup_{\bmu \in \Bc_{2,j}}  && O(\bmu) = 2\mud(2_t)\Omega_{2,2} + \sum_{3 \le g \le j}\mud(g_t){\bar u}_{g_t} \nonumber \\ \text{ such that } && \Fc_{2,j} = \left \{\mud({g_t}) \le \mud({(g-1)_t}) \text{ for all } 3 \le g \le j,    \text{ and } 2\mud(2_t) + \sum_{g=3}^j \mud(g_t) = 1 \right\}. \nonumber
\end{eqnarray}
We first show that given the constraints that $\bmu$ has to satisfy, it is clear that
  $\frac{1}{j} \le \mud(2_t) \le \frac{1}{2}$ (since $\mud(g_t) \le \mud(2_t) \text{ for all $g \ge 3$ and } 2\mud({2_t})+(j-2)\mud(2_t) = 1$); this also implies 
  $0\le \sum_{g = 3}^j \mud(g_t) \le 1 - \frac{2}{j}$ for any $\mud$ in feasible region, $\Fc_{2,j}$.  We will now show that $\mud(2_t) = \frac{1}{j}$ at optimality. Assume the contrary, i.e., 
$\mud(2_t) > \frac{1}{j}.$ 

Then, one can move mass from $\mud(2_t)$ to one of the $g_t$ with $3 \le g \le j$ and improve the objective function since ${\bar u}_{j_t} > \Omega_{2}$ without violating the constraints of $\Fc_{2,j}$, as 
at current $\bmu$, we have $\sum_{g=3}^j \mud(g_t) < 1- \frac{2}{j}$.  Thus such a $\mud$ is not optimal. 
The only $\bmu \in \mathcal{F}_{2,j}$ is as given in the theorem hypothesis. \eop}

\medskip

\noindent
\textbf{Proof of Theorem \ref{thm_SD}:} We begin with deriving the sub-optimizers in \eqref{eqn_gam_last}. While solving  \eqref{eqn_stationary_orig} with transitions as in Figure \ref{fig:TD_MC}, we obtain $\mud(2_t) = \mud(2_o)$. Further the stationary distribution (pmf) satisfies the following set of equations:
\begin{align}
   & 2\mud(2_t) + \sum_{g=3}^{j} \mud(g_t) + \sum_{g=3}^{i} \mud(g_o) = 1, \nonumber \\
\mud(g_o) &= \left(\prod_{l=1}^{g-2}\gam_l\right)\mud(2_t)\text{ for any } g < i, \text{ and } \mud(i_o) = \left(\frac{\prod_{l=1}^{i-2}\gam_l}{1-\gam_{i-1}}\right)\mud(2_t), \text{ and } \nonumber \\
\mud(g_t) &= \left(\prod_{l=1}^{g-2}\bet_l\right)\mud(2_t) \text{ for any } g < j, \text{ and } \mud(j_t) = \left(\frac{\prod_{l=1}^{j-2}\bet_l}{1-\bet_{j-1}}\right)\mud(2_t). \nonumber
\end{align}
Thus we have the following,
\begin{align}
\label{eqn_sd}
    \mud(2_t) = \frac{1}{ \sum_{g=2}^j  C_g + \sum_{g=2}^i D_g}, \text{ where } C_g = \prod_{l=1}^{g-2}\left[\bet_l\left(\mathds{1}_{\{g<j\}}+ \frac{1}{1-\bet_{j-1}}\ind_{\{g=j\}} \right)\right], 
    \text{ and } \nonumber \\
    D_g = \prod_{l=1}^{g-2}\left[\gam_l\left(\ind_{\{g<i\}}+\frac{1}{1-\gam_{i-1}}\ind_{\{g=i\}}\right)  \right].
\end{align}
One can substitute the above stationary distribution into the constraint of the optimization problem  given in \eqref{eqn_general_theta_fair} and can derive an equation between two variables (say $\gamma_{i-2}$, $\gamma_{i-1}$) such that the corresponding scheduler satisfies \eqref{eqn_general_theta_fair}: this relation is defined using $\gamma_{ij}^*$ defined in theorem hypothesis as below:
\begin{align*}
    \gs := \frac{\gam_{i-2}}{1-\gam_{i-1}} = -\frac{ \sum_{g=2}^{j} \varpi_{g_t}C_g+  \sum_{g=2}^{i-1} \varpi_{g_o}D_g}{\prod_{l=1}^{i-3}\gam_l} \text{ where } \varpi_{g_t} =  \frac{\ut_{g_t}-\theta\uo_{g_t}}{\ut_{i_o}-\theta\uo_{i_o}} \text{ and } \varpi_{g_o} =  \frac{\ut_{g_o}-\theta\uo_{g_o}}{\ut_{i_o}-\theta\uo_{i_o}}.
\end{align*}
Observe that $\rsbU_{g_t}^{(n)}$ and $\rsbU_{g_t}^{(n)}$ are constants for any $g$  and $n$, given the system parameters.
Now, using \eqref{eqn_sd} and the expression of $\gs$ from above, the objective function of optimisation problem \eqref{eqn_general_theta_fair} can be written as:
\begin{align}
& = \frac{ \sum_{g=2}^{j} {\bar u}_{g_t}C_g + \sum_{g=2}^{i-1} {\bar u}_{g_o}D_g - {\bar u}_{i_o}\left(\prod_{l=1}^{i-3}\gam_l\right)\left(\frac{ \sum_{g=2}^{j} \varpi_{g_t}C_g+  \sum_{g=2}^{i-1} \varpi_{g_o}D_g}{\prod_{l=1}^{i-3}\gam_l}\right) }{\sum_{g=2}^j  C_g + \sum_{g=2}^{i-1} D_g -\left(\prod_{l=1}^{i-3}\gam_l\right)\left(\frac{ \sum_{g=2}^{j} \varpi_{g_t}C_g+  \sum_{g=2}^{i-1} \varpi_{g_o}D_g}{\prod_{l=1}^{i-3}\gam_l}  \right)}, \nonumber \\
& = \frac{\sum_{g=2}^{j} \left({\bar u}_{g_t}- {\bar u}_{i_o}\varpi_{g_t}\right)C_g + \sum_{g=2}^{i-1} \left({\bar u}_{g_o}- {\bar u}_{i_o}\varpi_{g_o}\right)D_g }{ \sum_{g=2}^j  \left(1 - \varpi_{g_t}\right)C_g + \sum_{g=2}^{i-1} \left(1- \varpi_{g_o}\right)D_g }. 
\nonumber
\end{align}
Now consider any further sub-class of schedulers where (wlog) say all $\bet_l$ for any $1 < l \le j-2$ and $\gam_l$ for any $1 \le l \le i-3$ are fixed except for $\beta_1$.  When one considers optimizing among this class,  
from the above   the objective function is of the form $\frac{Qx+R}{Tx+U}$ with $x = \beta_1$ (for appropriate $Q,R,T$ and $U$). Using elementary (derivative-based) arguments this function is either increasing or is decreasing in $x$. Thus $\beta_1 = 0$ or $\beta_1 = 1$ is the optimizer in the considered sub-class.  
Similar arguments follow for other $\bet_l$ and $\gam_l$.

Since, we consider optimisation over uSMR policies, it is not possible to have
 $\bet_{\bar  g -1} = 1$ when some $\beta_g = 0$ for some $g < {\bar  g -1},$ (as otherwise we have disjoint communication classes and two distinct stationary distributions).

Substituting it in \eqref{eqn_sd}, we obtain the stationary distribution in \eqref{eqn_optimal_sd}. \eop 
\end{appendices}

\newpage
\setlength{\parskip}{5mm}
\titlespacing{\chapter}{0cm}{0mm}{0mm}
\titleformat{\chapter}[display]
  {\normalfont\huge\bfseries}
  {\chaptertitlename\ \thechapter}{20pt}{\Huge}

\bibliographystyle{unsrt}

\chapter*{List of Publications}
\label{ch:pub}
\addcontentsline{toc}{chapter}{\nameref{ch:pub}}


\section*{International Journals}
\begin{enumerate}
	
	\item \textbf{Shiksha Singhal}, and Veeraruna Kavitha. Coalition formation resource sharing games in networks. Performance Evaluation 152 (2021): 102239.
	
	\item \textbf{Shiksha Singhal}, Veeraruna Kavitha, and Jayakrishnan Nair. On the ubiquity of duopolies in constant sum congestion games. Operations Research (under review).

 \item \textbf{Shiksha Singhal}, and Veeraruna Kavitha. Dual Opportunistic Fair mmWave Scheduler: position-aided beam alignment and user assignment (to be submitted).

\rev{ \item Walunj, Tushar Shankar, \textbf{Shiksha Singhal}, Veeraruna Kavitha, and Jayakrishnan Nair. On the interplay between pricing, competition and QoS in ride hailing. Annals of Operations Research (ANOR) (under review).}

\end{enumerate}
\section*{International Conferences}
  \begin{enumerate}
   \item \textbf{Shiksha Singhal}, Veeraruna Kavitha, and Jayakrishnan Nair. Coalition formation in constant sum queueing games. In 2021 60th IEEE Conference on Decision and Control (CDC), pp. 3812-3817. IEEE, 2021.
    \item \textbf{Shiksha Singhal}, Veeraruna Kavitha, and Sreenath Ramanath. AoI-Based Opportunistic-Fair mmWave Schedulers. Accepted at  International Teletraffic Congress (ITC 34) - Teletraffic Engineering for Smart Networking, 2022. 
   \rev{ \item \textbf{Shiksha Singhal}, Veeraruna Kavitha, and Vidya Shankar. Social Optimal Freshness in Multi-Source, Multi-Channel Systems via MDP. Accepted at COMSNETS 2024.}
    \item Walunj, Tushar Shankar, \textbf{Shiksha Singhal}, Veeraruna Kavitha, and Jayakrishnan Nair. Pricing, competition and market segmentation in ride hailing. In 58th Annual Allerton Conference on Communication, Control, and Computing (Allerton), pp. 1-8. IEEE, 2022.

\end{enumerate}


\newpage
\setlength{\parskip}{5mm}
\titlespacing{\chapter}{0cm}{0mm}{0mm}
\titleformat{\chapter}[display]
  {\normalfont\huge\bfseries \centering}
  {\chaptertitlename\ \thechapter}{20pt}{\Huge}

\chapter*{Acknowledgments}
\label{ch:Acknowledgments}
\addcontentsline{toc}{chapter}{\nameref{ch:Acknowledgments}}
\vspace{10mm}
I would like to express my deepest gratitude to my supervisor, \textbf{Prof. Veeraruna Kavitha}, for her unwavering support, invaluable guidance, and immense patience throughout my doctoral journey. 
Her guidance has helped me navigate through complex challenges and overcome obstacles. 
I am particularly grateful for her open-door policy and willingness to engage in meaningful discussions. Her constructive feedback and suggestions have greatly enriched my research experience and broadened my perspectives. As my M.Tech. supervisor, she played a pivotal role in lifting me from an M.Tech. student to a Ph.D. student. Her training in critical thinking, paper writing, presentation skills, and various other aspects has been invaluable to my growth as a researcher.

Beyond her academic expertise, Prof. Veeraruna Kavitha possesses remarkable personal qualities. Her warmth, approachability, and genuine interest in the well-being of her students have fostered a nurturing and supportive environment. 

I am profoundly grateful to the members of my thesis committee, \textbf{Prof. Jayakrishnan Nair} and\textbf{ Prof. K.S. Mallikarjuna Rao}, for their valuable input, constructive criticism, and scholarly expertise. Their rigorous examination and valuable suggestions have significantly enriched the content of this thesis. 

My collaboration with Prof. Jayakrishnan Nair has proven to be immensely beneficial to my progress. Through our interactions, I have gained invaluable knowledge and insights. Prof. Nair's guidance has significantly enhanced my presentation skills, equipping me with a more effective approach to tackling complex problems. I am grateful for the valuable mentorship I have received from him.

 I also take this opportunity to thank \textbf{Prof. Jayendran Venkateswaran}, the Head of the
Department for his support in all academic matters.

I extend my heartfelt appreciation to the Indian Institute of Technology, Bombay, and the department for providing the resources, facilities, and academic environment that enabled me to pursue this research. I thank Mr. Abasaheb Molavane, Mr. Amlesh Kumar, Mr. Siddhartha Salve, and Mr. Pramod
Pawar for prompt, efficient, and friendly handling of all the administrative formalities. I would like to acknowledge the financial support of the \textbf{Ministry of Human Resource
Development (MHRD)}, Government of India.  The financial support in the form of the \textbf{Prime Minister's Research Fellowship (PMRF)} has been crucial in facilitating the completion of this work, and I am sincerely thankful for the opportunities it has provided.

I would like to extend my heartfelt gratitude and appreciation to \textbf{Prof. Shailendra Mishra}, whose  guidance played a crucial role in making my journey to IIT Bombay possible. It is with immense gratitude that I acknowledge the invaluable contribution Prof. Shailendra Mishra has made to my academic pursuits and overall development.

I am indebted to my colleagues and fellow researchers, whose discussions, collaborations, and friendships have been a source of inspiration and intellectual stimulation. Their support and camaraderie have made this challenging academic journey much more enjoyable and rewarding.

I would like to thank our `Stochastic Group’ members, for helping me in many academic
activities. 
I would like to take a moment to express a special thanks to my brother, Tushar, who has been a remarkable source of support and collaboration since the initial days of my Ph.D. journey. Our bond has grown stronger through our shared experiences, and I am incredibly grateful for his unwavering presence in my academic pursuit.

I extend my deepest gratitude to all the incredible individuals who have crossed paths with me during my Ph.D. journey.
Each of them has made an indelible impact on my life, and I would like to express my sincere gratitude to Shubham, Amit, Saumil, Anand, Sandesh, Chinmay, Tejal, Reena, Sachin, Adnan, and Mayuri. I am forever grateful for the friendships, collaborations, and support that have enriched my doctoral experience. 

I would like to take this opportunity to extend my deepest gratitude and appreciation to a truly special friend Shubham. 
I am grateful for the understanding and empathy you have shown me, particularly during times of stress and transition.

My thanks are also due to my M.Tech. (2020 batch) friends Sumit, Ankush, Mayur, Jayesh, Sandeep, Aman, Pankaj, Krishna, and Vijay for their company during my M.Tech. days (and after) at IIT Bombay.


I am deeply grateful for the  encouragement of my college friends Surbhi, Digvijay, and Nitish throughout my Ph.D. journey. Their friendship and support have been invaluable, and I would like to express my heartfelt appreciation to each and every one of them. In particular, I would like to extend a special mention to Surbhi, who has been a source of inspiration and strength, pushing me to overcome challenges and strive for excellence in my research

I would like to extend my gratitude to all my school friends Sonal, Megha, Priyanka, and Piyush who have been a part of my Ph.D. experience. Your friendship, understanding, and constant support have made a profound impact on my life. I am truly blessed to have such amazing individuals by my side. In particular, I would like to extend a special mention and heartfelt thanks to Sonal. Her constant support and willingness to lend an ear during challenging times have been truly remarkable. 

Last but not the least, I would like to thank my whole family, especially my parents, for
supporting and encouraging me always with their best wishes.
Their encouragement, love, and belief in my abilities have been a constant source of motivation throughout this endeavor.




\end{document}